\documentclass[12pt]{article}

\usepackage{amsmath,amssymb,graphicx,multirow,xspace}
\usepackage[colorlinks=true,urlcolor=blue,anchorcolor=blue,citecolor=blue,filecolor=blue,linkcolor=blue,menucolor=blue,pagecolor=blue]{hyperref}
\usepackage[compress,numbers]{natbib}
\usepackage{placeins}
\usepackage{booktabs}
\usepackage{blindtext, rotating}
\usepackage{afterpage}
\usepackage{enumitem}
\usepackage{ marvosym }
\usepackage{verbatim}
\usepackage{authblk}
\usepackage{booktabs,multirow}

\usepackage{amsfonts}
\usepackage{float}
\usepackage{colortbl}
\usepackage{ccaption}
\usepackage{todonotes}
\usepackage{cancel}
\usepackage{verbatim}
\usepackage{color}
\usepackage{caption}
\usepackage{subcaption}
\usepackage{placeins}
\usepackage{fancyhdr}
\usepackage{epstopdf}
\usepackage{arydshln}
\usepackage{pifont}
\usepackage{slashed}
\usepackage{psfrag}
\usepackage{epstopdf}
\usepackage{multicol}
\usepackage{enumerate}
\usepackage[top=1in,bottom=1in,left=1in,right=1in,footskip=0.5in,headheight=0.5in,footnotesep=0.2in,marginparwidth=0in,marginparsep=0in]{geometry}
\usepackage{titlesec}

\newcommand{\abs}[1]{\big | #1 \big |}
\newcommand\fverb{\setbox\fverbbox=\hbox\bgroup\verb}
\newcommand\fverbdo{\egroup\medskip\noindent%
            \fbox{\unhbox\fverbbox}\ }
\newcommand\fverbit{\egroup\item[\fbox{\unhbox\fverbbox}]}
\newbox\fverbbox

\numberwithin{equation}{section}

\long\def\symbolfootnote[#1]#2{\begingroup%
\def\thefootnote{\fnsymbol{footnote}}\footnote[#1]{#2}\endgroup}

\newcommand{\be}{\begin{equation}}
\newcommand{\ee}{\end{equation}}

\newcommand{\mat}{\begin{pmatrix}}
\newcommand{\rix}{\end{pmatrix}}

\renewcommand{\bar}{\overline}

\newcommand{\beqa}{\begin{eqnarray}}
\newcommand{\eeqa}{\end{eqnarray}}

\newcommand{\beq}{\begin{equation}}
\newcommand{\eeq}{\end{equation}}

\renewcommand{\arraystretch}{1.5}

\numberwithin{equation}{section}

\begin{document}
\author[1]{Daniel Egana-Ugrinovic}
\author[1]{Scott Thomas}

\affil[1]{ \small{New High Energy Theory Center, Rutgers University, Piscataway,
  NJ 08854}}
  \date{\vspace{-5ex}}
\title{Effective Theory of Higgs Sector Vacuum States}
\clearpage\maketitle
\thispagestyle{empty}

\begin{abstract}
The effective field theory description for modifications of 
Standard Model-like Higgs boson interactions arising from 
tree-level mixing with heavy Higgs sector vacuum states without 
conserved quantum numbers is presented. 
An expansion in terms of effective operator dimension based on powers of the heavy mass scale rather than operator dimension is utilized to 
systematically organize interactions within the effective theory. 
Vacuum states arising from electroweak singlet extensions of the 
Higgs sector yield at leading order only two effective dimension-six operators. 
One of these uniformly dilutes all the interactions of a single 
physical Higgs boson as compared with Standard Model expectations, 
while the combination of the two operators give more general modifications 
of all remaining interactions with two or more physical Higgs bosons.
Vacuum states arising from an additional electroweak doublet yield three types 
of effective dimension-six operators that modify 
physical Higgs boson couplings to fermion pairs, self-couplings, 
and introduce four-fermion interactions. 
However, in this case modification of physical Higgs boson interactions with 
massive gauge bosons arise at leading order only 
from three effective dimension-eight operators. 
If the underlying Yukawa couplings of the additional electroweak 
doublet satisfy the Glashow-Weinberg condition then time reversal violation 
at effective dimension-six depends on a single universal phase appearing 
only in couplings of physical Higgs bosons to fermion pairs. 
In all cases the couplings 
of any three physical states in the effective theory description 
must be equal to those in a unitary mixing description
in order for 
the long distance non-analytic components of 
physical on-shell scattering amplitudes to agree, while in contrast 
couplings of four or more physical states in general differ
due to short distance effects.   
\end{abstract}
\newpage
\setcounter{page}{1}
\section{Introduction}

The observation of a Higgs-like boson at the LHC \cite{Aad:2012tfa,Chatrchyan:2012ufa} opened a new era for discovery. The Higgs sector must now be probed, and it must be determined if it corresponds to the one of the Standard Model (SM).  The Higgs sector of the SM has a very particular phenomenology. In the SM there is a single vacuum state responsible for giving mass to all elementary fields, and its couplings are determined entirely by these masses and gauge invariance. Both ATLAS and CMS have an extensive program to search for scalar states that might be new vacuum states \cite{Khachatryan:2015tha,CMS:2015ooa,Khachatryan:2015tra,Aad:2014vgg,Aad:2015kna} and to test if Higgs couplings are SM like \cite{Aad:2015gba,Khachatryan:2014jba,ATLASCMSCOMB}. 

From the theory perspective, the interpretation of any deviation in the SM Higgs sector predictions requires an organization of the phenomenology of the possible extensions. This exercise cannot be done in full generality, but existing experimental data can provide strong motivation in favor of particular models. First, no significant deviations from the SM predictions for any of the Higgs couplings that have been measured at LHC have been found \cite{Aad:2015gba,Khachatryan:2014jba,ATLASCMSCOMB}. In the SM, the Higgs particle couples to fermions and gauge bosons with the same strength than the Higgs condensate itself. This property is called alignment, and it is not automatically fulfilled in extended Higgs sectors with multiple vacuum states, but it is recovered in the limit in which all the beyond the SM field content is decoupled \cite{Appelquist:1974tg}.  A second strong limitation comes from the $\rho$ parameter ($\rho=m_W^2/(m_Z^2\cos^2 \theta_W)$), which is measured in EW precision experiments to be very close to one \cite{Agashe:2014kda}, as predicted in the SM or in extensions  with an arbitrary number of Higgs doublets and/or singlets  \cite{Gunion:1989we}. Extensions with more complicated $SU(2)$ representations generally break this relation at tree level\footnote{For an example of an exception see \cite{Georgi:1985nv}.}.

The above considerations provide a strong motivation to study the low energy phenomenology of Higgs sectors extended only with heavy singlets and doublets. The main features of such extensions can be studied by considering the most general extension with a real singlet \cite{Veltman:1989vw, McDonald:1993ex} or a second doublet  \cite{Lee:1973iz,Lee:1974jb, Fayet:1974fj, Flores:1982pr}. These models are referred as the Singlet Higgs Standard Model (SHSM, also dubbed xSM) and the two Higgs doublet model (2HDM). Extensions with singlets arise in the NMSSM \cite{Ellis:1988er} and in models of EW baryogenesis \cite{Grojean:2004xa}. Theories with two Higgs doublets arise in different BSM scenarios and in particular, they are a fundamental part of all supersymmetric models \cite{Dimopoulos:1981zb}. The theory and phenomenology of both the SHSM  \cite{O'Connell:2006wi,Barger:2007im,Aad:2015pla} and the 2HDM \cite{Gunion:1989we,Gunion:2002zf,Branco:2011iw,Craig:2012vn,Craig:2013hca,Craig:2015jba} have been studied extensively in the literature in the so called mixing language, which relies in finding the vacuum mass eigenstates and its couplings using the full microscopic theories. 

In this work we follow an alternative approach to study the SHSM and 2HDM: we derive their tree level Wilsonian low energy effective theories (EFT)\cite{Wilson:1993dy} by integrating out the heavy SU(2) singlet or doublet. Near the decoupling limit, EFT is the most powerful available tool to study the low energy observables, since it automatically includes all the effects of UV physics and organizes them hierarchically in an expansion in terms suppressed by the heavy mass scale. This allows us to find patterns of deviations from the SM Higgs couplings for the different UV completions, and to clearly organize flavor and T violating effects. In order to get a consistent expansion, it is important to identify which operators must be kept in the EFT, and this is done by finding the correct concept of effective operator dimension.
We start by studying the SHSM. We first review the phenomenology in the mixing language and then we follow the alternative method using EFT. The correct concept of effective operator dimension in the SHSM EFT corresponds to just naive operator dimension. We work up to effective dimension six and we give only one example of an operator at effective dimension eight. We find that all the couplings of the Higgs are modified at effective dimension six with respect to their SM counterparts.  We also find that all the couplings of the Higgs to massive gauge bosons and fermions are generically smaller in magnitude than in the SM, while self couplings may be larger.  When comparing with the results obtained using the mixing language, we find that all trilinear couplings in the EFT and mixing languages coincide, but no coupling with more than three legs does. For instance, the coupling of two Higgses to two gauge bosons obtained in the EFT language differs from the one obtained in the mixing language. We provide a precise explanation on why this is the case - trilinear couplings in both languages must coincide, since they control long distance pieces of scattering amplitudes which cannot be modified by threshold corrections. This is a generic feature of EFT, and it is also valid in the UV completion with an SU(2) doublet. We also perform a thorough check of the EFT by comparing scattering amplitudes calculated in the mixing and EFT languages and we correct previous results in the literature \cite{Henning:2014gca}.
 %The 2HDM allows for a rich phenomenology, including modifications to the SM Higgs properties, new neutral and charged scalar states, and new sources of flavor and T violation. A complete analysis of the 2HDM is only available in the mixing language \cite{Gunion:2002zf,Gunion:1989we,Davidson:2005cw,Haber:2006ue}.
 
We then study the SM completed in the UV with an SU(2) doublet. Without loss of generality in this paper we work in the so called Higgs basis \cite{Branco:1999fs}, which is particularly useful near the alignment limit. In this basis, only one of the doublets contains the Higgs vacuum expectation value (vev), so in the exact alignment limit it must also contain the Higgs particle. The doublet with no vev must be interpreted as the doublet that is decoupled. We begin by reviewing the mixing language from a somewhat unconventional perspective, which relies on the use of the Higgs basis and background symmetry invariant eigenvectors and eigenstates of the mass matrix. These eigenvectors allow us to introduce a notion of complex alignment parameter, which is invariant under background symmetries, is a straightforward generalization of the alignment parameter of the T conserving 2HDM, and simplifies the analysis of the general T violating 2HDM in the mixing language. We then present the tree level low energy effective theory. We identify a  concept of operator effective dimension, which \textit{differs from naive operator dimension}. We work up to effective dimension six in operators involving fermions, and up to effective dimension eight in purely bosonic interactions. We find that operators of effective dimension six modify Higgs self-couplings, couplings to fermions and also lead to four fermion interactions, while Higgs interactions with massive gauge bosons are only modified by operators of effective dimension eight.  We also find that all the self-couplings of the Higgs and the couplings to massive gauge bosons are generically smaller in magnitude than the ones of the SM, while the couplings to fermions can be larger in magnitude.  Regarding flavor violation, Higgs Yukawas allow for $\Delta F=1$ chirality violating processes, while both $\Delta F=1$ and $\Delta F=2$, chirality preserving and chirality violating processes are allowed in four fermion operators. Regarding T violation, we find that at effective dimension six it only arises in Higgs interactions with fermions. The T violating phases that can be associated exclusively with couplings in the 2HDM potential \cite{Haber:2006ue}, including the one in the complex alignment parameter, are not relevant up to at least effective dimension eight. The consistency of the EFT is checked by comparing with scattering amplitudes calculated in the mixing language. We finally perform a detailed analysis of the 2HDM with Glashow-Weinberg conditions \cite{Glashow:1976nt}, namely the types I-IV 2HDM. We comment on the different $\tan\beta$ dependence on different types, and we find that at effective dimension six, there is a single universal T violating phase contained exclusively in the interactions of the Higgs boson with fermions. %We point out that this phase, which is common for all types of 2HDM, can be constrained from electron dipole moment experiments.
 
The paper is organized as follows. In section \ref{sec:realsinglet} we present the Standard Model extended with a real singlet. We first study the model using the mixing language, and then we derive the EFT and reinterpret all the results from this perspective. In section \ref{sec:2HDM} we present the most general extension of the Standard Model with an SU(2) doublet and derive the couplings of the Higgs in the decoupling limit using the mixing language. In section \ref{sec:2HDMEFT} we present the corresponding effective field theory. In section \ref{sec:analysis} we study the effective theories of the 2HDM with Glashow-Weinberg conditions. We conclude with comments on the phenomenology and a summary table of properties of the SHSM and 2HDM effective field theories.
\section{A Higgs and a heavy real scalar singlet}
\label{sec:realsinglet}
\subsection{The model}
\label{sec:realsingletmodel}
We begin by describing a Higgs sector containing a Higgs doublet $H$ and a real scalar gauge singlet $S$. $H$ is taken to have hypercharge $Y=1$. We consider canonically normalized fields and the most general potential. The Lagrangian density is given by
\begin{equation}
D_\mu H^\dagger D^\mu H
  - V(H) 
   - \bigg[
   ~  \lambda^u_{ij} ~ Q_i H \bar{u}_j
   - \lambda^d_{ij} Q_i H^c \bar{d}_j 
   - \lambda^\ell_{\ij} L_i H^c \bar{\ell}_j 
  +{\rm h.c.}~ 
  \bigg]
    \label{eq:actionsinglet}
\end{equation}
the covariant derivative acting on the doublet is
\begin{equation}
D_\mu H=\bigg[\partial_\mu-i\bigg(g_2W_{a\mu}T_a+\frac{1}{2}g_1 B_{\mu}\bigg)\bigg]H
\end{equation}
In the potential, any linear term in $S$ can be absorbed in a shift of $S$, so the most general potential at the renormalizable level is
\begin{eqnarray}
V(H)= \frac{\mu^2}{2} S^2 + \frac{\zeta}{3} S^3 +\frac{\lambda_S}{8} S^4   
+m^2 H^\dagger H + \frac{\lambda}{2} (H^\dagger H)^2  
 +\xi S H^\dagger H +\frac{\lambda'}{2} S^2 H^\dagger H 
 \label{eq:singletLagrangian}
\end{eqnarray}
%\begin{eqnarray}
%{\cal L} =(\partial S)^2 - \frac{\mu^2}{2} S^2 - \frac{\zeta}{3} S^3 - \frac{\lambda_S}{8} S^4    \nonumber \\
%+D H^\dagger D H-m^2 H^\dagger H - \frac{\lambda}{2} (H^\dagger H)^2  
% -\xi S H^\dagger H -\frac{\lambda'}{2} S^2 H^\dagger H 
% \label{eq:singletLagrangian}
%\end{eqnarray}
The potential is invariant under the $\mathbb{Z}_2$ background  symmetry specified in table \ref{eq:z2symmetry}. All measurable quantities must be invariant under the background symmetry. Note that the $\mathbb{Z}_2$ symmetry is explicitly broken by $\xi$ and $\zeta$.
\begin{table}[h]
\begin{center}
$$
\begin{array}{cc} 
\hline
  & \mathbb{Z}_2   \\
\hline
 H & +   \\
 S & -    \\
 m^2 , \mu^2 , \lambda, \lambda^\prime, \lambda_S & +   \\
 \xi, \zeta & -   \\
 \hline
\end{array}
$$
\end{center}
\caption{Charge assignments for the background $\mathbb{Z}_2$ symmetry of the potential \eqref{eq:singletLagrangian}.}
\label{eq:z2symmetry}
\end{table}

We consider perturbative marginal couplings. Stability of the potential requires $\lambda_S>0, \lambda>0, \lambda'>-\sqrt{\lambda \lambda_S}$. By redefining the sign of the singlet we can make either $\xi$ or $\zeta$ positive. We also consider a non tachyonic singlet mass $\mu^2>0$. For $H=0$ the potential is a quartic polynomial in $S$, with a stable and a metastable minimum and an unstable maximum. Without loss of generality we define the global minimum to be at $S=0$. In order for the minimum away from the origin to be the metastable minimum the potential parameters fulfill
\begin{equation}
\zeta^2 < \frac{9}{4}\lambda_S \mu^2
\end{equation}
The gauge invariant combination characterizing the Higgs condensate with symmetry breaking pattern $SU(2)_L \times U(1)_Y \to U(1)_Q$, and the $\mathbb{Z}_2$ invariant combination characterizing the singlet condensate are 
\begin{equation}
{v^2 \over 2} = \langle H^\dagger H \rangle  \quad \quad {v_s^2 \over 2} = \langle S^2 \rangle 
\end{equation}

The vacuum states $s,h$ are defined as 
\begin{equation}
S=v_s+s  \quad  \quad  H_0=\frac{1}{\sqrt{2}}(v+h)
\label{eq:fielddef}
\end{equation}
where $H_0$ is the neutral component of the doublet. Gauge invariance ensures $v \geq 0$. The Higgs vev is $v=246~\textrm{GeV}$. 
 $v_s$ has negative charge under the $\mathbb{Z}_2$ background symmetry of table \ref{eq:z2symmetry}.

The Higgs condensate gives mass to the gauge bosons corresponding to the broken gauge symmetries. The $W$ and $Z$ boson masses are
\begin{equation}
m_W=\frac{g_2 v}{2} ~~~~ ~~~~ m_Z=\frac{m_W}{\cos~\! \theta_W} ~~~~ ~~~~ \tan~\! \theta_W=\frac{g_1}{g_2}  
\end{equation}

%Yukawa interactions are given by
%\begin{equation}
%-{\cal L}_Y =  
%   ~  \lambda^u_{ij} ~ Q_i H \bar{u}_j
%   - \lambda^d_{ij} Q_i H^c \bar{d}_j 
%   - \lambda^\ell_{\ij} L_i H^c \bar{\ell}_j 
%  +{\rm h.c.}~ 
%\end{equation}
Fermions are defined in the mass eigenbasis so $\lambda^f_{ij}$ ($f=u,d,\ell$) are diagonal matrices in flavor space. The Yukawa matrices in terms of the fermion masses are 
%\begin{equation}
%m^f_{ij}=\frac{\lambda^f_{ij} v}{\sqrt{2}} 
%\end{equation}
\begin{equation}
\lambda^f_{ij}=\frac{\sqrt{2} m^f_{ij}}{v} =\frac{\sqrt{2}m^f_{i} }{v} ~\delta_{ij}
\end{equation}

%We will see in section \ref{sec:realsingletrotation} that the Higgs mass is of order $v^2$ plus corrections of order $v^2 \xi/\mu$ and $v^2 \zeta/\mu$. 
%This means that in order for the model to be natural (non-tuned) we consider, for order one couplings,
We are interested in the case in which there is a separation of scales $\mu^2 \gg \abs{m^2}$, which corresponds to the limit in which the mass of the singlet is much heavier than the EW scale. In the EFT language, $\mu$ will be the cutoff of the low energy theory. We allow the remaining two mass scales in the theory, $\xi$ and $\zeta$, to be as large as the cutoff
\begin{eqnarray}
\xi,\zeta\lesssim \mu
\label{eq:nontunedsinglet}
\end{eqnarray}

Finally, the decoupling limit is defined as 
\begin{equation}
\lambda \frac{v^2}{\mu^2}
~,~
\lambda' \frac{v^2}{\mu^2}
~,~
 \lambda_S \frac{v^2}{\mu^2}
 ~,~
  \frac{\xi^2}{\mu^2} \frac{v^2}{\mu^2}
  ~,~
   \frac{\zeta^2}{\mu^2} \frac{v^2}{\mu^2}
     ~,~
      \frac{\xi\zeta}{\mu^2} \frac{v^2}{\mu^2} 
    \ll 1
    \label{eq:decouplinglimitsinglet}
\end{equation}

\subsection{Mass eigenstates and couplings in the mixing language}
\label{sec:realsingletrotation}
In the mixing language, the couplings of the Higgs particle are found by identifying the mass eigenstates of the scalar potential \eqref{eq:singletLagrangian} in the Higgs-singlet condensate. The minimization conditions of the potential specify the Higgs and singlet vevs in terms of Lagrangian parameters
\begin{eqnarray}
\frac{\partial V}{\partial H}\Big |_{H=v/\sqrt{2}}
&=&
\sqrt{2} m^2 v+ \frac{\lambda }{\sqrt{2}}v^3+\frac{\lambda' }{\sqrt{2}}v v_s^2+\sqrt{2} \xi  v v_s
=0
\\
\label{eq:realsingletvev1}
\frac{\partial V}{\partial S}\Big |_{S=v_s}
&=&
\mu^2 v_s+\frac{\xi}{2}  v^2+\frac{\lambda'}{2} v^2 v_s+\zeta  v_s^2 +\frac{\lambda_s}{2}v_s^3
=0 
\label{eq:realsingletvev2}
\end{eqnarray}
We consider $\mu>0$, so the minimum at the origin of field space can be destabilized only by the Higgs mass at zero Higgs vev. From \eqref{eq:realsingletvev1}, the extrema for the Higgs are at
\begin{eqnarray}
v=0 ~~~~{\rm and } ~~~~
v^2 = - {2 \over \lambda} \bigg( m^2 + {1 \over 2} \lambda^\prime v_s^2 +  \xi v_s \bigg) 
\label{eq:veveq}
\end{eqnarray}
The term in parenthesis in the second expression is the Higgs mass at the origin. When positive, the potential has a global minimum at $v=0, v_s=0$. When negative, the global minimum is away from the origin and is a solution of the second expression in \eqref{eq:veveq}. The Higgs condensate induces a tadpole for the singlet, destabilizing the global minimum at the origin of the singlet field space and inducing a singlet condensate. The tadpole vanishes when $\xi \rightarrow 0$. The singlet vev is given by 
\eqref{eq:realsingletvev2}
\begin{eqnarray}
  v_s&=&-\frac{2 \zeta ^2-3 \lambda_S\left(\mu ^2+\frac{1}{2} \lambda'v^2\right)+2A^2+2 \zeta  A}{3 \lambda_SA}  \nonumber \\
A&=&
\left[\rule{0cm}{0.9cm} ~\! 
\zeta ^3-\frac{9}{4} \zeta  \lambda_S
\bigg( 
\mu ^2+\frac{1}{2}\lambda'v^2
 \bigg)
+{27\over 16} \lambda_S^2 \xi  v^2
\right.
 \nonumber \\
 && +  ~
  \left.
 \sqrt{
 \bigg( 
 \zeta ^3
 -\frac{9}{4} \zeta  \lambda_S
 \bigg( 
 \mu ^2+\frac{1}{2}\lambda'v^2
 \bigg)
 +\frac{27}{16} \lambda_S^2 \xi  v^2
 \bigg)^2
 - 
 \bigg(
 \zeta ^2-\frac{3}{2} \lambda_S 
 \bigg(
 \mu ^2+\frac{1}{2}\lambda'v^2
 \bigg)
 \bigg)^3}
 ~
\rule{0cm}{0.9cm} \right]
 ^{1/3} \nonumber \\
\end{eqnarray}
which is the solution of the cubic equation \eqref{eq:realsingletvev1} that vanishes at $\xi \rightarrow 0$.  The remaining two solutions are the unstable maximum and the metastable minimum which we do not study here. The factor $\mu ^2+\frac{1}{2}\lambda'v^2$ is the heavy singlet mass for a non-zero Higgs vev at the origin of the singlet field space. Expanding in $v^2/\left(\mu ^2+\frac{1}{2}\lambda'v^2\right)$ we get
\begin{eqnarray}
  v_s&=&
  -\frac{ \xi v^2}{ 2\mu ^2+\lambda' v^2}
\left[\rule{0cm}{0.85cm} ~\! 
  1+\frac{\zeta  \xi }
  {2 
  \big( \mu^2+\frac{1}{2}\lambda' v^2
  \big)}
  \Bigg(
  \frac{ v^2}
  { \mu ^2+\frac{\lambda' v^2}{2}}
  \Bigg)
  \right.
   \nonumber \\
   && +  ~
\Bigg(
\frac{\zeta^2\xi^2}{2\big(\mu ^2+\frac{1}{2}\lambda' v^2\big)^2}-
\frac{\lambda_S \xi^2}{8\big(\mu ^2+\frac{1}{2}\lambda' v^2\big)}
\Bigg)
\Bigg(
\frac{ v^2}{ \mu ^2+\frac{\lambda' v^2}{2}}
\Bigg)^2 
+{\cal O}\Bigg(\frac{ v^2}{ \mu ^2+\frac{\lambda' v^2}{2}}\Bigg)^3
\left.\rule{0cm}{0.85cm} ~\! \right] 
\nonumber \\
\end{eqnarray}
Note that $v_s$ vanishes in the limit $\xi \rightarrow 0$, as expected. Since near the decoupling limit $v\ll \mu$, we further expand the above expression in $(v/\mu)$,
\begin{eqnarray}
\label{eq:singletvev}
  v_s &=&-\frac{\xi  }{2}\left(\frac{v}{\mu}\right)^2\Bigg[1+\left(\frac{\zeta  \xi  }{2 \mu ^2}-\frac{\lambda' }{2 }\right)\left(\frac{v}{\mu}\right)^2 \nonumber \\
   &&  +  ~
  \left(\frac{\zeta ^2 \xi ^2}{2 \mu ^4}-\frac{3 \zeta  \lambda' \xi }{4 \mu ^2}+\frac{\lambda'^2}{4}-\frac{\lambda_S \xi ^2}{8 \mu ^2}\right)\left(\frac{v}{\mu}\right)^4+ {\cal O} \left(\frac{v^6}{\mu^6}\right)
\Bigg]
 \end{eqnarray}

The mass matrix for the vacuum states is specified by the second derivatives of the potential at the stable minimum. They are
\begin{eqnarray}
  \frac{\partial^2 V}{\partial h^2}
  &=&
  m_{hh}^2=
  m^2+\frac{3  }{2} \lambda v^2+\frac{1}{2} \lambda'v_s^2+\xi v_s   
  \nonumber \\
\frac{\partial^2 V}{\partial h\partial s}&=&
m_{hs}^2
= \lambda' v v_s+\xi  v  
 \nonumber \\
  \frac{\partial^2 V}{\partial s^2} 
  &=&
  m_{ss}^2 =\mu ^2+\frac{3 }{2} \lambda_S v_s^2+\frac{1}{2}\lambda'v^2+2 \zeta  v_s 
\label{eq:massmatrixsinglet}
\end{eqnarray}
Mixing is introduced between the singlet field and the Higgs in the mass matrix \eqref{eq:massmatrixsinglet} due to the condensates. The mixing matrix is
\begin{eqnarray}
\Bigg( 
{\def\arraystretch{1}\tabcolsep=10pt
\begin{array}{c} 
\varphi_1  \\ \varphi_2  \end{array} 
\Bigg) 
}
= 
\Bigg( 
{\def\arraystretch{1}\tabcolsep=10pt
\begin{array}{cc} \cos \gamma & \sin \gamma  
 \\ 
- \sin \gamma  & \cos \gamma
\end{array}   
}
\Bigg) 
\Bigg(
{\def\arraystretch{1}\tabcolsep=10pt 
\begin{array}{c}  h   
\\ s  
\end{array} 
}
\Bigg) 
        \label{eq:mixingmatrixsinglet}
\end{eqnarray}
where $\varphi_1$ is the lightest mass eigenstate and will be identified with the physical Higgs. The mixing angle is 
\begin{eqnarray}
\tan 2  \gamma = { -2 m_{hs}^2 \over m_{ss}^2 - m_{hh}^2}   
\label{eq:mixinganglesinglet}
\end{eqnarray}
or
\begin{equation}
\sin 2  \gamma = { -2 m_{hs}^2 \over m_+^2 - m_-^2} 
\end{equation}
where $m^2_+,m^2_-$ are the eigenvalues of the mass matrix. Using \eqref{eq:massmatrixsinglet} in \eqref{eq:mixinganglesinglet} we find
\begin{equation}
\tan 2  \gamma = {-2(\lambda'vv_s+\xi v) \over \mu^2-m^2+\frac{3}{2}(\lambda_Sv_s^2-\lambda v^2)+(2\zeta-\xi)v_s+\frac{1}{2}\lambda'(v^2-v_s^2)}   
\label{eq:tangammainterm}
\end{equation}
From \eqref{eq:veveq}, the soft mass is
\begin{equation}
m^2=-\frac{1}{2}\lambda v^2-\frac{1}{2}\lambda'v_s^2-\xi v_s
\label{eq:softmass}
\end{equation}
so \eqref{eq:softmass} in \eqref{eq:tangammainterm} we express the mixing angle as
\begin{eqnarray}
\tan 2  \gamma =\frac{-2(\lambda'vv_s+\xi v)}{\mu^2-\lambda v^2+2\zeta v_s+\frac{3}{2}\lambda_S v_s^2+\frac{1}{2}\lambda' v^2}
\label{eq:tangammainterm2}
\end{eqnarray}
A convenient expansion near the decoupling limit can be found for $\tan 2\gamma$ using the expansion for the singlet condensate \eqref{eq:singletvev} in \eqref{eq:tangammainterm2}  
\begin{equation}
  \tan 2\gamma = -\frac{2 \xi  }{\mu}\left(\frac{v}{\mu}\right)+\left(-\frac{2 \lambda  \xi }{\mu }+\frac{2 \lambda'
   \xi }{\mu }-\frac{2 \zeta  \xi ^2}{\mu ^3}\right)\left(\frac{v}{\mu}\right)^3+{\cal O}\left(\frac{v^5}{\mu^5}\right)  
   \end{equation}
   or, in terms of $\cos\gamma$ and $\sin\gamma$
\begin{eqnarray}
  \cos\gamma
  \!\!\!&=&\!\!\!
  1-\frac{\xi ^2}{2 \mu ^2}\left(\frac{v}{\mu}\right)^2+\left(-\frac{\lambda  \xi ^2}{\mu
   ^2}+\frac{\lambda' \xi ^2}{\mu ^2}-\frac{\zeta  \xi ^3}{\mu ^4}+\frac{11 \xi
   ^4}{8 \mu ^4}\right)\left(\frac{v}{\mu}\right)^4+{\cal O}\left(\frac{v^6}{\mu^6}\right) \nonumber \\
\sin  \gamma 
  \!\!\!&=&\!\!\!
-\frac{\xi }{\mu}\left(\frac{v}{\mu}\right)+ \left(-\frac{\lambda  \xi }{\mu }+\frac{\lambda'
   \xi }{\mu }-\frac{\zeta  \xi ^2}{\mu ^3}+\frac{3 \xi ^3}{2 \mu ^3}\right)\left(\frac{v}{\mu}\right)^3 +{\cal O}\left(\frac{v^5}{\mu^5}\right)
\label{eq:realsingletalign}
\end{eqnarray}

On the other hand, the two mass eigenvalues of the mass matrix \eqref{eq:massmatrixsinglet} are
\begin{eqnarray}
m^2_{\pm} = {1 \over 2} \Big( ~\!
  m_{hh}^2 + m_{ss}^2 \pm \sqrt{ (m_{ss}^2 - m_{hh}^2)^2 + 4m_{hs}^2 }   
    ~\! \Big) 
     \label{eq:exactmasses}
     \end{eqnarray}
We identify the Higgs with the lightest mass eigenstate $\varphi_1$. Using \eqref{eq:singletvev} and \eqref{eq:massmatrixsinglet} in \eqref{eq:exactmasses} we get the Higgs mass near the decoupling limit
\begin{eqnarray}
 m_{\varphi_1}^2
   \! \! \! & =  & \! \! \!   
    v^2 \Bigg[
       \bigg(   \lambda -\frac{\xi ^2}{\mu ^2}   \bigg)
       +\bigg(  \frac{3\lambda'\xi ^2}{2 \mu ^2}
       -\frac{\lambda  \xi ^2}{\mu ^2}
       -\frac{\zeta  \xi ^3}{\mu ^4}
       +\frac{\xi ^4}{\mu ^4}   \bigg)
       ~ \!  {v^2 \over \mu^2}
   \nonumber \\
   && ~ +  ~
   \bigg(  \! \! -\frac{3 \zeta ^2 \xi ^4}{2 \mu ^{6}}
   -\frac{2 \zeta  \lambda  \xi ^3}{\mu ^4}
   +\frac{3 \zeta \lambda'\xi ^3}{\mu ^4}
   +\frac{3 \zeta  \xi ^5}{\mu ^{6}}
   -\frac{\lambda ^2 \xi ^2}{\mu ^2}
   +    \frac{2 \lambda \lambda'\xi ^2}{\mu ^2}
   \nonumber \\ 
  &&  ~ +  ~
     \frac{3 \lambda  \xi ^4}{\mu ^4}
  -\frac{3 \lambda^{'2} \xi ^2}{2 \mu ^2}
  -\frac{7\lambda'\xi ^4}{2 \mu ^4}
  +\frac{3\lambda_S \xi ^4}{8 \mu ^4}
  -\frac{2 \xi ^6}{\mu ^{6}} \bigg)
  {v^4 \over \mu^4}
   +{\cal O}  \bigg( { v^6 \over \mu^6 }  
    \bigg)  
      ~ \!  \Bigg]
      \label{eq:Higgsmasssingletmixing}
%   \frac{3\lambda_S \xi ^4}{8 \mu ^4}\left(\frac{v}{\mu}\right)^4+\mathcal{O}\left(\lambda_S^2\left(\frac{v}{\mu}\right)^4,\frac{v^4}{\mu^4}\right)\Bigg)
\end{eqnarray}
while the mass of the heavy eigenstate is
\begin{eqnarray}
 m_{\varphi_2}^2
   \! \! \! & =  & \! \! \!   
    \mu^2 \Bigg[
      1
       +\bigg( -\frac{\zeta  \xi }{\mu ^2}+\frac{\lambda'}{2}+\frac{\xi ^2}{\mu ^2} \bigg)
       ~ \!  {v^2 \over \mu^2}
   \nonumber \\
   && ~ +  ~
   \bigg(  -\frac{\zeta ^2 \xi ^2}{2 \mu ^4}
   +\frac{\zeta  \lambda' \xi }{2 \mu ^2}
   +\frac{\zeta  \xi ^3}{\mu ^4}
   +\frac{\lambda  \xi ^2}{\mu ^2}
   -\frac{3 \lambda' \xi ^2}{2 \mu ^2}
   \nonumber \\ 
  && ~  +  ~
    \frac{3 \lambda_S \xi ^2}{8 \mu ^2}
    -\frac{\xi ^4}{\mu ^4}
   \bigg)
  {v^4 \over \mu^4}
   +{\cal O}  \bigg( { v^6 \over \mu^6 }  
    \bigg)  
      ~ \!  \Bigg]
%   \frac{3\lambda_S \xi ^4}{8 \mu ^4}\left(\frac{v}{\mu}\right)^4+\mathcal{O}\left(\lambda_S^2\left(\frac{v}{\mu}\right)^4,\frac{v^4}{\mu^4}\right)\Bigg)
\end{eqnarray}
 
We are now ready to study the couplings of the mass eigenstates. The Lagrangian for the mass eigenstates is obtained from \eqref{eq:actionsinglet} and \eqref{eq:mixingmatrixsinglet} and it is given by
\begin{eqnarray}
 && \frac{1}{2} \partial \varphi_a \partial \varphi_a + \frac{1}{2}Z^{\mu} Z_\mu \left(m_Z^2+ g_{\varphi_a ZZ}~ \varphi_a + \frac{1}{2}g_{\varphi_a^2ZZ}~ \varphi_a^2 +g_{\varphi_1 \varphi_2 ZZ}~\! \varphi_1 \varphi_2 \right)  \nonumber \\
   &&~ + 
   W^{+ \mu} W_{\mu}^- \left(m_W^2+ g_{\varphi_a WW}~ \varphi_a + \frac{1}{2}g_{\varphi_a^2WW}~ \varphi_a^2 +g_{\varphi_1 \varphi_2 WW}~\! \varphi_1 \varphi_2\right)   \nonumber \\
&&~- 
(m^f_{i} f_i\bar{f}_j~\delta_{ij}+\lambda^{f}_{\varphi_a ij} ~f_i  \varphi_a \bar{f}_j +\textrm{h.c.}) -V(\varphi_1,\varphi_2)
\label{eq:couplingdefmixing}
\end{eqnarray}
where we sum over repeated indices and
%+\lambda^{d}_{\varphi_a ij} ~d_i \varphi_a \bar{d}_j+\lambda^{\ell}_{\varphi_a ij} ~\ell_i \varphi_a \bar{\ell}_{j}
%\begin{eqnarray}
% && \frac{1}{2} \partial \varphi_a \partial \varphi_a +  g_{\varphi_a VV}~ \varphi_a V^{\dagger\mu} V_\mu + \frac{1}{2}g_{\varphi_a^2VV}~ \varphi_a^2 V^{\dagger\mu} V_\mu  \nonumber \\
%&&+( \lambda^{u\dagger}_{aij} ~ Q_i \varphi_a  \bar{u}_j+\lambda^{d}_{aij} ~Q_i \varphi_a \bar{d}_j+h.c.) -V(\varphi_1,\varphi_2)
%\label{eq:couplingdef}
%\end{eqnarray}
\begin{eqnarray}
V(\varphi_1,\varphi_2)
   \! \! \! & =  & \! \! \!   
\frac{1}{2} m_{\varphi_a}^2 \varphi_a^2-\sum_{n=0}^{n= 3} \frac{1}{n!(3-n)!}g_{\varphi_1^n \varphi_2^{3-n}}~\varphi_1^n\varphi_2^{3-n} \nonumber \\
&&-~\sum_{n=0}^{n=4} \frac{1}{n!(4-n)!}g_{\varphi_1^n \varphi_2^{4-n}}~\varphi_1^n\varphi_2^{4-n}
\label{eq:definitionpotentialsinglet}
\end{eqnarray}
We give explicit expressions for all couplings in \eqref{eq:couplingdefmixing} and \eqref{eq:definitionpotentialsinglet} below. As a consequence of mixing with the singlet the couplings of the Higgs are modified with respect to their SM expressions. We begin with the Higgs couplings to massive gauge bosons and fermions. Couplings of the Higgs to gauge bosons $V=Z,W$ and fermions $f=u,d,\ell$ are simply diluted by a factor of $\cos \gamma$ for each Higgs field, 
\begin{eqnarray}
  g_{\varphi_1 VV}
     \! \! \! & =  & \! \! \!   
     \frac{2 m_V^2}{v}  \cos\gamma \nonumber \\
  g_{\varphi_1^2VV}
     \! \! \! & =  & \! \! \!   
     \frac{2 m_V^2}{v^2}  \cos^2 \gamma  \nonumber \\
\lambda^f_{\varphi_1 ij}
   \! \! \! & =  & \! \! \!   
   \frac{m^f_{i}}{v}\cos\gamma ~\delta_{ij}
\label{eq:gaugecoupling1singlet}
\end{eqnarray}
%where $g_{\varphi_1 VV}^{SM}=\frac{2 m_V^2}{v}$, $g_{\varphi_1^2 VV}^{SM}=\frac{2 m_V^2}{v^2}$. 
The expressions for these couplings near the decoupling limit can be easily obtained using \eqref{eq:realsingletalign}. We omit the explicit expressions for brevity. The couplings of $\varphi_2$ to fermions and gauge bosons are inherited from the mixing between $h$ and $s$
\begin{eqnarray}
  g_{\varphi_2VV}
     \! \! \! & =  & \! \! \!   
     -\frac{2 m_V^2}{v}   \sin\gamma \nonumber \\
  g_{\varphi_2^2VV}
     \! \! \! & =  & \! \! \!   
     \frac{2 m_V^2}{v^2} \sin^2 \gamma  \nonumber \\
\lambda^f_{\varphi_2 ij}
   \! \! \! & =  & \! \! \!   
   -\frac{m^f_{i}}{v}\sin\gamma~ \delta_{ij}
\label{eq:gaugecoupling2singlet}
\end{eqnarray}
The couplings of $\varphi_1\varphi_2$ to massive gauge bosons are also inherited from mixing
\begin{eqnarray}
  g_{\varphi_1\varphi_2VV}
     \! \! \! & =  & \! \! \!   
     -\frac{2 m_V^2}{v}   \sin\gamma\cos\gamma 
\label{eq:gaugecoupling2singletmix}
\end{eqnarray}

Self couplings and Higgs-singlet couplings are given by using \eqref{eq:mixingmatrixsinglet} and the Higgs potential \eqref{eq:singletLagrangian}
\begin{eqnarray}
  g_{\varphi_1^3}
      \! \! & =  & \! \!  
     -3 \lambda  v \cos^3\gamma-3 \xi  \cos^2\gamma \sin\gamma-3 \lambda'v_s \cos^2\gamma \sin\gamma  \nonumber \\  &&-3 \lambda'v \cos\gamma \sin^2\gamma -2 \zeta    \sin^3\gamma-3 \lambda_Sv_s \sin^3\gamma \nonumber \\
  g_{\varphi_1^2\varphi_2}
   \! \! & =  & \! \!  
     -\xi \cos^3\gamma   -\lambda' v_s \cos^3\gamma+3\lambda v \cos^2\gamma   \sin\gamma -2\lambda'v \cos^2\gamma  \sin\gamma -2\zeta \cos\gamma\sin^2 \gamma \nonumber \\
  &&+2 \xi \cos\gamma   \sin^2\gamma+2\lambda'v_s \cos\gamma  \sin^2\gamma -3\lambda_Sv_s \cos\gamma  \sin^2\gamma +\lambda'v \sin^3\gamma  \nonumber \\
  g_{\varphi_1\varphi_2^2}
    \! \! & =  & \! \!   
      -v\lambda' \cos^3\gamma-2 \zeta \cos^2\gamma   \sin\gamma+2  \xi \cos^2\gamma  \sin\gamma+2\lambda'v_s \cos^2\gamma  \sin\gamma  \nonumber \\  &&-3 \lambda_Sv_s\cos^2\gamma\sin\gamma -3 \lambda v\cos\gamma   \sin^2\gamma+2\lambda'v \cos\gamma  \sin^2\gamma -\xi  \sin^3\gamma-\lambda' v_s \sin^3\gamma  \nonumber \\
  g_{\varphi_2^3}
   \! \! & =  & \! \!  
     -2 \zeta \cos^3\gamma -3\lambda_S v_s \cos^3\gamma +3 \lambda' v \cos^2\gamma \sin\gamma  \nonumber \\
  &&-3\xi \cos\gamma \sin^2\gamma-3\lambda'v_s \cos\gamma  \sin^2\gamma +3 \lambda v \sin^3\gamma  \nonumber \\
  g_{\varphi_1^4}
   \! \! & =  & \! \!  
     -3 \lambda\cos\gamma^4  -6 \lambda' \cos^2\gamma  \sin^2\gamma-3 \lambda_S \sin\gamma^4 \nonumber \\
  g_{\varphi_1^3\varphi_2}
   \! \! & =  & \! \!  
     3\lambda \cos^3\gamma   \sin\gamma-3\lambda' \cos^3\gamma  \sin\gamma+3 \lambda' \cos\gamma  \sin^3\gamma-3 \lambda_S \cos\gamma  \sin^3\gamma  \nonumber \\
  g_{\varphi_1^2\varphi_2^2}
   \! \! & =  & \! \!  
     -\lambda' \cos\gamma^4 -3 \lambda\cos^2\gamma   \sin^2\gamma+4\lambda'  \cos^2\gamma \sin^2\gamma-3\lambda_S \cos^2\gamma  \sin^2\gamma-\lambda' \sin\gamma^4  \nonumber \\
  g_{\varphi_1\varphi_2^3}
   \! \! & =  & \! \!  
     3 \lambda' \cos^3\gamma \sin\gamma-3 \lambda_S \cos^3\gamma \sin\gamma+3\lambda \cos\gamma   \sin^3\gamma-3\lambda' \cos\gamma  \sin^3\gamma \nonumber \\
 g_{\varphi_2^4}
   \! \! & =  & \! \!  
    -3\lambda_S \cos\gamma^4 -6 \lambda' \cos^2\gamma \sin^2\gamma-3 \lambda  \sin\gamma^4 \nonumber
\end{eqnarray}
%\begin{eqnarray}
%  g_{hVV}&=&g_V^2 v \left(1-\frac{\xi^2}{2\mu^2}\left(\frac{v}{\mu}\right)^2+{\cal O}\left(\lambda_S^2\left(\frac{v}{\mu}\right)^6,\frac{v^4}{\mu^4}\right)\right)  \nonumber \\
% g_{h^2VV}&=&g_V^2 \left(1-\frac{\xi^2}{\mu^2}\left(\frac{v}{\mu}\right)^2\right)+{\cal O}\left(\lambda_S^2\left(\frac{v}{\mu}\right)^6,\frac{v^4}{\mu^4}\right)
%\end{eqnarray}
%where $g_Z^2=m_Z^2/v^2$, $g_W^2=2m_W^2/v^2$. Couplings to fermions are also diluted
%\begin{equation}
%g_{hF\bar{F}}=-\frac{M_F}{v} \left(1-\frac{\xi^2}{2\mu^2}\left(\frac{v}{\mu}\right)^2+{\cal O}\left(\lambda_S^2\left(\frac{v}{\mu}\right)^6,\frac{v^4}{\mu^4}\right)
%\right)
%\end{equation}
The expressions for these couplings near the decoupling limit are obtained using \eqref{eq:singletvev} and \eqref{eq:realsingletalign}. Here for convenience we present the expansion for the Higgs self couplings 
\begin{eqnarray}
\label{eq:singletself1}
 { g_{\varphi_1^3}  \over v} 
 \! \! \!   &=& \! \! \!   
     -\frac{3m_{\varphi_1}^2}{v^2}+\bigg(  \frac{9 \lambda  \xi ^2}{2 \mu ^2}-\frac{3\lambda'\xi ^2}{\mu ^2}-\frac{9 \xi ^4}{2 \mu ^4}+\frac{2\zeta  \xi ^3}{\mu ^4}  \bigg) ~ \! 
     {v^2 \over \mu^2} +{\cal O} \bigg( {v^4 \over \mu^4}  \bigg)
   \nonumber \\
%-\frac{3\lambda_S \xi ^4}{8 \mu ^4}\left(\frac{v}{\mu}\right)^4+\mathcal{O}\left(\lambda_S^2\left(\frac{v}{\mu}\right)^6,\frac{v^4}{\mu^4}\right)\\
g_{\varphi_1^4} \! \! \!  &=& \! \! \!  
      -\frac{3m_{\varphi_1}^2}{v^2}-\frac{3 \xi ^2}{\mu ^2}
         +\bigg( \!  -\frac{3 \zeta  \xi ^3}{\mu ^4}+\frac{3 \lambda  \xi ^2}{\mu ^2}
           -\frac{3 \lambda' \xi ^2}{2 \mu ^2}+\frac{3 \xi ^4}{\mu ^4}  \bigg)
             ~ \! {v^2 \over \mu^2 }
             +{\cal O} \bigg(\frac{v^4}{\mu^4} \bigg)
%&&-\frac{15 \lambda_S \xi ^4}{8 \mu ^4}\left(\frac{v}{\mu}\right)^4+\mathcal{O}\left(\lambda_S^2\left(\frac{v}{\mu}\right)^6,\frac{v^4}{\mu^4}\right)
\end{eqnarray}
We will also make use of the coupling of two Higgses to a heavy mass eigenstate when calculating scattering amplitudes. Near the decoupling limit we get
\begin{eqnarray}
 { g_{\varphi_1^2\varphi_2}  } 
 \! \! \!   &=& \! \! \!   
     -\xi+\bigg(-3 \lambda  \xi   -\frac{2 \zeta  \xi ^2}{\mu ^2}+\frac{5 \lambda' \xi }{2}+\frac{7 \xi ^3}{2 \mu ^2}  \bigg) ~ \! 
     {v^2 \over \mu^2} +{\cal O} \bigg( {v^4 \over \mu^4}  \bigg)
    \label{eq:singletselftoheavy}
\end{eqnarray}
%where at order $\abs{m}^5/\mu^4$ and $m^4/\mu^4$ we dropped all terms not proportional to $\lambda_S$ for brevity. 

\subsubsection{Scattering amplitudes}
\label{sec:amplitudessingletmixing}
As an application of the results of the previous section we obtain some selected Higgs scattering amplitudes. This will prove to be useful as a consistency check of the EFT description, to be presented in section \ref{sec:realsingletEFT}, and to understand the difference between couplings in the mixing language and the effective theory language. All the scattering amplitudes are modified with respect to their SM values. The SM amplitudes can be read from this section by taking the limit $\xi\rightarrow 0$. We omit spinors in all amplitudes.

%The Higgs to diboson and difermion amplitudes are
%\begin{equation}
%{ \cal A} \left(\varphi_1 \rightarrow VV \right) 
%=g_{\mu\nu}~\!g_{\varphi_1 VV}
%=
%g_{\mu\nu}
%~\!
%\frac{2 m_V^2}{v} \bigg[1-\frac{\xi^2v^2}{2\mu^4}+ { \cal O} \bigg(  {v^4\over \mu^4} \bigg)\bigg]
%\label{eq:Higgsdibosonsingletmixing}
%\end{equation}
%%\begin{equation}
%%{ \cal A} \left(\varphi_1 \rightarrow f_i\bar{f}_j \right) =g_{\varphi_1 ij}=-\frac{m^f_{i}}{v}\bigg[1-\frac{\xi^2v^2}{2\mu^4}+ { \cal O} \bigg(  {v^4\over \mu^4} \bigg)\bigg]
%%\end{equation}
%\begin{equation} 
%{\cal A}  
%    \big(\varphi_1  \rightarrow  f_i \bar{f}_j  \big)
%  =-\lambda^f_{\varphi_1 ij}=   -{m^f_i \over v} \bigg[ ~ \!   1   -\frac{\xi ^2 v^2 }{2 \mu ^4}
%   + {\cal O} \bigg( { v^4 \over \mu^4 } \bigg) ~ \! 
%        \bigg]  ~ \delta_{ij}
%        \label{eq:Higgsdifermionsingletmixing}
%\end{equation}
%%\begin{eqnarray}
%%{ \cal A} \left(\varphi_1 \rightarrow \varphi_1\varphi_1 \right)  \! \! \! &=&   \! \! \!  g_{\varphi_1^3} =   -\frac{3m_{\varphi_1}^2}{v}\nonumber \\
%%     & &  \! \! \!  +~ v\bigg[\bigg(  \frac{9 \lambda  \xi ^2}{2 \mu ^2}-\frac{3\lambda'\xi ^2}{\mu ^2}-\frac{9 \xi ^4}{2 \mu ^4}+\frac{2\zeta  \xi ^3}{\mu ^4}  \bigg) ~ \! 
%%     {v^2 \over \mu^2}             +{\cal O} \bigg(\frac{v^4}{\mu^4} \bigg)\bigg]
%%\end{eqnarray}
%where we used the couplings \eqref{eq:gaugecoupling1singlet}.
The dihiggs scattering amplitude to two W bosons is given by the tree level coupling plus the contribution of diagrams with internal $\varphi_1$, $\varphi_2$ and $W$ boson propagators,
\begin{eqnarray}
\nonumber { \cal A} \left(\varphi_1 \varphi_1 \rightarrow W^{+ } W^-\right) 
  \! \! \! &=&   \! \! \! 
  g_{\mu\nu}
  ~\!
  \Bigg[
  g_{\varphi_1^2WW}-\frac{g_{\varphi_1^2 \varphi_a}g_{\varphi_a WW}}{s-m_{\varphi_a}^2}
  -g_{\varphi_1 WW}^2  \bigg(\frac{1}{t-m_{W}^2}+\frac{1}{u-m_{W}^2}\bigg) 
  \Bigg]\\
  \label{eq:exactdihiggsdibosonamplitudemixing}
 \end{eqnarray}
 where we sum over $a$. Using the couplings \eqref{eq:gaugecoupling1singlet}, \eqref{eq:gaugecoupling2singlet} and \eqref{eq:singletself1} in  \eqref{eq:exactdihiggsdibosonamplitudemixing} we get
 \begin{eqnarray}
{ \cal A} \left(\varphi_1 \varphi_1 \rightarrow W^{+} W^- \right) 
  \! \! \! &=&   \! \! \!   
  g_{\mu\nu}
  ~\!
  \Bigg[  
       \frac{2m_W^2}{v^2}  \bigg(\! 1-\frac{\xi^2v^2}{\mu^4} \bigg)
       +\frac{2m_W^2}{v^2}\bigg[ \frac{3m_{\varphi_1}^2}{v^2}
     +\bigg(  -\frac{2  \zeta  \xi ^3}{\mu ^4}   \nonumber \\
     & &  \! \! \!   -~ \frac{6 \lambda  \xi ^2}{\mu ^2}
       +\frac{3 \lambda' \xi ^2}{\mu ^2}
       +\frac{6 \xi ^4}{\mu ^4}\bigg)\frac{v^2}{\mu^2}
      \bigg]   
      \bigg( ~ \!  \frac{v^2}{s \! - \! m_{\varphi_1}^2}
          \bigg)   - ~ \!    \frac{2m_W^2\xi^2}{v^2\mu^2}\bigg(\frac{v^2}{\mu^2}\bigg)
\nonumber \\
 & & \! \! \!
  - ~\frac{4m_W^4}{v^4} \bigg[~\!1-\frac{\xi^2v^2}{\mu^4}~\!\bigg] \bigg(\frac{v^2}{t-m_{W}^2}+\frac{v^2}{u-m_{W}^2}\bigg) 
        \nonumber \\
  & & \! \! \! 
        + { \cal O}
               \bigg(  {sv^2\over \mu^4}, {s^2 v^2\over \mu^6},  {v^4\over \mu^4} \bigg)
               \Bigg]
             \label{eq:phiphiWWamplitudemixing}
% && + ~ { \cal O}
% \bigg( {x^2 \over \mu^2} , {v^2 \over \mu^2} \bigg)
%   % \left(\frac{s^2}{m_{\varphi_2}^4},\frac{t^2}{m_{\varphi_2}^4},
%    % \frac{u^2}{m_{\varphi_2}^4},\frac{v^4}{\mu^4}\right)
 %\nonumber \\
\end{eqnarray} 
The first term is the contact interaction. The second term is the long distance s-channel contribution mediated by the light mass eigenstate. The third term comes from an s-channel diagram mediated by the heavy state. The last term is the long distance contribution mediated by the $W$ boson. Note that all the long distance contributions are controlled by the trilinear couplings $g_{\varphi_1^3}$ and $g_{\varphi_1 WW}$. In general, all the long distance pieces of the amplitudes in this section are controlled exclusively by trilinear couplings. We will make use of this observation in section \ref{sec:realsingletEFT}.

The dihiggs to difermion chirality violating scattering amplitude is 
\begin{eqnarray}
{ \cal A} \left(\varphi_1 \varphi_1 \rightarrow f_i \bar{f}_j \right) 
  \! \! \! &=&   \! \! \! \frac{g_{\varphi_1^2 \varphi_{a}}\lambda^f_{\varphi_a ij}}{s-m_{\varphi_a}^2}
  \label{eq:exactdihiggsdifermionamplitudemixing}
 \end{eqnarray}
 where we sum over $a$. Using the couplings \eqref{eq:gaugecoupling1singlet}, \eqref{eq:gaugecoupling2singlet} and \eqref{eq:singletself1} in \eqref{eq:exactdihiggsdifermionamplitudemixing} we get
 \begin{eqnarray}
{ \cal A} \left(\varphi_1 \varphi_1 \rightarrow f_i \bar{f}_j \right) 
  \! \! \! &=&   \! \! \!  \frac{m_i^f~ \delta_{ij}}{v^2}\Bigg[\bigg( -\frac{3m_{\varphi_1}^2 }{v^2}
     +\bigg[ \frac{2 \zeta \xi ^3 }{\mu ^4} 
    -\frac{3 \lambda'\xi ^2}{\mu ^2}
    -\frac{6\xi ^4 }{\mu ^4} 
    \nonumber \\
     & &  \! \! \!   +~ \frac{6 \lambda  \xi ^2 }{\mu ^2} \bigg]\frac{v^2}{\mu^2}
      \bigg)
      \bigg( ~ \!  \frac{v^2}{s \! - \! m_{\varphi_1}^2}
          \bigg) 
          + \frac{\xi ^2 v^2}{\mu ^4} + { \cal O}
               \bigg(  {sv^2\over \mu^4}, {s^2 v^2\over \mu^6}, {v^4\over \mu^4},\frac{m_i^{f\,2}}{v^2} \bigg)
 \Bigg]
 \label{eq:dihiggsdifermionamplitudemixing}
 \end{eqnarray}
 The last term in the above expression comes from the s-channel amplitude mediated by the heavy state. The rest is the long distance contribution mediated by the light mass eigenstate. Note that the long distance contribution is controlled by the trilinear couplings $g_{\varphi_1^3}$ and $\lambda^f_{\varphi_1 ij}$

The dihiggs to dihiggs scattering amplitude is
\begin{eqnarray}
{ \cal A} \left(\varphi_1 \varphi_1 \rightarrow \varphi_1\varphi_1\right) 
  \! \! \! &=&   \! \! \! g_{\varphi_1^4}-g^2_{\varphi_1^2 \varphi_a}\left(\frac{1}{s-m_{\varphi_a}^2}+\frac{1}{t-m_{\varphi_a}^2}+\frac{1}{u-m_{\varphi_a}^2}\right)
    \label{eq:exactdihiggsdihiggsamplitudemixing}
 \end{eqnarray}
 where we sum over $a$. Using the couplings \eqref{eq:singletself1} in \eqref{eq:exactdihiggsdihiggsamplitudemixing} we get
 \begin{eqnarray}
{ \cal A} \left(\varphi_1 \varphi_1 \rightarrow \varphi_1\varphi_1\right) 
  \! \! \! &=&   \! \! \! 
       -\frac{3m_{\varphi_1}^2}{v^2} 
       -\bigg[{ 9 m_{\varphi_1}^4 \over v^4} 
       +\bigg(-\frac{12 \zeta  \lambda  \xi ^3}{\mu ^4}
       +\frac{12 \zeta  \xi ^5}{\mu ^6}  
       -\frac{27 \lambda ^2 \xi ^2}{\mu ^2} 
        + \frac{18 \lambda  \lambda' \xi ^2}{\mu ^2} \nonumber \\
     & &  \! \! \!   +~  \frac{54 \lambda  \xi ^4}{\mu ^4}
     -\frac{18 \lambda' \xi ^4}{\mu ^4}
     -\frac{27 \xi ^6}{\mu ^6} \bigg) \frac{v^2}{\mu^2}\bigg]  
      \bigg( ~ \!  \frac{v^2}{s \! - \! m_{\varphi_1}^2}
          +\frac{v^2}{t \! - \! m_{\varphi_1}^2}
          +\frac{v^2}{u \! - \! m_{\varphi_1}^2}
          \bigg) \nonumber \\
& &  \! \! \!       
       + ~ \! \bigg( \frac{12 \zeta  \xi ^3}{\mu ^4}
       +\frac{21 \lambda  \xi ^2}{\mu ^2}-\frac{18 \lambda' \xi ^2}{\mu ^2}
       -\frac{21 \xi ^4}{\mu ^4}  \bigg)
       ~ \! {v^2 \over \mu^2}
       %\nonumber 
       \\
   &&  \! \! \! 
     + ~ \!  \frac{\xi ^2}{\mu ^2}
        ~ \! 
         \bigg( {s+t+u \over \mu^2 } \bigg) 
         + { \cal O}
               \bigg( {x^2 \over \mu^4} , {x v^2 \over \mu^4} , {v^4\over \mu^4} \bigg)
     \nonumber 
% && + ~ { \cal O}
% \bigg( {x^2 \over \mu^2} , {v^2 \over \mu^2} \bigg)
%   % \left(\frac{s^2}{m_{\varphi_2}^4},\frac{t^2}{m_{\varphi_2}^4},
%    % \frac{u^2}{m_{\varphi_2}^4},\frac{v^4}{\mu^4}\right)
 %\nonumber \\
\end{eqnarray} 
where $ x = s,t,u$.  On shell we get
\begin{eqnarray}
{ \cal A} \left(\varphi_1 \varphi_1 \rightarrow \varphi_1\varphi_1\right) 
  \! \! \! &=&   \! \! \! -\frac{3m_{\varphi_1}^2}{v^2}
  +\left(\frac{12 \zeta  \xi ^3}{\mu ^4}
  +\frac{25 \lambda  \xi ^2}{\mu ^2}
  -\frac{18 \lambda' \xi ^2}{\mu ^2}
  -\frac{25\xi ^4}{\mu ^4}\right)\left(\frac{v}{\mu}\right)^2 \nonumber \\
  & &  \! \! \!       
  -~ \bigg[{ 9 m_{\varphi_1}^4 \over v^4} 
       +\bigg(-\frac{12 \zeta  \lambda  \xi ^3}{\mu ^4}
       +\frac{12 \zeta  \xi ^5}{\mu ^6}  
       -\frac{27 \lambda ^2 \xi ^2}{\mu ^2} 
        + \frac{18 \lambda  \lambda' \xi ^2}{\mu ^2} \nonumber \\
     & &  \! \! \!   +~  \frac{54 \lambda  \xi ^4}{\mu ^4}
     -\frac{18 \lambda' \xi ^4}{\mu ^4}
     -\frac{27 \xi ^6}{\mu ^6} \bigg) \frac{v^2}{\mu^2}\bigg]  
      \bigg( ~ \!  \frac{v^2}{s \! - \! m_{\varphi_1}^2}
          +\frac{v^2}{t \! - \! m_{\varphi_1}^2}
          +\frac{v^2}{u \! - \! m_{\varphi_1}^2}
          \bigg) \nonumber \\
  &&\! \! \!  +~ {\cal O}\left(\frac{x^2}{m_{\varphi_2}^4},\frac{v^4}{\mu^4}\right)  \label{eq:onshellamplitude}
\end{eqnarray}
Note that the long distance contribution to the amplitude is controlled by the trilinear coupling $g_{\varphi_1^3 }$. 

\subsection{The low energy effective theory}
\label{sec:realsingletEFT}

The mixing language presented in the last section provides a complete description of the Higgs particle in the SHSM. In this section we are interested in following an alternative approach in terms of EFT by integrating out the singlet. This approach is valid near the decoupling limit defined in \eqref{eq:decouplinglimitsinglet}. The cutoff of the EFT is the singlet mass $\mu$. No reference to mixing between the singlet and the Higgs is needed in the EFT approach to describe the properties of the Higgs boson, all the effects of mixing are automatically encoded in the low energy theory. In deriving the low energy theory we work at tree level. All the corrections to the SM Lagrangian in this case are threshold effects. This is enough to reproduce the tree level mixing effects presented in the last section. 
%The regime of parameters we work on are specified in section \ref{sec:realsingletmodel} and is briefly summarized here. We consider all the parameters in the potential to be perturbative. The dimensionful parameters in the potential are the heavy singlet mass $\mu$, the light doublet mass $m^2$, $\xi$ and $\zeta$. We allow $\xi$ and $\zeta$ to be as large as the singlet mass, which is the cutoff of the EFT.

EFT organizes the corrections to the SM properties hierarchically in terms of the small expansion parameter $v^2/\mu^2$.  In order to work consistently up to a particular order in the small expansion parameter, we must define a concept of \textit{effective operator dimension} which allows us to identify the operators that must be kept in the low energy theory. In the SHSM, the correct concept of effective dimension is just naive operator dimension. For instance, we will see that the effective Lagrangian contains a threshold correction to the quartic
 \begin{equation}
\frac{\xi^2}{\mu^2}(H^\dagger H)^2
\end{equation}
The coefficient of this operator contains a power of the heavy singlet mass in the denominator, but it must not be considered to be suppressed, since $\xi^2$ in the numerator is allowed to be as large as $\mu^2$, so this operator is of effective dimension four. In section \ref{sec:2HDMEFT} we will see that in the case of the effective field theory of the 2HDM, we will need to define a concept of effective dimension which does not coincide with naive operator dimension. In this paper, we build the low energy theory of the SHSM up to effective dimension six. Just for illustrative purposes, we also keep the leading effects in $\lambda_S$ at effective dimension eight.
%Motivated by the separation of scales, we define our concept of operator effective dimension by counting powers of the heavy scale in the operator's coefficient in the Lagrangian. We define the operator's effective dimension as
%\begin{equation}
%n_{E}=4-n_{m^2_{heavy}}
%\end{equation}
% where $n_{m^2_{heavy}}$ is the number of powers of the heavy mass scale in the operator's coefficient. 

\begin{figure}[ht!]
\centering
\begin{minipage}[t]{0.45 \textwidth}
\centering
\includegraphics[width=5cm]{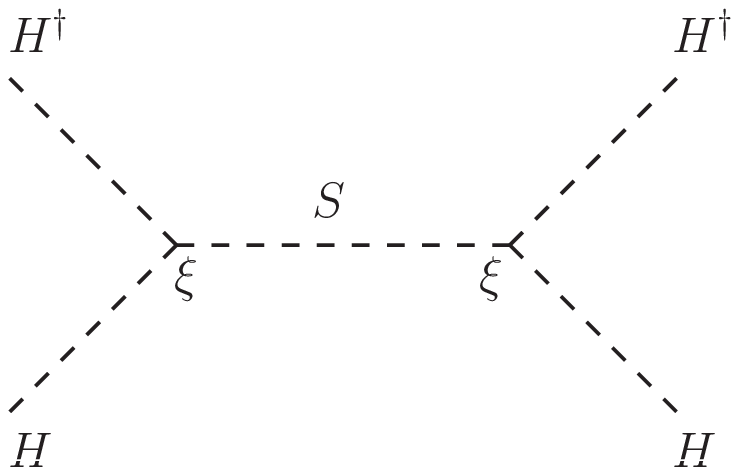}
\caption{At dimension four and six generates the operators $(H^\dagger H)^2$ and $\partial_\mu (H^\dagger H)\partial^\mu (H^\dagger H)$.}
\label{fig:realsinglet}
\end{minipage}%
~~~
\begin{minipage}[t]{0.45\textwidth}
\centering
\includegraphics[width=5cm]{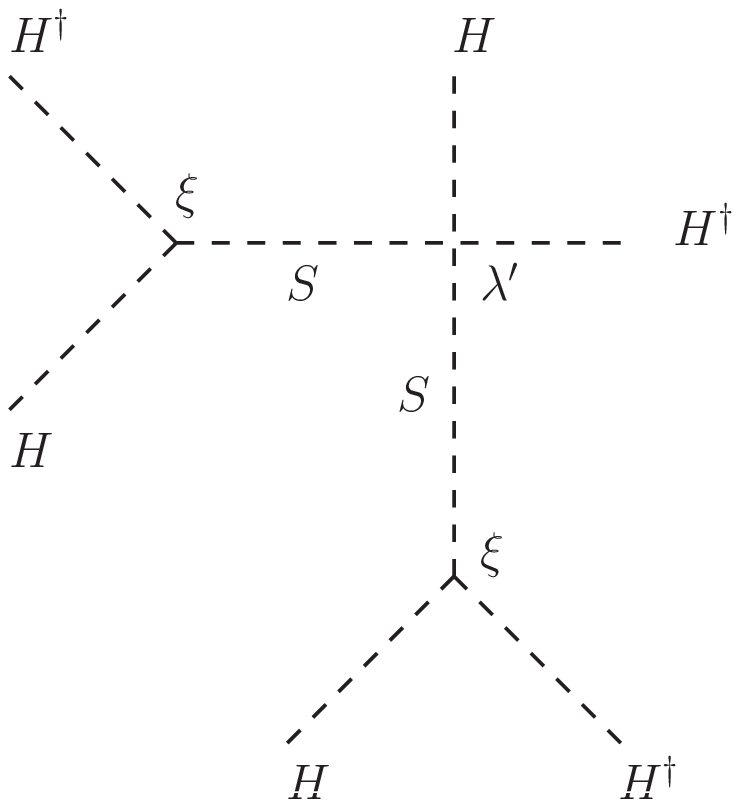}
\caption{At dimension six generates the operator $(H^\dagger H)^3$.}
\label{fig:H6diagramrealsinglet}
\end{minipage} \nonumber \\

\begin{minipage}[t]{0.45 \textwidth}
\centering
\includegraphics[width=7cm]{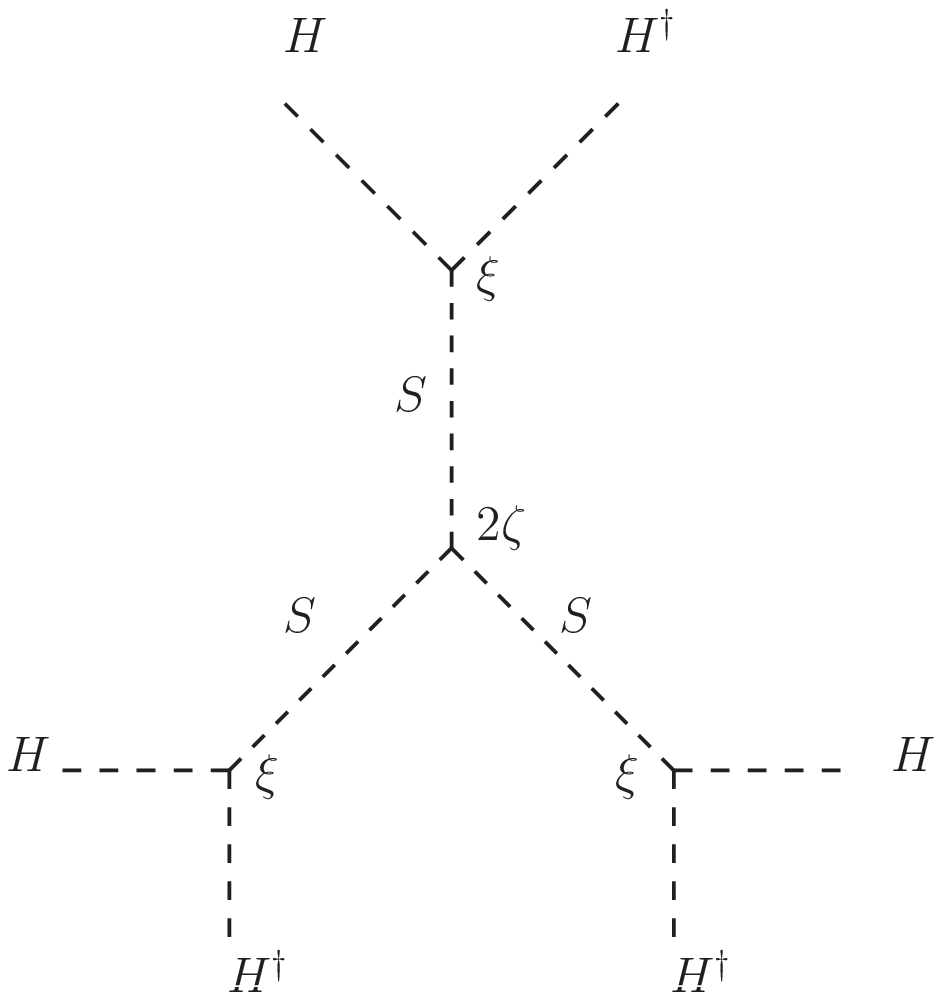}
\caption{At dimension six generates the operator $(H^\dagger H)^3$.}
\label{fig:realsinglezeta}
\end{minipage}%
~~~
\begin{minipage}[t]{0.45\textwidth}
\centering
\includegraphics[width=7cm]{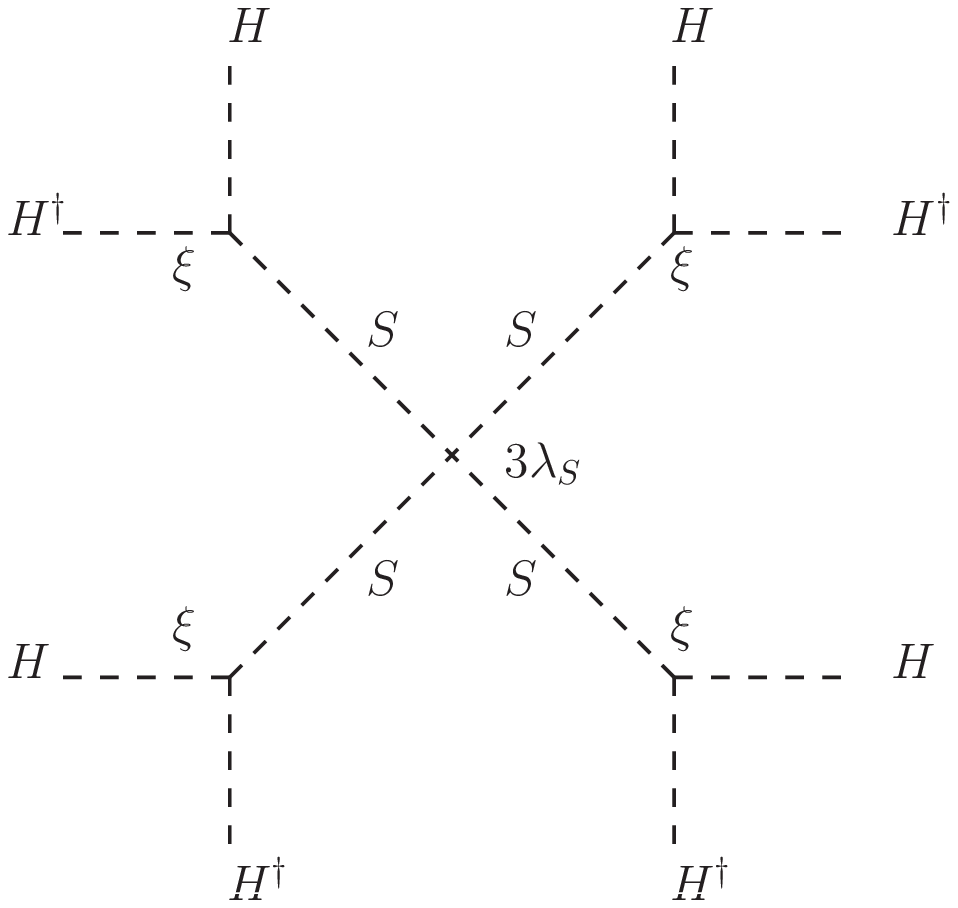}
\caption{At dimension eight generates the operator $(H^\dagger H)^4$. This diagrams represents the leading order correction in $\lambda_S$, and is considered only for illustrative purposes.}
\label{fig:H8diagramrealsinglet}
\end{minipage}
\end{figure}

The tree level low energy effective theory of the SHSM can be obtained by computing the diagrams of figures \ref{fig:realsinglet}-\ref{fig:H8diagramrealsinglet}. Diagrams  \ref{fig:realsinglet}-\ref{fig:realsinglezeta} give all the effective dimension six threshold corrections, and diagram \ref{fig:H8diagramrealsinglet} gives the leading correction in $\lambda_S$ at effective dimension eight. The resulting effective Lagrangian is
%We define all momenta corresponding to Higgs external legs to be incoming throughout the rest of this paper. 
\begin{eqnarray}
&& Z_H ~ \! D_\mu H^\dagger  D^\mu H 
 +  {1 \over 2} \zeta_{ H} ~ \!  \partial_\mu (H^\dagger H)\partial^\mu (H^\dagger H)
- V(H)
\nonumber \\
&&  \! \! \! 
 - \bigg[ 
   ~  \lambda^u_{ij} ~ Q_i H \bar{u}_j
   - \lambda^d_{ij} Q_i H^c \bar{d}_j 
   - \lambda^\ell_{\ij} L_i H^c \bar{\ell}_j 
  +{\rm h.c.}~ \bigg]
% $$
%  {\cal L}_{eff} \! \! \! &=& \! \! \! 
%  D_\mu H^\dagger D^\mu H
%  +\frac{\zeta_H}{2\mu^2}  \partial(H^\dagger H)\partial (H^\dagger H) \nonumber \\
%&&  \! \! \! - ~ m^2 H^\dagger H 
%-\frac{\tilde{\lambda}}{2}(H^\dagger H)(H^\dagger H)   
%-\frac{c_6}{\mu^2}(H^\dagger H)^3-\frac{c_8}{\mu^4}(H^\dagger H)^4  \nonumber \\
%&&  \! \! \! -
% ~  \bigg[ ~  \lambda^u_{ij} ~ Q_i H \bar{u}_j
%   - \lambda^d_{ij} Q_i H^c \bar{d}_j 
%   - \lambda^\ell_{\ij} L_i H^c \bar{\ell}_j 
%  +{\rm h.c.}~ \bigg]
\label{eq:summaryLETsinglet}
\end{eqnarray}
where
\begin{equation}
V(H) = m_H^2 H^\dagger H 
+ {1 \over 2} \lambda_H  (H^\dagger H)^2  
+ {1 \over 3}  \eta_6    (H^\dagger H)^3
+ {1 \over 4}  \eta_8     (H^\dagger H)^4
\end{equation}
with 
$$
Z_H = 1 ~~~~~~
\zeta_{H} =\frac{\xi^2}{ \mu^4} ~~~~ ~~~ 
m_H^2=m^2
$$
$$
\lambda_H=\lambda-\frac{\xi^2}{\mu^2} ~~~~ ~~~ 
\eta_6=   \frac{3 \lambda'\xi^2 }{2\mu^4}- \frac{\xi^{3}\zeta}{\mu^6}  ~~~~ ~~~ 
\eta_8=  \frac{\lambda_S\xi^{4}}{2\mu^8}  ~~~~ ~~~ 
\lambda^f_{ij}=\frac{\sqrt{2} m^f_{ij}}{v} =\frac{\sqrt{2}m^f_{i} }{v} ~\delta^f_{ij}
$$
%\begin{equation}
%\zeta_H=\frac{\xi^2}{\mu^2} ~~~~ , ~~~~ \tilde{\lambda}=\lambda-\frac{\xi^2}{\mu^2} ~~~~ , ~~~~ c_6=\frac{\lambda'\xi^2 }{2\mu^2}- \frac{\xi^{3}\zeta}{3\mu^4} ~~~~ , ~~~~ c_8=\frac{\lambda_S\xi^{4}}{8\mu^4} 
%\end{equation}
%The kinetic operator $\partial_\mu (H^\dagger H)\partial^\mu (H^\dagger H)$ and the shift in the Higgs quartic $(H^\dagger H)^2 $ come from figure \ref{fig:realsinglet}. The operator $(H^\dagger H)^3$ comes from figures \ref{fig:H6diagramrealsinglet} and \ref{fig:realsinglezeta}. The operator $(H^\dagger H)^4$ is our example of a higher order term, and it comes from figure \ref{fig:H8diagramrealsinglet}. 

The Higgs quartic in the low energy $\lambda_H=\lambda-\xi^2/\mu^2$ can be negative, since $\xi^2/\mu^2$ can be of order one. In that case, the Higgs potential can still be stabilized by the $(H^\dagger H)^3$ operator, as needed in some models of baryogenesis \cite{Grojean:2004xa}. Note that the coupling $\lambda_S$ is irrelevant at effective dimension six: it first enters as a coefficient of a dimension eight operator. As a consequence, its effects in the low energy theory are always subleading in the expansion in the small parameter $v^2/\mu^2$. The effective theory \eqref{eq:summaryLETsinglet} has the particularity that it does not contain operators with gauge boson or fermion fields beyond the ones already present in the SM, since the particle that is being integrated out is a singlet. All the modifications of the Higgs couplings to fermions and massive gauge bosons with respect to their SM values come from the operator $\partial_\mu (H^\dagger H)\partial^\mu (H^\dagger H)$, as we will see shortly.

% \footnote{This is not true anymore when working at loop level. For example, when working at two loops, there are diagrams with Higgs and singlet loops with external Higgs and gauge boson legs which must be computed.}
We now proceed to write the effective theory for the neutral component of the doublet in the unitary gauge, which corresponds to the Higgs in the EFT description. The extremum condition for the potential is
\begin{equation}
{ \partial V \over \partial v} \Big |_{h=0}= v \Big( 
m_H^2  
+ {1 \over 2} \lambda_H v^2 
+{1 \over 4} \eta_6 v^4
+{1 \over 8} \eta_8 v^6
 \Big) = 0 
\end{equation}
For a non-zero vev, the soft mass can be expressed in terms of the vev and couplings
\begin{equation}
m_H^2=- {1 \over 2} \lambda_H v^2 
-{1 \over 4} \eta_6 v^4
-{1 \over 8} \eta_8 v^6
\label{eq:softmasssinglettheory}
\end{equation}
The low energy Lagrangian density for the Higgs is
\begin{eqnarray}
{1 \over 2}  Z_H ~ \! \partial_\mu h ~ \! \partial^\mu h
+   \zeta_H 
\bigg(  
 ~ \!   {1 \over 2} v^2  ~ \! \partial_\mu  h ~ \! \partial^\mu h
   + v h  ~ \! \partial_\mu  h ~ \! \partial^\mu h
    +\frac{1}{2}  h^2  ~ \! \partial_\mu  h ~ \! \partial^\mu h 
\bigg)
-V(h)
 \notag \\
  +\frac{1}{2}  m_Z^2
    \bigg( 1 + {2 h \over v} + {h^2 \over v^2} ~ \! 
    \bigg)  Z^{\mu} Z_\mu 
  +
     m_W^2 
     \bigg(  1 + {2 h \over v} + {h^2 \over v^2} ~ \! 
     \bigg)  W^{+ \mu} W_{\mu}^-  \notag \\
      - \bigg(  1 + {h \over v}  \bigg) ~ \! 
  m^f_{ij} ~ \! f_i  \bar{f}_j +\textrm{h.c.} 
  \label{eq:lowenergyLagrangiansinglet1}
\end{eqnarray}
where
\begin{eqnarray}
  V(h)\! \! \! &=& \! \! \! 
  {1 \over 2}~ \!  m_H^2    
  %\left(  2 vh+h^2  \right)
     h^2
           +  {1 \over 2} ~ \! \lambda_H
        \bigg( %v^3 h
    ~ \!  \frac{3}{2}v^2h^2
     +vh^3+\frac{1}{4}h^4    \bigg)
      \nonumber \\  
&& \! \! \! 
    + ~ {1 \over 3} ~ \!  \eta_6 %\frac{\eta_6}{3}  
       \bigg(  %\frac{3  v^5}{4}h
    ~ \!   \frac{15 v^4}{8}h^2
    +\frac{5 v^3}{2}h^3+\frac{15v^2}{8}h^4
    +\frac{3v}{4}h^5+\frac{1}{8}h^6    \bigg)
    %\nonumber  
    \\  
&& \! \! \! + ~ {1 \over 4}~ \!  \eta_8
     \bigg(   %\frac{ v^7}{2}h
     ~ \!   \frac{7 v^6}{4}h^2
  +\frac{7 v^5}{2}h^3
  +\frac{35 v^4}{8}h^4 
  +\frac{7 v^3}{2} h^5
  +\frac{7 v^2}{4}h^6 
  +\frac{ v}{2}h^7
  +\frac{1}{16}h^8   \bigg) 
      \nonumber 
\label{eq:singletHiggspotential}
\end{eqnarray}
 The lowest order equation of motion for the Higgs field is
\begin{eqnarray}
Z_H \Box h \! \! \! &=& \! \! \!   -m_H^2 h
-\frac{1}{2}\lambda_H (3v^2 h
+3 v h^2
+h^3) \nonumber \\  
&& \! \! \!  + ~ \bigg(\frac{m_Z^2}{v}Z^\mu Z_\mu
+\frac{2m_W^2}{v}W^{+ \mu} W_{\mu}^- \bigg) \bigg(1+\frac{h}{v} \bigg)
 -\bigg(\frac{m^f_{ij}}{v} f_i \bar{f_j} +\textrm{h.c.}\bigg) +\dots
\label{eq:boxhsinglet}
\end{eqnarray}
%\begin{eqnarray}
%\Box h \! \! \! &=& \! \! \!   -m_h^2 h-\frac{3}{2}\lambda_Hv h^2
%-\frac{1}{2}\lambda_Hh^3
%+\bigg(\frac{m_Z^2}{v^2}Z^\mu Z_\mu
%+\frac{2m_W^2}{v^2}W^{+ \mu} W_{\mu}^- \bigg) \bigg(1+\frac{h}{v} \bigg)  \nonumber \\  
%&& \! \! \!  - ~ 
% \bigg(\frac{m^f_{ij}}{v} f_i \bar{f_j}  +\textrm{h.c.}\bigg) +\dots
%\label{eq:boxh}
%\end{eqnarray}
where the dots represent terms of higher effective dimension. From now on we commit to $Z_H=1$.

The operator $\partial_\mu (H^\dagger H)\partial^\mu (H^\dagger H)$ leading to the non canonical kinetic terms in \eqref{eq:lowenergyLagrangiansinglet1} has two effects in the couplings of the Higgs. First, it leads to wave function renormalization which dilutes all couplings. Second it gives additional irrelevant operators with derivatives which further modify the Higgs couplings, once they are replaced in favor of operators with no derivatives using integration by parts and the Higgs equation of motion. This second effect is usually not considered in the literature \cite{Henning:2014gca}, but it is important to obtain the correct couplings in the low energy theory. We start by discussing these irrelevant operators. First, using integration by parts, the operator $\partial_\mu (H^\dagger H)\partial^\mu (H^\dagger H)$ can be rewritten as
\begin{eqnarray}
%{1 \over 2}  Z_H ~ \! \partial_\mu h ~ \! \partial^\mu h
\frac{1}{2} \zeta_H \partial_\mu (H^\dagger H)\partial^\mu (H^\dagger H) \! \! \! &=& \! \! \!
\zeta_H 
\bigg(  
 ~ \!   {1 \over 2} v^2  ~ \! \partial_\mu  h ~ \! \partial^\mu h
   + v h  ~ \! \partial_\mu  h ~ \! \partial^\mu h
    +\frac{1}{2}  h^2  ~ \! \partial_\mu  h ~ \! \partial^\mu h 
\bigg) \nonumber \\
 \! \! \! &=& \! \! \!
\zeta_H \bigg(  
 ~ \!   {1 \over 2} v^2  ~ \! \partial_\mu  h ~ \! \partial^\mu h
   - \frac{1}{2} v h^2\! ~\Box h
   -\frac{1}{6}h^3 ~ \!\Box h
\bigg)
%-V(h)
% \notag \\
%  +\frac{1}{2}  m_Z^2
%    \bigg( 1 + {2 h \over v} + {h^2 \over v^2} ~ \! 
%    \bigg)  Z^{\mu} Z_\mu 
%  +
%     m_W^2 
%     \bigg(  1 + {2 h \over v} + {h^2 \over v^2} ~ \! 
%     \bigg)  W^{+ \mu} W_{\mu}^-  \notag \\
%      - \bigg(  1 + {h \over v}  \bigg) ~ \! 
%  m^f_{i} ~ \! f_i  \bar{f}_j ~\delta_{ij}+\textrm{h.c.} 
  \label{eq:lkinsinglet}
\end{eqnarray}
%\begin{equation}
%{\cal L}_{kin}= \frac{1}{2}\partial h \partial h
%+\frac{\zeta_H}{\mu^2}\bigg( \! \frac{1}{2}v^2\partial h \partial h- \frac{1}{2} v h^2 \Box h-\frac{1}{6}h^3 \Box h \bigg)
%\label{eq:lkinsinglet}
%\end{equation}
%\begin{eqnarray}
%\Box h \! \! \! &=& \! \! \!  \! 
%-\frac{\zeta_H}{\mu^2}\bigg[ v^2 \Box h
%+vh\Box h 
%+v \Box h^2 
%+\frac{1}{2}h^2\Box h 
%+\frac{1}{6}\Box h^3\bigg]
%- \frac{\partial V}{\partial h}
% \nonumber \\  
%&& \! \! \! ~ 
%%+m_h^2
%%+\frac{3}{2}\lambda_Hv h
%%+\frac{1}{2}\lambda_Hh^2\bigg) h
%+\bigg(\frac{m_Z^2}{v^2}Z^\mu Z_\mu
%+\frac{2m_W^2}{v^2}W^{+ \mu} W_{\mu}^- \bigg) \bigg(1+\frac{h}{v} \bigg)  - \bigg(\frac{m^f_{ij}}{v} f_i \bar{f_j}  +\textrm{h.c.}\bigg) %+
%%\nonumber \\  
%%&& \! \! \! ~ 
%%{\cal O}\left(\frac{\Box h v^2}{\mu^2},\frac{h^2 v^3}{\mu^2},\frac{h^3 v^2}{\mu^2},\frac{h^4 v}{\mu^2},\frac{h^5}{\mu^2},\lambda_S\right) 
%\end{eqnarray}
%where 
%\begin{eqnarray}
%  \frac{\partial V}{\partial h}\! \! \! &=& \! \! \! 
% m^2 h
%  +\frac{\lambda_H}{2}\bigg( \! 3v^2h+3vh^2+h^3 \bigg) \nonumber \\  
%&& \! \! \! + ~ \frac{c_6}{\mu^2}\bigg( \! \frac{15 v^4}{4}h
%+\frac{15 v^3}{2}h^2+\frac{15v^2}{2}h^3
%+\frac{15v}{4}h^4+\frac{3}{4}h^5\bigg)\nonumber  \\  
%&& \! \! \! + ~ \frac{c_8}{\mu^4}\bigg( \!  \frac{7 v^6}{2}h
%+\frac{21 v^5}{2}h^2
%+\frac{35 v^4}{2}h^3 
%+\frac{35 v^3}{2} h^4
%+\frac{21 v^2}{2}h^5 
%+\frac{ 7v}{2}h^6
%+\frac{1}{2}h^7\bigg) \nonumber \\
%\end{eqnarray}
The lowest order equation of motion \eqref{eq:boxhsinglet} can be used in \eqref{eq:lkinsinglet} to replace the operator $\Box h$ in favor of operators with no derivatives. The resulting Lagrangian is
\begin{eqnarray}
&& \! \! \!  {1 \over 2}  ~ \! \partial_\mu h ~ \! \partial^\mu h
+   \zeta_H 
\Bigg(
 ~ \!  \frac{1}{2}v^2~ \! \partial_\mu h ~ \! \partial^\mu h-
    \bigg(\frac{1}{2}v h^2  ~ 
    +\frac{1}{6}  h^3  \bigg)\bigg[-m_H^2 h  \nonumber \\ 
&& \! \! \!  -~ \frac{1}{2}\lambda_H (3v^2 h
+3 v h^2
+h^3) + \bigg(\frac{m_Z^2}{v}Z^\mu Z_\mu
+\frac{2m_W^2}{v}W^{+ \mu} W_{\mu}^- \bigg) \bigg(1+\frac{h}{v} \bigg)  \nonumber \\ 
&& \! \! \!  - ~
 \bigg(\frac{m^f_{ij}}{v} f_i \bar{f_j}+\textrm{h.c.}\bigg)  \bigg]
\Bigg)
-V(h)
 \nonumber \\
 && \! \! \!  + ~ \frac{1}{2}  m_Z^2
    \bigg( 1 + {2 h \over v} + {h^2 \over v^2} ~ \! 
    \bigg)  Z^{\mu} Z_\mu 
  +
     m_W^2 
     \bigg(  1 + {2 h \over v} + {h^2 \over v^2} ~ \! 
     \bigg)  W^{+ \mu} W_{\mu}^-  \notag \\
     && \! \! \!  -~\bigg(  1 + {h \over v}  \bigg) ~ \! 
  m^f_{ij} ~ \! f_i  \bar{f}_j +\textrm{h.c.} 
  \label{eq:secondstep}
\end{eqnarray}
%
%\begin{eqnarray}
%-\zeta_H  \bigg( \frac{1}{2} v h^2+\frac{1}{6}h^3\bigg)\Box h  \! \! \! &=& \! \! \!   \zeta_H  \bigg( \frac{1}{2} v h^2+\frac{1}{6}h^3\bigg) \bigg[ m_H^2 h
%+\frac{1}{2}\lambda_H (3v^2 h
%+3 v h^2
%+\frac{1}{2}\lambda_Hh^3) \nonumber \\  
%&& \! \! \!  - ~ \bigg(\frac{m_Z^2}{v^2}Z^\mu Z_\mu
%+\frac{2m_W^2}{v^2}W^{+ \mu} W_{\mu}^- \bigg)(v+h)  \nonumber \\  
%&& \! \! \!  + ~ 
% \bigg(\frac{m^f_{i}}{v}  f_i \bar{f_j} ~\delta_{ij} +\textrm{h.c.}\bigg)\bigg] 
% \label{eq:useeqmotionsinglet}
%\end{eqnarray}
We now discuss the wave function renormalization term $\! \zeta_H v^2 ~\! \partial_\mu h ~ \! \partial^\mu h$ in \eqref{eq:secondstep}. To canonicalize the kinetic Lagrangian, we perform a field redefinition
\begin{equation}
\varphi=( 1+\zeta_H v^2 )^{1/2} h=\bigg[ 1+\frac{\xi^2v^2}{\mu^4}  \bigg]^{1/2} h =\bigg[1+\sin^2\gamma+{\cal O}\left(\frac{v^4}{\mu^4}\right)\bigg]^{1/2}  h
\label{eq:WFrenormalizationsinglet}
\end{equation}
where using \eqref{eq:realsingletalign} we expressed the result in terms of $\sin^2 \gamma$ to point out the close relation between mixing and wave function renormalization. In the mixing language, all Higgs couplings are diluted democratically due to mixing with the singlet. In the EFT language, this dilution is represented as wave function renormalization of the Higgs field: the wave function renormalization constant is equal to $\cos^{-1}\gamma$ up to corrections that in principle could come from higher order derivative interactions. Note that in the SHSM EFT wave function renormalization is an effective dimension six effect. The full Lagrangian in its final form is
%%\begin{eqnarray}
%%Z_h=1+\zeta_H \left(\frac{v}{\mu}\right)^2=1+ \frac{\xi^2}{\mu^2}\left(\frac{v}{\mu}\right)^2=\sin^2(\gamma)\left(1+{\cal O}\left(\frac{v^2}{\mu^2}\right)\right)  
%%\end{eqnarray}
%
%a field redefinition
%\begin{equation}
%h = \frac{\varphi}{\sqrt{Z_h}}
%\label{eq:fieldredef}
%\end{equation}
%\begin{eqnarray}
%  {\cal L} \! \! \! &=& \! \! \! 
%  \frac{1}{2}\partial h \partial h 
%  -\frac{1}{2}m_h^2 h^2   
% + \sum_{n=3}^{n=8} \frac{1}{n!}g_{h^n}h^n
% -  (m^f_{ij} f_i\bar{f}_j+\lambda^{f}_{h ij} ~f_i  h \bar{f}_j +\textrm{h.c.})
%   \nonumber  \\
%&& \! \! \! 
% + ~ \frac{1}{2}Z^{\mu} Z_\mu \left( \! m_Z^2+ g_{h ZZ}~ h + \frac{1}{2}g_{h^2ZZ}~ h^2 \right)
% + W^{+ \mu} W_{\mu}^- \left( \! m_W^2+ g_{h WW}~ h + \frac{1}{2}g_{h^2WW}~ h^2 \right)   \nonumber \\
%\end{eqnarray}
\begin{eqnarray}
 \cal{L}\! \! \! &=& \! \! \! \frac{1}{2}\partial_\mu \varphi \partial^\mu \varphi
 -\frac{1}{2}m_\varphi^2 \varphi^2
 +\sum_{n=3}^{n=8} \frac{1}{n!}g_{\varphi^n}\varphi^n
 +\frac{1}{2}m_Z^2 Z^{\mu} Z_\mu 
 +m_W^2 W^{+ \mu} W_{\mu}^- \nonumber \\
&& \! \! \! + ~ \sum_{n=1}^{n=4}\frac{\varphi^n}{n!}\left[g_{\varphi^n ZZ}~  \frac{1}{2} Z^{\mu} Z_\mu  
+ g_{\varphi^n WW}~  W^{+ \mu} W_{\mu}^-\right] \nonumber   \nonumber  \\
&& \! \! \! - ~ m^{f }_{ij} f_i\bar{f}_j 
-\sum_{n=1}^{n=3} \frac{\varphi^n}{n! }\left[\lambda^{f}_{\varphi^n ij} ~f_i   \bar{f}_j +\textrm{h.c.}\right]   
%+\left(g_{h VV} h +\frac{1}{2}g_{h^2VV} h^2\right)V^{\dagger\mu} V_\mu  \nonumber 
%&+ &  h (\bar{Q}_L g_{h U\bar{U}} u_R+\bar{Q}_L g^{\dagger}_{h D \bar{D}} d_R+h.c.) 
\label{eq:LagrangianSHSMfinal}
\end{eqnarray}
where all the Higgs couplings can be expressed in terms of couplings of the UV completion using \eqref{eq:secondstep} and \eqref{eq:WFrenormalizationsinglet}, and are written down explicitly below. We start with the Higgs mass, which is given by
\begin{eqnarray}
 m_{\varphi}^2
   \! \! \! & =  & \! \! \!   
    v^2 \Bigg[
       \bigg(   \lambda -\frac{\xi ^2}{\mu ^2}   \bigg)
       +\bigg(  \frac{3\lambda'\xi ^2}{2 \mu ^2}
       -\frac{\lambda  \xi ^2}{\mu ^2}
       -\frac{\zeta  \xi ^3}{\mu ^4}
       +\frac{\xi ^4}{\mu ^4}   \bigg)
       ~ \!  {v^2 \over \mu^2}
   \nonumber \\
   && ~~ + ~ \!
  \frac{3\lambda_S \xi ^4}{8 \mu ^4}
  {v^4 \over \mu^4}
   +{\cal O}  \bigg({ v^4 \over \mu^4 } , \lambda_S^2 \frac{v^6}{\mu^6}
    \bigg)  
      ~ \!  \Bigg]
%   \frac{3\lambda_S \xi ^4}{8 \mu ^4}\left(\frac{v}{\mu}\right)^4+\mathcal{O}\left(\lambda_S^2\left(\frac{v}{\mu}\right)^4,\frac{v^4}{\mu^4}\right)\Bigg)
\label{eq:Higgssingletmass}
\end{eqnarray}
%
%\begin{eqnarray}
%  m_h^2&=&v^2 \Bigg[\left(\lambda -\frac{\xi ^2}{\mu ^2}\right)+\left(\frac{3\lambda'
%   \xi ^2}{2 \mu ^2}-\frac{\lambda  \xi ^2}{\mu ^2}-\frac{\zeta  \xi ^3}{\mu ^4}+\frac{\xi ^4}{\mu ^4}\right)\left(\frac{v}{\mu}\right)^2 \nonumber \\&+&\frac{3\lambda_S \xi ^4}{8 \mu ^4}\left(\frac{v}{\mu}\right)^4+{\cal O}\left(\lambda_S^2\left(\frac{v}{\mu}\right)^6,\frac{v^4}{\mu^4}\right)\Bigg]
%\end{eqnarray}
where we used \eqref{eq:softmasssinglettheory} to replace the Lagrangian mass $m_H$ in favor of the vacuum expectation value $v$. The term linear in $\lambda_S$ in \eqref{eq:Higgssingletmass} comes from our example of an effective dimension eight effect. We only keep this term for illustrative purposes for the Higgs mass, and we drop it from now on for the rest of the calculations. 

The couplings of the Higgs to gauge bosons are given by
\begin{eqnarray}
  g_{\varphi VV}  \! \! \! &=& \! \! \!  \frac{2m_V^2}{v} \bigg [   1-\frac{\xi^2}{2\mu^2}\frac{v^2}{\mu^2}+{\cal O} \bigg( {v^4 \over \mu^4}\bigg)  \bigg]  \nonumber \\
  g_{\varphi^2VV} \! \! \! &=& \! \! \!  \frac{2m_V^2}{v^2}  \bigg[ 1-2\frac{\xi^2}{\mu^2}\frac{v^2}{\mu^2}+{\cal O} \bigg( {v^4 \over \mu^4}\bigg)   \bigg] 
 \label{eq:HiggsgaugecouplingssingletEFT}
\end{eqnarray}
The modification of $ g_{\varphi VV} $  with respect to its SM value comes exclusively from wave function renormalization as defined in \eqref{eq:WFrenormalizationsinglet}. The modification of  $g_{\varphi^2 VV}$  with respect to its SM value comes from wave function renormalization and from the additional interaction terms obtained from using the Higgs equation of motion in going from \eqref{eq:lkinsinglet} to \eqref{eq:secondstep}. For completeness, the rest of the couplings to gauge bosons are
%\begin{equation}
%  vg_{\varphi^3VV} =  v^2g_{\varphi^4VV} =-\frac{8  \xi ^2 m_V^2 }{\mu ^4} +{\cal O} \bigg( {v^4 \over \mu^4}\bigg)  \nonumber \\
%  \end{equation}
  \begin{equation}
  vg_{\varphi^3VV} =  v^2g_{\varphi^4VV} =-\frac{m_V^2 }{v^2}\bigg[\frac{8  \xi ^2 v^2}{\mu ^4} +{\cal O} \bigg( {v^4 \over \mu^4}\bigg)\bigg]   
  \end{equation}
$g_{\varphi^3VV}$ and $g_{\varphi^4VV}$ are irrelevant couplings, and they vanish in the decoupling limit. They come from using the Higgs equation of motion in \eqref{eq:lkinsinglet}.

The couplings to fermions are 
  \begin{eqnarray}
   \lambda^f_{\varphi ij} \! \! \! &=& \! \! \!  \frac{m^f_{i}}{v} \bigg[1-\frac{\xi^2}{2\mu^2}\frac{v^2}{\mu^2}+{\cal O} \bigg( {v^4 \over \mu^4}\bigg)  \bigg] ~\delta_{ij} \nonumber \\
   v\lambda^f_{\varphi^2 ij} \! \! \! &=& \! \! \!   v^2\lambda^f_{\varphi^3 ij} = -\frac{m^f_{i}}{v}\bigg[\frac{\xi^2 v^2}{\mu^4} +{\cal O} \bigg( {v^4 \over \mu^4}\bigg) \bigg] ~\delta_{ij}
   \label{eq:HiggsfermioncouplingssingletEFT}
\end{eqnarray}
The modification of $ \lambda^f_{\varphi ij}$ with respect to its SM value comes exclusively from wave function renormalization.  $\lambda^f_{\varphi^2 ij}, \lambda^f_{\varphi^3 ij}$ are irrelevant couplings which come from using the Higgs equation of motion in \eqref{eq:lkinsinglet}. They vanish in the decoupling limit. 

The Higgs cubic and quartic self-couplings are modified by wave function renormalization and extra contributions from the rest of the irrelevant operators of the effective potential
\begin{eqnarray}
\label{eq:singletself2}
 { g_{\varphi^3}  \over v} 
 \! \! \!   &=& \! \! \!   
     -\frac{3m_{\varphi}^2}{v^2}+\bigg( \!  \frac{9 \lambda  \xi ^2}{2 \mu ^2}-\frac{3\lambda'\xi ^2}{\mu ^2}-\frac{9 \xi ^4}{2 \mu ^4}+\frac{2\zeta  \xi ^3}{\mu ^4}  \bigg) ~ \! 
     {v^2 \over \mu^2} +{\cal O} \bigg( {v^4 \over \mu^4}  \bigg)
   \nonumber \\
g_{\varphi^4} \! \! \!  &=& \! \! \!  
      -\frac{3m_{\varphi}^2}{v^2}
         +\bigg( \!  \frac{12 \zeta  \xi ^3}{\mu ^4}+\frac{25 \lambda  \xi ^2}{\mu ^2}
           -\frac{18 \lambda' \xi ^2}{ \mu ^2}-\frac{25 \xi ^4}{\mu ^4}  \bigg)
             ~ \! {v^2 \over \mu^2 }
             +{\cal O} \bigg(\frac{v^4}{\mu^4} \bigg)   
\label{eq:selfcouplings1}
\end{eqnarray}
The rest of the self-couplings are all irrelevant couplings, and they vanish in the decoupling limit. They are given by
\begin{eqnarray}
  vg_{\varphi^5} \! \! \!  &=& \! \! \!  \bigg(\! \frac{30 \zeta  \xi ^3}{\mu ^4}
  -\frac{45 \lambda' \xi ^2}{\mu ^2}
  +\frac{60 \lambda  \xi ^2}{\mu ^2}
  -\frac{60 \xi ^4}{\mu ^4}\bigg)~ \frac{v^2}{\mu^2}
  -\frac{105 \lambda_S \xi ^4}{2 \mu ^4}~ \frac{v^4}{\mu^4}
  +{\cal O}  \bigg({ v^4 \over \mu^4 } , \lambda_S^2 \frac{v^6}{\mu^6}
    \bigg)    \nonumber \\
  v^2g_{\varphi^6} \! \! \!  &=& \! \! \! \bigg(\! \frac{30 \zeta  \xi ^3}{\mu ^4}
  -\frac{45 \lambda' \xi ^2}{\mu ^2}
  +\frac{60 \lambda  \xi ^2}{\mu ^2}
  -\frac{60 \xi ^4}{\mu ^4}\bigg)~ \frac{v^2}{\mu^2}
  -\frac{315 \lambda_S \xi ^4}{2 \mu ^4}~ \frac{v^4}{\mu^4}
  +{\cal O}  \bigg({ v^4 \over \mu^4 } , \lambda_S^2 \frac{v^6}{\mu^6}
    \bigg) 
   \nonumber \\
  v^3g_{\varphi^7} \! \! \!  &=& \! \! \! -\frac{315 \lambda_S \xi ^4}{\mu ^4}~ \frac{v^4}{\mu^4}
  +{\cal O}  \bigg({ v^4 \over \mu^4 } , \lambda_S^2 \frac{v^6}{\mu^6}
    \bigg) 
 \nonumber \\
 v^4g_{\varphi^8} \! \! \!  &=& \! \! \! -\frac{315 \lambda_S \xi ^4}{\mu ^4}~ \frac{v^4}{\mu^4}
 +{\cal O}  \bigg({ v^4 \over \mu^4 } , \lambda_S^2 \frac{v^6}{\mu^6}
    \bigg) 
\end{eqnarray}

Note that in the SHSM EFT, all the modifications to Higgs couplings come at effective dimension six. Also, all the couplings of the Higgs to fermions and gauge bosons are reduced at leading order with respect to their SM counterparts, but Higgs self couplings may be larger. These features, together with the absence of novel flavor and T violating effects, are the main phenomenological features of the EFT of a Higgs sector completed in the UV with a real singlet.

We conclude this section by pointing out that all trilinear couplings involving at least one Higgs are the same in the EFT and in the mixing language. Integrating out heavy fields does not modify cubic interactions. The reason is that cubic interactions control the \textit{long distance}  (non analytic) pieces of the tree level four linear scattering amplitudes, which do not receive contributions from diagrams mediated by heavy fields. The equality of the cubic interactions in the EFT and mixing languages ensures the equality between the long distance pieces of these amplitudes in both languages. We give examples of calculations of these type of amplitudes in section \ref{sec:amplitudessingletmixing} in the mixing language and in section \ref{sec:amplitudessingletEFT} in the EFT language. This logic can be run backwards: the equality between the non analytic scattering amplitudes enforces that the trilinear couplings in both languages must match. The four linear couplings are not the same in the mixing and EFT languages. Integrating out heavy fields in general modifies quartic (and higher order) interactions. These couplings control \textit{short distance} pieces of the tree level four linear amplitudes, which do get contributions from diagrams mediated by the heavy mass eigenstate. In any case, even though couplings in general do not coincide in the mixing and EFT languages, the scattering amplitudes must coincide. Scattering amplitudes using the EFT language are obtained in the following section.

\subsubsection{Scattering amplitudes}
\label{sec:amplitudessingletEFT}
In this section we obtain the scattering amplitudes which were already computed in section \ref{sec:amplitudessingletmixing} using the mixing language, now using the low energy EFT summarized in equation \eqref{eq:LagrangianSHSMfinal}. This serves as a consistency check of the EFT, all amplitudes in the EFT and mixing language must match. We omit spinors in all amplitudes.
%
%The Higgs to diboson, difermion and dihiggs amplitudes are
%\begin{equation}
%{ \cal A} \left(\varphi \rightarrow VV \right) 
%=g_{\mu\nu}~\!g_{\varphi VV}
%=g_{\mu\nu}~\!\frac{2 m_V^2}{v} \bigg[1-\frac{\xi^2v^2}{2\mu^4}+ { \cal O} \bigg(  {v^4\over \mu^4} \bigg)\bigg]
%\label{eq:HiggsdibosonsingletEFT}
%\end{equation}
%\begin{equation}
%{ \cal A} \left(\varphi \rightarrow f_i\bar{f}_j \right) =g_{\varphi ij}=-\frac{m^f_{i}}{v}\bigg[1-\frac{\xi^2v^2}{2\mu^4}+ { \cal O} \bigg(  {v^4\over \mu^4} \bigg)\bigg] ~ \delta_{ij}
%\label{eq:HiggsdifermionsingletEFT}
%\end{equation}
%where we used the couplings of the effective theory \eqref{eq:HiggsgaugecouplingssingletEFT} and \eqref{eq:HiggsfermioncouplingssingletEFT}. Note that the amplitudes \eqref{eq:HiggsdibosonsingletEFT} and  \eqref{eq:HiggsdifermionsingletEFT} are trivially identical to the corresponding amplitudes obtained in the mixing language, \eqref{eq:Higgsdibosonsingletmixing} and   \eqref{eq:Higgsdifermionsingletmixing}.

%\begin{eqnarray}
%{ \cal A} \left(\varphi \rightarrow \varphi \varphi \right)  \! \! \! &=&   \! \! \!  g_{\varphi_1^3} =   -\frac{3m_{\varphi_1}^2}{v}\nonumber \\
%     & &  \! \! \!  +~ v\bigg[\bigg(  \frac{9 \lambda  \xi ^2}{2 \mu ^2}-\frac{3\lambda'\xi ^2}{\mu ^2}-\frac{9 \xi ^4}{2 \mu ^4}+\frac{2\zeta  \xi ^3}{\mu ^4}  \bigg) ~ \! 
%     {v^2 \over \mu^2}             +{\cal O} \bigg(\frac{v^4}{\mu^4} \bigg)\bigg]
%\end{eqnarray}
The dihiggs scattering amplitude to two W bosons is 
\begin{eqnarray}
{ \cal A} \left(\varphi \varphi \rightarrow W^{+} W^-\right) 
  \! \! \! &=&   \! \! \! g_{\mu\nu}~\! 
  \Bigg[g_{\varphi^2WW}-\frac{g_{\varphi^3}g_{\varphi WW}}{s-m_{\varphi}^2}  -g_{\varphi WW}^2  \bigg(\frac{1}{t-m_{W}^2}+\frac{1}{u-m_{W}^2}\bigg) \Bigg]
 \label{eq:exactdihiggsdibosonamplitudeEFT}
 \end{eqnarray}
Note that the same amplitude calculated in the mixing language (see \eqref{eq:exactdihiggsdibosonamplitudemixing}) also contains a contribution from a diagram mediated by the heavy mass eigenstate. In the EFT language the heavy state has already been integrated out, so its contribution is already included in the low energy effective couplings. Using the couplings of the effective theory \eqref{eq:HiggsgaugecouplingssingletEFT} and \eqref{eq:selfcouplings1} in \eqref{eq:exactdihiggsdibosonamplitudeEFT}, we get
 \begin{eqnarray}
{ \cal A} \left(\varphi \varphi \rightarrow W^{+} W^- \right) 
  \! \! \! &=&   \! \! \!     
     g_{\mu\nu}~\!  \Bigg[
       \frac{2m_W^2}{v^2}  \bigg(\! 1-\frac{2\xi^2v^2}{\mu^4}  \bigg)
       +  \frac{2m_W^2}{v^2}\bigg[ \frac{3m_\varphi^2}{v^2}
     +\bigg(  -\frac{2  \zeta  \xi ^3}{\mu ^4}   \nonumber \\
     & &  \! \! \!   -~ \frac{6 \lambda   \xi ^2}{\mu ^2}
       +\frac{3 \lambda'  \xi ^2}{\mu ^2}
       +\frac{6 \xi ^4}{\mu ^4}\bigg)\frac{v^2}{\mu^2}
      \bigg]   
      \bigg( ~ \!  \frac{v^2}{s \! - \! m_{\varphi}^2}
          \bigg) \nonumber \\
               & &  \! \! \!   
               -~  \frac{4m_W^4}{v^4} \bigg[1-\frac{\xi^2v^2}{\mu^4} \bigg] \bigg(\frac{v^2}{t-m_{W}^2}+\frac{v^2}{u-m_{W}^2}\bigg) 
        + { \cal O}
               \bigg({v^4\over \mu^4} \bigg)
               \Bigg]
% && + ~ { \cal O}
% \bigg( {x^2 \over \mu^2} , {v^2 \over \mu^2} \bigg)
%   % \left(\frac{s^2}{m_{\varphi_2}^4},\frac{t^2}{m_{\varphi_2}^4},
%    % \frac{u^2}{m_{\varphi_2}^4},\frac{v^4}{\mu^4}\right)
 %\nonumber \\
\end{eqnarray} 
which coincides with the corresponding calculation in the mixing language \eqref{eq:phiphiWWamplitudemixing}.

The chirality violating dihiggs to difermion scattering amplitude is 
\begin{eqnarray}
{ \cal A} \left(\varphi \varphi \rightarrow f_i \bar{f}_j \right) 
  \! \! \! &=&   \! \! \! -\lambda^f_{\varphi^2 ij}+\frac{g_{\varphi^3}\lambda^f_{\varphi ij}}{s-m_{\varphi}^2}
  \label{eq:exactdihiggsdifermionamplitudeEFT}
 \end{eqnarray}
The same amplitude calculated in the mixing language \eqref{eq:exactdihiggsdifermionamplitudemixing} also contains a diagram mediated by the heavy mass eigenstates, which in the EFT is already included in the low energy effective couplings. Note that in the EFT language there is a four linear effective coupling $\lambda^f_{\varphi^2 ij}$ that contributes to the amplitude. Using the effective couplings of the EFT \eqref{eq:HiggsfermioncouplingssingletEFT} and \eqref{eq:selfcouplings1} in \eqref{eq:exactdihiggsdifermionamplitudeEFT} we get
% \begin{eqnarray}
%{ \cal A} \left(\varphi \varphi \rightarrow f_i \bar{f}_j \right) 
%  \! \! \! &=&   \! \! \! \frac{m^f_{i} \xi ^2 }{\mu ^4}~\delta_{ij}  +\bigg[ -\frac{3m_\varphi^2 m^f_{i}}{v^3}
%     +\bigg( \frac{2 m^f_{i} \zeta \xi ^3 }{v\mu ^4} 
%    -\frac{3m^f_{i} \lambda'\xi ^2}{v\mu ^2}
%    -\frac{6 m^f_{i} \xi ^4 }{v\mu ^4} 
%    \nonumber \\
%     & &  \! \! \!   +~ \frac{6  m^f_{i} \lambda  \xi ^2 }{v\mu ^2} \bigg)\frac{v^2}{\mu^2}
%      \bigg]   
%      \bigg( ~ \!  \frac{v}{s \! - \! m_{\varphi}^2}
%          \bigg) ~ \delta_{ij}
% \end{eqnarray}
 \begin{eqnarray}
{ \cal A} \left(\varphi \varphi \rightarrow f_i \bar{f}_j \right) 
  \! \! \! &=&   \! \! \!  \frac{m_i^f~ \delta_{ij}}{v^2}\Bigg[\bigg( -\frac{3m_{\varphi}^2 }{v^2}
     +\bigg[ \frac{2 \zeta \xi ^3 }{\mu ^4} 
    -\frac{3 \lambda'\xi ^2}{\mu ^2}
    -\frac{6\xi ^4 }{\mu ^4} 
    \nonumber \\
     & &  \! \! \!   +~ \frac{6 \lambda  \xi ^2 }{\mu ^2} \bigg]\frac{v^2}{\mu^2}
      \bigg)
      \bigg( ~ \!  \frac{v^2}{s \! - \! m_{\varphi}^2}
          \bigg) 
          + \frac{\xi ^2 v^2}{\mu ^4} + { \cal O}
               \bigg( {v^4\over \mu^4},\frac{m_i^{f\,2}}{v^2} \bigg)
 \Bigg]
 \label{eq:dihiggsdifermionamplitudeEFT}
 \end{eqnarray}
which coincides with the result obtained using the mixing language \eqref{eq:dihiggsdifermionamplitudemixing}.

The dihiggs to dihiggs amplitude is
\begin{eqnarray}
{ \cal A} \left(\varphi \varphi\rightarrow \varphi\varphi\right) 
  \! \! \! &=&   \! \! \! g_{\varphi^4}-g^2_{\varphi^3}\left(\frac{1}{s-m_{\varphi}^2}+\frac{1}{t-m_{\varphi}^2}+\frac{1}{u-m_{\varphi}^2}\right)
  \label{eq:exactdihiggsdihiggsamplitudeEFT}
 \end{eqnarray}
 This amplitude calculated in the mixing language \eqref{eq:exactdihiggsdihiggsamplitudemixing} also contains a diagram mediated by the heavy mass eigenstates, which in the EFT are already included in the low energy effective couplings. Using the couplings of the effective theory \eqref{eq:selfcouplings1} in \eqref{eq:exactdihiggsdihiggsamplitudeEFT} and expanding in $v^2/\mu^2$ we get
\begin{eqnarray}
{ \cal A} (\varphi \varphi \rightarrow \varphi\varphi ) 
  \! \! \! &=&   \! \! \! -\frac{3m_{\varphi}^2}{v^2}
  +\bigg(\frac{12 \zeta  \xi ^3}{\mu ^4}
  +\frac{25 \lambda  \xi ^2}{\mu ^2}
  -\frac{18 \lambda' \xi ^2}{\mu ^2}
  -\frac{25\xi ^4}{\mu ^4}\bigg)\frac{v^2 }{\mu^2 } \nonumber \\
  & &  \! \! \!       
  -~ \Bigg[{ 9 m_{\varphi}^4 \over v^4} 
       +\bigg(-\frac{12 \zeta  \lambda  \xi ^3}{\mu ^4}
       +\frac{12 \zeta  \xi ^5}{\mu ^6}  
       -\frac{27 \lambda ^2 \xi ^2}{\mu ^2} 
        + \frac{18 \lambda  \lambda' \xi ^2}{\mu ^2} \nonumber \\
     & &  \! \! \!   +~ \!  \frac{54 \lambda  \xi ^4}{\mu ^4}
     -\frac{18 \lambda' \xi ^4}{\mu ^4}
     -\frac{27 \xi ^6}{\mu ^6} \bigg) \frac{v^2}{\mu^2}\Bigg]  
      \bigg( ~ \!  \frac{v^2}{s \! - \! m_{\varphi}^2}
          \! +\! \frac{v^2}{t \! - \! m_{\varphi}^2}
          \! +\! \frac{v^2}{u \! - \! m_{\varphi}^2}
          \bigg) \nonumber \\
  &&\! \! \!  
    + ~ \! { \cal O}
               \bigg(  {v^4\over \mu^4} \bigg)
%  +~ {\cal O}\left(\frac{s^2}{m_{\varphi_2}^4},\frac{t^2}{m_{\varphi_2}^4},
  %   \frac{u^2}{m_{\varphi_2}^4},\frac{v^4}{\mu^4}\right) \nonumber \\
  \label{eq:amplitude2}
\end{eqnarray}
Note that the equivalence between this result and the corresponding amplitude calculated in the mixing language \eqref{eq:onshellamplitude} is only on-shell, since we used equations of motion to write our EFT in its final form. 

\section{The two Higgs doublet model}
\label{sec:2HDM}
%TODO: Make sure what is the bar over fermions in Haber's notation, I think it's just a dagger. Check if relation between WF renormalization and mixing angle is exact, and modify expressions accordingly (remove {\cal O}()). 
%\cite{Davidson:2005cw,Haber:2006ue}. 
%In this work we are interested in studying the 2HDM near the decoupling limit, in which the mass of one of the doublets is much larger than the electroweak scale. In this section, however, we present the model in full generality. Whenever convenient and in addition to the general results, we present convenient expansions valid near the decoupling limit. 
The 2HDM is a theory with a Higgs sector composed by two identical complex scalar fields $\Phi_a$, $a=1,2$, charged under $SU(2)_L \times U(1)_Y $, with hypercharge one. 
The two Higgs doublets span a two dimensional complex plane. A $U(2)$ rotation in this complex plane does not affect the canonically normalized kinetic terms and leads to a different choice of Higgs fields, which we call the choice of basis. The ungauged $SU(2)=U(2)/U(1)_Y$ subgroup of the $U(2)$ is the full background symmetry of the model \cite{Davidson:2005cw,Haber:2006ue}. The Lagrangian density of the Higgs doublets in a generic basis is
\begin{equation}
%D_\mu \Phi_1^\dagger D^\mu \Phi_1 + D_\mu \Phi_2^\dagger D^\mu \Phi_2 
% \sum_{a=1}^2 
D_\mu \Phi_a^\dagger D^\mu \Phi_a
  - V(\Phi_1, \Phi_2) 
   - 
   %\sum_{a=1}^2~
   \bigg[ ~
     \lambda^u_{aij} ~ Q_i \Phi_a \bar{u}_j
   - \lambda^{d\dagger}_{aij} Q_i \Phi_a^c \bar{d}_j 
   - \lambda^{\ell\dagger}_{aij} L_i \Phi_a^c \bar{\ell}_j 
  +{\rm h.c.}~ \!  \bigg]
  \label{eq:2HDMactiongeneric}
\end{equation}
where we sum over repeated indices and the covariant derivative acting on the doublets is
\begin{equation}
D_\mu \Phi_{1,2}=\bigg[\partial_\mu-i\bigg(g_2W_{a\mu}T_a+\frac{1}{2}g_1 B_{\mu}\bigg)\bigg]H_{1,2}
\end{equation}
The most general potential at the renormalizable level is given by
\begin{eqnarray}
V( \Phi_1 , \Phi_2) &=&m_{1}^2 \Phi_1^\dagger \Phi_1+m_{2}^2 \Phi_2^\dagger \Phi_2
 -\Big(   m_{12}^2 \Phi_1^\dagger \Phi_2 +{\rm h.c.} \Big) 
  + \frac{1}{2}\lambda_1 ( \Phi_1^\dagger  \Phi_1)^2
  \nonumber 
  \nonumber \\
&+&\frac{1}{2}\lambda_2 ( \Phi_2^\dagger  \Phi_2)^2+\lambda_3 ( \Phi_1^\dagger  \Phi_1)(\Phi_2^\dagger  \Phi_2)+\lambda_4 ( \Phi_1^\dagger  \Phi_2)(\Phi_2^\dagger  \Phi_1)
  \nonumber
   \nonumber \\
&+& \bigg[ ~
      \frac{1}{2}\lambda_5(\Phi_1^\dagger \Phi_2)^2
      +\lambda_6(\Phi_1^\dagger \Phi_1)(\Phi_1^\dagger \Phi_2)
      +\lambda_7(\Phi_2^\dagger \Phi_2)(\Phi_1^\dagger \Phi_2)+{\rm h.c.} 
      ~\bigg]
      \label{eq:2HDMpotentialgeneric}
\end{eqnarray}
The parameters $m_1^2 , m_2^2 , \lambda_1 , \lambda_2 , \lambda_3 , \lambda_4$ are 
real, while 
$m_{12}^2 , \lambda_5, \lambda_6, \lambda_7$ are in general complex.

%$$
%\begin{array}{cc} 
%%  & U(1)_{{1 \over 2}(Y + PQ)}  \nonumber \\
% & U(1)_{\textrm{PQ}}   \\
%   &     \\
% \Phi_1 & +1~~\! ~    \\
% \Phi_2 & -1~~\! ~    \\
% m_1^2 , m_2^2 , \lambda_1, \lambda_2, \lambda_3, \lambda_4 & 0   \\
% m_{12}^2, \lambda_6, \lambda_7 & + 2 ~~\! ~   \\
% \lambda_5 &  +4~~\!~   \\
% e^{i \xi} & -2 ~~ \! ~   \\
% \lambda^f_{1ij} & -1 ~~ \! ~   \\ 
%  \lambda^f_{2ij} & 1 ~~ \! ~   \\ 
%\end{array}
%$$

The gauge invariant combinations of the fields that characterize the Higgs condensate 
with $SU(2)_L \times U(1)_Y \to U(1)_Q$ are 
\begin{equation}
{v_1^2 \over 2} = \langle \Phi_1^\dagger \Phi_1 \rangle  ~~~~~~ 
{v_2^2 \over 2} = \langle \Phi_2^\dagger \Phi_2 \rangle ~~~~~~
v_1^2+v_2^2=v^2 
\label{eq:defvev}
\end{equation}
$$
{v_{12}^2 \over 2} = \langle \Phi_1^\dagger \Phi_2 \rangle 
$$
while the gauge invariant combination that measures charge breaking is 
$$
{v_c^2 \over 2}= \langle \Phi_1^c \Phi_2 \rangle 
$$
$v_1$ and $v_2$ are real, while $v_{12}$ and $v_c$ are in general complex. The Higgs vacuum expectation value takes the value $v=246~\textrm{GeV}$. We also define the ratio of the vacuum expectation values $\tan\beta$ and the relative phase of the condensate $\xi$
\begin{equation}
\tan \beta = {v_2 \over v_1} \quad \quad \quad \quad \xi = {\rm Arg} \langle \Phi_1^\dagger \Phi_2 \rangle
\label{eq:deftanb}
\end{equation}
The Higgs condensate gives mass to the gauge bosons corresponding to the broken gauge symmetries. The $W$ and $Z$ boson masses are
\begin{equation}
m_W=\frac{g_2 v}{2} ~~~~ ~~~~ m_Z=\frac{m_W}{\cos~\! \theta_W} ~~~~ ~~~~ \tan~\! \theta_W=\frac{g_1}{g_2}  
\end{equation}
%{\color{gray} (***\textbf{we do not make use of this, should I drop it?} The potential for the Higgs condensate with breaking pattern XXX in terms gauge invariant quantities is 
%\begin{eqnarray}
%V(v_1 , v_2, \xi) \! \! \! &=& \! \! \! {1 \over 2} m_1^2 v_1^2
%+ {1 \over 2} m_2^2 v_2^2
%+ {1 \over 8} \lambda_1 v_1^4 
%+ {1 \over 8} \lambda_2 v_2^4 
%+ {1 \over 4} \big( \lambda_3+ \lambda_4) v_1^2 v_2^2 
% \nonumber \\
%&&+~  \! {\rm Re} \Big( 
% - v_1 v_2 m_{12}^2 e^{i \xi} + 
% {1 \over 4} v_1^2 v_2^2 \lambda_5 e^{2i \xi} 
% +  {1 \over 2} v_1^3 v_2 \lambda_6 e^{i \xi} 
%  +  {1 \over 2} v_1 v_2^3 \lambda_7 e^{i \xi} \Big)
%\end{eqnarray}
%Spontaneous T violation occurs when $\xi\neq 0, \pi/2, \pi$ at the potential minimum. This happens when 
%
%\begin{equation}
%0\neq \abs{m_{12}^2-\frac{1}{2}\lambda_6v_1^2-\frac{1}{2}\lambda_7 v_2^2} \leq \lambda_5 v_1 v_2  ~~~~ \textrm{and} ~~~~  \lambda_5 >0
%\end{equation}
%*****)}
The magnitudes of the Higgs condensates $v_1,v_2$ break the $SU(2)$ background symmetry down to a $U(1)_{\textrm{PQ}}$ subset. The $U(1)_{\textrm{PQ}}$ charges of fields and couplings are specified in table \ref{tab:U1PQ1}. 
\begin{table}[h]
\begin{center}
$$
\begin{array}{cc} 
%  & U(1)_{{1 \over 2}(Y + PQ)}  \nonumber \\
\hline
 & U(1)_{\textrm{PQ}}   \\
\hline
 \Phi_1 & +1~~\! ~    \\
 \Phi_2 & -1~~\! ~    \\
 m_1^2 , m_2^2 , \lambda_1, \lambda_2, \lambda_3, \lambda_4 & 0   \\
 m_{12}^2, \lambda_6, \lambda_7 & + 2 ~~\! ~   \\
 \lambda_5 &  +4~~\!~   \\
v_1,v_2 & 0 ~~ \! ~   \\
 e^{i \xi} & -2 ~~ \! ~   \\
 \lambda^f_{1ij} & -1 ~~ \! ~   \\ 
  \lambda^f_{2ij} & +1 ~~ \! ~   \\ 
  \hline
\end{array}
$$
\end{center}
\caption{$U(1)_{\textrm{PQ}}$ charges. $v_1,v_2$ and $\xi$ specify the Higgs condensate as defined in \eqref{eq:defvev} and \eqref{eq:deftanb}. The $U(1)_{\textrm{PQ}}$ is the part of the background symmetry that is left unbroken by $v_1$ and $v_2$. The label $f$ in the Yukawa couplings corresponds to $u,d$ or $\ell$. Fermions and gauge bosons are $U(1)_{\textrm{PQ}}$ invariant fields.}
\label{tab:U1PQ1}
\end{table}

%{\color{gray}(**** \textbf{we will never make use of these invariants in this basis, we will only make use of the invariants in the Higgs basis. Should I drop them?} There are three complex combinations of potential parameters which are invariant 
%under the background Peccei-Quinn symmetry, which may be taken to be 
%$$
%( m_{12}^2)^2 \lambda_5^* ~~~, ~~~
%m_{12}^2\lambda_6^*~~~, ~~~
%m_{12}^2\lambda_7^*
%$$
%The imaginary components of these combinations of parameters are odd under 
%the anti-unitary time reversal symmetry. ***)}
%The most general two Higgs doublet model contains no feature that allows us to distinguish between the two doublets, either in the Higgs potential or Higgs fermionic interactions. As such, $\tan\beta$ is a parameter that contains no physical information. 
In a general 2HDM with the EWSB pattern \eqref{eq:defvev}, the only feature that allows us to distinguish between the two doublets is the direction of the vev in the space of the neutral components of the doublets. This motivates the introduction of the \textit{Higgs basis}, in which the vev is contained exclusively in one of the doublets. The Higgs basis is obtained by a rotation in the doublet space into a new set of doublets $H_a$, $a=1,2$
\begin{eqnarray}
%e^{-i\xi /2 }
e^{-i \xi /2} 
H_1 &=& \cos  \beta ~ \Phi_1+\sin \beta ~e^{-i\xi} ~\Phi_2 
  \nonumber \\
H_2 &=& -\sin \beta ~e^{i\xi} ~ \Phi_1+\cos \beta ~\Phi_2
\label{eq:rotationtoHiggsbasis}
\end{eqnarray}
such that in the new basis, only one of the doublets is responsible for EWSB
\begin{equation}
{v^2 \over 2} = \langle H_1^\dagger H_1 \rangle  ~~~~~~
0 = \langle H_2^\dagger H_2 \rangle ~~~~~~
\label{eq:HiggscondensatesHiggsbasis}
\end{equation}
Note that the Higgs basis is not unique. A $U(1)_{\textrm{PQ}}$ transformation leaves the magnitudes of the Higgs condensates invariant. As a consequence, the condition \eqref{eq:HiggscondensatesHiggsbasis} is $U(1)_{\textrm{PQ}}$ invariant, and different Higgs basis can be obtained by performing a $U(1)_{\textrm{PQ}}$ transformation on the doublets $H_a$. 

We now describe the 2HDM in the Higgs basis. The Lagrangian density in the Higgs basis is
\begin{equation}
%D_\mu \Phi_1^\dagger D^\mu \Phi_1 + D_\mu \Phi_2^\dagger D^\mu \Phi_2 
% \sum_{a=1}^2 
D_\mu H_a^\dagger D^\mu H_a
  - V(H_1, H_2) 
   - 
   %\sum_{a=1}^2~
   \bigg[ ~
     \tilde{\lambda}^u_{aij} ~ Q_i H_a \bar{u}_j
   - \tilde{\lambda}^{d\dagger}_{aij} Q_i H_a^c \bar{d}_j 
   - \tilde{\lambda}^{\ell\dagger}_{aij} L_i H_a^c \bar{\ell}_j 
  +{\rm h.c.}~ \!  \bigg]
  \label{eq:actionHiggsbasis}
\end{equation}
Since only $H_1$ carries a vev, the quark mass matrices are given by
%\begin{equation}
%M_F=\frac{\kappa_F v}{\sqrt{2}} \quad , \quad F=U,D,L
%\label{eq:kUkD}
%\end{equation}
\begin{equation}
\tilde{\lambda}^f_{1ij} = \frac{\sqrt{2}m^f_{ij} }{v}
\label{eq:kUkD}
\end{equation}
where the mass matrices for the quarks and leptons are defined in the Lagrangian as
\begin{equation}
-m^f_{ij} u_i \bar{u}_j-m^{d \dagger}_{ij} d_i \bar{d}_j-m^{\ell \dagger}_{ij} \ell_i \bar{\ell}_j   + \textrm{h.c.}
\end{equation}
In the above definition and for the rest of this paper, note that there is a dagger in the definition of the down type mass matrices with respect to the up type mass matrix.

The potential in \eqref{eq:actionHiggsbasis} is given by
\begin{eqnarray}
 V(H_1,H_2) &=&{\tilde{m}_1}^2  H_1^\dagger  H_1+{\tilde{m}_2}^2  H_2^\dagger  H_2+\Big({\tilde{m}_{12}}^2  H_1^\dagger  H_2 +\textrm{h.c.}\Big) \nonumber  \nonumber \\
&+&\frac{1}{2} {\tilde{\lambda}}_1 ( H_1^\dagger  H_1)^2+ \frac{1}{2} {\tilde{\lambda}}_2( H_2^\dagger  H_2)^2+ {\tilde{\lambda}}_3( H_2^\dagger  H_2)( H_1^\dagger  H_1)+ {\tilde{\lambda}}_4 ( H_2^\dagger  H_1)( H_1^\dagger  H_2) \nonumber  \nonumber \\
&+&\bigg[ ~ \frac{1}{2}{\tilde{\lambda}}_5( H_1^\dagger  H_2)^2+{\tilde{\lambda}}_6  H_1^\dagger  H_1  H_1^\dagger  H_2 +{\tilde{\lambda}}_7( H_2^\dagger  H_2)( H_1^\dagger  H_2)+\textrm{h.c.}~\bigg]
\label{eq:2HDMLagrangian}
\end{eqnarray}
The parameters $\tilde{m}_1^2 , \tilde{m}_2^2 , \tilde{\lambda}_1 , \tilde{\lambda}_2 , \tilde{\lambda}_3 , \tilde{\lambda}_4$ are 
real, while $\tilde{m}_{12}^2 , \tilde{\lambda}_5, \tilde{\lambda}_6, \tilde{\lambda}_7$ are in general complex. The relations between the couplings in the Higgs basis and the generic basis are \cite{Davidson:2005cw}
\begin{eqnarray}
\tilde{\lambda}^f_{1ij} \! \! \! &=& \! \! \! e^{-i\xi /2  } \lambda_{1ij}^f  \cos   \beta 
			+ e^{i\xi /2 } \lambda_{2ij}^f \sin   \beta  \notag \\ 
\tilde{\lambda}^f_{2ij} \! \! \!   &=& \! \! \! -e^{-i \xi} \lambda_{1ij}^f  \sin   \beta 
			+ \lambda_{2ij}^f \cos   \beta  
\label{eq:relationYukawas}
\end{eqnarray}
\begin{eqnarray}
 {\tilde{m}_1}^2 &=& m^2_{1} c^2_\beta +m^2_{2} s^2_\beta -\textrm{Re}(m^2_{12}e^{i\xi})s_{2\beta} \nonumber  \nonumber \\
{\tilde{m}_2}^2 &=& m^2_{1} s^2_\beta +m^2_{2} c^2_\beta +\textrm{Re}(m^2_{12}e^{i\xi})s_{2\beta} \nonumber  \nonumber \\
 {\tilde{m}_{12}}^2e^{i\xi/2}&=&-\frac{1}{2}(m^2_{1}-m^2_{2}) s_{2\beta} -\textrm{Re}(m_{12}^2e^{i\xi})c_{2\beta} -i\textrm{Im}(m^2_{12}e^{i\xi})
 \label{eq:relationmass}
\end{eqnarray}
\begin{eqnarray}
 {\tilde{\lambda}}_1&=&\lambda_1 c^4_\beta+\lambda_2 s^4_\beta+\frac{1}{2}\lambda_{345}s^2_{2\beta}+2s_{2\beta}\left(c^2_\beta\textrm{Re}(\lambda_6 e^{i\xi})+s^2_\beta \textrm{Re}(\lambda_7 e^{i\xi})  \right)  \nonumber \\
 {\tilde{\lambda}}_2&=&\lambda_1 s^4_\beta+\lambda_2 c^4_\beta+\frac{1}{2}\lambda_{345}s^2_{2\beta}-2s_{2\beta}\left(s^2_\beta\textrm{Re}(\lambda_6 e^{i\xi})+c^2_\beta \textrm{Re}(\lambda_7 e^{i\xi})  \right)  \nonumber \\
 {\tilde{\lambda}}_3&=&\frac{1}{4}s^2_{2\beta}(\lambda_1+\lambda_2-2\lambda_{345})+\lambda_3-s_{2\beta}c_{2\beta}\textrm{Re}((\lambda_6-\lambda_7)e^{i\xi}) \nonumber \\
 {\tilde{\lambda}}_4&=&\frac{1}{4}s^2_{2\beta}(\lambda_1+\lambda_2-2\lambda_{345})+\lambda_4-s_{2\beta}c_{2\beta}\textrm{Re}((\lambda_6-\lambda_7)e^{i\xi}) \nonumber  \nonumber \\
 {\tilde{\lambda}}_5 e^{i\xi}&=&\frac{1}{4}s^2_{2\beta}(\lambda_1+\lambda_2-2\lambda_{345})+\textrm{Re}(\lambda_5 e^{2i\xi})+ic_{2\beta} \textrm{Im}(\lambda_5 e^{2i\xi}) \nonumber \\
 &-&s_{2\beta}c_{2\beta}\textrm{Re}\left((\lambda_6-\lambda_7)e^{i\xi})\right)-is_{2\beta}\textrm{Im}\left((\lambda_6-\lambda_7)e^{i\xi})\right)\nonumber  \nonumber \\
 {\tilde{\lambda}}_6 e^{i\xi/2}&=&-\frac{1}{2}s_{2\beta}\left(\lambda_1 c_{\beta}^2-\lambda_2 s_{\beta}^2-\lambda_{345}c_{2\beta}-i\textrm{Im}(\lambda_5 e^{2i\xi})\right)\nonumber \\
 &+&c_\beta c_{3\beta} \textrm{Re}(\lambda_6 e^{i\xi})+s_\beta s_{3\beta}\textrm{Re}(\lambda_7 e^{i\xi})+i c^2_\beta\textrm{Im}(\lambda_6 e^{i\xi})+i s^2_\beta\textrm{Im}(\lambda_7 e^{i\xi})\nonumber \\
 {\tilde{\lambda}}_7 e^{i\xi /2 }&=&-\frac{1}{2}s_{2\beta}\left(\lambda_1 s_{\beta}^2-\lambda_2 c_{\beta}^2+\lambda_{345}c_{2\beta}+i\textrm{Im}(\lambda_5 e^{2i\xi})\right)\nonumber \\
 &+&s_\beta s_{3\beta} \textrm{Re}(\lambda_6 e^{i\xi})+c_\beta c_{3\beta}\textrm{Re}(\lambda_7 e^{i\xi})+i s^2_\beta\textrm{Im}(\lambda_6 e^{i\xi})+i c^2_\beta\textrm{Im}(\lambda_7 e^{i\xi})
 \label{eq:relationcouplings}
\end{eqnarray}
where $c_{n \beta} \equiv \cos n \beta$ and 
$s_{n \beta} \equiv \sin n \beta$ and we defined
\begin{equation}
\lambda_{345} \equiv \lambda_3+\lambda_4+\textrm{Re}(\lambda_5 e^{2i\xi})
\end{equation} 
In a general 2HDM, $\tan\beta$ is only required to map between couplings in the original generic basis and the Higgs basis, but it is irrelevant for any calculation, since the generic basis is arbitrary. In other words, $\tan\beta$ is a parameter that contains no physical information in a general 2HDM, since there is no feature in the model that allows us to distinguish the direction of the doublets $\Phi_1,\Phi_2$ as special directions in field space. For this reason, we will make no further reference to $\tan\beta$ until section \ref{sec:analysis}, where we study particular cases of 2HDM with features that allow us to specify a preferred basis which is different from the Higgs basis. The fields and couplings in the Higgs basis are charged under a $U(1)_{\textrm{PQ}}$ background symmetry as specified in table \ref{tab:U1PQ2}. Note that all the relations \eqref{eq:relationYukawas}, \eqref{eq:relationmass} and \eqref{eq:relationcouplings} are covariant under the $U(1)_{\textrm{PQ}}$. 

The decoupling limit in the Higgs basis can be very simple defined as the limit in which 
\begin{equation}
{\tilde{m}_2}^2  \gg  \abs{\tilde{\lambda}_i}v^2
\label{eq:decouplinglimitdef}
\end{equation}
where $i=1..7$. This definition is motivated by the fact that in the decoupling limit, the Higgs aligns with the vacuum expectation value which is contained entirely in $H_1$, so $H_2$ must be the decoupled doublet. Near the decoupling limit alignment is not exact, and there are corrections that can be understood in two ways. In the mixing language, corrections arise due to mixing between the neutral components of $H_1$ and $H_2$, which is suppressed by powers of the electroweak scale over the heavy mass of $H_2$, as we will see in section \ref{sec:2HDMrotation}. In the effective field theory language, corrections arise in the form of higher dimensional operators that modify the SM Lagrangian, which are induced when $H_2$ is integrated out, as we will see in section \ref{sec:2HDMEFT}.
\begin{table}[h]
\begin{center}
$$
\begin{array}{cc}
\hline 
%  & U(1)_{{1 \over 2}(Y + PQ)}  \nonumber \\
 & U(1)_{\textrm{PQ}}   \\
 \hline
 H_1 & 0~~\! ~    \\
 H_2 & -1~~\! ~    \\
 {\tilde{m}_1}^2 , {\tilde{m}_2}^2 , \tilde{\lambda}_1, {\tilde{\lambda}}_2, {\tilde{\lambda}}_3, {\tilde{\lambda}}_4 & 0   \\
{\tilde{m}_{12}}^2, {\tilde{\lambda}}_6, {\tilde{\lambda}}_7 & + 1 ~~\! ~   \\
{\tilde{\lambda}}_5 &  +2~~\!~   \\
v & 0 ~~ \! ~   \\
\tilde{\lambda}^f_{1ij} &  0~~\!~   \\
\tilde{\lambda}^f_{2ij} &  +1~~\!~   \\
\hline
\end{array}
$$
\end{center}
\caption{$U(1)_{\textrm{PQ}}$ charges in the Higgs basis. The label $f$ in the Yukawa couplings corresponds to $u,d$ or $\ell$. Fermions and gauge bosons are $U(1)_{\textrm{PQ}}$ invariant fields.}
\label{tab:U1PQ2}
\end{table}

% In these types of 2HDM the relations \eqref{eq:relationmass} and \eqref{eq:relationcouplings} between the couplings in the Higgs basis and a different preferred basis are useful, because constraints on the couplings in the preferred basis are imposed.

The EWSB conditions in the Higgs basis are simplified by the fact that only one of the doublets contains a vev. They are given by
\begin{eqnarray}
\frac{\partial V}{\partial H_1}\Big |_{H_1=v/\sqrt{2},H_2=0}&\!\!\!=\!\!\!&\frac{1}{\sqrt{2}}\tilde{m}_1^2 v+\frac{1}{2\sqrt{2}} \tilde{\lambda}_1 v^3=0
\nonumber \\
\frac{\partial V}{\partial H_2}\Big |_{H_1=v/\sqrt{2},H_2=0}&\!\!\!=\!\!\!&\frac{1}{\sqrt{2}}\tilde{m}_{12}^2 v+\frac{1}{2\sqrt{2}}\tilde{\lambda}_6 v^3=0
\end{eqnarray}
These conditions can be rewritten as
\begin{equation}
v^2=-2\frac{{\tilde{m}_1}^2}{ {\tilde{\lambda}}_1}
\label{eq:EWSBscale}
\end{equation}
\begin{equation}
{\tilde{m}_{12}}^2=-\frac{1}{2}  {\tilde{\lambda}}_6  v^2
\label{eq:notadpolecondition}
\end{equation}
where \eqref{eq:EWSBscale} involves only real parameters and \eqref{eq:notadpolecondition} is a complex equation. The expressions \eqref{eq:EWSBscale} and \eqref{eq:notadpolecondition} relate $\tilde{m}_1$ and $\tilde{m}_{12}$ with the electroweak scale. The relation \eqref{eq:notadpolecondition} can be interpreted as a no tadpole condition for the $H_2$ doublet. 

The Higgs potential \eqref{eq:2HDMLagrangian} contains fourteen parameters. However, not all of these parameters are physical or independent of each other.  Only thirteen of the parameters are invariants under the background $U(1)_{\textrm{PQ}}$, namely the six real couplings, the four magnitudes of the complex couplings and $ {\rm Arg}({\tilde{m}}^4_{12}{\tilde{\lambda}}^*_5), {\rm Arg}({\tilde{m}_{12}}^2{\tilde{\lambda}}^*_6),{\rm Arg}({\tilde{m}_{12}}^2{\tilde{\lambda}}^*_7)$. On the other hand, the EWSB relations \eqref{eq:EWSBscale} and \eqref{eq:notadpolecondition} impose three conditions on the couplings, so the final number of physical and independent invariants in the potential is eleven. One of the invariants can be chosen to be the Higgs vev. Some of the parameters specifying the Higgs potential and Yukawa interactions in a general 2HDM are complex, and allow for T violation. There are two independent physical T violating phases in the bosonic sector in the case in which $\tilde{\lambda}_5 \neq 0$  \cite{Haber:2006ue}. We take them to be the  $U(1)_{\textrm{PQ}}$ invariant combinations
\begin{eqnarray}
\theta_1
&\!\!\!=\!\!\!&
\frac{1}{2}
{\rm Arg} \big({\tilde{\lambda}}_6^2 {\tilde{\lambda}}_5^* \big)=\frac{1}{2}{\rm Arg} \big(\tilde{m}_{12}^4 {\tilde{\lambda}}_5^* \big) 
\nonumber \\
 \theta_2
 &\!\!\!=\!\!\!&
 \frac{1}{2}
 {\rm Arg} \big({\tilde{\lambda}}_7^2 {\tilde{\lambda}}_5^* \big)
\label{eq:cpviolatingphases}
\end{eqnarray}
where in the first line we made use of \eqref{eq:notadpolecondition}. In the case in which $\tilde{\lambda}_5=0$, there is only one independent T violating phase given by $\theta_3={\rm Arg} \big({\tilde{\lambda}}_7 {\tilde{\lambda}}_6^* \big)$. When considering also fermionic interactions, there are several additional T violating phases. The $U(1)_{\textrm{PQ}}$ invariant phases are
\begin{equation}
{\tilde{\lambda}}_5^* (\tilde{\lambda}^f_{2ij})^2 ~~~~ ~~~~ {\tilde{\lambda}}_6^* \tilde{\lambda}^f_{2ij}~~~~ ~~~~ {\tilde{\lambda}}_7^* \tilde{\lambda}^f_{2ij}
\label{eq:cpviolatingphases2}
\end{equation}
All the $U(1)_{\textrm{PQ}}$ invariant phases in the most general 2HDM can be expressed in terms of the complete set of independent phases
\begin{equation}
\theta_1 \textrm{(or $\theta_3$ if $\tilde{\lambda}_5$ vanishes)}  ~~~~   ~~~~  \theta_2   ~~~~   ~~~~  \textrm{Arg} \big({\tilde{\lambda}}_6^* \tilde{\lambda}^f_{2ij}\big) ~~,~~ f=u,d,\ell
\label{eq:completesetCP}
\end{equation}
In this work we allow for T violation both in the bosonic sector and the fermionic interactions, and we will specify when we specialize to the T conserving case. In this paper we assume that there is a low energy solution for the strong CP problem and we do not worry about possible contributions from complex phases to $\theta_{QCD}$.

We now turn to the components of the Higgs doublets. The neutral and charged components of the Higgs basis doublets are
\begin{eqnarray} 
H_1^0 \! \! \! &=& \! \! \! {1 \over \sqrt{2}} \Big( v + h_1 + i G^0 \Big) 
\nonumber \\
H_2^0 \! \! \! &=&  \! \! \! {1 \over \sqrt{2}} 
    %e^{(-i/2) {\rm Arg}(\lambda^H_5) } 
    %e^{-i \theta_5 / 2} 
     ~ e^{ i~ \! \! {\rm Arg}({\tilde{\lambda}}_5^*) /2}    
          \Big( h_2 + i h_3 \Big)
  \nonumber
\\
H_1^+ \! \! \! &=&  \! \! \! G^+ 
\nonumber \\
H_2^+ \! \! \! &=&  \! \! \! 
    %e^{(-i/2) {\rm Arg}(\lambda^H_5) } 
    %e^{-i \theta_5 / 2} 
      e^{ i~ \! \! {\rm Arg}({\tilde{\lambda}}_5^*) /2}    
        ~ H^+ 
    \label{eq:Higgsbasis}
 \end{eqnarray}
where gauge invariance ensures $v\geq 0$. As advertised, the second doublet $H_2$ in the Higgs basis does not participate in EWSB. The phase $ e^{i ~ \! \! {\rm Arg}({\tilde{\lambda}}_5^*)/2}$ is added in the definition of the components of $H_2$ because it makes the mass matrix automatically block diagonal in the T conserving case, as we will see in section \ref{sec:2HDMrotation}. In this case the fields $h_1$ and $h_2$ do not mix with $h_3$, which will automatically be a T odd eigenstate. Note that the fields $h_1, h_2$ and $h_3$ are invariant under the $U(1)_{\textrm{PQ}}$ background symmetry. $G^0$, $G^+$ are Goldstone bosons. From now on we work in the unitary gauge.

\subsection{Mass eigenstates}
\label{sec:2HDMrotation}
In the unitary gauge, the two doublets contain four Higgs particles, three of which are neutral and one is charged. The  charged Higgs boson is a mass eigenstate with mass
\begin{equation}
m^2_{H\pm}={\tilde{m}_2}^2+\frac{1}{2}{\tilde{\lambda}}_3v^2
\end{equation}
The three neutral states mix in the mass matrix
\begin{equation}
V \supset \frac{1}{2}{\cal M}^2_{ij}h_i h_j
\end{equation}
where we sum over $i,j=1,2,3$. This mass matrix is $U(1)_{\textrm{PQ}}$ invariant, real and symmetric. It is specified by five real numbers and it is given by
\begin{equation}
{\cal M}^2=\left(\begin{array}{ccc} v^2 {\tilde{\lambda}}_1 &  v^2 \abs{{\tilde{\lambda}}_6} \cos \theta_1   & -v^2 \abs{{\tilde{\lambda}}_6}\sin \theta_1    \\  
v^2 \abs{{\tilde{\lambda}}_6} \cos\theta_1   &{\tilde{m}_2}^2+ \frac{1}{2}v^2 \Big( {\tilde{\lambda}}_3+ {\tilde{\lambda}}_4+ \abs{{\tilde{\lambda}}_5}\Big) & 0   \\
 -v^2 \abs{{\tilde{\lambda}}_6}\sin \theta_1 & 0 & {\tilde{m}_2}^2+\frac{1}{2}v^2\Big( {\tilde{\lambda}}_3+ {\tilde{\lambda}}_4-\abs{{\tilde{\lambda}}_5}\Big)
\end{array}\right)
\label{eq:massmatrixHiggsbasis}
\end{equation}
where $\theta_1$ is one of the bosonic CP violating phases defined in \eqref{eq:cpviolatingphases}. In this paper we identify the Higgs particle with the lightest of the three neutral mass eigenstates, which are labeled in increasing order by their mass as $(\varphi_1,\varphi_2,\varphi_3)$. We are interested in analyzing the decoupling limit \eqref{eq:decouplinglimitdef}, in which mixing in the mass matrix is suppressed.

The time reversal transformation properties of the mass eigenstates can be derived by inspecting their couplings to gauge bosons and fermions. In general we expect the mass eigenstates to be fields with indefinite T properties, due to the T violating phases \eqref{eq:cpviolatingphases} and \eqref{eq:cpviolatingphases2} of the 2HDM. Note that the T violating effect of $\theta_1$ is encoded in mixing of $h_1$ with $h_3$. On the other hand, the T violating phase $\theta_2$ does not appear in the mass matrix.  Note that the phase  $e^{i ~ \! \! {\rm Arg}({\tilde{\lambda}}_5^*)/2}$ in the definition \eqref{eq:Higgsbasis} guarantees that the mass matrix is automatically block diagonal in the T conserving case. 

The mass matrix has two zeroes, which have their origin in the underlying $U(1)_{\textrm{PQ}}$ background symmetry. To understand the zeroes, consider the limit in which we suppress the mixing between $h_1$ and $h_2$ by taking $\tilde{\lambda}_6$ and $\tilde{m}^2_{12}$ equal to zero. In this case, the only $U(1)_{\textrm{PQ}}$ breaking term that enters in the mass matrix is $\tilde{\lambda}_5$, which breaks the mass degeneracy of the two heavy eigenstates. The phase $e^{ i~ \! \! {\rm Arg}({\tilde{\lambda}}_5^*) /2}$ included in our choice of component fields $h_2,h_3$ in \eqref{eq:Higgsbasis}, makes sure that $h_2$ and $h_3$ do not mix in the mass matrix,  giving rise to the aforementioned zeroes. In this case, $h_2$ and $h_3$ are the heavy neutral mass eigenstates, with mass splitting controlled by $\abs{\tilde{\lambda}_5}$.

\subsubsection{The T conserving case}
Before analyzing the neutral mass matrix in generality, let us start by studying the mass eigenstates in the simpler, T conserving case. This is defined as the case in which all the T violating phases \eqref{eq:cpviolatingphases} and \eqref{eq:cpviolatingphases2} vanish \cite{Haber:2006ue}. In the case that $\tilde{\lambda}_5=0$, one must also check that ${\rm Arg} \big({\tilde{\lambda}}_7 {\tilde{\lambda}}_6^* \big)$ vanishes. In this section we can take all couplings to be real. In the T conserving case the mass matrix simplifies to a block diagonal form, mixing only the fields $h_1$ and $h_2$, which are now T even. The field $h_3$ is T odd and corresponds to the mass eigenstate $\varphi_2$ 
\begin{equation}
m_{\varphi_2}^2={\cal M}^2_{33}= {\tilde{m}_2}^2+\frac{1}{2}v^2\Big( {\tilde{\lambda}}_3+ {\tilde{\lambda}}_4-\abs{{\tilde{\lambda}}_5}\Big)
\end{equation}
The masses of the remaining neutral eigenstates are
\begin{eqnarray}
m^2_{\varphi_1,\varphi_3} = {1 \over 2} \bigg( 
  {\cal M}^2_{11} + {\cal M}^2_{22} \pm \sqrt{ ({\cal M}^2_{22} - {\cal M}^2_{11})^2 + 4{\cal M}^2_{12} }   
     \bigg) 
\end{eqnarray}
In terms of couplings in the potential, the Higgs mass near the decoupling limit is given by
\begin{eqnarray}
\label{eq:HiggsmassTconserving}
 m^2_{\varphi_1}   \! \! \! & =  & \! \! \!   
 v^2\Bigg[{\tilde{\lambda}}_1 
 -{\tilde{\lambda}}_6^2  ~ \! \frac{ v^2}{{\tilde{m}_2}^2}   
 - 
 \frac{1}{2} {\tilde{\lambda}}_6^2~\!  \big[ 2{\tilde{\lambda}}_1  
  -{\tilde{\lambda}}_3 -{\tilde{\lambda}}_4-{\tilde{\lambda}}_5
  \big]
  \frac{v^4}{{\tilde{m}_2}^4}
  +{\cal O}\bigg ( { v^6 \over {\tilde{m}_2}^6} \bigg) 
  \Bigg] 
\end{eqnarray}
For completeness, the mass of the remaining neutral eigenstate is
\begin{equation}
m_{\varphi_2}^2= {\tilde{m}_2}^2\Bigg[1+\frac{1}{2}\Big( {\tilde{\lambda}}_3+ {\tilde{\lambda}}_4+\abs{{\tilde{\lambda}}_5}\Big) {v^2 \over {\tilde{m}_2}^2}
+ 
{\tilde{\lambda}}_6^2 ~\!
 {v^4 \over {\tilde{m}_2}^4}
   +{\cal O}\bigg ( { v^6 \over {\tilde{m}_2}^6} \bigg) 
  \Bigg]
\end{equation}
Note that the splitting between the two heavy neutral eigenstates is 
\begin{equation}
m_{\varphi_3}^2-m_{\varphi_2}^2=v^2\Bigg[\abs{{\tilde{\lambda}}_5}
+{\tilde{\lambda}}_6^2 ~\! { v^2 \over {\tilde{m}_2}^2}
+{\cal O}\bigg({ v^4 \over {\tilde{m}_2}^4}\bigg) \Bigg]
\end{equation}

The mixing angle between the $h_1$ and $h_2$ is generally named in the literature $\beta-\alpha$, where $\alpha$ is the rotation angle that diagonalizes the T even part of the mass matrix in the generic basis \cite{Gunion:1989we}. The relation between the mass eigenstates and $h_1$ and $h_2$ is given by 
\begin{eqnarray}
{\def\arraystretch{1}\tabcolsep=10pt
\left( 
\begin{array}{c}  \varphi_1  \\ \varphi_3  \end{array} \right) = 
\left( \begin{array}{cc} \sin (\beta-\alpha) &  \cos (\beta-\alpha)    \\ 
                         - \cos (\beta-\alpha)   &  \sin (\beta-\alpha)
        \end{array}  
                   \right)
    \left( \begin{array}{c} h_1  \\ h_2  \end{array} \right)  }
        \label{eq:rotationTconserving}
        \end{eqnarray}
$\cos(\beta-\alpha)$ is called the alignment parameter \cite{Craig:2013hca}, and the alignment limit is defined as the limit in which $\cos(\beta-\alpha)=0$. In this limit, the Higgs lies along the direction of $h_1$, which by definition of the Higgs basis is the direction of the vev in field space. The alignment parameter controls the couplings of the Higgs to gauge bosons and fermions \cite{Gunion:1989we}. In the alignment limit the couplings of the Higgs are Standard Model like.  The alignment parameter is given by
%\begin{eqnarray}
%\tan 2  (\beta-\alpha) = { -2 {\cal M}^2_{12} \over {\cal M}^2_{22} - {\cal M}^2_{11}}   
%\end{eqnarray}
\begin{eqnarray}
\cos (\beta-\alpha) ={ -2 {\cal M}^2_{12} 
\over \sqrt{
4
\big(
{\cal M}^2_{12}
\big)^2
+\Big[
{\cal M}^2_{22}
-{\cal M}^2_{11}
+\sqrt{
\big({\cal M}^2_{22}-{\cal M}^2_{11}\big)^2
+4{\cal M}^2_{12}
}
\Big]^2
}   
}
\end{eqnarray}
which at first order in $v^2/\tilde{m}_2^2$ is
\begin{equation}
\cos (\beta-\alpha) =-\abs{{\tilde{\lambda}}_6} ~\! \frac{v^2}{{\tilde{m}_2}^2}+{\cal O}\bigg({v^4 \over {\tilde{m}_2}^4}\bigg)
\label{eq:cosbaeval}
\end{equation}

\subsubsection{The general T violating case}
\label{sec:generalTviolating}
%As usual, the way to proceed is to minimize the scalar potential, and solve for $m_{aa}$ and $m_{bb}$ in terms of the Higgs vevs. However, in the Higgs basis there is a small subtlety. When we solve for $m_{bb}$ we get, using $\bra  H_1 \ket =v$
%
%\begin{equation}
%{\tilde{m}_2}^2=-\frac{{\tilde{m}_{12}}^2 v +v^3 {\tilde{\lambda}}_6 +\bra  H_2 \ket (...) }{\bra  H_2 \ket}
%\end{equation}
%
%so we immediately see that when we use the no tadpole condition \eqref{eq:notadpolecondition2hdm} and $\bra  H_2 \ket=0$, we get an indeterminate expression. This happens since ${\tilde{m}_2}^2$ controls the decoupling of $ H_2$. Therefore, let us fix ${\tilde{m}_2}^2$ to a certain value, and only solve for ${\tilde{m}_1}^2$ in terms of $v$. This leads to the mass matrix

%\begin{equation}
%{\cal M}^4=\left(\begin{array}{cc} 4\frac{1}{2}\lambda v^2 & 2 {\tilde{\lambda}}_6 v^2  \nonumber \\2 {\tilde{\lambda}}_6 v^2 & {\tilde{m}_2}^2+v^2( {\tilde{\lambda}}_3+ {\tilde{\lambda}}_4+2 {\tilde{\lambda}}_5) \end{array}\right)
%\end{equation}
We now proceed to study the neutral mass eigenstates in the general T violating case, in which all three fields $h_1,h_2$ and $h_3$ mix in the mass matrix. Phases must now be included in the mixing discussion \cite{Asner:2013psa}. The characteristic equation of the mass matrix is 
\begin{equation}
  -\det({\cal M}^2-xI)= x^3 + a x^2
  +bx
  +c=0  
\end{equation}
with
\begin{eqnarray}
  a\! \! \! &=&\! \! \!   
  -2 {\tilde{m}_2}^2
  -v^2({\tilde{\lambda}}_1
  +{\tilde{\lambda}}_3
  +{\tilde{\lambda}}_4)
   \nonumber \\
  b\! \! \! &=&\! \! \!   
  {\tilde{m}_2}^4
  +v^2{\tilde{m}_2}^2( 2 {\tilde{\lambda}}_1
  +  {\tilde{\lambda}}_3
  + {\tilde{\lambda}}_4)
  +v^4\bigg[~\! {\tilde{\lambda}}_1( {\tilde{\lambda}}_3
  + {\tilde{\lambda}}_4) +\frac{1}{4}( {\tilde{\lambda}}_3+{\tilde{\lambda}}_4)^2 -\frac{1}{4}{\tilde{\lambda}}_5 {\tilde{\lambda}}_5^*-{\tilde{\lambda}}_6 {\tilde{\lambda}}_6^*  ~\! \bigg]
 \nonumber \\
  c\! \! \! &=&\! \! \!   
  -v^2 {\tilde{m}_2}^4 {\tilde{\lambda}}_1
  +v^4 {\tilde{m}_2}^2 \bigg[{\tilde{\lambda}}_6 {\tilde{\lambda}}_6^*
  -{\tilde{\lambda}}_1 ({\tilde{\lambda}}_3+ {\tilde{\lambda}}_4)
  \bigg]
   \nonumber \\
  &&-~
  \frac{1}{4}v^6 \bigg[~\!  {\tilde{\lambda}}_1( {\tilde{\lambda}}_3  +{\tilde{\lambda}}_4)^2 
  -{\tilde{\lambda}}_1 {\tilde{\lambda}}_5 {\tilde{\lambda}}_5^*
  -2{\tilde{\lambda}}_6 {\tilde{\lambda}}_6^*({\tilde{\lambda}}_3 +{\tilde{\lambda}}_4) 
  + ({\tilde{\lambda}}_5 {\tilde{\lambda}}_6^{* 2}+\textrm{h.c.}) ~\!\bigg]
\end{eqnarray}

The three solutions to the characteristic equation are the masses of the neutral Higgs particles, and they are given by
\begin{eqnarray}
  m_{\varphi_a}^2\! \! \! &=&\! \! \!  -\frac{1}{3 C}\bigg[~\! aC+\omega_aC^2+\frac{A}{\omega_a} ~\! \bigg]  \nonumber \\
  A\! \! \! &=&\! \! \!  a^2-3b  \nonumber \\
B\! \! \! &=&\! \! \!  2a^3 - 9ab + 27c \nonumber \\
  C\! \! \! &=&\! \! \!  \bigg[~\!  \frac{B}{2}+ \frac{1}{2}\sqrt{B^2 - 4A^3} ~\! \bigg]^{1/3}  \nonumber \\
  \omega_1\! \! \! &=&\! \! \!  1 \quad , \quad \omega_2=-\frac{1}{2}+i\frac{\sqrt{3}}{2} \quad , \quad \omega_3=\omega_2^*
  \label{eq:exactmasseigenstates}
\end{eqnarray}
We identify the smallest root of the characteristic polynomial with the Higgs mass. 

Near the decoupling limit, the mass eigenstates \eqref{eq:exactmasseigenstates} have simple expressions in terms of expansions in $v^2/\tilde{m}_2^2$. In this case, the Higgs mass is given by
\begin{eqnarray}
\label{eq:Higgsmass}
 m^2_{\varphi_1}   \! \! \! & =  & \! \! \!   
 v^2\Bigg[{\tilde{\lambda}}_1 
 -{\tilde{\lambda}}_6^* {\tilde{\lambda}}_6 ~ \! \frac{ v^2}{{\tilde{m}_2}^2}   
 - 
 \frac{1}{2}  \Big[ (2{\tilde{\lambda}}_1  
  -{\tilde{\lambda}}_3 -{\tilde{\lambda}}_4) {\tilde{\lambda}}_6^* {\tilde{\lambda}}_6
  -\frac{1}{2}( {\tilde{\lambda}}_5^*  {\tilde{\lambda}}_6^2+\textrm{h.c}) 
  \Big]
  \frac{v^4}{{\tilde{m}_2}^4}
  +{\cal O}\bigg ( { v^6 \over {\tilde{m}_2}^6} \bigg) 
  \Bigg] \nonumber \\
\end{eqnarray}
while the masses of the heavy eigenstates are
\begin{eqnarray}
m_{\varphi_2,\varphi_3}^2
={\tilde{m}_2}^2
\Bigg[1+ \frac{1}{2} \big({\tilde{\lambda}}_3+{\tilde{\lambda}}_4\pm\abs{{\tilde{\lambda}}_5}\big)~\! \frac{v^2}{{\tilde{m}_2}^2} 
+ \frac{1}{2}
{\tilde{\lambda}}_6^* {\tilde{\lambda}}_6
\big[1 \pm\cos~\!2\theta_1\big] {v^4 \over {\tilde{m}_2}^4}
+{\cal O}\bigg({v^6 \over {\tilde{m}_2}^6}\bigg)\Bigg]
\label{eq:massheavyeigenstates}
\end{eqnarray}
Note that the mass splitting between the heavy states is
\begin{equation}
m_{\varphi_3}^2-m_{\varphi_2}^2=v^2\Bigg[\abs{{\tilde{\lambda}}_5}
+{\tilde{\lambda}}_6^* {\tilde{\lambda}}_6 ~\! \cos ~\! 2\theta_1~\! { v^2 \over {\tilde{m}_2}^2}
+{\cal O}\bigg({ v^4 \over {\tilde{m}_2}^4}\bigg) \Bigg]
\end{equation}

The eigenvectors $V^a$ corresponding to the three mass eigenvalues $m^2_{\varphi_a}$ characterize the mixing and satisfy
\begin{equation}
{\cal M}^2_{ij} V^{a}_j=m^2_{\varphi_a}V^{a}_i
\end{equation}
where only a sum over $j$ is intended. The eigenvectors can be chosen to be real, and they are given by \cite{Kopp:2006wp}\footnote{This expression is not valid when the eigenvalues are not degenerate. In practice, we will use it only for the Higgs eigenstate, which in the decoupling limit is not degenerate. For the degenerate case, see \cite{Kopp:2006wp}.}
\begin{equation}
V_i^{a}= \varepsilon_{ijk} ({\cal M}^2_{Aj}-m_{\varphi_a}^2 \delta_{Aj})  ({\cal M}^2_{Bk} -m_{\varphi_a}^2\delta_{Bk}) 
%({\cal M}^2_1-m_i^2 \hat{e}_1) \times ({\cal M}^2_2 -m_i^2 \hat{e}_2) 
\label{eq:Vi}
\end{equation}
where we sum over $j,k$. $A$ and $B$ can take the values $1,2$ or $3$ subject to $A\neq B$. Different choices lead to different sign conventions for the eigenvectors. We commit to the choice $A=2, B=3$ for the rest of this paper. 
 The eigenvectors are invariant under the $U(1)_{\textrm{PQ}}$ background symmetry, so they are physical, measurable quantities. The normalized eigenvectors are defined as 
\begin{equation}
\widehat{V}_i^{a}=\frac{V_i^{a}}{\abs{V^a}}
\end{equation}
The normalized eigenvectors can be used to diagonalize the mass matrix
\begin{equation}
\widehat{V}_i^{ a}{\cal M}_{ij}^2 \widehat{V}_j^b = \delta_{ab} m_{\varphi_a}^2
\label{eq:rotationTviolating}
\end{equation}
Each normalized eigenvector has only two independent real components due to the normalization condition 
\begin{equation}
 \big(\widehat{V}_1^{a}\big)^2+\big(\widehat{V}_2^{a}\big)^2+\big(\widehat{V}_3^{a}\big)^2=1
 \label{eq:sumrule}
\end{equation}
Furthermore, the three eigenvectors are orthogonal to each other for non degenerate mass eigenvalues, so all together they can be specified by three real numbers. Now, using \eqref{eq:Vi}, the normalized eigenvectors are
\begin{eqnarray}
\widehat{V}_1^{a}
&\!\!\!=\!\!\!&
\frac{({\cal M}_{22}^2-m_{\varphi_a}^2) ({\cal M}_{33}^2-m_{\varphi_a}^2)}{\sqrt{({\cal M}_{33}^2-m_{\varphi_a}^2)^2 \left({\cal M}_{12}^4+({\cal M}_{22}^2-m_{\varphi_a}^2)^2\right)+{\cal M}_{13}^4 ({\cal M}_{22}^2-m_{\varphi_a}^2)^2}}
\nonumber \\
\widehat{V}_2^{a}
&\!\!\!=\!\!\!&
-\frac{{\cal M}_{12}^2 ({\cal M}_{33}^2-m_{\varphi_a}^2)}
{\sqrt{({\cal M}_{33}^2-m_{\varphi_a}^2)^2 \left({\cal M}_{12}^4+({\cal M}_{22}^2-m_{\varphi_a}^2)^2\right)+{\cal M}_{13}^4 ({\cal M}_{22}^2-m_{\varphi_a}^2)^2}}
 \nonumber\\
\widehat{V}_3^{a}
&\!\!\!=\!\!\!&
-\frac{{\cal M}_{13}^2 ({\cal M}_{22}^2-m_{\varphi_a}^2)}
{\sqrt{({\cal M}_{33}^2-m_{\varphi_a}^2)^2 \left({\cal M}_{12}^4+({\cal M}_{22}^2-m_{\varphi_a}^2)^2\right)+{\cal M}_{13}^4 ({\cal M}_{22}^2-m_{\varphi_a}^2)^2}}
\label{eq:exactalignmentparameter}
\end{eqnarray}
% Note that near the decoupling limit ${\cal M}_{22}^2$ ${\cal M}_{33}^2$ are of order  $\tilde{m}_2^2 \gg m_{\varphi_1}^2$, so $\widehat{V}_1^{1}>0$. In the general case
 $\widehat{V}_1^{1}$ can always be chosen to be positive by a sign redefinition of the eigenvector. The definition chosen in \eqref{eq:exactalignmentparameter} leads to $\widehat{V}_1^{1}>0$ near the decoupling limit, since in that case ${\cal M}_{22}^2$ and ${\cal M}_{33}^2$ are of order  $\tilde{m}_2^2 \gg m_{\varphi_1}^2$. Without loss of generality we can write the normalized eigenvector corresponding to the Higgs in terms of a \textit{complex alignment parameter} $\Xi$
\begin{eqnarray}
\widehat{V}^1_1=\sqrt{1-\abs{\Xi } ^2}  ~~~ ~~~ \widehat{V}^1_2=\textrm{Re}~\!\Xi  ~~~ ~~~ \widehat{V}^1_3=\textrm{Im}~\!\Xi  ~~~,  ~~~ \Xi  \in \mathbb{C}^1
% \left(\widehat{V}_h\right)_a=\delta_{1a}\sqrt{1-\abs{\Xi } ^2}+\delta_{2a}V_{\perp 1}+\delta_{3a}{\Xi  2} \nonumber \\
%\widehat{V}^1=\left(\begin{array}{c}\sqrt{1-\abs{\Xi } ^2}   \\ \textrm{Re}\Xi   \\  \textrm{Im}\Xi \end{array}\right)  \quad \quad , \quad \quad \Xi  \in \mathbb{C}^1
\label{eq:Vh}
\end{eqnarray}
$\Xi $ specifies the projections of the Higgs particle $\varphi_1$ along $h_1,h_2$ and $h_3$. Note that $\sqrt{1-\abs{\Xi } ^2}$ gives the component of the Higgs in the direction of $h_1$, which by definition in the Higgs basis is the direction of the vev in field space. $\Xi$ is the complex generalization of the alignment parameter $\cos(\beta-\alpha)$ for the general T violating 2HDM.  In the T conserving case, ${\rm Im} ~\! \Xi=0$ and $\Xi$ reduces to $\cos(\beta-\alpha)$.  We will see in the next section that $\Xi$ controls the deviations of the Higgs couplings to fermions and gauge bosons from their SM counterparts. In the alignment limit, $\Xi=0$, and the couplings of the Higgs are SM like. The complex alignment parameter is invariant under the $U(1)_{\textrm{PQ}}$ background symmetry and both its magnitude and its phase are physical, measurable quantities. 

Near the decoupling limit it is useful to find the complex alignment parameter as an expansion in $v^2/\tilde{m}_2^2$ in terms of parameters of the Higgs potential. The real and imaginary components of the complex alignment parameter can be obtained using 
\eqref{eq:exactalignmentparameter}. The calculation is straightforward, using the elements of the mass matrix \eqref{eq:massmatrixHiggsbasis} and the expression for the Higgs mass \eqref{eq:Higgsmass} we get
\begin{equation}
% \Xi =-\abs{{\tilde{\lambda}}_6} e^{-i ~ \! \! \theta_1}~\! \frac{v^2}{{\tilde{m}_2}^2}+{\cal O}\bigg({v^4 \over {\tilde{m}_2}^4}\bigg)
\Xi =
-\abs{{\tilde{\lambda}}_6} e^{-i ~ \! \! \theta_1}~\! \frac{v^2}{{\tilde{m}_2}^2}
\Bigg[1
-\frac{1}{2}\big(2\tilde{\lambda}_1-\tilde{\lambda}_3-\tilde{\lambda}_4-\abs{\tilde{\lambda}_5} \big)~\! \frac{v^2}{{\tilde{m}_2}^2}
+
{\cal O}\bigg({v^4 \over {\tilde{m}_2}^4}\bigg)
\Bigg]
\label{eq:vperpapprox}
\end{equation}
where $\theta_1=1/2~\!{\rm Arg} \big({\tilde{\lambda}}_6^2 {\tilde{\lambda}}_5^* \big)$ is one of the CP violating phases defined in \eqref{eq:cpviolatingphases}. In calculating the couplings of the light Higgs, an important combination will be
\begin{equation}
\Xi ~\!  e^{ i ~ \! \! {\rm Arg}({\tilde{\lambda}}_5^*)/2}=-{{\tilde{\lambda}}_6^*} ~\! \frac{v^2}{{\tilde{m}_2}^2}+{\cal O}\bigg({v^4 \over {\tilde{m}_2}^4}\bigg)
\label{eq:convenientxi}
\end{equation}
Also, note from \eqref{eq:massheavyeigenstates} that at first order, $\tilde{m}_2$ corresponds to the mass of the heavy mass eigenstates, so we can write
\begin{equation}
\Xi ~\!  e^{ i ~ \! \! {\rm Arg}({\tilde{\lambda}}_5^*)/2}=-{{\tilde{\lambda}}_6^*} ~\! \frac{v^2}{m^2_{\varphi_{2,3}}}+{\cal O}\bigg({v^4 \over {\tilde{m}_2}^4}\bigg)
\end{equation}
In the next section we will also make use of the projection of the heavy neutral mass eigenstates in the direction of the vev, which are given by $\widehat{V}^2_1$ and $\widehat{V}^3_1$. Using the mass matrix elements \eqref{eq:massmatrixHiggsbasis} and the expression for the mass of the heavy eigenstates \eqref{eq:massheavyeigenstates} in \eqref{eq:exactalignmentparameter}, to first order in $v^2/\tilde{m}_2^2$ we get
\begin{eqnarray}
\nonumber
\widehat{V}^2_1
&=&  
-\textrm{Im}~\!\Xi
\Bigg[1 +\textrm{Re}~\!\Xi~\! {\cal O}\bigg({v^2 \over {\tilde{m}_2}^2}\bigg)
\Bigg]
=
-\abs{{\tilde{\lambda}}_6} \sin~\! \theta_1~\! \frac{v^2}{{\tilde{m}_2}^2}
\Bigg[
1
+\, {\cal O}\bigg({v^2 \over {\tilde{m}_2}^2}\bigg)
\Bigg]
\\ 
\widehat{V}^3_1 
&=&
-\textrm{Re}~\!\Xi 
\Bigg[1 +\textrm{Im}~\!\Xi~\! {\cal O}\bigg({v^2 \over {\tilde{m}_2}^2}\bigg)
\Bigg]
=
 \abs{{\tilde{\lambda}}_6} \cos~\! \theta_1~\! \frac{v^2}{{\tilde{m}_2}^2}
 \Bigg[
 1
 +{\cal O}\bigg({v^2 \over {\tilde{m}_2}^2}\bigg)
 \Bigg]
\label{eq:restofvectors}
\end{eqnarray}
Note that in the T conserving limit, $\textrm{Im}~\!\Xi=0$ and $\varphi_2=h_3$ is a pseudoscalar mass eigenstate that does not mix with the CP even fields $h_1$ or $h_2$, so $\widehat{V}^2_1$ vanishes and $\widehat{V}^3_1$ reduces to the usual alignment parameter $-\cos(\beta-\alpha)$.

%\begin{eqnarray}
%\nonumber
%\widehat{V}^2_1
%&=&  -\abs{{\tilde{\lambda}}_6} \sin~\! \theta_1~\! \frac{v^2}{{\tilde{m}_2}^2}+{\cal O}\bigg({v^4 \over {\tilde{m}_2}^4}\bigg)
%=-\textrm{Im}~\!\Xi +{\cal O}\bigg({v^4 \over {\tilde{m}_2}^4}\bigg)
%\\ \nonumber 
%\widehat{V}^2_2 &=& 
%\frac{1}{2} \tilde{\lambda}_6 \tilde{\lambda}_6^* \abs{\tilde{\lambda}_5}^{-1}
%\sin~\! 2\theta_1 
%~\! \frac{v^2}{{\tilde{m}_2}^2} 
%+{\cal O}\bigg({v^4 \over {\tilde{m}_2}^4}\bigg)
%\\ \nonumber
%\widehat{V}^2_3&=&
%1 +{\cal O}\bigg({v^4 \over {\tilde{m}_2}^4}\bigg)
%\\ \nonumber 
%\widehat{V}^3_1 &=&
%-\textrm{Re}~\!\Xi +{\cal O}\bigg({v^4 \over {\tilde{m}_2}^4}\bigg)
%\\ \nonumber 
%\widehat{V}^3_2
%&=& 
%1 +{\cal O}\bigg({v^4 \over {\tilde{m}_2}^4}\bigg)
%\\  
%\widehat{V}^3_3
%&=&
%\label{eq:restofvectors}
%\end{eqnarray}

\subsection{Couplings of the Higgs boson}
\label{sec:couplings2HDMmixing}
\label{sec:couplingsmixing}
We now study the couplings of the Higgs particle $\varphi_1$ in the 2HDM. In this section, the complex alignment parameter \eqref{eq:vperpapprox} is used extensively. Working near the decoupling limit is not a necessary condition to make use of the complex alignment parameter.

In this section, when working near the decoupling limit, we express the deviations from the SM couplings of the light Higgs as an expansion in $v^2/\tilde{m}_2^2$. We work up to first order in $v^2/\tilde{m}_2^2$ for the fermionic interactions, and second order in the bosonic interactions. We will see that this captures all the leading order deviations from the SM predictions near the decoupling limit. 

We start by defining the Lagrangian density for the light Higgs 
\begin{eqnarray}
 && \frac{1}{2} \partial \varphi_1 \partial \varphi_1 + \frac{1}{2}Z^{\mu} Z_\mu \left(m_Z^2+ g_{\varphi_a ZZ}~ \varphi_a + \frac{1}{2}g_{\varphi_1^2ZZ}~ \varphi_1^2 \right)  \nonumber \\
   && ~ + 
   W^{+ \mu} W_{\mu}^- \left(m_W^2+ g_{\varphi_a WW}~ \varphi_a + \frac{1}{2}g_{\varphi_1^2WW}~ \varphi_1^2 \right)   \nonumber \\
     && ~ + \Big[g_{\varphi_1 H^{\pm} W }  (W^{+ \mu} \partial_\mu H^- \varphi_1-W^{+ \mu} H^- \partial_\mu \varphi_1) +\textrm{h.c.}\Big] \nonumber \\
          && ~ + ( g_{\varphi_1 \varphi_2 Z } Z^{\mu} \partial_\mu \varphi_2 ~\!\varphi_1-Z^{\mu} \varphi_2 \partial_\mu \varphi_1 )+ g_{\varphi_1 \varphi_3 Z } ( Z^{\mu} \partial_\mu \varphi_3 ~\!\varphi_1-Z^{\mu} \varphi_3 \partial_\mu \varphi_1 )
     \nonumber \\
&&~-
(m^f_{ij} f_i\bar{f}_j+\lambda^{f}_{\varphi_a ij} ~f_i  \varphi_a \bar{f}_j +\textrm{h.c.}) -V(\varphi_1,\varphi_2,\varphi_3)
\label{eq:couplingdef}
\end{eqnarray}
where we sum over repeated indices and we included the Yukawa and gauge couplings involving one heavy Higgs, for later use. We leave out interactions with more than one heavy Higgs, which will not be used in this paper. The potential for the light Higgs is
\begin{eqnarray}
V(\varphi_1,\varphi_2,\varphi_3)  \! \! \! &=&\! \! \!   
\frac{1}{2} m_{\varphi_1}^2 \varphi_1^2
-\frac{1}{3!}g_{\varphi_1^3}\varphi_1^3 
-\frac{1}{4!}g_{\varphi_1^4}\varphi_1^4 
-\frac{1}{2}g_{\varphi_1^2 \varphi_2 }\varphi_1^2 \varphi_2
%-\frac{1}{2}g_{\varphi_1 \varphi_2^2 }\varphi_1 \varphi_2^2
%-\frac{1}{3!}g_{\varphi_1 \varphi_2^3 }\varphi_1 \varphi_2^3 
\nonumber \\
   & & - ~  % \frac{1}{2}g_{\varphi_1 \varphi_3^2 }\varphi_1 \varphi_3^2
%-\frac{1}{3!}g_{\varphi_1 \varphi_3^3 }\varphi_1 \varphi_3^3
%-\frac{1}{4}g_{\varphi_1^2 \varphi_2^2 }\varphi_1^2 \varphi_2^2
%\nonumber \\
 %  & & - 
   \frac{1}{2}g_{\varphi_1^2 \varphi_3 }\varphi_1^2 \varphi_3
%-\frac{1}{4}g_{\varphi_1^2 \varphi_3^2 }\varphi_1^2 \varphi_3^2
-\frac{1}{3!}g_{\varphi_1^3 \varphi_2 }\varphi_1^3 \varphi_2
-\frac{1}{3!}g_{\varphi_1^3 \varphi_3 }\varphi_1^3 \varphi_3
\end{eqnarray}
where again we included the couplings involving at most one heavy Higgs.
%+\lambda^{d}_{\varphi_a ij} ~d_i \varphi_a \bar{d}_j+\lambda^{\ell}_{\varphi_a ij} ~\ell_i \varphi_a \bar{\ell}_{j}
%\begin{eqnarray}
% && \frac{1}{2} \partial \varphi_a \partial \varphi_a +  g_{\varphi_a VV}~ \varphi_a V^{\dagger\mu} V_\mu + \frac{1}{2}g_{\varphi_a^2VV}~ \varphi_a^2 V^{\dagger\mu} V_\mu  \nonumber \\
%&&+( \lambda^{u\dagger}_{aij} ~ Q_i \varphi_a  \bar{u}_j+\lambda^{d}_{aij} ~Q_i \varphi_a \bar{d}_j+h.c.) -V(\varphi_1,\varphi_2)
%\label{eq:couplingdef}
%\end{eqnarray}
%\begin{eqnarray}
%V(h)&=&\frac{1}{2} m_h^2 h^2-\frac{1}{3!}g_{h^3}h^3-\frac{1}{4!}g_{h^4}h^4
%\end{eqnarray}
%\begin{equation}
%{\cal L}_{n}=  \frac{\phi_1^n}{n!}\left(g_{\phi_1^nVV} V^{\dagger\mu} V_\mu+ (\bar{Q}_L g_{\phi_1^n U\bar{U}} u_R+\bar{Q}_L g^{\dagger}_{\phi_1^n D \bar{D}} d_R+h.c.) + g_{\phi_1^n}\right)
%\label{eq:couplingdef}
%\end{equation}

Due to mixing of the doublets, the couplings of the Higgs are modified with respect to their SM values. An exception is the coupling between two Higgses and two gauge bosons since it is determined by gauge invariance and both $H_1$ and $H_2$ have the same quantum numbers. This coupling is given by
\begin{equation}
g_{\varphi_1^2 VV}=\frac{2m_V^2}{v^2} 
\label{eq:varphi1varphi1VVmixing}
\end{equation}
The coupling of one Higgs to two gauge bosons comes exclusively from the projection of the Higgs into the direction of the vev in field space $\widehat{V}^1_1=\sqrt{1-\abs{\Xi } ^2} $, so it is diluted with respect to its SM value by the complex alignment parameter as
\begin{equation}
g_{\varphi_1 VV}=\frac{2m_V^2}{v}\sqrt{1-\abs{\Xi }^2} 
\label{eq:varphi1VVmixing0}
\end{equation}
Note that when the Higgs potential is T conserving, this leads to the familiar relation $g_{\varphi_1 VV}=2m_V^2/v ~\! \sin(\beta-\alpha)$ \cite{Gunion:1989we}. Near the decoupling limit, using the complex alignment parameter to lowest order in $v^2/\tilde{m}_2^2$ given in \eqref{eq:vperpapprox}, we get
\begin{equation}
g_{\varphi_1 VV}=\frac{2m_V^2}{v} \Bigg[1-\frac{1}{2}{\tilde{\lambda}}_6^* {\tilde{\lambda}}_6 ~ \! \frac{v^4}{{\tilde{m}_2}^4}+{\cal O}\bigg({v^6 \over {\tilde{m}_2}^6}\bigg) \Bigg]
\label{eq:varphi1VVmixing}
\end{equation}

Yukawa couplings to up and down type fermions can be found by using the Higgs doublet Yukawa couplings defined in \eqref{eq:actionHiggsbasis} and the projections of the Higgs into the neutral components of the doublets \eqref{eq:Vh}. They are given by
 \begin{equation}
 \lambda^f_{\varphi_1  ij}=\delta_{ij}~\!\frac{m^f_{i}}{v}\sqrt{1-\abs{\Xi }^2} +\frac{1}{\sqrt{2}}\tilde{\lambda}^f_{2ij}~\!\Xi ~\!  e^{ i ~ \! \! {\rm Arg}({\tilde{\lambda}}_5^*)/2}  
 \label{eq:varphi1ffmixing0}
  \end{equation}
This coupling violates the flavor and T symmetries due to the correction proportional to $\tilde{\lambda}^f_{2ij}~\!\Xi$, which is inherited from the mixing with the heavy doublet. Near the decoupling limit, using \eqref{eq:vperpapprox} we get
\begin{eqnarray}
 \lambda^f_{\varphi_1  ij}=\delta_{ij}~\!\frac{m^f_{i}}{v} -\frac{1}{\sqrt{2}}\tilde{\lambda}^f_{2ij}{\tilde{\lambda}}_6^* ~\! \frac{v^2}{{\tilde{m}_2}^2}
 +{\cal O}\bigg({v^4 \over {\tilde{m}_2}^4}\bigg)
  \label{eq:varphi1ffmixing}
\end{eqnarray}
Note that if the Yukawa of the heavy doublet vanishes $\tilde{\lambda}^f_{2ij}=0$, the Higgs Yukawas are not modified with respect to their SM values at first order in $v^2/\tilde{m}_2^2$. In this case, the modifications come exclusively from dilution due to the complex alignment parameter in \eqref{eq:varphi1ffmixing0}, which is a second order effect. Note that in this case, the modifications to the SM predictions for the Higgs Yukawas and to the coupling of a Higgs to two gauge bosons are identical: both couplings are diluted by $\sqrt{1-\abs{\Xi}^2}$. This limit resembles the results of the SHSM studied in section \ref{sec:realsinglet}.

The Higgs self-couplings are obtained from the 2HDM potential \eqref{eq:2HDMLagrangian} and the projections of the Higgs into the neutral components of the doublets \eqref{eq:Vh}. They are given by
\begin{eqnarray}
  \frac{1}{v }g_{\varphi_1^3} \! \! \! &=&\! \! \!   
   -3 {\tilde{\lambda}}_1\left(1-\abs{\Xi }^2\right)^{3/2}
   -3({\tilde{\lambda}}_3+{\tilde{\lambda}}_4)\abs{\Xi }^2\left(1-\abs{\Xi }^2\right)^{1/2} 
   \nonumber \\
&& - ~ 3\abs{{\tilde{\lambda}}_5}\textrm{Re}~ \! \Xi ^2\left(1-\abs{\Xi }^2\right)^{1/2}
 -\frac{9}{2}\bigg[~\! \abs{{\tilde{\lambda}}_6}e^{i \theta_1}~\! \Xi  (1-\abs{\Xi }^2)    +\textrm{h.c.}~\!  \bigg]
 \nonumber \\
&& -~ \frac{3}{2}\bigg[~\! \abs{{\tilde{\lambda}}_7}e^{i \theta_2}~\! \Xi  \abs{\Xi }^2 + \textrm{h.c.}  ~\! \bigg]
\label{eq:Higgsself10}
\end{eqnarray}
\begin{eqnarray}
  g_{\varphi_1^4}\! \! \! &=&\! \! \!   
   -3 {\tilde{\lambda}}_1\left(1-\abs{\Xi }^2\right)^{2}
   -3{\tilde{\lambda}}_2 \abs{\Xi }^4
   -6({\tilde{\lambda}}_3+{\tilde{\lambda}}_4)\abs{\Xi }^2\left(1-\abs{\Xi }^2\right) 
   \nonumber \\
  &&-~6\abs{{\tilde{\lambda}}_5}\textrm{Re}~\!\Xi ^2\left(1-\abs{\Xi }^2\right)-6\bigg[~\! \abs{{\tilde{\lambda}}_6}e^{i \theta_1 }~\! \Xi  \left(1-\abs{\Xi }^2\right)^{3/2} +\textrm{h.c.} ~\!\bigg] \nonumber \\
&&-~6\bigg[ ~\! \abs{{\tilde{\lambda}}_7}e^{i \theta_2} ~\!\Xi  \abs{\Xi }^2 \left(1-\abs{\Xi }^2\right)^{1/2}+ \textrm{h.c.} ~\! \bigg]
\label{eq:Higgsself20}
\end{eqnarray}
%For the cubic Higgs self-coupling, we give the expression to order $v^4/{\tilde{m}_2}^4~\!$. This will be useful later as a consistency check of the results obtained using effective field theory: in the EFT language we will work up to order $v^4/{\tilde{m}_2}^4~\!$ in all bosonic interactions. The cubic Higgs self coupling is particularly useful for checking the consistency of the EFT, since it contains information about the Higgs potential, and since as discussed in section \ref{sec:realsingletEFT}, trilinear couplings in the mixing and EFT language must match, while this is not true for four-linear couplings. To get the cubic self coupling to second order in $v^2/{\tilde{m}_2}^2~\!$ we need the complex alignment parameter to the same order, which is given by
%\begin{equation}
%\Xi =-\abs{{\tilde{\lambda}}_6} e^{-i ~ \! \! \theta_1}~\! \frac{v^2}{{\tilde{m}_2}^2}-\frac{1}{2}\abs{\tilde{\lambda}_6}e^{-i\theta_1}\bigg[2\tilde{\lambda}_1-\tilde{\lambda}_3-\tilde{\lambda}_4-\abs{\tilde{\lambda}_5} \bigg]~\! \frac{v^4}{{\tilde{m}_2}^4}+
%{\cal O}\bigg({v^6 \over {\tilde{m}_2}^6}\bigg)
%\label{eq:vperpapprox2}
%\end{equation}
Using \eqref{eq:vperpapprox} in the exact expression for the cubic Higgs self-coupling \eqref{eq:Higgsself20}, we get
\begin{eqnarray}
\frac{1}{v}g_{\varphi_1^3} \! \! \! &=&\! \! \!    
-\frac{3m_{\varphi_1}^2}{v^2}
+6 {\tilde{\lambda}}_6^* {\tilde{\lambda}}_6~ \!\frac{v^2}{{\tilde{m}_2}^2} \nonumber \\ 
  &&+~ \frac{1}{2}\bigg[~\! 
  \big[ 21{\tilde{\lambda}}_1
  -12  ({\tilde{\lambda}}_3 +{\tilde{\lambda}}_4)
  \big] 
  {\tilde{\lambda}}_6^* {\tilde{\lambda}}_6
  -\big( 6{\tilde{\lambda}}_5 {\tilde{\lambda}}_6^{* 2}
  +\textrm{h.c.} 
  \big) 
  \bigg] 
  ~\!\frac{v^4}{{\tilde{m}_2}^4}
  +{\cal O}\bigg({v^6 \over {\tilde{m}_2}^6}\bigg) 
\label{eq:Higgsself2}
\end{eqnarray}
\begin{eqnarray}
g_{\varphi_1^4}\! \! \! &=&\! \! \!   
-\frac{3m_{\varphi_1}^2}{v^2}
+9 {\tilde{\lambda}}_6^* {\tilde{\lambda}}_6~ \!\frac{v^2}{{\tilde{m}_2}^2} \nonumber \\ 
  &&+~ \frac{1}{4}\bigg[~\! 
  \big[ 60{\tilde{\lambda}}_1
  -42  ({\tilde{\lambda}}_3 +{\tilde{\lambda}}_4)
  \big] 
  {\tilde{\lambda}}_6^* {\tilde{\lambda}}_6
  -\big( 21{\tilde{\lambda}}_5 {\tilde{\lambda}}_6^{* 2}
  +\textrm{h.c.} 
  \big) 
  \bigg] 
  ~\!\frac{v^4}{{\tilde{m}_2}^4}
  +{\cal O}\bigg({v^6 \over {\tilde{m}_2}^6}\bigg) 
\label{eq:Higgsself1}
\end{eqnarray}
%from and \eqref{eq:varphi1varphi1VVmixing}, \eqref{eq:varphi1ffmixing0}, \eqref{eq:Higgsself10} and \eqref{eq:Higgsself20} we see that 
Note that all the deviations from the SM values of the Higgs couplings are parametrized by the complex alignment parameter, such that if $\Xi=0$ all the coupling of the Higgs are SM like. 

Let us now consider couplings involving also one heavy Higgs state. They will be used in the next section to calculate scattering amplitudes of the Higgs to gauge bosons, fermions and Higgses. The couplings of one neutral heavy Higgs to two gauge bosons are obtained from the projection of these states into the field direction of the vev
\begin{eqnarray}
\nonumber \frac{1}{v}  g_{\varphi_2 VV} 
\! \! \! &=&\! \! \! 
\frac{2m_V^2}{v^2}
\widehat{V}_1^{2}
=-\frac{2m_V^2}{v^2} \abs{{\tilde{\lambda}}_6} \sin\theta_1~\! \frac{v^2}{{\tilde{m}_2}^2}
+{\cal O}\bigg({v^4 \over {\tilde{m}_2}^4}\bigg) 
\\ 
\frac{1}{v }g_{\varphi_3 VV} 
\! \! \! &=&\! \! \! 
\frac{2m_V^2}{v^2} \widehat{V}_1^{3}
=
\frac{2m_V^2}{v^2}\abs{{\tilde{\lambda}}_6} \cos\theta_1 ~\! \frac{v^2}{{\tilde{m}_2}^2}
+{\cal O}\bigg({v^4 \over {\tilde{m}_2}^4}\bigg) 
\label{eq:varphi2VV}
\end{eqnarray}
where we used \eqref{eq:restofvectors} . The $g_{\varphi_1 H^{\pm}W}$ coupling is obtained from the projection of the Higgs into the neutral components of the heavy doublet $H_2$, since the charged Higgs resides entirely in $H_2$, so its couplings to gauge bosons come exclusively from the kinetic term $D_\mu H_2^\dagger \, D^\mu H_2$.  In terms of the complex alignment parameter this coupling can be expressed in a remarkably simple way
\begin{equation}
g_{\varphi_1 H^{\pm}W} = -\frac{i m_W}{v} ~\! \Xi
\label{eq:hHpmW0}
\end{equation}
Using \eqref{eq:vperpapprox}, to lowest order in $v^2/\tilde{m}_2^2$ we get
\begin{equation}
g_{\varphi_1 H^{\pm}W}=  \frac{im_W}{v}\abs{{\tilde{\lambda}}_6} e^{-i ~ \! \! \theta_1}~\! \frac{v^2}{{\tilde{m}_2}^2}+{\cal O}\bigg({v^4 \over {\tilde{m}_2}^4}\bigg)
\label{eq:hHpmW}
\end{equation}

The couplings involving a light Higgs, a heavy Higgs and the $Z$ boson come exclusively from the kinetic term of the second doublet $D_\mu H_2^\dagger \, D^\mu H_2$. They are given by
 %, obtained from the projections of the Higgs into $h_2$ and $h_3$
\begin{eqnarray}
\nonumber g_{\varphi_1 \varphi_2 Z } 
\! \! \! &=&\! \! \! 
\frac{m_Z}{v} \textrm{Re}~\!\Xi\Bigg[ ~\!
 1
 + 
 \textrm{Im}~\! \Xi \,{\cal O}\bigg({v^2 \over {\tilde{m}_2}^2}\bigg)\Bigg]
=
-\frac{m_Z}{v}\abs{{\tilde{\lambda}}_6}\cos\theta_1~\! \frac{v^2}{{\tilde{m}_2}^2} +{\cal O}\bigg({v^4 \over {\tilde{m}_2}^4}\bigg) 
\\ 
g_{\varphi_1 \varphi_3 Z } 
\! \! \! &=&\! \! \! 
- \frac{m_Z}{v}\textrm{Im}~\!\Xi \Bigg[~\!
1
+
 \textrm{Re}~\!\Xi \,{\cal O}\bigg({v^2 \over {\tilde{m}_2}^2}\bigg)
\Bigg] 
 =
 -\frac{m_Z}{v} \abs{{\tilde{\lambda}}_6} \sin\theta_1~\! \frac{v^2}{{\tilde{m}_2}^2}+{\cal O}\bigg({v^4 \over {\tilde{m}_2}^4}\bigg)
\end{eqnarray}

The couplings of the heavy neutral states to fermions are to lowest order in $v^2/\tilde{m}_2^2$ given by the Yukawas of the heavy doublet,
\begin{eqnarray}
\lambda^f_{\varphi_2 ij}= \frac{1}{\sqrt{2}}~\!i~\! \tilde{\lambda}^f_{2ij}e^{ i ~ \! \! {\rm Arg} ({\tilde{\lambda}}_5^*)/2} +{\cal O}\bigg({v^2 \over {\tilde{m}_2}^2}\bigg)
 \label{eq:varphi2ff3}
\end{eqnarray}
\begin{eqnarray}
 \lambda^f_{\varphi_3 ij}=\frac{1}{\sqrt{2}}\tilde{\lambda}^f_{2ij}e^{ i ~ \! \! {\rm Arg} ({\tilde{\lambda}}_5^*)/2} +{\cal O}\bigg({v^2 \over {\tilde{m}_2}^2}\bigg)
 \label{eq:varphi2ff2}
\end{eqnarray}
where the factor $e^{ i ~ \! \! {\rm Arg} ({\tilde{\lambda}}_5^*)/2} $ comes from the definition of the component fields \eqref{eq:Higgsbasis}. Note that these couplings are not controlled by the complex alignment parameter $\Xi$, and do not vanish in the alignment limit. 

The dihiggs couplings to one heavy Higgs are obtained from the Higgs potential, and they are
 \begin{eqnarray}
 \nonumber \frac{1}{v}g^2_{\varphi_1 ^2 \varphi_2}  \! \! \! &=& 3 \abs{\tilde{\lambda}_6} \sin \theta_1  +{\cal O}\bigg({v^2 \over {\tilde{m}_2}^2}\bigg)
\\
 \frac{1}{v}g^2_{\varphi_1^2 \varphi_3}  \! \! \! &=&-3\abs{\tilde{\lambda}_6} \cos \theta_1 +{\cal O}\bigg({v^2 \over {\tilde{m}_2}^2}\bigg) 
 \label{eq:varphi1varphi1varphi2}
\end{eqnarray}
Note that these couplings are not controlled by the complex alignment parameter. In particular, $\Xi$ could be small due to a large separation of scales $v^2/{\tilde{m}_2}^2 \ll 1$, while $g^2_{\varphi_1^2 \varphi_2}, g^2_{\varphi_1 ^2 \varphi_3}$ could be of order $v$, if $\abs{\tilde{\lambda}_6}$ is not small.

\subsection{Scattering amplitudes}
\label{sec:amplitudesmixing2HDM}
We now work out some examples of tree level dihiggs scattering amplitudes. They will be used as a consistency check for the 2HDM low energy EFT that will be presented in section \ref{sec:physicalHiggs2HDMEFT} and to compare couplings in the EFT and mixing languages. The Standard Model results for the amplitudes can be read from this section by taking the limit $\tilde{\lambda}_6\rightarrow 0$. We omit spinors in all amplitudes.

The tree level dihiggs to di-W boson scattering amplitude is 
\begin{eqnarray}
{ \cal A} \left(\varphi_1 \varphi_1 \rightarrow W^+  W^- \right) 
  \! \! \! &=&  \!\!\!
  g_{\mu\nu}~\!
   \Bigg[ 
   g_{\varphi_1^2W^2} 
  -\frac{g_{\varphi_1^2 \varphi_{a}}g_{\varphi_a W^2}}{s-m_{\varphi_a}^2}
 -g_{\varphi_1 W^2}^2  \bigg(\frac{1}{t-m_{W}^2}+\frac{1}{u-m_{W}^2}\bigg) 
 \Bigg]
  \notag \\
  && 
  - ~ \abs{g_{\varphi_1 H^{\pm}W}}^2
  \Bigg[
  \frac{(2p_1-p_+)_\mu  (p_1+p_2+p_+)_\nu }{t-m_{H^{\pm}}^2} 
  + ~(p_+ \leftrightarrow p_-)  
  \Bigg]\notag \\
    \label{eq:exactdihdiWmixing}
 \end{eqnarray}
where we sum over $a$ and $p_\pm$ are the momenta of the $W^\pm$ bosons. In the last line, exchanging $p_+$ with  $p_-$ changes the Mandelstam variable $t$ to $u$. Using the expressions for the couplings \eqref{eq:varphi1varphi1VVmixing}, \eqref{eq:varphi1VVmixing}, \eqref{eq:Higgsself1},  \eqref{eq:varphi2VV} \eqref{eq:hHpmW} and \eqref{eq:varphi1varphi1varphi2}, we get 
\begin{eqnarray}
{ \cal A} \left(\varphi_1\varphi_1\rightarrow W^+  W^- \right) 
  \! \! \! &=&   \! \! \!
g_{\mu\nu} 
~\!
\Bigg[ \frac{2m_W^2}{v^2} 
  -6\tilde{\lambda}_6\tilde{\lambda}_6^*\frac{m_W^2}{v^2} ~\! \frac{v^4}{\tilde{m}_2^4} \nonumber \\
    && - ~ \frac{2m_W^2}{v^2} \bigg[~\!
    -\frac{3m_{\varphi_1}^2}{v^2}
+6 {\tilde{\lambda}}_6^* {\tilde{\lambda}}_6~ \!\frac{v^2}{{\tilde{m}_2}^2} 
+ \frac{1}{2}\bigg(~\! 
  \big[ 21{\tilde{\lambda}}_1
  -12  ({\tilde{\lambda}}_3 +{\tilde{\lambda}}_4)
  \big] 
  {\tilde{\lambda}}_6^* {\tilde{\lambda}}_6
\nonumber \\ 
  &&-~ 
 \big( 6{\tilde{\lambda}}_5 {\tilde{\lambda}}_6^{* 2}
  +\textrm{h.c.} 
  \big) 
  \bigg)
  ~\!\frac{v^4}{{\tilde{m}_2}^4}
  +\frac{3}{2} {\tilde{\lambda}}_6^* {\tilde{\lambda}}_6 \frac{m_{\varphi_1}^2v^2}{{\tilde{m}_2}^4}
     ~\!\bigg]\bigg(\frac{v^2}{s-m_{\varphi_1}^2} \bigg) \notag \\
       && - ~\frac{4m_W^4}{v^4} \bigg[1-{\tilde{\lambda}}_6^\dagger {\tilde{\lambda}}_6 ~ \! \frac{v^4}{{\tilde{m}_2}^4} \bigg] \bigg(\frac{v^2}{t-m_{W}^2}+\frac{v^2}{u-m_{W}^2}\bigg) 
       + {\cal O}\bigg(\frac{s v^2}{{\tilde{m}_2}^4},\frac{v^6}{{\tilde{m}_2}^6}\bigg)       \Bigg]\notag \\
       \label{eq:dihiggsdiWmixing}
 \end{eqnarray}
where the first term is the contact interaction, the second term comes from the short distance contribution of the heavy neutral states, and the two last terms are the long distance contributions from Higgs and $W$ boson mediated diagrams. Note that the contributions mediated by the charged Higgs written in the last line of \eqref{eq:exactdihdiWmixing} drop out of the calculations - they do not contribute to the order we work to.

The dihiggs to difermion chirality violating scattering amplitude is  
\begin{eqnarray}
{ \cal A} \left(\varphi_1 \varphi_1 \rightarrow f_i \bar{f}_j \right) 
  \! \! \! &=&   \! \! \! \frac{g_{\varphi_1^2 \varphi_{a}}\lambda^f_{\varphi_a ij}}{s-m_{\varphi_a}^2}
  \label{eq:dihiggsdifexactmixing}
 \end{eqnarray}
 where we sum over $a$. Using the expressions for the couplings \eqref{eq:varphi1ffmixing}, \eqref{eq:Higgsself1}, \eqref{eq:varphi2ff3}, \eqref{eq:varphi2ff2} and \eqref{eq:varphi1varphi1varphi2} we get  
 \begin{eqnarray}
{ \cal A} \left(\varphi_1 \varphi_1 \rightarrow f_i \bar{f}_j \right) 
  \! \! \! &=&   \! \! \! \frac{1}{v }
  ~\!\Bigg[
  ~\!
  % Long Distance Piece
  \bigg[
  \delta_{ij} \frac{m^f_{i}}{v}
  \bigg(-\frac{3m_{\varphi_1}^2}{v^2}
  +6\tilde{\lambda}_6\tilde{\lambda}_6^* ~\!\frac{v^2}{\tilde{m}^2_2}
   \bigg)
   % Short distance piece
  +\frac{3}{\sqrt{2}}~\! \tilde{\lambda}^f_{2ij} \tilde{\lambda}_6^*~\! \frac{ m_{\varphi_1}^2}{\tilde{m}^2_2}   
  \bigg] 
  \bigg(\frac{v^2}{s-m_{\varphi_1}^2}\bigg) \notag \\
    && \! \! \!  
    +~ 
    \bigg[ \frac{3}{\sqrt{2}} \tilde{\lambda}^f_{2ij}e^{i\textrm{Arg}(\tilde{\lambda}_5^*)/2} \abs{\tilde{\lambda}_6}e^{-i\theta_1}~\! 
    \frac{v^2}{\tilde{m}_2^2}\bigg] 
    % Error
    + {\cal O}\bigg(\frac{s v^2}{{\tilde{m}_2}^4},\frac{v^4}{{\tilde{m}_2}^4},\frac{m^{f\,2}_{i}}{v^2}\bigg)
  ~\!  \Bigg]
    \notag \\
    % Second equality
     \! \! \! &=&   \! \! \! \frac{1}{v }
     ~\!
     % Long distance Piece
     \Bigg[
     ~\!
     \bigg[
  \delta_{ij} \frac{m^f_{i}}{v}
  \bigg(-\frac{3m_{\varphi_1}^2}{v^2}
  +6\tilde{\lambda}_6\tilde{\lambda}_6^* ~\!\frac{v^2}{\tilde{m}^2_2}
   \bigg)
  +\frac{3}{\sqrt{2}}~\! \tilde{\lambda}^f_{2ij} \tilde{\lambda}_6^*~\! \frac{ m_{\varphi_1}^2}{\tilde{m}^2_2}   
  \bigg] 
  \bigg(\frac{v^2}{s-m_{\varphi_1}^2}\bigg) \notag \\
    && \! \! \!  
    % Short distance piece
    +~ \frac{3}{\sqrt{2}} \tilde{\lambda}^f_{2ij}\tilde{\lambda}_6^*~\! 
    \frac{v^2}{\tilde{m}_2^2}
    % Error
    + {\cal O}\bigg(\frac{s v^2}{{\tilde{m}_2}^4},\frac{v^4}{{\tilde{m}_2}^4},\frac{m^{f\,2}_{i}}{v^2}\bigg)
    ~\!\Bigg]
    \label{eq:dihiggsdifmixing}
 \end{eqnarray}
where in the last step we made use of the definition of the CP violating phase $\theta_1=\frac{1}{2}{\rm Arg} \big({\tilde{\lambda}}_6^2 {\tilde{\lambda}}_5^* \big)$  (see \eqref{eq:cpviolatingphases}). The first term in square brackets is the long distance contribution mediated by the Higgs, and the last term corresponds to the short distance contributions mediated by the heavy Higgses. Note that this amplitude has a flavor and T violating term both in the long and short distance pieces, inherited from $\tilde{\lambda}^f_{2ij}$, the coupling of the heavy doublet to the fermions. 

The four Higgs scattering amplitude is
\begin{eqnarray}
{ \cal A} \left(\varphi_1 \varphi_1 \rightarrow \varphi_1\varphi_1\right) 
  \! \! \! &=&   \! \! \! g_{\varphi_1^4} -g^2_{\varphi_1^2 \varphi_a}\left(\frac{1}{s-m_{\varphi_a}^2}+\frac{1}{t-m_{\varphi_a}^2}+\frac{1}{u-m_{\varphi_a}^2}\right)
 \end{eqnarray}
where we sum over $a$. Using \eqref{eq:Higgsself1} and \eqref{eq:varphi1varphi1varphi2} we get
\begin{eqnarray}
{ \cal A} \left(\varphi_1 \varphi_1 \rightarrow \varphi_1\varphi_1\right) \! \! \! &=&   
% Contact term
\! \! \! g_{\varphi_1^4}
% Short distance contributions
-  \frac{g^2_{\varphi_1^2 \varphi_2}}{v^2}\bigg[-\frac{3v^2}{m_{\varphi_2}^2} +{\cal O}\bigg({xv^2 \over {\tilde{m}_{\varphi_2}}^4}\bigg) \bigg]
-  \frac{g^2_{\varphi_1^2 \varphi_3}}{v^2}\bigg[-\frac{3v^2}{m_{\varphi_3}^2} +{\cal O}\bigg({xv^2 \over {\tilde{m}_{\varphi_3}}^4}\bigg) \bigg]
\notag \\
&& \! \! \!  
% Long distance contributions
- ~ \frac{g^2_{\varphi_1^3}}{v^2}\bigg(\frac{v^2}{s-m_{\varphi_1}^2}
+\frac{v^2}{t-m_{\varphi_1}^2}
+\frac{v^2}{u-m_{\varphi_1}^2}\bigg)
\notag \\
% Second equality
&=&   
% Contact term and short distance contributions
\! \! \!-\frac{3m_{\varphi_1}^2}{v^2}
+9{\tilde{\lambda}}_6^* {\tilde{\lambda}}_6 ~ \! \frac{v^2}{{\tilde{m}_2}^2}  
+27{\tilde{\lambda}_6}{\tilde{\lambda}_6^*} \cos^2 \theta_1 ~ \frac{v^2}{\tilde{m}_{2}^2}
+27{\tilde{\lambda}_6}{\tilde{\lambda}_6^*} \sin^2 \theta_1 ~ \frac{v^2}{\tilde{m}_{2}^2}
\notag \\
% Long distance pieces
&&  \! \! \! - ~\bigg(\frac{9m_{\varphi_1}^4}{v^4}
  -36{\tilde{\lambda}}_1 {\tilde{\lambda}}_6^* {\tilde{\lambda}}_6 ~  \frac{v^2}{{\tilde{m}_2}^2}
   \bigg)\bigg(
\frac{v^2}{s-m_{\varphi_1}^2}+\frac{v^2}{t-m_{\varphi_1}^2}+\frac{v^2}{u-m_{\varphi_1}^2}
\bigg) \notag \\
&& \! \! \! 
% Error
+ ~{\cal O}\bigg(\frac{v^4}{\tilde{m}_{2}^4},\frac{xv^2}{\tilde{m}_{2}^4}\bigg) \notag\\
  \! \! \! &=&   \! \! \! -\frac{3m_{\varphi_1}^2}{v^2}+36 {\tilde{\lambda}}_6^* {\tilde{\lambda}}_6\left(\frac{v^2}{{\tilde{m}_2}^2}\right)
  - \bigg(
  \frac{9m_{\varphi_1}^2}{v^4}
  -36{\tilde{\lambda}}_1 {\tilde{\lambda}}_6^* {\tilde{\lambda}}_6 ~  \frac{v^2}{{\tilde{m}_2}^2}
   \bigg)
   \notag \\
&&  \! \! \!
\bigg(\frac{v^2}{s-m_{\varphi_1}^2}+\frac{v^2}{t-m_{\varphi_1}^2}+\frac{v^2}{u-m_{\varphi_1}^2}\bigg) + {\cal O}\bigg(\frac{x v^2}{{\tilde{m}_2}^4},\frac{v^4}{{\tilde{m}_2}^4}\bigg)
\label{eq:fourfermionamplitude2HDMmixing}
 \end{eqnarray}
where the first two terms come from the contact interaction and the short distance diagrams mediated by the Heavy Higgses, and the last term is the long distance contribution mediated by the Higgs particle.

 \section{The low energy effective theory of the 2HDM}
\label{sec:2HDMEFT}

In this section we present the tree level low energy effective theory of the Standard Model extended in the UV with a heavy Higgs doublet. Near the decoupling limit the heavy doublet can be integrated out, and the remaining Higgs doublet in the low energy theory contains the vev and the Higgs particle. The cutoff of the EFT is the heavy doublet mass $\tilde{m}_2$. The EFT description is an alternative to the mixing language, and no reference to mixing between the neutral components of the doublets is needed in deriving the EFT. Working at tree level is enough to reproduce the mixing effects described in the previous section. The effective theory can be derived most easily by working in the Higgs basis. In this basis one of the doublets carries no vev and must be identified with the heavy doublet, as discussed in the previous section. The remaining doublet contains the particle that must be identified with the Higgs in the low energy theory.  In the derivation of the EFT we keep track of the $U(1)_{\textrm{PQ}}$ symmetry defined in table \ref{tab:U1PQ2}, which can be used at any point of the calculations as a consistency check, since all the fields in the low energy theory (fermions, gauge bosons and the Higgs doublet $H_1$) are PQ invariants, so no coupling of the low energy theory can be charged under the $U(1)_{\textrm{PQ}}$.

The regime of the 2HDM parameters we work in is the following. We consider all the marginal couplings of the 2HDM in the Higgs basis $\tilde{\lambda}_i$ $i=1..7$ to be perturbative. The 2HDM has three dimensionful couplings: $\tilde{m}_1, \tilde{m}_2$ and $\tilde{m}_{2} $. However, there are only two mass scales, $\tilde{m}_1$  $\tilde{m}_2$; due to the no tadpole condition \eqref{eq:notadpolecondition}, $\tilde{m}_{12}$ is of the order of $\tilde{m}_1$. Due to the EWSB condition \eqref{eq:EWSBscale}, the mass of the remaining doublet $\tilde{m}_1$ must be identified with the EWSB scale. Near the decoupling limit, ${\tilde{m}_2}^2  \gg  \abs{\tilde{\lambda}_i}v^2$. Due to the separation of scales, we organize the corrections to the SM predictions as an expansion in the small parameter $v^2/\tilde{m}_2^2$. To obtain a consistent expansion in $v^2/\tilde{m}_2^2$ in the 2HDM EFT, we need to define a concept of effective operator dimension which \textit{differs from naive operator dimension}. The separation of scales motivates us to define our concept of effective operator dimension by counting powers of the heavy scale in the operator's coefficient in the Lagrangian. Since $\tilde{m}_2$ is the only heavy mass term, we define the operator's effective dimension as
\begin{equation}
n_{E}=4-n_{\tilde{m}_2^2}
\end{equation}
 where $n_{\tilde{m}_2^2}$ is the number of powers of $\tilde{m}_2^2$ in the operator's coefficient. For instance, the operators
 \begin{equation}
\frac{\tilde{m}^2_{12}}{\tilde{m}_2^2} (H_1^\dagger H_1)^2 \quad \quad \frac{1}{\tilde{m}_2^2} (H_1^\dagger H_1)^3
\end{equation}
are both of effective dimension six, even though the first operator is of naive operator dimension four. In this work, we choose to work up to effective dimension eight in the Higgs potential and kinetic terms (including gauge interactions), and up to effective dimension six in operators involving fermions. We will see that this captures all the lowest order deviations to the SM predictions for the Higgs bosonic and fermionic interactions.

%%Any renormalizable coupling $\eta_h$ of mass dimension $d$ of the field $h$ to itself or to other fields  can be written as 
%%\begin{equation}
%%\eta_h=v^d \sum_{a \geq 0} c_a \left(\frac{v^2}{{\tilde{m}_2}^2} \right)^{a} 
%%\end{equation}
%%where $c_a$ are coefficients that depend on the couplings of the 2HDM. To obtain $\eta_h$ in effective field theory to order $a$, we must keep operators up to effective dimension $n_E=2a+4$. 
%\begin{equation}
%n_E=n_D+2n_{{\tilde{m}_{12}}^d{equation}
%where $n_E$ is the operator dimension and $n_{{\tilde{m}_{12}}^2}$ is the number of ${\tilde{m}_{12}}^2$ couplings \textit{in the operator's coefficient}. 

\subsection{Diagrams and derivation of the EFT}
In this section we present all the needed diagrams to derive the low energy EFT up to effective dimension eight in the Higgs potential and kinetic terms and up to effective dimension six in operators involving fermions. Since the maximum effective dimension we work up to is eight, all diagrams can have at most two inverse powers of $\tilde{m}_2^2$, \textit{i.e.}, they can have at most two $H_2$ propagators. Note that of all couplings in the 2HDM Lagrangian \eqref{eq:2HDMLagrangian}, ${\tilde{m}_{12}}^2$, $({\tilde{m}_{12}}^{2})^*~$, ${\tilde{\lambda}}_6$, ${\tilde{\lambda}}_6^*$, $\tilde{\lambda}_{2ij}$ and $\tilde{\lambda}_{2ij}^*~\!$ are the only ones that involve only one heavy doublet $H_2$, so using these interactions it is possible to draw diagrams with only one $H_2$ propagator. These will be the diagrams that, at zero momentum, induce the irrelevant operators with the lowest effective dimension (six). As such, leading deviations to the SM predictions will be controlled by this small subset of parameters of the UV completion. We organize the presentation of the diagrams by number of insertions of Higgs potential or fermionic couplings involving the heavy Higgs doublet $H_2$. 

\begin{itemize}[leftmargin=*]
\item {\bf Diagrams with two ${\tilde{m}_{12}}^2$ insertions}:

%\begin{figure}[h]
%\begin{center}
%%\psfrag{Y}[Bc]{\small $\cos(\theta)$ \normalsize}
%%\psfrag{X}[lc]{ \small $\kappa$  \normalsize}
%\includegraphics[width=16cm]{2hdmdiagram1.eps}
%\caption{Diagrams with two $m_{12}^2$ insertions up to effective dimension eight. The upper diagram must be expanded up to quadratic order in the momentum of the external leg to work up to effective dimension eight. There are also diagrams with one gauge boson attached to the internal heavy Higgs propagator which are also considered. The upper diagram at zero momentum leads to the operator $H_1^\dagger  H_1$. The upper diagram at quadratic order in the momentum expansion, together with the diagrams with gauge bosons lead to  $\left(D_\mu H_1 \right)^\dagger \left(D^\mu H_1 \right)$.}
%\label{fig:2HDM4}
%\end{center}
%\end{figure}

\begin{figure}[h]
\begin{center}
%\psfrag{Y}[Bc]{\small $\cos(\theta)$ \normalsize}
%\psfrag{X}[lc]{ \small $\kappa$  \normalsize}
\includegraphics[width=14cm]{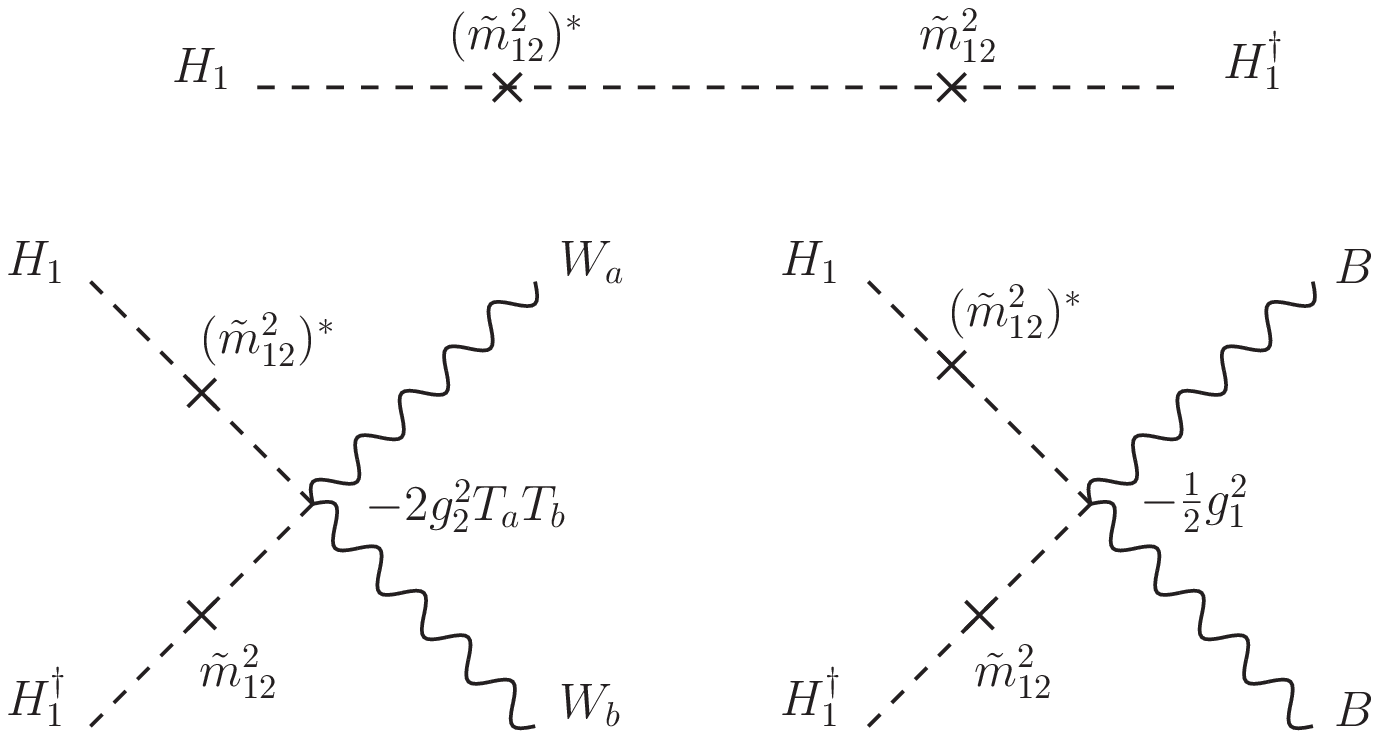}
\caption{Diagrams with two $m_{12}^2$ insertions and gauge boson legs leading to operators up to effective dimension eight. By gauge invariance there are also diagrams with only one gauge boson attached to the internal heavy Higgs propagator, which we do not draw for brevity, but that are considered when deriving the low energy theory. The upper diagram must be expanded up to quadratic order in the momentum of the external leg to work up to effective dimension eight. The upper diagram at zero momentum leads to the operator $H_1^\dagger  H_1$. The upper diagram at quadratic order in the momentum expansion, together with the diagrams with gauge bosons lead to the operator $\left(D_\mu H_1 \right)^\dagger \left(D^\mu H_1 \right)$.}
\label{fig:2HDM4}
\end{center}
\end{figure}

%\begin{figure}[h]
%\begin{center}
%%\psfrag{Y}[Bc]{\small $\cos(\theta)$ \normalsize}
%%\psfrag{X}[lc]{ \small $\kappa$  \normalsize}
%\includegraphics[width=17cm]{2hdmdiagram4.eps}
%\caption{}
%\label{fig:2HDM4}
%\end{center}
%\end{figure}

Let us start with the diagrams of figure \ref{fig:2HDM4}. Up to effective dimension eight they lead to
%\begin{equation}
%\bra T  H^\dagger  H \ket \supset i \frac{\abs{{\tilde{m}_{12}}^2}^2}{{\tilde{m}_2}^2}
%\end{equation}
\begin{eqnarray}
 && \frac{\abs{{\tilde{m}_{12}}^2}^2}{{\tilde{m}_2}^2}  H^\dagger  H
+ \frac{\abs{{\tilde{m}_{12}}^2}^2}{{\tilde{m}_2}^4} \partial_\mu  H^\dagger \partial^\mu  H
+\frac{\abs{{\tilde{m}_{12}}^2}^2}{{\tilde{m}_2}^4} 
\bigg[
g_2^2  H_1^\dagger  T_a T_b W_{a\mu} W_b ^\mu  H_1
\nonumber \\
&& \! \! \! - ~g_2 \left(i\partial_\mu  H_1^\dagger T_a W_a^\mu  H_1 
+\textrm{h.c.}
\right) 
+ 
\frac{1}{4}g_1^2  H_1^\dagger B_{\mu} B^\mu  H_1
-\frac{1}{2}g_1 \left(i\partial_\mu  H_1^\dagger B^\mu  H_1 
+\textrm{h.c.}
\right) 
\bigg]
\nonumber \\
\label{eq:diagram10}
\end{eqnarray}
which can be rearranged into 
\begin{equation}
\frac{\abs{{\tilde{m}_{12}}^2}^2}{{\tilde{m}_2}^2}  H_1^\dagger  H_1+\frac{\abs{{\tilde{m}_{12}}^2}^2}{{\tilde{m}_2}^4} \left(D_\mu H_1 \right)^\dagger \left(D^\mu H_1 \right) 
\label{eq:diagram1}
\end{equation}

\item  {\bf Diagrams with two $ {\tilde{\lambda}}_6$ insertions:}

%\begin{figure}[h]
%\begin{center}
%%\psfrag{Y}[Bc]{\small $\cos(\theta)$ \normalsize}
%%\psfrag{X}[lc]{ \small $\kappa$  \normalsize}
%\includegraphics[width=16cm]{2hdmdiagram3joined.eps}
%\caption{Diagrams with two $\lambda_6$ insertions up to effective dimension eight.  The upper diagram must be expanded up to quadratic order in the momentum of the external leg to work up to effective dimension eight. There are also diagrams with one gauge boson attached to the internal heavy Higgs propagator which are also considered. The upper diagram at zero momentum leads to the operator $( H_1^\dagger  H_1)^3$. The upper diagram at quadratic order in momentum of the external legs leads to $\partial_\mu( H_1^\dagger  H_1)
%\partial^\mu( H_1^\dagger  H_1)( H_1^\dagger  H_1)$ and also, together with the diagrams with gauge bosons, to $(D_\mu H_1)^\dagger  (D^\mu  H_1) ( H_1^\dagger  H_1)^2 $.}
%\label{fig:2HDM3}
%\end{center}
%\end{figure}

\begin{figure}[h]
\begin{center}
%\psfrag{Y}[Bc]{\small $\cos(\theta)$ \normalsize}
%\psfrag{X}[lc]{ \small $\kappa$  \normalsize}
\includegraphics[width=14cm]{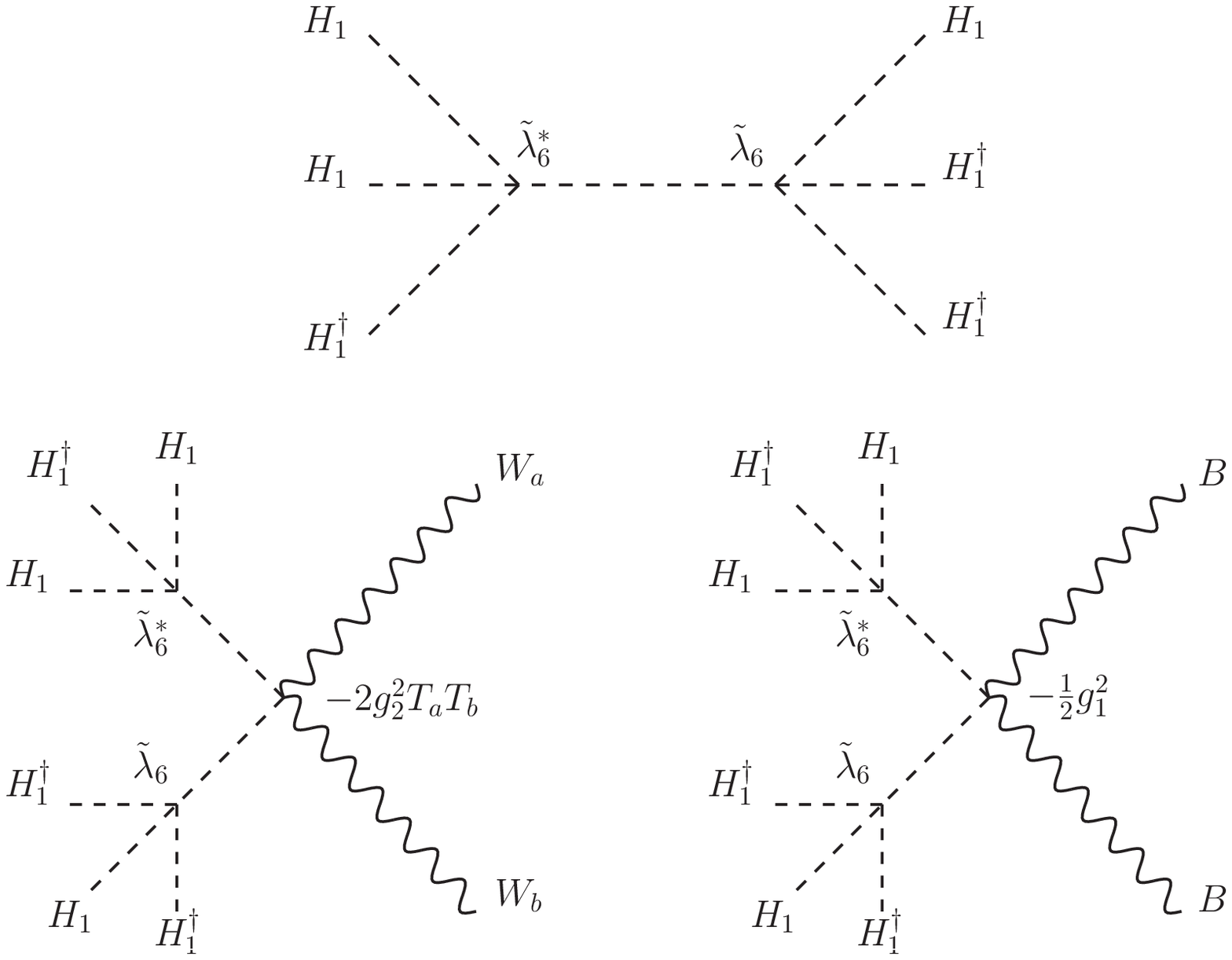}
\caption{Diagrams with two $\lambda_6$ insertions and gauge boson legs leading to operators up to effective dimension eight. By gauge invariance there are also diagrams with only one gauge boson attached to the internal heavy Higgs propagator, which we do not draw for brevity, but that are considered when deriving the low energy theory. The upper diagram must be expanded up to quadratic order in the momentum of the external leg to work up to effective dimension eight. The upper diagram at zero momentum leads to the operator $( H_1^\dagger  H_1)^3$. The upper diagram at quadratic order in the momentum of the external legs leads to $\partial_\mu( H_1^\dagger  H_1)
\partial^\mu( H_1^\dagger  H_1)( H_1^\dagger  H_1)$ and also, together with the diagrams with gauge bosons, to $(D_\mu H_1)^\dagger  (D^\mu  H_1) ( H_1^\dagger  H_1)^2 $.}
\label{fig:2HDM3}
\end{center}
\end{figure}

The diagrams of figure \ref{fig:2HDM3} lead to
\begin{eqnarray}
&& \frac{ {\tilde{\lambda}}_6^*  {\tilde{\lambda}}_6}{{\tilde{m}_2}^2} ( H_1^\dagger  H_1)^3+ \frac{ {\tilde{\lambda}}_6^*  {\tilde{\lambda}}_6}{{\tilde{m}_2}^4}
\Bigg[~\!
- 
\frac{3}{2}( H_1^\dagger  H_1)^2 
(~\! \Box H_1^\dagger  H_1+\textrm{h.c})
\nonumber \\
&& \! \! \! - ~
 H_1^\dagger  H_1\Big(
\partial_\mu  H_1^\dagger\partial^\mu  H_1  H_1^\dagger  H_1
+\partial_\mu H_1^\dagger H_1\partial^\mu H_1^\dagger   H_1
+ H_1^\dagger \partial_\mu  H_1 \partial^\mu H_1^\dagger   H_1 
+\textrm{h.c} \Big)
\nonumber \\
&& \! \! \! + ~
( H_1^\dagger  H_1)^2
\bigg[
~\!
  g_2^2  H_1^\dagger  T_a T_b W_{a\mu} W_{b}^\mu   H_1
-g_2 \Big(i\partial_\mu  H_1^\dagger T_a W_a^\mu  H_1 
+\textrm{h.c.}
\Big)
\nonumber \\
&& \! \! \! + ~
\frac{1}{4}g_1^2  H_1^\dagger B_{\mu} B^\mu  H_1
-\frac{1}{2}g_1 \Big(i\partial_\mu  H_1^\dagger B^\mu  H_1 
+\textrm{h.c.}
\Big) 
~\!
\bigg]
~\!
\Bigg]
\label{eq:diagram21}
\end{eqnarray}
%We can also plug in 1 or 2 outgoing gauge bosons in the $ H_2$ propagator, in a similar way as we did in figure \ref{fig:2HDM4}. Those diagrams lead to
%\begin{equation}
%\frac{ {\tilde{\lambda}}_6^*  {\tilde{\lambda}}_6}{2 {\tilde{m}_2}^4}
%( H_1^\dagger  H_1)^2
%\bigg[
%2  g_2^2  H_1^\dagger  T_a T_b W_{a\mu} W_{b}^\mu   H_1
%-2g_2 \left(i\partial_\mu  H_1^\dagger T_a W_a^\mu  H_1 
%+\textrm{h.c.}
%\right)
%\bigg]
%\label{eq:diagram22}
%\end{equation}
Rearranging the derivatives and the interactions with gauge bosons into covariant derivatives we get
\begin{equation}
\frac{ {\tilde{\lambda}}_6^*  {\tilde{\lambda}}_6}{{\tilde{m}_2}^2}( H_1^\dagger  H_1)^3+\frac{ {\tilde{\lambda}}_6^*  {\tilde{\lambda}}_6}{{\tilde{m}_2}^4}
\bigg[
~\!2~\!\partial_\mu( H_1^\dagger  H_1)
\partial^\mu( H_1^\dagger  H_1)( H_1^\dagger  H_1)
+ (D_\mu H_1)^\dagger  (D^\mu  H_1) ( H_1^\dagger  H_1)^2 
\bigg]
\label{eq:diagram2}
\end{equation} 

\item   {\bf Diagrams with one $ {\tilde{\lambda}}_6$ and one ${\tilde{m}_{12}}^2$ insertions}

%\begin{figure}[h]
%\begin{center}
%%\psfrag{Y}[Bc]{\small $\cos(\theta)$ \normalsize}
%%\psfrag{X}[lc]{ \small $\kappa$  \normalsize}
%\includegraphics[width=16cm]{2hdmdiagram2.eps}
%\caption{Diagrams with one $\lambda_6$ and one $(\tilde{m}_{12}^2)^*$ insertion up to effective dimension eight. The upper diagram must be expanded up to quadratic order in the momentum of the external leg to work up to effective dimension eight. There are also hermitic conjugate versions of these diagrams and diagrams with one gauge boson attached to the internal heavy Higgs propagator, which are also considered. The upper diagram at zero momentum leads to the operator $( H_1^\dagger  H_1)^2$. The upper diagram at quadratic order in momentum of the external legs leads to $\partial_\mu( H_1^\dagger  H_1)
%$ and also, together with the diagrams with gauge bosons, to $(D_\mu H_1)^\dagger  (D^\mu  H_1) H_1^\dagger  H_1 $.}
%\label{fig:2HDM2}
%\end{center}
%\end{figure}

\begin{figure}[h]
\begin{center}
%\psfrag{Y}[Bc]{\small $\cos(\theta)$ \normalsize}
%\psfrag{X}[lc]{ \small $\kappa$  \normalsize}
\includegraphics[width=14cm]{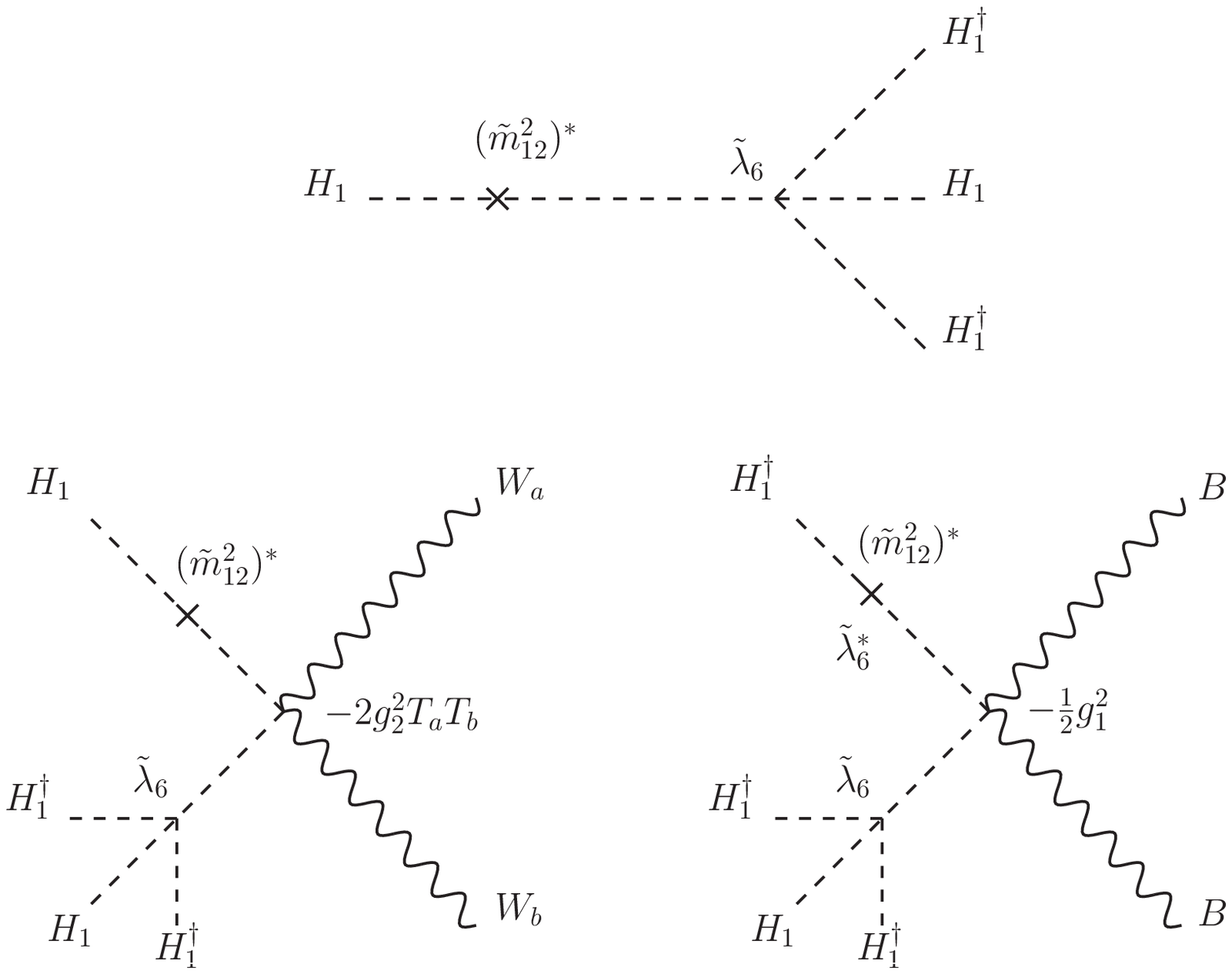}
\caption{Diagrams with one $\lambda_6$ and one $(\tilde{m}_{12}^2)^*$ insertion up to effective dimension eight. There are also hermitic conjugate versions of these diagrams and diagrams with only one gauge boson attached to the internal heavy Higgs propagator, which are not drawn for brevity but considered when deriving the low energy theory. The upper diagram must be expanded up to quadratic order in the momentum of the external leg to work up to effective dimension eight. The upper diagram at zero momentum leads to the operator $( H_1^\dagger  H_1)^2$. The upper diagram at quadratic order in momentum of the external legs leads to $H_1^\dagger  H_1 H_1^\dagger \Box  H_1
$ and also, together with the diagrams with gauge bosons, to $(D_\mu H_1)^\dagger  (D^\mu  H_1) H_1^\dagger  H_1 $.}
\label{fig:2HDM2}
\end{center}
\end{figure}
%The first effect is to induce a correction to the Higgs quartic
%\begin{equation}
%\bra T ( H_1^\dagger  H_1)^2\ket = 2!2! \frac{i ( {\tilde{\lambda}}_6^* {\tilde{m}_{12}}^2 +h.c.) }{{\tilde{m}_2}^2}
%\end{equation}
The diagrams in figure \ref{fig:2HDM2} lead to
\begin{eqnarray}
&& 
\frac{ {\tilde{\lambda}}_6 ({\tilde{m}_{12}}^2)^*}{{\tilde{m}_2}^2}  ( H_1^\dagger  H_1)^2
+
\frac{ {\tilde{\lambda}}_6 ({\tilde{m}_{12}}^2)^*}{{\tilde{m}_2}^4} 
\Bigg[~\!
-  H_1^\dagger  H_1 ~\! H_1^\dagger \Box  H_1
\nonumber \\
&& \! \! \! + ~
H_1^\dagger  H_1
\bigg[
~\!
  g_2^2  H_1^\dagger  T_a T_b W_{a\mu} W_{b}^\mu   H_1
-g_2 \Big(i\partial_\mu  H_1^\dagger T_a W_a^\mu  H_1 
+\textrm{h.c.}
\Big)
\nonumber \\
&& \! \! \! + ~
\frac{1}{4}g_1^2  H_1^\dagger B_{\mu} B^\mu  H_1
-\frac{1}{2}g_1 \Big(i\partial_\mu  H_1^\dagger B^\mu  H_1 
+\textrm{h.c.}
\Big) 
~\!
\bigg]
~\!
\Bigg]+\textrm{h.c.}
\label{eq:diagram31}
\end{eqnarray}

%\begin{equation}
%\frac{ ( {\tilde{\lambda}}_6^* {\tilde{m}_{12}}^2 +\textrm{h.c.})}{{\tilde{m}_2}^2} ~\!( H_1^\dagger  H_1)^2
%\label{eq:diagram31}
%\end{equation}
%while the terms at quadratic order in the momentum expansion are
%\begin{equation}
%\bra T  H_1^\dagger  H_1 (p) H_1^\dagger  H_1 \ket \supset 2!2! \frac{i   {\tilde{\lambda}}_6 {\tilde{m}_{12}}^{2\dagger}}{{\tilde{m}_2}^4}p^2
%\end{equation}
%and the corresponding correlation function hermitic conjugate field combination, which lead to
%\begin{equation}
%-\frac{1}{{\tilde{m}_2}^4}(~\! {\tilde{\lambda}}_6 {\tilde{m}_{12}}^{2\dagger} ~\! H_1^\dagger ~\!\Box H_1+\textrm{h.c.})
%\label{eq:diagram32}
%\end{equation}
%We can also plug in gauge bosons in the internal propagator, leading to  
% \begin{equation}
%\frac{  {\tilde{\lambda}}_6^* {\tilde{m}_{12}}^2
%+\textrm{h.c.}}{2 {\tilde{m}_2}^4}  
%\bigg[~\!
%g^2  H_1^\dagger  T_a T_b W_{a\mu} W_{b}^\mu 
%H_1
%-g 
%\bigg(
%i\partial_\mu H_1^\dagger 
%T_aW_a^\mu  H_1 
%+\textrm{h.c.}
%\bigg)
%\bigg] H_1^\dagger  H_1
%\label{eq:diagram33}
%\end{equation}
Rearranging the derivatives
\begin{equation}
\frac{2{\tilde{\lambda}}_6^* {\tilde{m}_{12}}^2}{{\tilde{m}_2}^2}  ( H_1^\dagger  H_1)^2
 +
\frac{{\tilde{\lambda}}_6^* {\tilde{m}_{12}}^2}{{\tilde{m}_2}^4} 
\Bigg[
 \partial_\mu
 ( H_1^\dagger  H_1)
 \partial^\mu( H_1^\dagger  H_1)
 +2(D_\mu H_1)^\dagger (D^\mu H_1) H_1^\dagger  H_1
 ~\Bigg]
 \label{eq:diagram3}
 \end{equation}
where we used ${\tilde{\lambda}}_6^* {\tilde{m}_{12}}^2={\tilde{\lambda}}_6 {\tilde{m}_{12}}^{2*}$ due to the no tadpole condition \eqref{eq:notadpolecondition}. 

\item { \bf  Diagrams with ${\tilde{\lambda}}_3,{\tilde{\lambda}}_4$ and ${\tilde{\lambda}}_5$ insertions}
%\begin{figure}[h]
%\begin{center}
%%\psfrag{Y}[Bc]{\small $\cos(\theta)$ \normalsize}
%%\psfrag{X}[lc]{ \small $\kappa$  \normalsize}
%\includegraphics[width=14cm]{2hdmdiagram5.eps}
%\caption{Diagrams with a $\tilde{\lambda}_3$ vertex up to effective dimension eight. All diagrams are at zero momentum. There are similar diagrams replacing the $\tilde{\lambda}_3$ vertex with $\tilde{\lambda}_4$ and $\tilde{\lambda}_5$, which are also considered. The diagrams with a $\tilde{\lambda}_5$ vertex are slightly different. For instance, they contain two $\tilde{m}_{12}^2$ vertices instead of one $\tilde{m}_{12}^2$ and one $(\tilde{m}_{12}^2)^*$ vertex. The upper left diagram leads to the operator $( H_1^\dagger  H_1)^2$. The upper right diagram leads to the operator $( H_1^\dagger  H_1)^3$. The lower diagram leads to the operator $( H_1^\dagger  H_1)^4$.}
%\label{fig:etadiagrams}
%\end{center}
%\end{figure}

\begin{figure}[h]
\begin{center}
%\psfrag{Y}[Bc]{\small $\cos(\theta)$ \normalsize}
%\psfrag{X}[lc]{ \small $\kappa$  \normalsize}
\includegraphics[width=16cm]{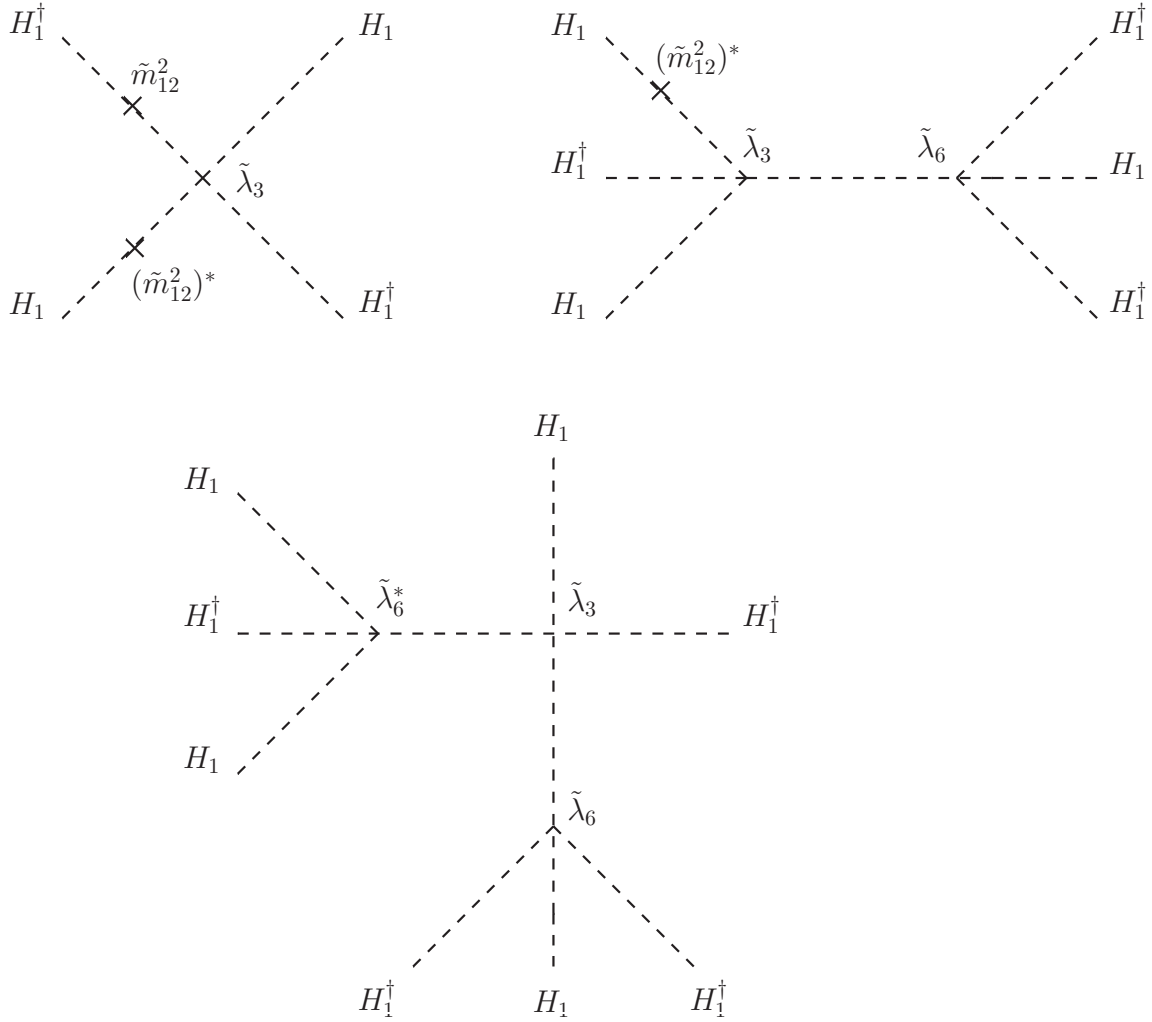}
\caption{Diagrams with a $\tilde{\lambda}_3$ vertex up to effective dimension eight. All diagrams are at zero momentum. There are similar diagrams replacing the $\tilde{\lambda}_3$ vertex with $\tilde{\lambda}_4$ and $\tilde{\lambda}_5$, which are also considered to derive the low energy EFT. The diagrams with a $\tilde{\lambda}_5$ vertex are slightly different, they contain two $\tilde{m}_{12}^2$ vertices instead of one $\tilde{m}_{12}^2$ and one $(\tilde{m}_{12}^2)^*$ vertex. The upper left diagram leads to the operator $( H_1^\dagger  H_1)^2$. The upper right diagram leads to the operator $( H_1^\dagger  H_1)^3$. The lower diagram leads to the operator $( H_1^\dagger  H_1)^4$.}
\label{fig:etadiagrams}
\end{center}
\end{figure}

\label{sec:etaoperators}
The diagrams of figure \ref{fig:etadiagrams} lead to
\begin{eqnarray}
&&
  -\frac{ {\tilde{\lambda}}_3 \abs{{\tilde{m}_{12}}^2}^2}{{\tilde{m}_2}^4} ( H_1^\dagger  H_1)^2
 -  \frac{2  {\tilde{\lambda}}_3  {\tilde{\lambda}}_6^* {\tilde{m}_{12}}^2 }{{\tilde{m}_2}^4} ( H_1^\dagger  H_1)^3 
 - \frac{ {\tilde{\lambda}}_3  {\tilde{\lambda}}_6^*  {\tilde{\lambda}}_6}{{\tilde{m}_2}^4}( H_1^\dagger  H_1)^4 
  \nonumber \\
&&\!\!\! -~ 
\frac{ {\tilde{\lambda}}_4 \abs{{\tilde{m}_{12}}^2}^2}{{\tilde{m}_2}^4} ( H_1^\dagger  H_1)^2
-  \frac{2  {\tilde{\lambda}}_4  {\tilde{\lambda}}_6^* {\tilde{m}_{12}}^2}{{\tilde{m}_2}^4} ( H_1^\dagger  H_1)^3 
- \frac{ {\tilde{\lambda}}_4  {\tilde{\lambda}}_6^*  {\tilde{\lambda}}_6}{{\tilde{m}_2}^4}( H_1^\dagger  H_1)^4  
\nonumber \\
&&\!\!\! -~ 
\frac{  ({\tilde{\lambda}}_5^* {\tilde{m}_{12}}^4+\textrm{h.c.})}{2{\tilde{m}_2}^4} ( H_1^\dagger  H_1)^2
-  \frac{  ({\tilde{\lambda}}_5^* {\tilde{\lambda}}_6 {\tilde{m}_{12}}^2+\textrm{h.c.})}{{\tilde{m}_2}^4} ( H_1^\dagger  H_1)^3 
- \frac{ ({\tilde{\lambda}}_5^* {\tilde{\lambda}}_6^2+\textrm{h.c.})}{2{\tilde{m}_2}^4}( H_1^\dagger  H_1)^4
\nonumber 
\\
\label{eq:diagram4}
\end{eqnarray} 
where we used ${\tilde{\lambda}}_6^* {\tilde{m}_{12}}^2={\tilde{\lambda}}_6 {\tilde{m}_{12}}^{2*}$ due to the no tadpole condition \eqref{eq:notadpolecondition}. 

\item  { \bf Diagrams with Yukawas and an $\tilde{m}_{12}^2$ or $\tilde{\lambda}_6$ insertion:}

\begin{figure}[h]
\begin{center}
%\psfrag{Y}[Bc]{\small $\cos(\theta)$ \normalsize}
%\psfrag{X}[lc]{ \small $\kappa$  \normalsize}
\includegraphics[width=16cm]{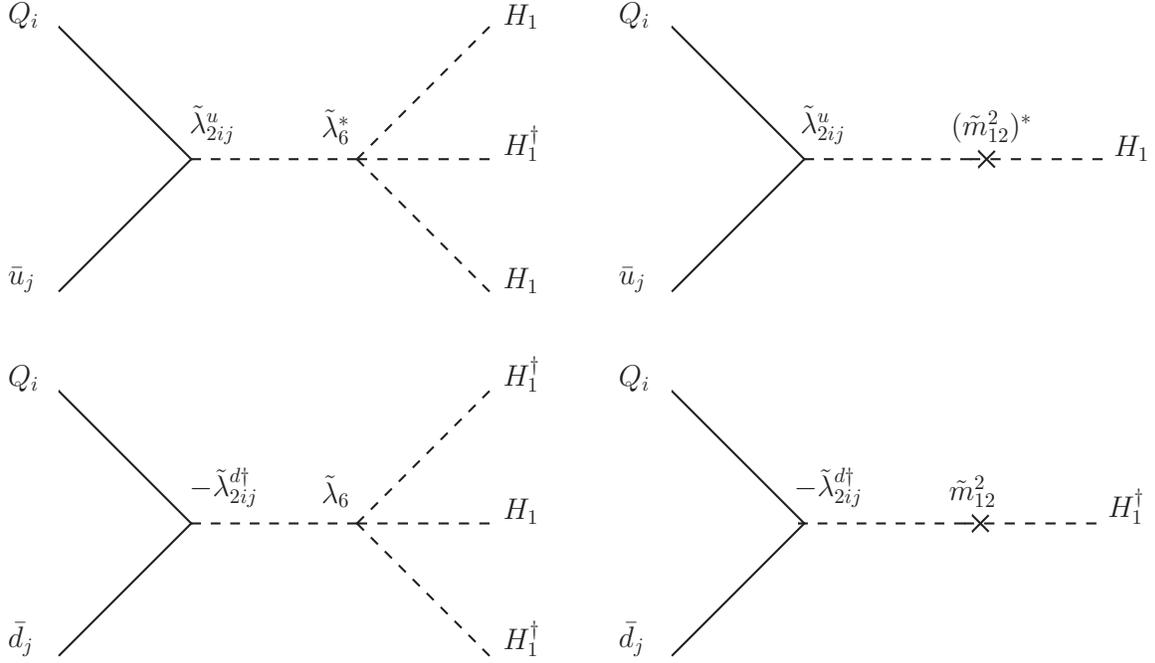}
\caption{Diagrams with one Yukawa and one $\tilde{\lambda}_6$ or $\tilde{m}_{12}^2$ coupling up to effective dimension six. All diagrams are at zero momentum. There are also complex conjugate versions of these diagrams and diagrams with leptonic Yukawas, which are also considered to derive the low energy EFT.
The upper left diagram leads to the operator $Q_i H_1 \bar{u}_j$. The upper right diagram leads to the operator $Q_i H_1 \bar{u}_j~\!H_1^\dagger H_1$. The lower left diagram leads to the operator $Q_i H_1^c \bar{d}_j$. The lower right diagram leads to the operator $Q_i H_1^c \bar{d}_j ~\!H_1^\dagger H_1$.}
\label{fig:2HDMyukup}
\end{center}
\end{figure}
%
%\begin{figure}[h]
%\begin{center}
%%\psfrag{Y}[Bc]{\small $\cos(\theta)$ \normalsize}
%%\psfrag{X}[lc]{ \small $\kappa$  \normalsize}
%\includegraphics[width=12cm]{2hdmYukawadown.eps}
%\caption{}
%\label{fig:2HDMyudown}
%\end{center}
%\end{figure}

The diagrams of figure \ref{fig:2HDMyukup} lead to %where $y_u$ and $y_d$ are SM Yukawas (for instance, $y_t v/\sqrt{2}=m_t$), lead to 
\begin{eqnarray}
&&
\frac{\tilde{\lambda}_{2ij}^u}{{\tilde{m}_2}^2}
Q_i H_1 \bar{u}_j
\bigg[({\tilde{m}_{12}}^2)^*
+ {\tilde{\lambda}}_6^*  H_1^\dagger  H_1
\bigg] 
-
\frac{\tilde{\lambda}_{2ij}^{d\dagger}}{{\tilde{m}_2}^2}
Q_i H_1^c \bar{d}_j
\bigg[{\tilde{m}_{12}}^2
+ {\tilde{\lambda}}_6  H_1^\dagger  H_1
\bigg] 
\nonumber \\
&&\!\!\!-~
\frac{\tilde{\lambda}_{2ij}^{\ell\dagger}}{{\tilde{m}_2}^2}
L_i H_1^c \bar{\ell}_j
\bigg[{\tilde{m}_{12}}^2
+ {\tilde{\lambda}}_6  H_1^\dagger  H_1
\bigg] 
+
\textrm{h.c.}
\label{eq:diagram5}
\end{eqnarray}
We do not present the operators at effective dimension 8 that modify the Yukawas. 

\item   {\bf Diagrams with Yukawas only (four fermion operators):}

The diagrams of figure \ref{fig:fourfermion} lead to

\begin{eqnarray}
&&
 \frac{\tilde{\lambda}_{2ij}^u \tilde{\lambda}_{2mn}^{u\dagger}}{{\tilde{m}_2}^2}  (Q_i \bar{u}_j)(\bar{u}_m^\dagger Q_n^\dagger)
+ \frac{\tilde{\lambda}^{d\dagger}_{2ij}\tilde{\lambda}^{d}_{2mn}}{{\tilde{m}_2}^2}  (Q_i \bar{d}_j)(\bar{d}_m^\dagger Q_n^\dagger)
+\frac{\tilde{\lambda}^{\ell\dagger}_{2ij}\tilde{\lambda}^{\ell}_{2mn}}{{\tilde{m}_2}^2}(L_i \bar{\ell}_j)(\bar{\ell}_m^\dagger L_n^\dagger) \nonumber \\
&& \! \! \! + ~
\left[ \frac{\tilde{\lambda}^{d\dagger}_{2ij}\tilde{\lambda}^{\ell}_{2mn}}{{\tilde{m}_2}^2} (Q_i \bar{d}_j)(\bar{\ell}_m^\dagger L_n^\dagger)
+\frac{\tilde{\lambda}_{2ij}^u \tilde{\lambda}^{d\dagger}_{2mn}}{{\tilde{m}_2}^2} (Q_i \bar{u}_j)(Q_m \bar{d}_n) 
+\frac{\tilde{\lambda}_{2ij}^u \tilde{\lambda}^{\ell\dagger}_{2mn}}{{\tilde{m}_2}^2}(Q_i \bar{u}_j)(L_m \bar{\ell}_n)+\textrm{h.c.} \right] \nonumber \\
\label{eq:fourfermionderiv}
\end{eqnarray}

\begin{figure}[h]
\begin{center}
%\psfrag{Y}[Bc]{\small $\cos(\theta)$ \normalsize}
%\psfrag{X}[lc]{ \small $\kappa$  \normalsize}
\includegraphics[width=14cm]{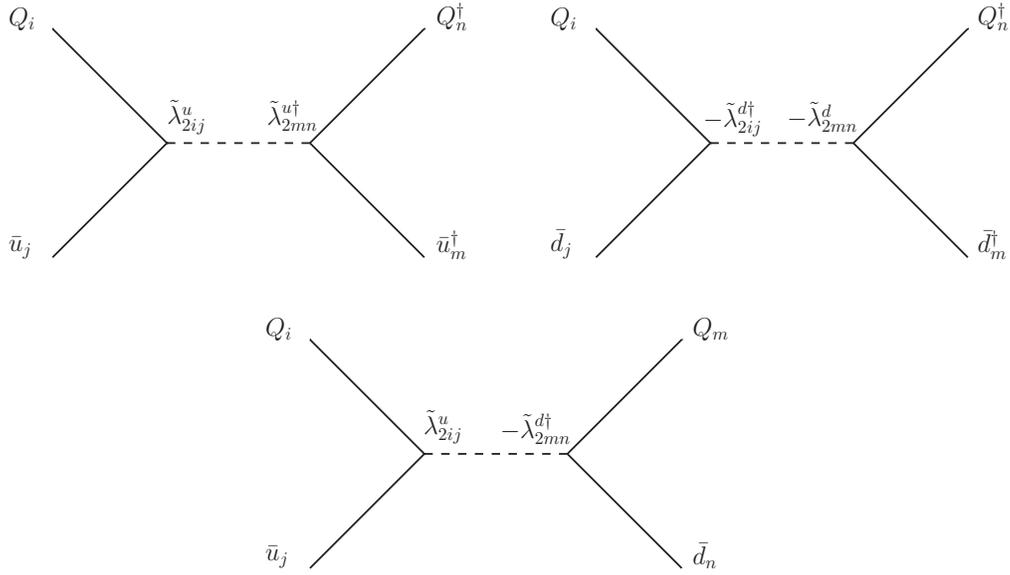}
\caption{Diagrams with two Yukawas up to effective dimension six.  The lower diagram also has a complex conjugate version. There are also diagrams with leptonic Yukawas. The upper left diagram leads to the operator $(Q_i\bar{u}_j)(\bar{u}_m^\dagger Q_n^\dagger)$. The upper right diagram leads to the operator $(Q_i\bar{d}_j)(\bar{d}_m^\dagger Q_n^\dagger)$. The lower diagram leads to the operator $-(Q_i\bar{u}_j)(Q_i\bar{d}_j)$, where the minus sign comes from reordering the contraction of fields charged under $SU(2)$.}
\label{fig:fourfermion}
\end{center}
\end{figure}
\end{itemize}

\FloatBarrier
\subsection{The low energy theory of the Higgs doublet}

Collecting the results \eqref{eq:diagram1}, \eqref{eq:diagram2}, \eqref{eq:diagram3}, \eqref{eq:diagram4}, \eqref{eq:diagram5} and \eqref{eq:fourfermionderiv} we are ready to write down the low energy theory. For the remainder of this paper we drop the subscripts in the doublet of the low energy theory $H_1$ and its neutral component $h_1$. The effective Lagrangian density for the Higgs doublet is
\begin{eqnarray}
  {\cal L}\! \! \!&=&\! \! \! Z_H ~ \! D_\mu H^\dagger  D^\mu H 
  +\zeta_H
  \bigg[~ \! \frac{1}{2} \partial_\mu (H^\dagger H)\partial^\mu (H^\dagger H)+H^\dagger  H~\! D_\mu H^\dagger D^\mu H ~ \!\bigg] 
  \nonumber \\
&& \! \! \! + ~
 \zeta'_H
 \bigg[~ \! 2H^\dagger  H ~\! \partial_\mu( H^\dagger  H)\partial^\mu( H^\dagger  H)
 + ( H^\dagger  H)^2~\!D_\mu  H^\dagger D^\mu  H  ~ \! \bigg]
 -V(H) 
 \nonumber  \\
&& \! \! \!-~ \bigg[
Q_i H \big( \lambda^u_{ij} 
+\eta^{u }_{ij} H^\dagger  H \big) 
\bar{u}_j 
-Q_i H^c \big(  \lambda^{d\dagger}_{ij}  
+\eta^{d\dagger}_{ij}H^\dagger H \big)
\bar{d}_j
%\nonumber \\
%&& \! \! \! 
%~~~~~ - 
-L_i H^c \big( \lambda^{\ell\dagger}_{ij}  
+\eta^{\ell\dagger}_{ij}H^\dagger H \big)
\bar{\ell}_j 
+\textrm{h.c.} \bigg] \nonumber \\
 \label{eq:2HDMsummary}
\end{eqnarray}
\begin{equation}
V(H)=m_H^2 H^\dagger  H
+\frac{1}{2} \lambda_H ( H^\dagger  H)^2
+\frac{1}{3}\eta_6( H^\dagger  H)^3
+\frac{1}{4}\eta_8( H^\dagger  H)^4 
\end{equation}
where

\begin{eqnarray}
  \zeta'_H\! \! \!&=&\! \! \! 
  \frac{{\tilde{\lambda}}_6^* {\tilde{\lambda}}_6  }{\tilde{m}_2^4} 
  \nonumber \\
  Z_H\! \! \!&=& \! \! \!
  1+\Delta_{Z_H}
  =1+  \frac{\abs{{\tilde{m}_{12}}^2}^2}{{\tilde{m}_2}^4}
  \nonumber \\
  \zeta_H\! \! \!&=&\! \! \!
  \frac{2 {\tilde{m}_{12}}^2  {\tilde{\lambda}}_6^*}{{\tilde{m}_2}^4}
  \nonumber \\
    \Delta_{Z_H} \! \! \!&=& \! \! \! \frac{1}{4}\zeta_H'v^4=-\frac{1}{4}\zeta_H v^2
  \label{eq:parameterrelations2HDM}
  \end{eqnarray}
  \begin{eqnarray}
  m_H^2\! \! \!&=&\! \! \!
  {\tilde{m}_1}^2
  - \frac{\abs{{\tilde{m}_{12}}^2}^2}{{\tilde{m}_2}^2}
  \nonumber \\
  \lambda_H\! \! \!&=&\! \! \!
 {\tilde{\lambda}}_1
  -\frac{4{\tilde{m}_{12}}^2  {\tilde{\lambda}}_6^*}{{\tilde{m}_2}^2}
  +\frac{2 ({\tilde{\lambda}}_3 +{\tilde{\lambda}}_4 ) \abs{{\tilde{m}_{12}}^2}^2}{{\tilde{m}_2}^4}
  +\frac{ ({\tilde{\lambda}}_5^* {\tilde{m}_{12}}^4
  +\textrm{h.c.})}{{\tilde{m}_2}^4}  \nonumber \\
  \eta_6\! \! \!&=&\! \! \! 
  -\frac{3 {\tilde{\lambda}}_6^\dagger {\tilde{\lambda}}_6}{{\tilde{m}_2}^2}
  + \frac{6 ( {\tilde{\lambda}}_3 + {\tilde{\lambda}}_4){\tilde{m}_{12}}^2  {\tilde{\lambda}}_6^*}{{\tilde{m}_2}^4} 
  +\frac{3 ({\tilde{\lambda}}_5^* {\tilde{\lambda}}_6 {\tilde{m}_{12}}^2+\textrm{h.c})}{{\tilde{m}_2}^4}  
  \nonumber \\
 \eta_8\! \! \!&=&\! \! \!
\frac{4({\tilde{\lambda}}_3 +{\tilde{\lambda}}_4 ) {\tilde{\lambda}}_6^* {\tilde{\lambda}}_6}{{\tilde{m}_2}^4} 
 + \frac{2({\tilde{\lambda}}_5^* {\tilde{\lambda}}_6^2+\textrm{h.c.})}{{\tilde{m}_2}^4}
\label{eq:coefficients2HDMEFT}
\end{eqnarray}
\begin{eqnarray}
\nonumber \eta^f_{ij}
\! \! \!&=&\! \! \!
-\frac{\tilde{\lambda}_{2ij}^f {\tilde{\lambda}}_6^*}
{{\tilde{m}_2}^2}  
\\
\nonumber \lambda^f_{ij}
\! \! \!&=&\! \! \!
\tilde{\lambda}_{1ij}^f
- \frac{\tilde{\lambda}_{2ij}^f \tilde{m}_{12}^{*2}}{{\tilde{m}_2}^2}
%\\
%\nonumber
%\! \! \!&=&\! \! \!
%\frac{\sqrt{2}m^f_{ij} }{v}
%+
%\frac{1}{2} 
%\tilde{\lambda}_{2ij}^f 
%{\tilde{\lambda}_6}^*
%~\!
%\frac{v^2}{\tilde{m}_2^2}
%\\
%\! \! \!&=&\! \! \!
%\frac{\sqrt{2}m^f_{ij} }{v}
%-
%\frac{1}{2}\eta^f_{ij} v^2
\quad \quad \quad \quad
f=u,d,\ell 
  \label{eq:coefficients2HDMEFT2}
\end{eqnarray}
Note that due to the no tadpole condition \eqref{eq:notadpolecondition}, the operator coefficients $\Delta_{Z_H}$, $\zeta_H'$ and $\zeta_H$ are not independent, as we see in the last lines of \eqref{eq:parameterrelations2HDM}. Note also that in the effective theory there are no threshold corrections to the Higgs kinetic Lagrangian at effective dimension six, the first corrections are of effective dimension eight. The Higgs potential is modified at effective dimension six. The EFT contains cubic Yukawa operators at effective dimension six, for all types of SM fermions. We do not present operators with naive dimension six and eight involving the Higgs and gauge field strength tensors since they come at effective dimension higher than eight. We also drop all operator dimension six Higgs-fermion interactions with derivatives since they come at effective dimension eight. Note that T violation in purely bosonic operators (operators involving no SM fermions) is irrelevant to the effective dimension we work to; all  the bosonic operators in \eqref{eq:2HDMsummary} are T conserving. Up to the order we work to, T violation only arises in the Higgs-fermion interactions. 
The fermion mass matrices are given by the linear and cubic Yukawas by
  \begin{equation}
\frac{\lambda^f_{ij}v}{\sqrt{2}}+\frac{\eta_{ij}^f v^3}{2\sqrt{2}}=m_{ij}^f
\label{eq:massandeta}
\end{equation}

%The operators on the first two lines in \eqref{eq:2HDMsummary}, which besides the canonical kinetic term are all of effective dimension 8, lead to wave function renormalization of the physical Higgs and modifications to the Higgs couplings. The rest of the operators correspond to modifications of the Higgs potential at effective dimension six and eight, and they affect the Higgs mass and self couplings. 

%The Higgs-fermion interaction effective Lagrangian is 
%\begin{equation}
%-{\cal L}_Y= Q_i H \left( \lambda^u_{ij} +\eta^{u }_{ij} H^\dagger  H \right) \bar{u}_j -Q_i H^c \left(  y^{d\dagger}_{ij}  +\eta^{d\dagger}_{ij}H^\dagger H \right)\bar{d}_j-L_i H^c \left( y^{\ell\dagger}_{ij}  +\eta^{\ell\dagger}_{ij}H^\dagger H \right)\bar{\ell}_j +\textrm{h.c.}
%\end{equation}
%%\begin{equation}
%%-{\cal L}_Y= \bar{Q}_L \tilde{H} \left(y_U + \eta_U H^\dagger  H \right) u_R +\bar{Q}_L  H \left(y_D^\dagger +\eta_D^\dagger  H^\dagger  H\right)d_R+h.c. 
%%\end{equation}
%where
%\begin{eqnarray}
%y^f=\kappa^f- \frac{\rho^f {\tilde{m}_{12}}^{2\dagger}}{{\tilde{m}_2}^2} \quad , \quad \eta^f=-\frac{\rho^f {\tilde{\lambda}}_6^*}{{\tilde{m}_2}^2}  \quad , \quad f=u,d,l
%\end{eqnarray}
%%\begin{eqnarray}
%%y_F=\kappa_F- \frac{\rho_F{\tilde{m}_{12}}^2}{{\tilde{m}_2}^2} \quad , \quad \eta_F=-\frac{\rho_F{\tilde{\lambda}}_6}{{\tilde{m}_2}^2}  \quad , \quad F=U,D,L
%%\end{eqnarray}

%The mass matrices are given by
%\begin{equation}
%M_F=y_F \left(\frac{v}{\sqrt{2}}\right)+\eta_F \left(\frac{v}{\sqrt{2}}\right)^3  \quad , \quad F=U,D,L
%\end{equation} 

The four fermion operators of the low energy theory are given by \eqref{eq:fourfermionderiv}. Here we write them as
\begin{eqnarray}
&& \Omega^{u u ~\!\! (0)}_{ijmn }(Q_i \bar{u}_j)(\bar{u}_m^\dagger Q_n^\dagger)
+\Omega^{dd~\!\! (0)}_{ijmn }(Q_i \bar{d}_j)(\bar{d}_m^\dagger Q_n^\dagger)
+\Omega^{\ell\ell ~\!\! (0)}_{ijmn }(L_i \bar{\ell}_j)(\bar{\ell}_m^\dagger L_n^\dagger) \nonumber \\
&& \! \! \! + ~
\left[ \Omega^{d\ell ~\!\! (0)}_{ijmn }(Q_i \bar{d}_j)(\bar{\ell}_m^\dagger L_n^\dagger)
+\Omega^{u d ~\!\! (2)}_{ijmn }(Q_i \bar{u}_j)(Q_m \bar{d}_n) 
+\Omega^{u \ell ~\!\! (2)}_{ijmn }(Q_i \bar{u}_j)(L_m \bar{\ell}_n)+\textrm{h.c.} \right] \nonumber \\
\label{eq:fourfermion}
\end{eqnarray}
or, expanding in components of the fermion doublets
\begin{eqnarray}
&& \Omega^{uu ~\!\! (0)}_{ijmn }
\bigg[
(u_i \bar{u}_j)(\bar{u}_m^\dagger u_n^\dagger)
+(d_i \bar{u}_j)(\bar{u}_m^\dagger d_n^\dagger) 
\bigg]
+
\Omega^{dd~\!\! (0)}_{ijmn }
\bigg[
(u_i \bar{d}_j)(\bar{d}_m^\dagger u_n^\dagger)
+
(d_i \bar{d}_j)(\bar{d}_m^\dagger d_n^\dagger)
\bigg]
\nonumber \\
&&\! \! \! + ~
\Omega^{\ell\ell ~\!\! (0)}_{ijmn }
\bigg[ 
(\nu_i \bar{\ell}_j)(\bar{\ell}_m^\dagger \nu_n^\dagger) 
+
(\ell_i \bar{\ell}_j)(\bar{\ell}_m^\dagger \ell_n^\dagger) 
\bigg]
\nonumber \\
&&\! \! \! + ~
\Bigg[ ~
\Omega^{d\ell ~\!\! (0)}_{ijmn }
\bigg[
(u_i \bar{d}_j)(\bar{\ell}_m^\dagger \nu_n^\dagger)
+
(d_i \bar{d}_j)(\bar{\ell}_m^\dagger \ell_n^\dagger)
\bigg]
+
\Omega^{u d ~\!\! (2)}_{ijmn }
\bigg[
(u_i \bar{u}_j)(d_m \bar{d}_n) 
-
(d_i \bar{u}_j)(u_m \bar{d}_n) 
\bigg]
\nonumber \\
&&\! \! \! + ~
\Omega^{u \ell ~\!\! (2)}_{ijmn }
\bigg[
(u_i \bar{u}_j)(\ell_m \bar{\ell}_n)
-(d_i \bar{u}_j)(\nu_m \bar{\ell}_n)
\bigg]
+\textrm{h.c.} 
~
\Bigg] 
\label{eq:fourfermiontype}
\end{eqnarray}

These are all the four fermion operators that are obtained at effective dimension six. The superindices $(0)$ and $(2)$ represent operators that violate chirality by zero or two units respectively. The labels $uu, dd, \ell\ell, d\ell, ud, u\ell$ label the right handed quarks appearing in the four fermion operator. The coefficients of the operators are
%Note that they are also all the gauge invariant four fermion operators that can be written at effective dimension six (\textit{I take this back, what about $(Q_i Q_j) (Q_k Q_m)^\dagger$? It's not generated here, but it is gauge invariant. }).
\begin{eqnarray}
&& \! \! \!  ~ 
\Omega^{uu~\!\! (0)}_{ijmn }  
= \frac{\tilde{\lambda}_{2ij}^u \tilde{\lambda}_{2mn}^{u\dagger}}{{\tilde{m}_2}^2} 
 ~~~, ~~~ 
 \Omega^{dd ~\!\! (0)}_{ijmn } 
 =  \frac{\tilde{\lambda}^{d\dagger}_{2ij}\tilde{\lambda}^{d}_{2mn}}{{\tilde{m}_2}^2} 
 ~~~, ~~~ 
 \Omega^{\ell\ell ~\!\! (0)}_{ijmn } 
 =  \frac{\tilde{\lambda}^{\ell\dagger}_{2ij}\tilde{\lambda}^{\ell}_{2mn}}{{\tilde{m}_2}^2}
 \nonumber \\
&& \! \! \!  ~ 
\Omega^{d\ell ~\!\! (0)}_{ijmn}
=\frac{\tilde{\lambda}^{d\dagger}_{2ij}\tilde{\lambda}^{\ell}_{2mn}}{{\tilde{m}_2}^2} 
~~~, ~~~
\Omega^{u d ~\!\! (2)}_{ijmn}
=\frac{\tilde{\lambda}_{2ij}^u \tilde{\lambda}^{d\dagger}_{2mn}}{{\tilde{m}_2}^2} 
~~~, ~~~
\Omega^{u \ell ~\!\! (2)}_{ijmn}
=\frac{\tilde{\lambda}_{2ij}^u \tilde{\lambda}^{\ell\dagger}_{2mn}}{{\tilde{m}_2}^2}
\label{eq:fourfermionoperators}
\end{eqnarray}
For general $\tilde{\lambda}_{2ij}^f$ matrices, these four fermion interactions are flavor and T violating.
%We summarize the effective dimensions at which the Higgs sector is modified in table \ref{tab:summaryED}.
%\begin{table}[h]
%$\begin{array}{ccc} 
%  & \textrm{Is modified at effective dimension:}   \quad & \quad \textrm{Order we work to:}   \\
%  &     & \\
%    \textrm{Higgs kinetic Lagrangian} &  \geq 8  & 8 \\
%        \textrm{Higgs-gauge boson interactions} &  \geq 8  & 8 \\
% \textrm{Higgs potential} & \geq  6  & 8 \\
%   \textrm{Higgs-fermion interactions *} & \geq 6 & 6   \\
%\textrm{Four-fermion interactions *} & \geq 6  & 6  \\
%\end{array}$
%\caption{Summary of effective dimensions. * indicates T and flavor violating terms starting at effective dimension 6.}
%\label{tab:summaryED}
%\end{table}

\subsection{The Low energy theory of the Higgs boson}
\label{sec:physicalHiggs2HDMEFT}
We now write down the effective theory for the neutral component of the Higgs doublet in the unitary gauge. The no tadpole condition \eqref{eq:notadpolecondition} will be used extensively in the rest of the discussion. Writing the Lagrangian density \eqref{eq:2HDMsummary} in terms of $h$ we get
\begin{eqnarray}
&& Z_H \Bigg[ \frac{1}{2}  \partial_\mu h \partial^\mu h  +\frac{1}{2}  m_Z^2
    Z^{\mu} Z_\mu  \bigg( 1 + {2 h \over v} + {h^2 \over v^2} ~ \! 
    \bigg) 
  +
     m_W^2 
   W^{+ \mu} W_{\mu}^-   \bigg(  1 + {2 h \over v} + {h^2 \over v^2} ~ \! 
     \bigg)  \Bigg]  \notag \\
  && + ~ \zeta_H\Bigg[  
   \frac{1}{2} \partial_\mu h \partial^\mu h  \bigg(\frac{3 }{2}v^2+3 v h+\frac{3 }{2}h^2\bigg) +
 \notag \\
 && +~ 
  \bigg(\frac{1}{2}m_Z^2 Z^\mu Z_\mu+m_W^2 W^{+\mu}W^-_\mu  \bigg)
  \left(\frac{1}{2} v^2 
 +  2 v h 
 + 3  h^2 
 +   \frac{2h^3}{v} 
 + \frac{1}{2}\frac{h^4}{v^2} \right)
   \Bigg] 
   \notag \\
   && +~  \zeta'_H\Bigg[ 
   \frac{1}{2}  \partial_\mu h \partial^\mu h  
   \bigg(\frac{9}{4}v^4+9 v^3 h+\frac{27}{2} v^2 h^2+9 v h^3+\frac{9 }{4}h^4\bigg) 
     \notag \\
    && +~   \bigg(\frac{1}{2}m_Z^2 Z^\mu Z_\mu+m_W^2 W^{+\mu}W^-_\mu  \bigg)
     \left( \frac{1}{4} v^4
   +  \frac{3}{2}  v^3 h
   + \frac{15}{4} v^2 h^2  
   + 5    v h^3
   + \frac{15}{4}   h^4 
   + \frac{3}{2}    \frac{h^5}{v} 
   +\frac{1}{4} \frac{h^6 }{v^2} 
 \right) 
    \Bigg] \notag \\
   &&  - ~ V(h) 
   - \Bigg[ u_i\bar{u}_j  \bigg[m^u_{ij}
+\bigg(\frac{m^u_{ij}}{v}+2\bigg(\frac{\eta^u_{ij} v^2}{2\sqrt{2}}\bigg)
\bigg)h
+3\bigg(\frac{\eta_{ij}^u v}{2\sqrt{2}}\bigg)h^2
+\bigg(\frac{\eta_{ij}^u}{2\sqrt{2}}\bigg)h^3
\bigg]  
\nonumber \\
   &&-~ 
   d_i\bar{d}_j  
   \bigg[m^{d\dagger}_{ij}+
   \bigg(\frac{m^{d\dagger}_{ij}}{v}
   +2\bigg(\frac{\eta^{d\dagger}_{ij} v^2}{2\sqrt{2}}\bigg)\bigg)h
   +3\bigg(\frac{\eta_{ij}^{d\dagger} v}{2\sqrt{2}}\bigg)h^2
   +\bigg(\frac{\eta_{ij}^{d\dagger}}{2\sqrt{2}}\bigg)h^3
   \bigg]
    \nonumber \\
    &&- ~ 
    \ell_i\bar{\ell}_j  \bigg[
    m^{\ell\dagger}_{ij}+
    \bigg(\frac{m^{\ell\dagger}_{ij}}{v}
    +2\bigg(\frac{\eta^{\ell\dagger}_{ij} v^2}{2\sqrt{2}}\bigg)\bigg)h
    +3\bigg(\frac{\eta_{ij}^{\ell\dagger} v}{2\sqrt{2}}\bigg)h^2
    +\bigg(\frac{\eta_{ij}^{\ell\dagger}}{2\sqrt{2}}\bigg)h^3
    \bigg]
    +\textrm{h.c.} \Bigg]  
    \label{eq:intermediatestep0}
  \end{eqnarray}
  %  the no tadpole condition \eqref{eq:notadpolecondition} and
The cubic Yukawa for the doublet contributes a term $3\eta^f_{ij} v^2/(2\sqrt{2})$ to the linear Yukawas of the Higgs $f_i \bar{f}_j h$, but one factor of $\eta^f_{ij} v^2/(2\sqrt{2})$ is already included in the definition of the fermion mass matrix $m^{f}_{ij}/v$, Eq. \eqref{eq:massandeta}. The Higgs potential is given by
  \begin{eqnarray}
  V(h)\! \! \! &=& \! \! \! 
  {1 \over 2}~ \!  m_H^2    
  %\left(  2 vh+h^2  \right)
     h^2
           +  {1 \over 2} ~ \! \lambda_H
        \bigg( %v^3 h
    ~ \!  \frac{3}{2}v^2h^2
     +vh^3+\frac{1}{4}h^4    \bigg)
      \nonumber \\  
&& \! \! \! 
    + ~ {1 \over 3} ~ \!  \eta_6 %\frac{\eta_6}{3}  
       \bigg(  %\frac{3  v^5}{4}h
    ~ \!   \frac{15 v^4}{8}h^2
    +\frac{5 v^3}{2}h^3+\frac{15v^2}{8}h^4
    +\frac{3v}{4}h^5+\frac{1}{8}h^6    \bigg)
    %\nonumber  
    \\  
&& \! \! \! + ~ {1 \over 4}~ \!  \eta_8
     \bigg(   %\frac{ v^7}{2}h
     ~ \!   \frac{7 v^6}{4}h^2
  +\frac{7 v^5}{2}h^3
  +\frac{35 v^4}{8}h^4 
  +\frac{7 v^3}{2} h^5
  +\frac{7 v^2}{4}h^6 
  +\frac{ v}{2}h^7
  +\frac{1}{16}h^8   \bigg) 
      \nonumber 
\end{eqnarray}

The irrelevant kinetic operators in \eqref{eq:intermediatestep0} can be replaced in favor of operators with no derivatives using integration by parts and equations of motion. We first use integration by parts to rewrite the kinetic terms proportional to $\zeta_H$  as
\begin{eqnarray}
&& 
\frac{1}{2} \partial_\mu (H^\dagger H)\partial^\mu (H^\dagger H)+H^\dagger  H~\! D_\mu H^\dagger D^\mu H 
  \nonumber \\
&& =  
   \frac{3}{4}  v^2~\! \partial_\mu h \partial^\mu h 
   + \frac{3}{2}v h ~\! \partial_\mu h \partial^\mu h 
   + \frac{3}{4}  h^2~\! \partial_\mu h \partial^\mu h  +
 \notag \\
 && +~ 
  \bigg(\frac{1}{2}m_Z^2 Z^\mu Z_\mu+m_W^2 W^{+\mu}W^-_\mu  \bigg)
  \left(\frac{1}{2} v^2 
 +  2 v h 
 + 3  h^2 
 +   \frac{2h^3}{v} 
 + \frac{1}{2}\frac{h^4}{v^2} \right)
   \notag \\
    && =
   \frac{3}{4} \partial_\mu h \partial^\mu h~\! v^2
   - \frac{3}{4} v h^2 ~\! \Box h
   - \frac{1}{4} h^3 ~\! \Box h+
 \notag \\
 && +~ 
  \bigg(\frac{1}{2}m_Z^2 Z^\mu Z_\mu
  +m_W^2 W^{+\mu}W^-_\mu  \bigg)
  \left(\frac{1}{2} v^2 
 +  2 v h 
 + 3  h^2 
 +   \frac{2h^3}{v} 
 + \frac{1}{2}\frac{h^4}{v^2} \right)
    \label{eq:intermediatestepkineticLagrangian1}
\end{eqnarray}
We also use integration by parts to rewrite the kinetic terms proportional to $\zeta_H'$,
\begin{eqnarray}
&& 
2H^\dagger  H ~\! \partial_\mu( H^\dagger  H)\partial^\mu( H^\dagger  H)
 + ( H^\dagger  H)^2~\!D_\mu  H^\dagger D^\mu  H  \notag  \\
&& =
   \frac{9}{8}  v^4~\! \partial_\mu h \partial^\mu h 
   +\frac{9}{2}  v^3 h ~\! \partial_\mu h \partial^\mu h 
   +   \frac{27}{4}v^2 h^2~\! \partial_\mu h \partial^\mu h 
   +\frac{9}{2} v h^3 ~\! \partial_\mu h \partial^\mu h 
   +\frac{9 }{8} h^4 ~\! \partial_\mu h \partial^\mu h 
     \notag \\
    && +~   \bigg(\frac{1}{2}m_Z^2 Z^\mu Z_\mu+m_W^2 W^{+\mu}W^-_\mu  \bigg)
     \left( \frac{1}{4} v^4
   +  \frac{3}{2}  v^3 h
   + \frac{15}{4} v^2 h^2  
   + 5    v h^3
   + \frac{15}{4}   h^4 
   + \frac{3}{2}    \frac{h^5}{v} 
   +\frac{1}{4} \frac{h^6 }{v^2} 
 \right) 
 \notag \\
    %%%%%
    && =
   \frac{9}{8}  \partial_\mu h \partial^\mu h ~\! v^4
   -\frac{9}{4}   v^3 h^2 ~\! \Box h
   -   \frac{9}{4}  v^2 h^3 ~\! \Box h
   - \frac{9}{8} v h^4 ~\! \Box h
   -\frac{9 }{40} h^5 ~\! \Box h
     \notag \\
    && +~   \bigg(\frac{1}{2}m_Z^2 Z^\mu Z_\mu+m_W^2 W^{+\mu}W^-_\mu  \bigg)
     \left( \frac{1}{4} v^4
   +  \frac{3}{2}  v^3 h
   + \frac{15}{4} v^2 h^2  
   + 5    v h^3
   + \frac{15}{4}   h^4 
   + \frac{3}{2}    \frac{h^5}{v} 
   +\frac{1}{4} \frac{h^6 }{v^2} 
 \right) 
\ \notag \\
    \label{eq:intermediatestepkineticLagrangian2}
\end{eqnarray}
The lowest order (effective dimension four) equation of motion for the Higgs field is
\begin{eqnarray}
\Box h \! \! \! &=& \! \! \!   -m_H^2 h
-\frac{1}{2}\lambda_H (3v^2 h
+3 v h^2
+h^3) \nonumber \\  
&& \! \! \!  + ~ \bigg(\frac{m_Z^2}{v}Z^\mu Z_\mu
+\frac{2m_W^2}{v}W^{+ \mu} W_{\mu}^- \bigg) \bigg(1+\frac{h}{v} \bigg)
\nonumber \\  
&& \! \! \! 
 -\bigg(\frac{m^u_{ij}}{v} u_i \bar{u}_j +\frac{m^{d \dagger}_{ij}}{v} d_i \bar{d}_j +\frac{m^{\ell \dagger}_{ij}}{v} \ell_i \bar{\ell}_j+\textrm{h.c.}\bigg) +\dots
\label{eq:boxh}
\end{eqnarray}
where the dots represent terms of higher effective dimension. Using \eqref{eq:boxh} in \eqref{eq:intermediatestepkineticLagrangian1} and in  \eqref{eq:intermediatestepkineticLagrangian2}, we can rewrite the Higgs Lagrangian \eqref{eq:intermediatestep0} with no irrelevant operators with derivatives
\begin{eqnarray}
&& Z_H \Bigg[ \frac{1}{2}  \partial_\mu h \partial^\mu h  +\frac{1}{2}  m_Z^2
    \bigg( 1 + {2 h \over v} + {h^2 \over v^2} ~ \! 
    \bigg)  Z^{\mu} Z_\mu 
  +
     m_W^2 
     \bigg(  1 + {2 h \over v} + {h^2 \over v^2} ~ \! 
     \bigg)  W^{+ \mu} W_{\mu}^-\Bigg]  \notag \\
&& + ~ \zeta_H\Bigg[  
   \frac{3}{4} \partial_\mu h \partial^\mu h~\! v^2
   - \bigg(\frac{3}{4} v h^2+\frac{1}{4} h^3 \bigg) ~\! \bigg[ -m_H^2 h
-\frac{1}{2}\lambda_H (3v^2 h
+3 v h^2
+h^3) \nonumber \\  
&& \! \! \!  + ~ \bigg(\frac{m_Z^2}{v}Z^\mu Z_\mu
+\frac{2m_W^2}{v}W^{+ \mu} W_{\mu}^- \bigg) \bigg(1+\frac{h}{v} \bigg)
 -\bigg(
\frac{m^u_{ij}}{v} u_i \bar{u}_j +\frac{m^{d \dagger}_{ij}}{v} d_i \bar{d}_j +\frac{m^{\ell \dagger}_{ij}}{v} \ell_i \bar{\ell}_j+\textrm{h.c.}
 \bigg) \bigg]
 \notag \\
 && +~ 
  \bigg(\frac{1}{2}m_Z^2 Z^\mu Z_\mu
  +m_W^2 W^{+\mu}W^-_\mu  \bigg)
  \left(\frac{1}{2} v^2 
 +  2 v h 
 + 3  h^2 
 +   \frac{2h^3}{v} 
 + \frac{1}{2}\frac{h^4}{v^2} \right)
   \Bigg] 
   \notag \\
   && +~  \zeta'_H\Bigg[ 
   \frac{9}{8}  \partial_\mu h \partial^\mu h ~\! v^4
   -\bigg( \frac{9}{4}   v^3 h^2 
    +  \frac{9}{4}  v^2 h^3 
   + \frac{9}{8} v h^4 
   +\frac{9 }{40} h^5 
   \bigg) 
   \bigg[ -m_H^2 h
-\frac{1}{2}\lambda_H (3v^2 h
+3 v h^2
+h^3) \nonumber \\  
&& \! \! \!  + ~ \bigg(\frac{m_Z^2}{v}Z^\mu Z_\mu
+\frac{2m_W^2}{v}W^{+ \mu} W_{\mu}^- \bigg) \bigg(1+\frac{h}{v} \bigg)
 -\bigg(
\frac{m^u_{ij}}{v} u_i \bar{u}_j +\frac{m^{d \dagger}_{ij}}{v} d_i \bar{d}_j +\frac{m^{\ell \dagger}_{ij}}{v} \ell_i \bar{\ell}_j+\textrm{h.c.}
 \bigg) \bigg]
     \notag \\
    && +~   \bigg(\frac{1}{2}m_Z^2 Z^\mu Z_\mu+m_W^2 W^{+\mu}W^-_\mu  \bigg)
     \left( \frac{1}{4} v^4
   +  \frac{3}{2}  v^3 h
   + \frac{15}{4} v^2 h^2  
   + 5    v h^3
   + \frac{15}{4}   h^4 
   + \frac{3}{2}    \frac{h^5}{v} 
   +\frac{1}{4} \frac{h^6 }{v^2} 
 \right) 
    \Bigg] \notag \\
     &&  - ~ V(h) 
   - \Bigg[ u_i\bar{u}_j  \bigg[m^u_{ij}
+\bigg(\frac{m^u_{ij}}{v}+2\bigg(\frac{\eta^u_{ij} v^2}{2\sqrt{2}}\bigg)
\bigg)h
+3\bigg(\frac{\eta_{ij}^u v}{2\sqrt{2}}\bigg)h^2
+\bigg(\frac{\eta_{ij}^u}{2\sqrt{2}}\bigg)h^3
\bigg]  
\nonumber \\
   &&-~ 
   d_i\bar{d}_j  
   \bigg[m^{d\dagger}_{ij}+
   \bigg(\frac{m^{d\dagger}_{ij}}{v}
   +2\bigg(\frac{\eta^{d\dagger}_{ij} v^2}{2\sqrt{2}}\bigg)\bigg)h
   +3\bigg(\frac{\eta_{ij}^{d\dagger} v}{2\sqrt{2}}\bigg)h^2
   +\bigg(\frac{\eta_{ij}^{d\dagger}}{2\sqrt{2}}\bigg)h^3
   \bigg]
    \nonumber \\
    &&- ~ 
    \ell_i\bar{\ell}_j  \bigg[
    m^{\ell\dagger}_{ij}+
    \bigg(\frac{m^{\ell\dagger}_{ij}}{v}
    +2\bigg(\frac{\eta^{\ell\dagger}_{ij} v^2}{2\sqrt{2}}\bigg)\bigg)h
    +3\bigg(\frac{\eta_{ij}^{\ell\dagger} v}{2\sqrt{2}}\bigg)h^2
    +\bigg(\frac{\eta_{ij}^{\ell\dagger}}{2\sqrt{2}}\bigg)h^3
    \bigg]
    +\textrm{h.c.} \Bigg] 
        \label{eq:intermediatestepLagrangian}
\end{eqnarray}
Note that using equations of motion leads to new interaction terms for the Higgs with fermions and gauge bosons - irrelevant operators involving derivatives of the Higgs cannot be neglected when studying Higgs couplings. The wave function renormalization term in \eqref{eq:intermediatestepLagrangian} can be made canonical by performing a field redefinition
\begin{eqnarray}
\varphi& \!\!\!=\!\!\!\ & \bigg( 1+\Delta Z_H+\frac{3}{2} \zeta_H v^2+ \frac{9}{4}\zeta_H' v^4\bigg)^{1/2} h \notag \\
& \!\!\!=\!\!\!\ & 
\bigg[ 1
+{\tilde{\lambda}}_6^\dagger  {\tilde{\lambda}}_6 
\left(\frac{ v^2}{{\tilde{m}_2}^2}\right)^2\bigg]^{1/2} 
h 
=\bigg[1
+\abs{\Xi }^2
+{\cal O}\bigg(\frac{v^6}{{\tilde{m}_2}^6}\bigg) 
\bigg]^{1/2}  
h
\label{eq:fieldredef}
\end{eqnarray}
where in going from the first to the second line we made use of \eqref{eq:parameterrelations2HDM}.
%\begin{equation}
%\Delta Z_h={\tilde{\lambda}}_6^\dagger  {\tilde{\lambda}}_6 \left(\frac{ v^2}{{\tilde{m}_2}^2}\right)^2=\abs{\Xi }^2+{\cal O}\bigg(\frac{v^4}{{\tilde{m}_2}^4}\bigg)
%\label{eq:2HDMWF}
%\end{equation}
As in section \ref{sec:realsingletEFT}, we see that there is a close relation between the mixing parameter (in this case, the complex alignment parameter $\Xi$) in the mixing language and wave function renormalization in the EFT language. 

Finally, note that the extremum condition for the potential is
\begin{equation}
{ \partial V \over \partial v} \Big |_{h=0}= v \Big( 
m_H^2  
+ {1 \over 2} \lambda_H v^2 
+{1 \over 4} \eta_6 v^4
+{1 \over 8} \eta_8 v^6
 \Big) = 0 
\end{equation}
so the Lagrangian mass parameter $m_H$ can be expressed in terms of Higgs potential couplings and the vev
\begin{equation}
m_H^2=- {1 \over 2} \lambda_H v^2 
-{1 \over 4} \eta_6 v^4
-{1 \over 8} \eta_8 v^6
\label{eq:softmass2HDM}
\end{equation}
Using \eqref{eq:fieldredef} and \eqref{eq:softmass2HDM} in \eqref{eq:intermediatestepLagrangian} we write the effective field theory for the Higgs boson in its final form 
\begin{eqnarray}
 \cal{L}\! \! \! &=& \! \! \! \frac{1}{2}\partial_\mu \varphi \partial^\mu \varphi
 -\frac{1}{2}m_\varphi^2 \varphi^2
 +\sum_{n=3}^{n=8} \frac{1}{n!}g_{\varphi^n}\varphi^n
 +\frac{1}{2}m_Z^2 Z^{\mu} Z_\mu 
 +m_W^2 W^{+ \mu} W_{\mu}^- \nonumber \\
&& \! \! \! + ~ \sum_{n=1}^{n=6}\frac{\varphi^n}{n!}\left[g_{\varphi^n ZZ}~  \frac{1}{2} Z^{\mu} Z_\mu  
+ g_{\varphi^n WW}~  W^{+ \mu} W_{\mu}^-\right] \nonumber   \nonumber  \\
 && -m^{u }_{ij} u_i\bar{u}_j -m^{d\dagger }_{ij} d_i\bar{d}_j-m^{\ell \dagger}_{ij} \ell_i\bar{\ell}_j  \nonumber  \nonumber \\
&& \! \! \!
 -\sum_{n=1}^{n=3} \frac{\varphi^n}{n! }\left[\lambda^{u}_{\varphi^n ij} ~u_i   \bar{u}_j +\lambda^{d\dagger}_{\varphi^n ij} ~d_i   \bar{d}_j +\lambda^{\ell \dagger}_{\varphi^n ij} ~\ell_i   \bar{\ell}_j +\textrm{h.c.}\right]   
 \label{eq:2HDMfinalLagrangian}
\end{eqnarray}
We now give explicit expressions for all the couplings of the effective theory. All the SM values of the couplings of the Higgs can be read from this section by taking the limit $\tilde{\lambda}_6 \rightarrow 0$. The no tadpole condition is crucial for the final results. For instance, note that in \eqref{eq:intermediatestepLagrangian} there are several mass terms for the W and Z bosons at effective dimension eight. The no tadpole condition imposes conditions between the coefficients of these operators, Eq. \eqref{eq:parameterrelations2HDM}. These conditions ensure that all the effective dimension eight mass terms cancel out, such that the mass of the W and Z boson coincide with their values in the renormalizable SM, $m_W=g_2v/2$ and $m_Z=g_2v/(2 \cos\theta_W)$. Note that this ensures that $\rho=m_W^2/(m_Z^2\cos^2 \theta_W)=1$.  

We start by the Higgs mass, which is given by
\begin{eqnarray}
\label{eq:2HDMmass}
 m^2_{\varphi}   \! \! \! & =  & \! \! \!   
 v^2\Bigg[{\tilde{\lambda}}_1 
 -{\tilde{\lambda}}_6^* {\tilde{\lambda}}_6 ~ \! \frac{ v^2}{{\tilde{m}_2}^2}   
 - 
 \frac{1}{2}  \Big[ (2{\tilde{\lambda}}_1  
  -{\tilde{\lambda}}_3 -{\tilde{\lambda}}_4) {\tilde{\lambda}}_6^* {\tilde{\lambda}}_6
  -\frac{1}{2}( {\tilde{\lambda}}_5^*  {\tilde{\lambda}}_6^2+\textrm{h.c}) 
  \Big]
  \frac{v^4}{{\tilde{m}_2}^4}
  +{\cal O}\bigg ( { v^6 \over {\tilde{m}_2}^6} \bigg) 
  \Bigg] \nonumber \\
\end{eqnarray}
The second term in \eqref{eq:2HDMmass} comes from the threshold correction to the operators $( H^\dagger  H)^2$ and $( H^\dagger  H)^3$  at effective dimension six. The rest of the corrections come from operators of effective dimension eight: the term proportional to $\tilde{\lambda}_1{\tilde{\lambda}}_6^* {\tilde{\lambda}}_6$ is due to wave function renormalization \eqref{eq:intermediatestepLagrangian}, and the rest of the terms come from threshold corrections to the operators $( H^\dagger  H)^n$, $n=2,3,4$ at effective dimension eight.

The couplings of one Higgs to two massive gauge bosons $V=W,Z$ are
\begin{equation}
g_{\varphi VV}=\frac{2m_V^2}{v} \Bigg[1-\frac{1}{2}{\tilde{\lambda}}_6^* {\tilde{\lambda}}_6 ~ \! \frac{v^4}{{\tilde{m}_2}^4}+{\cal O}\bigg({v^6 \over {\tilde{m}_2}^6}\bigg) \Bigg]
\label{eq:varphiVVEFT}
\end{equation}
The difference with respect to the SM values comes exclusively from dilution by wave function renormalization \eqref{eq:fieldredef} at effective dimension eight. The rest of the contributions to these couplings coming from the terms with covariant derivatives proportional to $\zeta_H$ and $\zeta_H'$ in \eqref{eq:2HDMsummary} cancel out, thanks to \eqref{eq:parameterrelations2HDM}. The result \eqref{eq:varphiVVEFT} coincides with the corresponding couplings obtained in the mixing language \eqref{eq:varphi1VVmixing}. The couplings of two Higgses to two massive gauge bosons are
\begin{equation}
g_{\varphi^2VV}
=\frac{2m_V^2}{v^2}\Bigg[1-3{\tilde{\lambda}}_6^* {\tilde{\lambda}}_6~\!\frac{ v^4}{{\tilde{m}_2}^4}
+{\cal O}\bigg({v^6 \over {\tilde{m}_2}^6}\bigg)\Bigg]
\label{eq:varphivarphiVVEFT}
\end{equation}
The difference with respect to the SM prediction comes exclusively from the terms that were generated by using the equations of motion in \eqref{eq:intermediatestepLagrangian}. This is an effective dimension eight effect. The rest of the effective dimension eight terms cancel out. Note that \eqref{eq:varphivarphiVVEFT} does not coincide with the result in the mixing language \eqref{eq:varphi1varphi1VVmixing}. As discussed in the SHSM EFT, four linear couplings in general do not coincide in the mixing and EFT languages, we comment on the difference in the end of this section. For completeness, the irrelevant couplings to gauge bosons come at effective dimension eight and are given by
\begin{eqnarray}
  vg_{\varphi^3VV}&=&
  -\frac{m_V^2}{v^2}
  \bigg[ 24 {\tilde{\lambda}}_6^* {\tilde{\lambda}}_6 \!~\frac{v^4}{{\tilde{m}_2}^4}
  +{\cal O}\bigg({v^6 \over {\tilde{m}_2}^6}\bigg)\bigg] 
  \nonumber \\
  v^2g_{\varphi^4VV}&=&
  -\frac{m_V^2}{v^2}
  \bigg[ 72 {\tilde{\lambda}}_6^* {\tilde{\lambda}}_6 \!~\frac{v^4}{{\tilde{m}_2}^4}
  +{\cal O}\bigg({v^6 \over {\tilde{m}_2}^6}\bigg)\bigg] 
  \nonumber \\
v^3g_{\varphi^5VV}&=&
-\frac{m_V^2}{v^2}
  \bigg[ 144 {\tilde{\lambda}}_6^* {\tilde{\lambda}}_6 \!~\frac{v^4}{{\tilde{m}_2}^4}
  +{\cal O}\bigg({v^6 \over {\tilde{m}_2}^6}\bigg)\bigg] 
  \nonumber \\
  v^4g_{\varphi^6VV}&=&
  -\frac{m_V^2}{v^2}
  \bigg[ 144 {\tilde{\lambda}}_6^* {\tilde{\lambda}}_6 \!~\frac{v^4}{{\tilde{m}_2}^4}
  +{\cal O}\bigg({v^6 \over {\tilde{m}_2}^6}\bigg)\bigg] 
  \end{eqnarray}
These couplings vanish in the decoupling limit.
%\begin{eqnarray}
%  vg_{h^3VV}&=&9 g_V^2 {\tilde{\lambda}}_6^* {\tilde{\lambda}}_6\left(\frac{v^2}{{\tilde{m}_2}^2}\right)^2+{\cal O}(v^6/{\tilde{m}_2}^6)  \nonumber \\
%  v^2g_{h^4VV}&=&39 g_V^2 {\tilde{\lambda}}_6^* {\tilde{\lambda}}_6\left(\frac{v^2}{{\tilde{m}_2}^2}\right)^2+{\cal O}(v^6/{\tilde{m}_2}^6)  \nonumber \\
%v^3g_{h^5VV}&=&90 g_V^2 {\tilde{\lambda}}_6^* {\tilde{\lambda}}_6\left(\frac{v^2}{{\tilde{m}_2}^2}\right)^2+{\cal O}(v^6/{\tilde{m}_2}^6)  \nonumber \\
%  v^4g_{h^6VV}&=&90 g_V^2 {\tilde{\lambda}}_6^* {\tilde{\lambda}}_6\left(\frac{v^2}{{\tilde{m}_2}^2}\right)^2+{\cal O}(v^6/{\tilde{m}_2}^6)
%\end{eqnarray}

Higgs-fermion interactions are modified by the cubic Yukawas of the Higgs doublet, which come at effective dimension six. They are given by
\begin{eqnarray}
 \lambda^f_{\varphi ij}=\frac{m^f_{i}}{v}~\! \delta_{ij}-2\left(\frac{\tilde{\lambda}^f_{2ij}{\tilde{\lambda}}_6^* }{2\sqrt{2}}\right)\left(\frac{v^2}{{\tilde{m}_2}^2}\right)+{\cal O}\bigg({v^4 \over {\tilde{m}_2}^4}\bigg) \nonumber  \nonumber \\
 v\lambda^f_{\varphi^2 ij}= -3\left(\frac{\tilde{\lambda}^f_{2ij}{\tilde{\lambda}}_6^* }{\sqrt{2}}\right)\left(\frac{v^2}{{\tilde{m}_2}^2}\right)+{\cal O}\bigg({v^4 \over {\tilde{m}_2}^4}\bigg) 
 \nonumber \\
 v^2 \lambda^f_{\varphi^3 ij}=-3\left(\frac{\tilde{\lambda}^f_{2ij}{\tilde{\lambda}}_6^* }{\sqrt{2}}\right)\left(\frac{v^2}{{\tilde{m}_2}^2}\right)+{\cal O}\bigg({v^4 \over {\tilde{m}_2}^4}\bigg) 
\label{eq:varphiffEFT}
\end{eqnarray}
%\begin{eqnarray}
%% g_{hF\bar{F}}=-\frac{M_F}{v} +\frac{ \rho_F {\tilde{\lambda}}_6}{ \sqrt{2}}\left(\frac{v^2}{{\tilde{m}_2}^2}\right)+{\cal O}\bigg(\frac{v^4}{{\tilde{m}_2}^4}\bigg)  \nonumber \\
%g_{hF\bar{F}}=-\frac{M_F}{v} +2 \left( \frac{\rho_F {\tilde{\lambda}}_6}{2\sqrt{2}}\left(\frac{v^2}{{\tilde{m}_2}^2}\right)\right)+{\cal O}\bigg(\frac{v^4}{{\tilde{m}_2}^4}\bigg)  \nonumber \\
%vg_{h^2F\bar{F}}=v^2 g_{h^3F\bar{F}}=\frac{ 3\rho_F {\tilde{\lambda}}_6}{\sqrt{2}}\left(\frac{v^2}{{\tilde{m}_2}^2}\right)+{\cal O}\bigg(\frac{v^4}{{\tilde{m}_2}^4}\bigg)
%\label{eq:2HDMyuk}
%\end{eqnarray}
Note that we neglect the effective dimension eight terms coming from wave function renormalization \eqref{eq:fieldredef} and from the use of the equations of motion in \eqref{eq:intermediatestepkineticLagrangian2}, since we only work up to effective dimension six in interactions involving fermions. All the flavor violating terms in the fermionic couplings \eqref{eq:varphiffEFT} are inherited from the heavy doublet Yukawa matrix, $\tilde{\lambda}^f_{2ij}$. Non zero phases in $\tilde{\lambda}^f_{2ij}{\tilde{\lambda}}_6^*$ lead to T violating processes mediated by Higgs boson exchange, since they induce a relative phase between the Yukawas and the quark mass matrices. Note that the irrelevant couplings $\lambda^f_{\varphi^2 ij}$ and $\lambda^f_{\varphi^3 ij}$ vanish in the decoupling limit.

Higgs self couplings are modified by the operators of effective dimension six and eight in \eqref{eq:2HDMsummary}. They are given by
\begin{eqnarray}
\frac{1}{v}g_{\varphi_1^3} \! \! \! &=&\! \! \!    
-\frac{3m_{\varphi_1}^2}{v^2}
+6 {\tilde{\lambda}}_6^* {\tilde{\lambda}}_6~ \!\frac{v^2}{{\tilde{m}_2}^2} \nonumber \\ 
  &&+~ \frac{1}{2}\bigg[~\! 
  \big[ 21{\tilde{\lambda}}_1
  -12  ({\tilde{\lambda}}_3 +{\tilde{\lambda}}_4)
  \big] 
  {\tilde{\lambda}}_6^* {\tilde{\lambda}}_6
  -\big( 6{\tilde{\lambda}}_5 {\tilde{\lambda}}_6^{* 2}
  +\textrm{h.c.} 
  \big) 
  \bigg] 
  ~\!\frac{v^4}{{\tilde{m}_2}^4}
  +{\cal O}\bigg({v^6 \over {\tilde{m}_2}^6}\bigg) 
\label{eq:Higgsself2EFT}
\end{eqnarray}
\begin{eqnarray}
g_{\varphi_1^4} \! \! \! &=&\! \! \!    
-\frac{3m_{\varphi_1}^2}{v^2}
+36 {\tilde{\lambda}}_6^* {\tilde{\lambda}}_6~ \!\frac{v^2}{{\tilde{m}_2}^2} \nonumber \\ 
  &&+~ \bigg[~\! 
  \big[ 105{\tilde{\lambda}}_1
  -60  ({\tilde{\lambda}}_3 +{\tilde{\lambda}}_4)
  \big] 
  {\tilde{\lambda}}_6^* {\tilde{\lambda}}_6
  -\big( 30{\tilde{\lambda}}_5 {\tilde{\lambda}}_6^{* 2}
  +\textrm{h.c.} 
  \big) 
  \bigg] 
  ~\!\frac{v^4}{{\tilde{m}_2}^4}
  +{\cal O}\bigg({v^6 \over {\tilde{m}_2}^6}\bigg) 
\label{eq:Higgsself1EFT}
\end{eqnarray}
For completeness, the irrelevant self couplings are given by
\begin{eqnarray}
  vg_{\varphi^5} \! \! \!  &=& \! \! \!  
  90 \tilde{\lambda}_6 \tilde{\lambda}_6^* ~\! {v^2 \over {\tilde{m}_2}^2}+
   \bigg(~\! 
  \big[ 585{\tilde{\lambda}}_1
  -330  ({\tilde{\lambda}}_3 +{\tilde{\lambda}}_4)
  \big] 
  {\tilde{\lambda}}_6^* {\tilde{\lambda}}_6
  -\big( 165{\tilde{\lambda}}_5 {\tilde{\lambda}}_6^{* 2}
  +\textrm{h.c.} 
  \big) 
  \bigg)
  ~\!\frac{v^4}{{\tilde{m}_2}^4}
  +{\cal O}\bigg({v^6 \over {\tilde{m}_2}^6}\bigg) 
    \nonumber \\
  v^2g_{\varphi^6} \! \! \!  &=& \! \! \!
  90 \tilde{\lambda}_6 \tilde{\lambda}_6^* ~\! {v^2 \over {\tilde{m}_2}^2}+
   \bigg(~\! 
  \big[ 2097{\tilde{\lambda}}_1
  -1170  ({\tilde{\lambda}}_3 +{\tilde{\lambda}}_4)
  \big] 
  {\tilde{\lambda}}_6^* {\tilde{\lambda}}_6
  -\big( 585{\tilde{\lambda}}_5 {\tilde{\lambda}}_6^{* 2}
  +\textrm{h.c.} 
  \big) 
  \bigg)
  ~\!\frac{v^4}{{\tilde{m}_2}^4}
  +{\cal O}\bigg({v^6 \over {\tilde{m}_2}^6}\bigg) 
     \nonumber \\
  v^3g_{\varphi^7} \! \! \!  &=& \! \! \! 
   \bigg(~\! 
  \big[ 4536{\tilde{\lambda}}_1
  -2520  ({\tilde{\lambda}}_3 +{\tilde{\lambda}}_4)
  \big] 
  {\tilde{\lambda}}_6^* {\tilde{\lambda}}_6
  -\big( 1260{\tilde{\lambda}}_5 {\tilde{\lambda}}_6^{* 2}
  +\textrm{h.c.} 
  \big) 
  \bigg) 
  ~\!\frac{v^4}{{\tilde{m}_2}^4}
  +{\cal O}\bigg({v^6 \over {\tilde{m}_2}^6}\bigg) 
   \nonumber \\
 v^4g_{\varphi^8} \! \! \!  &=& \! \! \! 
   \bigg(~\! 
  \big[ 4536{\tilde{\lambda}}_1
  -2520  ({\tilde{\lambda}}_3 +{\tilde{\lambda}}_4)
  \big] 
  {\tilde{\lambda}}_6^* {\tilde{\lambda}}_6
  -\big( 1260{\tilde{\lambda}}_5 {\tilde{\lambda}}_6^{* 2}
  +\textrm{h.c.} 
  \big) 
  \bigg)
  ~\!\frac{v^4}{{\tilde{m}_2}^4}
  +{\cal O}\bigg({v^6 \over {\tilde{m}_2}^6}\bigg) 
 \end{eqnarray}
All these irrelevant self couplings vanish in the decoupling limit.

Note that all trilinear couplings coincide in the EFT and mixing languages, while four-linear interactions do not coincide, as already discussed in section \ref{sec:realsingletEFT} for the case of the SM extended with a heavy singlet. The equality of the trilinear couplings in the EFT and mixing languages ensures the equality between the non analytic (long distance) physical scattering amplitudes in the EFT and mixing languages. Integrating out heavy fields cannot modify cubic interactions. Four-linear interactions can be different in both languages, since four linear couplings in the mixing language do not contain all the short distance pieces of the corresponding amplitudes, but EFT couplings do. Scattering amplitudes calculated in both languages coincide, as we will see in section \ref{sec:amplitudesEFT}.

We conclude this section emphasizing a series of particular phenomenological characteristics of the 2HDM EFT. First, deviations to the SM predictions of the Higgs-fermion couplings come at effective dimension six, while deviations to the Higgs-gauge boson couplings come first at effective dimension eight. This hierarchy is one of the main features of the 2HDM EFT. Deviations from the SM predictions in a measurement of fermionic Higgs couplings at colliders, together with no corresponding modifications to the Higgs-gauge boson couplings, would be a possible indication of a Higgs sector completed with a second doublet in the UV. Second, all the deviations to the SM predictions for the couplings of the Higgs at effective dimension six are controlled exclusively by a small subset of the couplings of the UV completion: the PQ invariant combinations $\tilde{\lambda}^f_{2ij}{\tilde{\lambda}}_6$ and ${\tilde{\lambda}}_6^*{\tilde{\lambda}}_6$. This significantly reduces the complexity of the analysis of the most general 2HDM. In the mixing language, $\tilde{\lambda}_6$ is related to the complex alignment parameter through \eqref{eq:vperpapprox}. The modifications to Higgs-gauge boson couplings with respect to their SM predictions at effective dimension eight, are also controlled by ${\tilde{\lambda}}_6^*{\tilde{\lambda}}_6$. Another special characteristic of the 2HDM EFT is that \textit{all} the bosonic couplings of the Higgs are smaller in magnitude than their SM counterparts. This is evident in the EFT, but less obvious in the mixing language, where the coupling of two Higgses to two gauge bosons is left unmodified with respect to its SM value.

T violation has a particular structure in the 2HDM EFT. All T violation at effective dimension six comes exclusively from the PQ invariant phases in the fermionic interactions proportional to $\tilde{\lambda}^f_{2ij}{\tilde{\lambda}}_6$. The T violating phases $\theta_1$ and $\theta_2$ defined in \eqref{eq:cpviolatingphases} which come from the 2HDM potential do not show up at effective dimension six. Note that $\theta_1$ is the phase of the complex alignment parameter \eqref{eq:vperpapprox}. As a consequence, measuring the phase of the complex alignment parameter is challenging for a 2HDM near the decoupling limit, since the T violating effects of $\theta_1$ are subleading. The T violating effects of $\theta_1$ might first appear in fermionic interactions involving an insertion of the coupling ${\tilde{\lambda}}_5$ at effective dimension eight. Finally, there is no CP violation in purely bosonic interactions up to at least effective dimension ten in the 2HDM EFT. For instance, the operator $H^\dagger H F_{\mu \nu} \tilde{F}^{\mu \nu}$ is not induced up to at least effective dimension ten.

Regarding the four fermion operators \eqref{eq:fourfermion}, note that they are the only operators which are not controlled by $\tilde{\lambda}_6$, and they are the only operators at effective dimension six which do not vanish in the exact alignment limit, when $\Xi \propto \tilde{\lambda}_6=0$. In the exact alignment limit and near the decoupling limit, the only hope to get hints of a second doublet would be at flavor experiments sensitive to flavor violation in the four fermion operators \eqref{eq:fourfermion}.

Finally, note that in our results there is no reference to $\tan\beta$ anywhere, since in a general 2HDM, $\tan\beta$ has no physical meaning.

 \subsection{Scattering amplitudes}
\label{sec:amplitudesEFT}

As a consistency check of the EFT we compute scattering amplitudes, and compare with the corresponding results obtained in the mixing language in section \ref{sec:amplitudesmixing2HDM}. We omit spinors in the amplitudes.

The dihiggs to di-W boson scattering amplitude is 
\begin{eqnarray}
{ \cal A} \left(\varphi_1 \varphi_1\rightarrow W^+  W^- \right) 
  \! \! \! &=&g_{\mu\nu}\Bigg[ g_{\varphi^2W^2} 
  -\frac{g_{\varphi^3}g_{\varphi W^2}}{s-m_{\varphi}^2}
 -g_{\varphi W^2}^2  \bigg(\frac{1}{t-m_{W}^2}+\frac{1}{u-m_{W}^2}\bigg) 
 \Bigg]
    \label{eq:exactdihdiWEFT}
 \end{eqnarray} 
Using the EFT couplings \eqref{eq:varphiVVEFT}, \eqref{eq:varphivarphiVVEFT} and \eqref{eq:Higgsself2EFT} the resulting amplitude is

\begin{eqnarray}
{ \cal A} \left(\varphi\varphi\rightarrow W^+  W^- \right) 
  \! \! \! &=& 
g_{\mu\nu} 
\Bigg[ \frac{2m_W^2}{v^2} 
  -6\tilde{\lambda}_6\tilde{\lambda}_6^*\frac{m_W^2}{v^2}   ~\! \frac{v^4}{\tilde{m}_2^4} \nonumber \\
    && - ~ \frac{2m_W^2}{v^2} 
    \bigg[~\!
    -\frac{3m_{\varphi}^2}{v^2}
+6 {\tilde{\lambda}}_6^* {\tilde{\lambda}}_6~ \!\frac{v^2}{{\tilde{m}_2}^2} 
+ \frac{1}{2}\bigg(~\! 
  \big[ 21{\tilde{\lambda}}_1
  -12  ({\tilde{\lambda}}_3 +{\tilde{\lambda}}_4)
  \big] 
  {\tilde{\lambda}}_6^* {\tilde{\lambda}}_6
\nonumber \\ 
  &&-~ 
 \big( 6{\tilde{\lambda}}_5 {\tilde{\lambda}}_6^{* 2}
  +\textrm{h.c.} 
  \big) 
  \bigg)
  ~\!\frac{v^4}{{\tilde{m}_2}^4}
  +\frac{3}{2} {\tilde{\lambda}}_6^* {\tilde{\lambda}}_6 \frac{m_{\varphi}^2v^2}{{\tilde{m}_2}^4}
     ~\!\bigg]\bigg(\frac{v^2}{s-m_{\varphi}^2} \bigg) \notag \\
       && - ~\frac{4m_W^4}{v^4} \bigg[1-{\tilde{\lambda}}_6^\dagger {\tilde{\lambda}}_6 ~ \! \frac{v^4}{{\tilde{m}_2}^4} \bigg] \bigg(\frac{v^2}{t-m_{W}^2}+\frac{v^2}{u-m_{W}^2}\bigg) 
       + {\cal O}\bigg(\frac{v^6}{{\tilde{m}_2}^6}\bigg)
       \Bigg]
        \notag \\
        \label{eq:dihiggsdiWEFT}
 \end{eqnarray}
which coincides with  the result obtained in the mixing language \eqref{eq:dihiggsdiWmixing}. Note that the second term in \eqref{eq:dihiggsdiWEFT} comes from the modification to the dihiggs di-W boson coupling with respect to its SM value. 

The dihiggs to difermion chirality violating scattering amplitude is  
\begin{eqnarray}
{ \cal A} \left(\varphi \varphi \rightarrow f_i \bar{f}_j \right) 
  \! \! \! &=&   \! \! \! -\lambda^f_{\varphi^2 ij}+ \frac{g_{\varphi^3}\lambda^f_{\varphi ij}}{s-m_{\varphi}^2}
 \end{eqnarray}
Using the EFT couplings \eqref{eq:varphiffEFT} and \eqref{eq:Higgsself2EFT}  we get 
 \begin{eqnarray}
{ \cal A} \left(\varphi \varphi \rightarrow f_i \bar{f}_j \right) 
  \! \! \! &=&   ~ \frac{1}{v}
  ~\!\Bigg[
    %%% Contact Piece
    ~\! \frac{3}{\sqrt{2}} \tilde{\lambda}^f_{2ij}\tilde{\lambda}_6^*~\! 
    \frac{v^2}{\tilde{m}_2^2} 
    %%% Long Distance Piece
+ \bigg[
  \delta_{ij} \frac{m^f_{i}}{v}
  \bigg(-\frac{3m_{\varphi}^2}{v^2}
  +6\tilde{\lambda}_6\tilde{\lambda}_6^* ~\!\frac{v^2}{\tilde{m}^2_2}
   \bigg) \nonumber \\
  && +~ 
  \frac{3}{\sqrt{2}}~\! \tilde{\lambda}^f_{2ij} \tilde{\lambda}_6^*~\! \frac{ m_{\varphi}^2}{\tilde{m}^2_2}   
  \bigg] 
  \bigg(\frac{v^2}{s-m_{\varphi}^2}\bigg)
  %Error
  + {\cal O}\bigg(\frac{v^4}{{\tilde{m}_2}^4}\bigg)
        \Bigg]
 \end{eqnarray}
which coincides with the result obtained in the mixing language \eqref{eq:dihiggsdifmixing}. Note that this amplitude has a flavor and T violating term both in the long and short distance pieces, inherited from the Yukawas of the heavy doublet $\tilde{\lambda}^f_{2ij}$.

The dihiggs to dihiggs scattering amplitude is
 \begin{equation}
{ \cal A} \left(\varphi \varphi \rightarrow \varphi \varphi \right) =g_{\varphi^4}-g^2_{\varphi^3}\left(\frac{1}{s-m_{\varphi}^2}+\frac{1}{t-m_{\varphi}^2}+\frac{1}{u-m_{\varphi}^2}\right) 
\end{equation}
Using the EFT coupling \eqref{eq:Higgsself2EFT} we get
\begin{eqnarray}
{ \cal A} \left(\varphi \varphi \rightarrow \varphi \varphi \right)  \! \! \! &=& \! \! \! -\frac{3m_{\varphi}^2}{v^2}+36 {\tilde{\lambda}}_6^* {\tilde{\lambda}}_6\left(\frac{v^2}{{\tilde{m}_2}^2}\right)
  - \bigg(
  \frac{9m_{\varphi}^2}{v^4}
  -36{\tilde{\lambda}}_1 {\tilde{\lambda}}_6^* {\tilde{\lambda}}_6 ~  \frac{v^2}{{\tilde{m}_2}^2}
   \bigg)
   \notag \\
&&  \! \! \!
\bigg(\frac{v^2}{s-m_{\varphi}^2}+\frac{v^2}{t-m_{\varphi}^2}+\frac{v^2}{u-m_{\varphi}^2}\bigg) 
+ {\cal O}\bigg(\frac{v^4}{{\tilde{m}_2}^4}\bigg)
\end{eqnarray}
which coincides with  the result obtained in the mixing language \eqref{eq:fourfermionamplitude2HDMmixing}.

\section{The EFT of the 2HDM with Glashow-Weinberg conditions}
\label{sec:analysis}

In the SM, the fermion mass matrices and the Higgs Yukawas are aligned in flavor space, so they can be simultaneously diagonalized by performing quark and lepton field redefinitions. As a consequence, there are no flavor changing neutral currents (FCNC). FCNC are experimentally strongly constrained. 

In a general 2HDM, it is not possible to simultaneously diagonalize the fermion mass matrices and the Yukawas of both Higgs doublets simultaneously by performing field redefinitions. This leads to FCNC mediated by neutral Higgs states. In the 2HDM, FCNCs can be avoided by imposing the Glashow-Weinberg (GW) conditions \cite{Glashow:1976nt}. The conditions consist in giving mass to all fermions in a particular representation by allowing them to couple with only one of the two doublets. These conditions can be imposed by discrete symmetries or supersymmetry. The Glashow-Weinberg (GW) conditions are satisfied by four discrete choices of the Yukawa couplings \eqref{eq:2HDMactiongeneric} of the doublets $\Phi_1$, $\Phi_2$
\begin{eqnarray}
\textrm{Type I: } \lambda_1^u=\lambda_1^d=\lambda_1^{\ell}=0 \nonumber \\
\textrm{Type II: } \lambda_1^u=\lambda_2^d=\lambda_2^{\ell}=0  \nonumber\\
\textrm{Type III: } \lambda_1^u=\lambda_1^d=\lambda_2^{\ell}=0 \nonumber\\
\textrm{Type IV: } \lambda_1^u=\lambda_2^d=\lambda_1^{\ell}=0
\label{eq:GWconditions}
\end{eqnarray}
In the 2HDM with GW conditions, the ratio of the vevs of the doublets  $v_2/v_1=\tan\beta$ defined in \eqref{eq:deftanb} contains physical information, since it is the ratio of the vevs in the preferred basis of the Higgs doublets in which one of the conditions \eqref{eq:GWconditions} holds. In other words, in a 2HDM with GW conditions, $\tan\beta$ is physical, since it is the rotation angle relating the preferred basis with the Higgs basis \eqref{eq:rotationtoHiggsbasis} \footnote{In the MSSM, $\tan\beta$ can be defined independently of the Yukawas, since there is a flat direction $H_u=H_d$ which specifies the preferred basis.}.

The objective of this section is to study the types I-IV 2HDM using the EFT presented in section \ref{sec:2HDMEFT}. GW conditions impose a particular structure on the Yukawas and four fermion operators of the low energy theory, but we impose no other restrictions in addition to one of the conditions \eqref{eq:GWconditions}, \textit{i.e.}, we consider the Higgs potential to be the most general one at the renormalizable level, and we allow for all the possible T violating phases. Since the GW conditions only refer to the fermionic interactions, we do not present the bosonic interactions in this section. All the bosonic interactions are the ones of a general 2HDM, and were presented in section \ref{sec:2HDMEFT}. Each type of 2HDM will be presented in sections \ref{sec:typeI}, \ref{sec:typeII}, \ref{sec:typeIII} and \ref{sec:typeIV}. The detailed discussion of T violation in the types I-IV 2HDM is left to section \ref{sec:Tviolationtypes}. 

%Recall that in a generic basis, the most general Yukawa couplings are given by 
%\begin{equation}
%\label{eq:generalYukawas}
% - 
%   %\sum_{a=1}^2~
%  ~
%     \lambda^u_{aij} ~ Q_i \Phi_a \bar{u}_j
%   + \lambda^{d\dagger}_{aij} Q_i \Phi_a^c \bar{d}_j 
%   + \lambda^{\ell\dagger}_{aij} L_i \Phi_a^c \bar{\ell}_j 
%  +{\rm h.c.}~ \!  
%\end{equation}
%where quarks are in the mass eigenbasis and $\Phi_1$ and $\Phi_2$ are Higgs doublets. The doublets are related to the Higgs basis doublets by a rotation matrix specified by an angle $\beta$.  \sidecomment{Half of this page is repetition of what was already said. Should I repeat it for clarity?}
%\begin{equation}
%e^{-i\xi /2 }H=\cos  \beta\Phi_1+\sin  \beta e^{-i\xi} \Phi_2 \quad \quad  H_2=-\sin  \beta e^{i\xi}\Phi_1+\cos  \beta \Phi_2
%\end{equation}
%such that the Yukawa coupling the Higgs basis doublets to fermions are
%\begin{eqnarray}
%\tilde{\lambda}^f_{1ij} \! \! \! &=& \! \! \! e^{-i\xi /2  } \lambda_{1ij}^f  \cos   \beta 
%			+ e^{i\xi /2 } \lambda_{2ij}^f \sin   \beta  \notag \\ 
%\tilde{\lambda}^f_{2ij} \! \! \!   &=& \! \! \! -e^{-i \xi} \lambda_{1ij}^f  \sin   \beta 
%			+ \lambda_{2ij}^f \cos   \beta  
%			\end{eqnarray}
%where the fermion mass matrix is given by
%\begin{equation}
% m^{f}_{ij}=\frac{v}{\sqrt{2}} \tilde{\lambda}^f_{1ij}   \quad \quad f  = u,d,\ell 
% \label{eq:massrelation}
%\end{equation}
%
%We immediately see that the Yukawas of the heavy doublet $\tilde{\lambda}^f_{2ij}$, which are in general not aligned with $ m^{f}_{ij}$ 

\subsection{ Type I}
\label{sec:typeI}
%%% build section: low energy effective theory of four fermi interactions. stress out GIM suppression and T violation, especially in the GW case.
In the type I 2HDM the Yukawas of the doublet $\Phi_1$ are set to zero
\begin{equation}
\lambda_1^u=\lambda_1^d=\lambda_1^{\ell}=0
\label{eq:2HDMI}
\end{equation}
so all the fermions get their mass from their coupling to the second doublet $\Phi_2$.

The Yukawas of the doublets in the Higgs basis are obtained by using \eqref{eq:2HDMI} in \eqref{eq:relationYukawas}
\begin{eqnarray}
\label{eq:UVYukawas1I}
 \tilde{\lambda}^{u,d,\ell}_{1ij}   \! \! \! &=& \! \! \! 
 e^{{i\xi \over 2}} \lambda_{2ij}^{u,d,\ell}\sin\beta
  \\
\tilde{\lambda}^{u,d,\ell}_{2ij}  \! \! \! &=& \! \! \! \lambda_{2ij}^{u,d,\ell}\cos\beta
\label{eq:UVYukawasI}
\end{eqnarray}
The fermion mass matrices are related to $ \tilde{\lambda}^{u,d,\ell}_{1ij}$ through \eqref{eq:kUkD}, so using \eqref{eq:UVYukawas1I} we rewrite  $\lambda_{2ij}^{u,d,\ell}$ as
\begin{equation}
\lambda_{2ij}^{u,d,\ell}= e^{{-i\xi \over 2}}  \frac{\sqrt{2}m^{u,d,\ell}_{ij} }{v } \csc \beta 
\label{eq:secondyukawaI}
\end{equation}
Using \eqref{eq:secondyukawaI} in \eqref{eq:UVYukawasI}, we write the Yukawas of the second doublet in the Higgs basis as
\begin{equation}
\tilde{\lambda}^{u,d,\ell}_{2ij} =  \sqrt{2}e^{-{i\xi \over 2}} \cot\beta ~\!\frac{m^{u,d,\ell}_{ij}}{v}
\label{eq:heavyyukI}
\end{equation}
Using \eqref{eq:heavyyukI} in the Higgs Yukawas \eqref{eq:varphiffEFT}, we get 
\begin{eqnarray}
 \lambda^{u,d,\ell}_{\varphi ij}
  \! \! \! &=& \! \! \!
\frac{m_{ij}^{u,d,\ell}}{v}
 \bigg[1
 -{\tilde{\lambda}}_6^* e^{-i\xi /2 }\cot\beta
 ~\! \frac{v^2}{{\tilde{m}_2}^2}
  +{\cal O}\bigg(\frac{v^4}{{\tilde{m}_2}^4}\bigg)  
 \bigg]
  \nonumber \\
v  \lambda^{u,d,\ell}_{\varphi^2 ij}  \! \! \! &=& \! \! \! 
  -~\! \frac{m^{u,d,\ell}_{ij}}{v} ~\! 
 \bigg[ 3  {\tilde{\lambda}}_6^* e^{-{i\xi \over 2}} \cot\beta 
~\! \frac{v^2}{{\tilde{m}_2}^2}
+{\cal O}\bigg(\frac{v^4}{{\tilde{m}_2}^4}\bigg) 
\bigg]
\nonumber \\
v^2\lambda^{u,d,\ell}_{\varphi^3 ij}
 \! \! \! &=& \! \! \! 
  -~\! \frac{m^{u,d,\ell}_{ij}}{v} ~\! 
 \bigg[ 3  {\tilde{\lambda}}_6^* e^{-{i\xi \over 2}} \cot\beta 
~\! \frac{v^2}{{\tilde{m}_2}^2}
+{\cal O}\bigg(\frac{v^4}{{\tilde{m}_2}^4}\bigg) 
\bigg]
\label{eq:YukawastypeIpre}
\end{eqnarray}
Note that using \eqref{eq:convenientxi} the Higgs Yukawas can also be expressed in terms of the complex alignment parameter
\begin{eqnarray}
\lambda^{u,d,\ell}_{\varphi ij}  \! \! \! &=& \! \! \! 
\delta_{ij}~\!\frac{m_{ij}^{u,d,\ell}}{v} 
\bigg[1+  e^{ i ~ \! \! {\rm Arg}({\tilde{\lambda}}_5^*)/2} e^{-{i\xi \over 2}}\,  \Xi \, \cot\beta   +{\cal O}\Big(\Xi^2\Big ) 
\bigg] 
  \nonumber \\
v  \lambda^{u,d,\ell}_{\varphi^2 ij}  \! \! \! &=& \! \! \! 
  ~\! \frac{m^{u,d,\ell}_{ij}}{v} ~\! 
 \bigg[ 3  
 e^{ i ~ \! \! {\rm Arg}({\tilde{\lambda}}_5^*)/2} e^{-{i\xi \over 2}}\,  \Xi \, \cot\beta   +{\cal O}\Big(\Xi^2\Big )
\bigg]
\nonumber \\
v^2\lambda^{u,d,\ell}_{\varphi^3 ij}
 \! \! \! &=& \! \! \! 
  ~\! \frac{m^{u,d,\ell}_{ij}}{v} ~\! 
 \bigg[ 3  
 e^{ i ~ \! \! {\rm Arg}({\tilde{\lambda}}_5^*)/2} e^{-{i\xi \over 2}}\,  \Xi \, \cot\beta   +{\cal O}\Big(\Xi^2\Big )
\bigg]
\end{eqnarray}

The coefficients of the four fermion operators are found using \eqref{eq:heavyyukI} in \eqref{eq:fourfermion} 
\begin{eqnarray}
\Omega^{uu~\!\! (0)}_{ijmn }  
  \! \! \! &=& \! \! \!
 2~
 \! \frac{1}{{\tilde{m}_2}^2} \frac{m^{u}_{ij} m^{u\dagger}_{mn}}{v^2} \cot^2 \beta 
 \nonumber \\
 \Omega^{dd ~\!\! (0)}_{ijmn } 
  \! \! \! &=& \! \! \!
 2 ~ 
 \! \frac{1}{{\tilde{m}_2}^2} \frac{m^{d\dagger}_{ij} m^{d}_{mn}}{v^2} \cot^2 \beta  
  \nonumber \\
 \Omega^{\ell\ell ~\!\! (0)}_{ijmn } 
   \! \! \! &=& \! \! \!
 2~ 
 \! \frac{1}{{\tilde{m}_2}^2} \frac{m^{\ell\dagger}_{ij} m^{\ell}_{mn}}{v^2} \cot^2 \beta 
 \nonumber \\
\Omega^{d\ell ~\!\! (0)}_{ijmn}
  \! \! \! &=& \! \! \!
 2 ~ 
 \! \frac{1}{{\tilde{m}_2}^2} \frac{m^{d\dagger}_{ij} m^{\ell}_{mn}}{v^2} \cot^2 \beta 
 \nonumber \\
 \Omega^{u d ~\!\! (2)}_{ijmn}
  \! \! \! &=& \! \! \!
 2 ~ 
 \! \frac{1}{{\tilde{m}_2}^2} \frac{m^{u}_{ij} m^{d\dagger}_{mn}}{v^2} \cot^2 \beta 
 \nonumber \\
\Omega^{u \ell ~\!\! (2)}_{ijmn}
  \! \! \! &=& \! \! \!
2 ~ 
 \! \frac{1}{{\tilde{m}_2}^2} \frac{m^{u}_{ij} m^{\ell\dagger}_{mn}}{v^2} \cot^2 \beta 
 \label{eq:fourfermioncoefftypeI}
\end{eqnarray}

In the EFT of a type I 2HDM at effective dimension six, all the modifications to the SM predictions for the fermionic couplings of the Higgs \eqref{eq:YukawastypeIpre}, and all the four fermion operators \eqref{eq:fourfermioncoefftypeI} vanish in the large $\tan\beta$ limit. The reason is that in this limit, the doublet $H_2$ that is integrated out is aligned with the doublet $\Phi_1$, which does not couple to fermions. The leading effects on the Higgs Yukawas in this case are of effective dimension eight, and come from kinetic operators in the effective Lagrangian. Alternatively, in the mixing language, at large $\tan\beta$ the only modifications to the Higgs Yukawa couplings come from dilution due to the complex alignment parameter as discussed in section \ref{sec:couplings2HDMmixing}. In this limit, the modifications to the SM predictions for the Higgs Yukawas and Higgs couplings to gauge bosons are identical: both couplings are diluted by $\sqrt{1-\abs{\Xi}^2}$.

%The quark mass matrices can be expressed in terms of its eigenvalues using the unitary rotations
%\begin{equation}
%m^u_{ij}=
%\end{equation}
We now perform field redefinitions to present the Yukawa couplings and four fermion operators in the quark mass eigenbasis. Without loss of generality, the quark mass matrices in a general flavor basis are given by
\begin{eqnarray}
m^{u}  & = & U_{Q_u}~\! \textrm{diag}(m_u,m_c,m_t) ~\! U_{\bar{u}}^\dagger \\
m^{d\dagger}  & = & U_{Q_d} ~\! \textrm{diag}(m_d,m_s,m_b) ~\! U_{\bar{d}}^\dagger 
  \label{eq:lambdau}
\end{eqnarray}
where all the mass eigenvalues are real and non negative. We define the CKM matrix following the conventions of \cite{Agashe:2014kda}
\begin{equation}
V=U_{Q_u}^T U_{Q_d}^*
\label{eq:VCKM}
\end{equation}
The quark mass matrices are diagonalized by the field redefinitions
\begin{eqnarray}
&& u'_i=u_j \big[U_{Q_u}\big]_{ji}  
\quad\quad 
d'_i=d_j \big[U_{Q_d} \big]_{ji} 
\quad\quad 
\bar{u}'_i =\big[~\! U^\dagger_{\bar{u}} ~\!  \big]_{ij}  \bar{u}_j  
\quad\quad 
\bar{d}'_i =\big[~\!  U_{\bar{d}}^\dagger ~\!  \big]_{ij} \bar{d}_j 
%&& L'_i=L _j(U_\ell^\dagger)_{ji} \quad\quad \bar{\ell}'_i =(U_{\bar{\ell}})_{ij} \bar{\ell}_j 
\label{eq:flavortransf}
\end{eqnarray}
We drop the primes in the quark fields in the mass eigenbasis for the remainder of this paper. Using the field redefinitions \eqref{eq:flavortransf} in the effective lagrangian \eqref{eq:2HDMfinalLagrangian} and the Yukawas \eqref{eq:YukawastypeIpre}, we find the Yukawas in the quark mass eigenbasis
\begin{eqnarray}
 \lambda^{u,d,\ell}_{\varphi ij}
  \! \! \! &=& \! \! \!
\delta_{ij}~\!\frac{m_{i}^{u,d,\ell}}{v}
 \bigg[1
 -{\tilde{\lambda}}_6^* e^{-i\xi /2 }\cot\beta
 ~\! \frac{v^2}{{\tilde{m}_2}^2}
  +{\cal O}\bigg(\frac{v^4}{{\tilde{m}_2}^4}\bigg)  
 \bigg]
  \nonumber \\
v  \lambda^{u,d,\ell}_{\varphi^2 ij}  \! \! \! &=& 
\! \! \! 
 -\delta_{ij}~\! \frac{m^{u,d,\ell}_{i}}{v} ~\! 
 \bigg[ 3  {\tilde{\lambda}}_6^* e^{-{i\xi \over 2}} \cot\beta 
~\! \frac{v^2}{{\tilde{m}_2}^2}
+{\cal O}\bigg(\frac{v^4}{{\tilde{m}_2}^4}\bigg) 
\bigg]
\nonumber \\
v^2\lambda^{u,d,\ell}_{\varphi^3 ij}
 \! \! \! &=& 
\! \! \! 
  -\delta_{ij}~\! \frac{m^{u,d,\ell}_{i}}{v} ~\! 
 \bigg[ 3  {\tilde{\lambda}}_6^* e^{-{i\xi \over 2}} \cot\beta 
~\! \frac{v^2}{{\tilde{m}_2}^2}
+{\cal O}\bigg(\frac{v^4}{{\tilde{m}_2}^4}\bigg) 
\bigg]
\label{eq:YukawastypeI}
\end{eqnarray}
Note that the Yukawas in the mass eigenbasis are diagonal, as expected, since the GW conditions ensure that there are no FCNCs. 
%Note that in the mixing language, by using \eqref{eq:varphi1ffmixing0} and \eqref{eq:UVYukawasI} one could also express these Yukawas in terms of the complex alignment parameter, but for brevity, we omit explicit expression in terms of $\Xi$ in what follows.

We define the coefficients of the four fermion operators for the components of the quark and lepton doublets in the quark mass eigenbasis as
\begin{eqnarray}
&&  
\omega^{uu ~\!\! (0)}_{ijmn}~\!
(u_i \bar{u}_j)(\bar{u}_m^\dagger u_n^\dagger)
+
\omega^{uu \pm ~\!\! (0)}_{ijmn}~\!
(d_i \bar{u}_j)(\bar{u}_m^\dagger d_n^\dagger) 
\nonumber \\
&&\! \! \! + ~
\omega^{dd ~\!\! (0)}_{ijmn}~\!
 (d_i \bar{d}_j)(\bar{d}_m^\dagger d_n^\dagger)
 +
 \omega^{dd\pm ~\!\! (0)}_{ijmn}~\!
(u_i \bar{d}_j)(\bar{d}_m^\dagger u_n^\dagger)
\nonumber \\
&&\! \! \! + ~
\omega^{\ell\ell ~\!\! (0)}_{ijmn}~\!
(\nu_i \bar{\ell}_j)(\bar{\ell}_m^\dagger \nu_n^\dagger) 
+
\omega^{\ell\ell \pm ~\!\! (0)}_{ijmn}~\!
 (\ell_i \bar{\ell}_j)(\bar{\ell}_m^\dagger \ell_n^\dagger) 
\nonumber \\
&&\! \! \! + ~
\Bigg[
\omega^{d\ell  ~\!\! (0)}_{ijmn}~\!
 (d_i \bar{d}_j)(\bar{\ell}_m^\dagger \ell_n^\dagger)
 +
\omega^{d\ell \pm ~\!\! (0)}_{ijmn}~\!
  (u_i \bar{d}_j)(\bar{\ell}_m^\dagger \nu_n^\dagger)
\nonumber \\
&& \! \! \! + ~ 
\omega^{ud  ~\!\! (2)}_{ijmn}~\!
(u_i \bar{u}_j)(d_m \bar{d}_n) 
+
\omega^{ud\pm  ~\!\! (2)}_{ijmn}~\!
 (d_i \bar{u}_j)(u_m \bar{d}_n) 
\nonumber \\
&& \! \! \! + ~
\omega^{u\ell  ~\!\! (2)}_{ijmn}~\!
(u_i \bar{u}_j)(\ell_m \bar{\ell}_n)
+
\omega^{u\ell\pm  ~\!\! (2)}_{ijmn}~\!
(d_i \bar{u}_j)(\nu_m \bar{\ell}_n)
+\textrm{h.c.} 
~
\Bigg]
\label{eq:fourfermiontypeI}
\end{eqnarray}
where the $\pm$ superindex indicates four fermion operators generated by integrating out the charged Higgs, while the operators with no $\pm$ superindex are induced by integrating out neutral heavy Higgs states. In the original gauge interaction basis for the quark fields, the coefficients of these two types of operators are identical, as in \eqref{eq:fourfermiontype}, due to gauge invariance of the operators \eqref{eq:fourfermion}. In the quark mass eigenbasis these two types of operators have different coefficients, since the field redefinitions \eqref{eq:flavortransf} act differently on the two components of the quark doublet $Q$. 
The coefficients in \eqref{eq:fourfermiontypeI} are found by using \eqref{eq:fourfermioncoefftypeI} in  \eqref{eq:fourfermiontype}, and performing the field redefinitions \eqref{eq:flavortransf}. We get 

%
%Note that in the original gauge interaction basis for the quark fields, the four fermion operators involving quark fields \eqref{eq:fourfermiontype} satisfy 
%\begin{eqnarray}
%\omega^{ff ~\!\! (0)}_{ijmn}&=&\omega^{ff \pm ~\!\! (0)}_{ijmn}=\Omega^{ff~\!\! (0)}_{ijmn }   \quad \quad \quad f=u,d 
%  \nonumber \\ 
%\omega^{d\ell  ~\!\! (0)}_{ijmn}&=&\omega^{d\ell \pm ~\!\! (0)}_{ijmn}={\Omega}^{d\ell  ~\!\! (0)}_{ijmn}
%  \nonumber \\ 
% \omega^{ud  ~\!\! (2)}_{ijmn}&=&\omega^{ud\pm  ~\!\! (2)}_{ijmn}= {\Omega}^{ud  ~\!\! (2)}_{ijmn}
%   \nonumber \\ 
% \omega^{u\ell  ~\!\! (2)}_{ijmn}&=&\omega^{u\ell\pm  ~\!\! (2)}_{ijmn}= {\Omega}^{u\ell  ~\!\! (2)}_{ijmn}
%\end{eqnarray}
\begin{eqnarray}
&&
\omega^{uu ~\!\! (0)}_{ijmn}
=2~ \! \frac{1}{{\tilde{m}_2}^2}\delta_{ij}\delta_{mn} \frac{m^{u}_{i} m^{u*}_{m}}{v^2} \cot^2 \beta 
\quad \quad 
\omega^{uu\pm ~\!\! (0)}_{ijmn}
=2~ \! \frac{1}{{\tilde{m}_2}^2} \frac{V^T_{ij} m^{u}_{j} m^{u*}_{m}V^*_{mn}}{v^2} \cot^2 \beta 
\nonumber \\ &&
\omega^{dd ~\!\! (0)}_{ijmn}
=2 ~ \! \frac{1}{{\tilde{m}_2}^2}\delta_{ij}\delta_{mn}  \frac{m^{d*}_{i} m^{d}_{m}}{v^2} \cot^2 \beta 
\quad\quad 
\omega^{dd\pm ~\!\! (0)}_{ijmn}
=2 ~ \! \frac{1}{{\tilde{m}_2}^2} \frac{V^*_{ij}m^{d*}_{j} m^{d}_{m}V^T_{mn}}{v^2} \cot^2 \beta  
\nonumber \\ &&
 \omega^{\ell\ell ~\!\! (0)}_{ijmn}= \omega^{\ell\ell \pm~\!\! (0)}_{ijmn}
 =
 2~  \! \frac{1}{{\tilde{m}_2}^2}\delta_{ij}\delta_{mn}  \frac{m^{\ell*}_{i} m^{\ell}_{m}}{v^2} \cot^2 \beta  \nonumber \\
% \quad \quad 
% \omega^{\ell\ell \pm~\!\! (0)}_{ijmn}
% =
% 2~  \! \frac{1}{{\tilde{m}_2}^2}\delta_{ij}\delta_{mn}  \frac{m^{\ell*}_{i} m^{\ell}_{m}}{v^2} \cot^2 \beta 
%\nonumber \\
 &&
\omega^{d\ell~\!\! (0)}_{ijmn}
=
 2 ~ \! \frac{1}{{\tilde{m}_2}^2} \delta_{ij}\delta_{mn}  \frac{m^{d*}_{i} m^{\ell}_{m}}{v^2} \cot^2 \beta  
 \quad\quad
\omega^{d\ell\pm ~\!\! (0)}_{ijmn}
=
 2 ~ \! \frac{1}{{\tilde{m}_2}^2} \delta_{mn} \frac{V^*_{ij}m^{d*}_{j} m^{\ell}_{m}}{v^2} \cot^2 \beta 
\nonumber \\ &&
 \omega^{ud ~\!\! (2)}_{ijmn}=
 2 ~  \! \frac{1}{{\tilde{m}_2}^2}\delta_{ij}\delta_{mn}  \frac{m^{u}_{i} m^{d*}_{m}}{v^2} \cot^2 \beta 
\quad\quad
 \omega^{ud\pm ~\!\! (2)}_{ijmn}=
-
 2 ~  \! \frac{1}{{\tilde{m}_2}^2} \frac{V^T_{ij} m^{u}_{j} V_{mn}^*m^{d*}_{n}}{v^2} \cot^2 \beta
\nonumber \\ &&
 \omega^{u\ell ~\!\! (2)}_{ijmn}=
 2 ~  \! \frac{1}{{\tilde{m}_2}^2} \delta_{ij}\delta_{mn} \frac{m^{u}_{i} m^{\ell*}_{m}}{v^2} \cot^2 \beta 
 \quad \quad
 \omega^{u\ell\pm ~\!\! (2)}_{ijmn}=
 -2 ~  \! \frac{1}{{\tilde{m}_2}^2} \delta_{mn} \frac{V^T_{ij} m^{u}_{j} m^{\ell*}_{m}}{v^2} \cot^2 \beta  
\nonumber \\ &&
\label{eq:fourfermiontypeI2}
\end{eqnarray}
where no sum over any index is intended. We keep the stars in the quark mass terms for generality, but in the quark mass eigenbasis they are real. All flavor and T violation in the four fermion operators comes from the CKM matrix, and it is induced by integrating out the charged heavy Higgses. We leave a more detailed discussion of T violation for section \ref{sec:Tviolationtypes}.

\subsection{Type II: $\lambda_1^u=\lambda_2^d=\lambda_2^{\ell}=0$}
\label{sec:typeII}

The type II 2HDM is defined by setting $\lambda_1^u=\lambda_2^d=\lambda_2^{\ell}=0$. Here we follow closely the calculations and the discussion performed in section \ref{sec:typeI}, and we omit repeating some of the details. The Yukawa matrices for the Higgs basis doublets are

\begin{eqnarray}
 \tilde{\lambda}_{1ij}^u\! \! \! &=& \! \! \! 
  e^{{i\xi\over 2}} \lambda_{2ij}^u\sin\beta 
   \nonumber \\
\label{eq:UVYukawas1II}   \tilde{\lambda}^{d,\ell}_{1ij}\! \! \! &=& \! \! \! 
  e^{-{i\xi\over 2}}    \lambda_{1ij}^{ d,\ell}\cos\beta 
   \end{eqnarray}
   \begin{eqnarray}
 \tilde{\lambda}_{2ij}^u\! \! \! &=& \! \! \! 
 \lambda_{2ij}^u\cos\beta
 \nonumber \\
\tilde{\lambda}^{d,\ell}_{2ij}\! \! \! &=& \! \! \! 
-e^{-i\xi}  \lambda_{1ij}^{d,\ell} \sin\beta  
\label{eq:UVYukawasII}
\end{eqnarray}
Using \eqref{eq:kUkD}, \eqref{eq:UVYukawas1II} and \eqref{eq:UVYukawasII} we rewrite  $\lambda_{2ij}^{u,d,\ell}$ in terms of the fermion mass matrices
\begin{eqnarray}
\nonumber \lambda_{2ij}^{u} \! \! \! &=& \! \! \!  e^{{-i\xi \over 2}}  \frac{\sqrt{2}m^{u}_{ij} }{v } \csc \beta \\
\lambda_{2ij}^{d,\ell}\! \! \! &=& \! \! \!  e^{{i\xi \over 2}}  \frac{\sqrt{2}m^{d,\ell}_{ij} }{v } \sec \beta 
\label{eq:secondyukawaII}
\end{eqnarray}
Using \eqref{eq:secondyukawaII} in \eqref{eq:UVYukawasII}, the Yukawas for the heavy doublet can be expressed in terms of the quark mass matrices as
\begin{eqnarray}
&& \tilde{\lambda}^{u}_{2ij} =  \sqrt{2}e^{-{i\xi \over 2}} \cot\beta ~\!\frac{m^{u}_{ij}}{v}
\nonumber \\
&& \tilde{\lambda}^{d,\ell}_{2ij} =  -\sqrt{2}e^{-{i\xi \over 2}} \tan\beta ~\!\frac{m^{d,\ell}_{ij}}{v}
\label{eq:heavyyukII}
\end{eqnarray}

Using \eqref{eq:heavyyukII} in \eqref{eq:varphiffEFT}, the resulting couplings of the physical Higgs to up type fermions in the low energy theory are
\begin{eqnarray}
 \lambda^u_{\varphi ij}
  \! \! \! &=& \! \! \!
~\!\frac{m_{ij}^u}{v}
 \bigg[1
 -{\tilde{\lambda}}_6^* e^{-i\xi /2 }\cot\beta
 ~\! \frac{v^2}{{\tilde{m}_2}^2}
  +{\cal O}\bigg(\frac{v^4}{{\tilde{m}_2}^4}\bigg)  
 \bigg]
  \nonumber \\
v  \lambda^u_{\varphi^2 ij}  \! \! \! &=& \! \! \!
  -~\! \frac{m^u_{ij}}{v} ~\! 
 \bigg[ 3  {\tilde{\lambda}}_6^* e^{-{i\xi \over 2}} \cot\beta 
~\! \frac{v^2}{{\tilde{m}_2}^2}
+{\cal O}\bigg(\frac{v^4}{{\tilde{m}_2}^4}\bigg) 
\bigg]
  \nonumber \\
 v^2\lambda^u_{\varphi^3 ij}
 \! \! \! &=& \! \! \!
  -~\! \frac{m^u_{ij}}{v} ~\! 
 \bigg[ 3  {\tilde{\lambda}}_6^* e^{-{i\xi \over 2}} \cot\beta 
~\! \frac{v^2}{{\tilde{m}_2}^2}
+{\cal O}\bigg(\frac{v^4}{{\tilde{m}_2}^4}\bigg) 
\bigg]
\end{eqnarray}
while down type quark Yukawas are
\begin{eqnarray}
 \lambda^{d,\ell}_{\varphi ij}
  \! \! \! &=& \! \! \!
 ~\!\frac{m_{ij}^{d,\ell}}{v}
 \bigg[1
 +{\tilde{\lambda}}_6^* e^{-i\xi /2 }\tan\beta
 ~\! \frac{v^2}{{\tilde{m}_2}^2}
  +{\cal O}\bigg(\frac{v^4}{{\tilde{m}_2}^4}\bigg)  
 \bigg]
  \nonumber \\
v  \lambda^{d,\ell}_{\varphi^2 ij}  \! \! \! &=& \! \! \! 
  ~\! \frac{m^{d,\ell}_{ij}}{v} ~\! 
 \bigg[ 3  {\tilde{\lambda}}_6^* e^{-{i\xi \over 2}} \tan\beta 
~\! \frac{v^2}{{\tilde{m}_2}^2}
+{\cal O}\bigg(\frac{v^4}{{\tilde{m}_2}^4}\bigg) 
\bigg]
  \nonumber \\
v^2\lambda^{d,\ell}_{\varphi^3 ij}
\! \! \! &=& \! \! \! 
  ~\! \frac{m^{d,\ell}_{ij}}{v} ~\! 
 \bigg[ 3  {\tilde{\lambda}}_6^* e^{-{i\xi \over 2}} \tan\beta 
~\! \frac{v^2}{{\tilde{m}_2}^2}
+{\cal O}\bigg(\frac{v^4}{{\tilde{m}_2}^4}\bigg) 
\bigg]
\end{eqnarray}
Note that using \eqref{eq:convenientxi} the Higgs Yukawas can be expressed in terms of the complex alignment parameter
\begin{eqnarray}
\lambda^{u}_{\varphi ij}  \! \! \! &=& \! \! \! 
\delta_{ij}~\!\frac{m_{ij}^{u}}{v} 
\bigg[1+  e^{ i ~ \! \! {\rm Arg}({\tilde{\lambda}}_5^*)/2} e^{-{i\xi \over 2}}\,  \Xi \, \cot\beta   +{\cal O}\Big(\Xi^2\Big ) 
\bigg] 
  \nonumber \\
v  \lambda^{u}_{\varphi^2 ij}  \! \! \! &=& \! \! \! 
  ~\! \frac{m^{u}_{ij}}{v} ~\! 
 \bigg[ 3  
 e^{ i ~ \! \! {\rm Arg}({\tilde{\lambda}}_5^*)/2} e^{-{i\xi \over 2}}\,  \Xi \, \cot\beta   +{\cal O}\Big(\Xi^2\Big )
\bigg]
\nonumber \\
v^2\lambda^{u}_{\varphi^3 ij}
 \! \! \! &=& \! \! \! 
  ~\! \frac{m^{u}_{ij}}{v} ~\! 
 \bigg[ 3  
 e^{ i ~ \! \! {\rm Arg}({\tilde{\lambda}}_5^*)/2} e^{-{i\xi \over 2}}\,  \Xi \, \cot\beta   +{\cal O}\Big(\Xi^2\Big )
\bigg]
\end{eqnarray}
%%%
\begin{eqnarray}
\lambda^{d,\ell}_{\varphi ij}  \! \! \! &=& \! \! \! 
\delta_{ij}~\!\frac{m_{ij}^{d,\ell}}{v} 
\bigg[1-  e^{ i ~ \! \! {\rm Arg}({\tilde{\lambda}}_5^*)/2} e^{-{i\xi \over 2}}\,  \Xi \, \tan\beta   +{\cal O}\Big(\Xi^2\Big ) 
\bigg] 
  \nonumber \\
v  \lambda^{d,\ell}_{\varphi^2 ij}  \! \! \! &=& \! \! \! 
  ~\!- \frac{m^{d,\ell}_{ij}}{v} ~\! 
 \bigg[ 3  
 e^{ i ~ \! \! {\rm Arg}({\tilde{\lambda}}_5^*)/2} e^{-{i\xi \over 2}}\,  \Xi \, \tan\beta   +{\cal O}\Big(\Xi^2\Big )
\bigg]
\nonumber \\
v^2\lambda^{d,\ell}_{\varphi^3 ij}
 \! \! \! &=& \! \! \! 
  ~\! -\frac{m^{d,\ell}_{ij}}{v} ~\! 
 \bigg[ 3  
 e^{ i ~ \! \! {\rm Arg}({\tilde{\lambda}}_5^*)/2} e^{-{i\xi \over 2}}\,  \Xi \, \tan\beta   +{\cal O}\Big(\Xi^2\Big )
\bigg]
\end{eqnarray}
In the EFT the type II 2HDM at large $\tan\beta$ the the only modifications to the up type Higgs Yukawas are subleading effects coming from dilution due to the complex alignment parameter by a factor $\sqrt{1-\abs{\Xi}^2}$ as discussed in section \ref{sec:couplings2HDMmixing}.  In this limit, the modifications to the SM predictions for the up type Higgs Yukawas and Higgs couplings to gauge bosons are identical: both couplings are diluted by $\sqrt{1-\abs{\Xi}^2}$. In this limit down type Yukawas suffer the largest deviations with respect to their SM values, since they are modified at effective dimension six, and the modifications are enhanced by $\tan\beta$.

The coefficients of the four fermion interactions \eqref{eq:fourfermion} are
\begin{eqnarray}
\Omega^{uu~\!\! (0)}_{ijmn }  
  \! \! \! &=& \! \! \!
 2 ~ 
 \! \frac{1}{{\tilde{m}_2}^2} \frac{m^{u}_{ij}m^{u\dagger}_{mn}}{v^2} \cot^2 \beta 
 \nonumber \\
 \Omega^{dd ~\!\! (0)}_{ijmn } 
  \! \! \! &=& \! \! \!
 2 ~ 
 \! \frac{1}{{\tilde{m}_2}^2} \frac{m^{d\dagger}_{ij} m^{d}_{mn}}{v^2} \tan^2 \beta  
  \nonumber \\
 \Omega^{\ell\ell ~\!\! (0)}_{ijmn } 
   \! \! \! &=& \! \! \!
 2 ~ 
 \! \frac{1}{{\tilde{m}_2}^2} \frac{m^{\ell\dagger}_{ij} m^{\ell}_{mn}}{v^2} \tan^2 \beta 
 \nonumber \\
\Omega^{d\ell ~\!\! (0)}_{ijmn}
  \! \! \! &=& \! \! \!
 2 ~ 
 \! \frac{1}{{\tilde{m}_2}^2} \frac{m^{d\dagger}_{ij} m^{\ell}_{mn}}{v^2} \tan^2 \beta 
 \nonumber \\
 \Omega^{u d ~\!\! (2)}_{ijmn}
  \! \! \! &=& \! \! \!
- 2 ~ 
 \! \frac{1}{{\tilde{m}_2}^2} \frac{m^{u}_{ij}m^{d\dagger}_{mn}}{v^2}
 \nonumber \\
\Omega^{u \ell ~\!\! (2)}_{ijmn}
  \! \! \! &=& \! \! \!
- 2 ~ 
 \! \frac{1}{{\tilde{m}_2}^2} \frac{m^{u}_{ij}m^{\ell\dagger}_{mn}}{v^2} 
\end{eqnarray}

We now rotate to the quark mass eigenbasis. The up type Yukawas are
\begin{eqnarray}
 \lambda^u_{\varphi ij}
  \! \! \! &=& \! \! \!
 \delta_{ij}~\!\frac{m_{i}^u}{v}
 \bigg[1
 -{\tilde{\lambda}}_6^* e^{-i\xi /2 }\cot\beta
 ~\! \frac{v^2}{{\tilde{m}_2}^2}
  +{\cal O}\bigg(\frac{v^4}{{\tilde{m}_2}^4}\bigg)  
 \bigg]
  \nonumber \\
v  \lambda^u_{\varphi^2 ij}  \! \! \! &=& \! \! \! 
  -\delta_{ij}~\! \frac{m^u_{i}}{v} ~\! 
 \bigg[ 3  {\tilde{\lambda}}_6^* e^{-{i\xi \over 2}} \cot\beta 
~\! \frac{v^2}{{\tilde{m}_2}^2}
+{\cal O}\bigg(\frac{v^4}{{\tilde{m}_2}^4}\bigg) 
\bigg]
  \nonumber \\
v^2\lambda^u_{\varphi^3 ij}
\! \! \! &=& \! \! \! 
  -\delta_{ij}~\! \frac{m^u_{i}}{v} ~\! 
 \bigg[ 3  {\tilde{\lambda}}_6^* e^{-{i\xi \over 2}} \cot\beta 
~\! \frac{v^2}{{\tilde{m}_2}^2}
+{\cal O}\bigg(\frac{v^4}{{\tilde{m}_2}^4}\bigg) 
\bigg]
\label{eq:YukawastypeIIA}
\end{eqnarray}
while the down type Yukawas are
\begin{eqnarray}
 \lambda^{d,\ell}_{\varphi ij}
  \! \! \! &=& \! \! \!
 \delta_{ij}~\!\frac{m_{i}^{d,\ell}}{v}
 \bigg[1
 +{\tilde{\lambda}}_6^* e^{-i\xi /2 }\tan\beta
 ~\! \frac{v^2}{{\tilde{m}_2}^2}
  +{\cal O}\bigg(\frac{v^4}{{\tilde{m}_2}^4}\bigg)  
 \bigg]
  \nonumber \\
v  \lambda^{d,\ell}_{\varphi^2 ij}  \! \! \! &=& \! \! \!
\delta_{ij}  ~\! \frac{m^{d,\ell}_{i}}{v} ~\! 
 \bigg[ 3  {\tilde{\lambda}}_6^* e^{-{i\xi \over 2}} \tan\beta 
~\! \frac{v^2}{{\tilde{m}_2}^2}
+{\cal O}\bigg(\frac{v^4}{{\tilde{m}_2}^4}\bigg) 
\bigg]
  \nonumber \\
 v^2\lambda^{d,\ell}_{\varphi^3 ij}
 \! \! \! &=& \! \! \!
\delta_{ij}  ~\! \frac{m^{d,\ell}_{i}}{v} ~\! 
 \bigg[ 3  {\tilde{\lambda}}_6^* e^{-{i\xi \over 2}} \tan\beta 
~\! \frac{v^2}{{\tilde{m}_2}^2}
+{\cal O}\bigg(\frac{v^4}{{\tilde{m}_2}^4}\bigg) 
\bigg]
\label{eq:YukawastypeIIB}
\end{eqnarray}

Note that in a type II 2HDM $\tan \beta$ can be measured from comparing the deviations of up type Yukawas versus down type Yukawas from their SM values. Using  \eqref{eq:YukawastypeIIA} and \eqref{eq:YukawastypeIIB} we get
%%% put words around this. deviation with respect to SM. also put absolute value. 
\begin{equation}
\tan^2\beta=\Bigg |
\frac{m^u_{i}}{m^{d,\ell}_{i}} \bigg(
 \frac{\lambda^{d,\ell}_{\varphi ij}-m^{d,\ell}_{i}/v}
 {\lambda_{\varphi ij}^u-m^u_{i}/v} 
 \Bigg)\Bigg |
 +{\cal O}\bigg(\frac{v^2}{{\tilde{m}_2}^2}\bigg) 
\end{equation}

The four fermion interactions in the quark mass  eigenbasis are defined as in \eqref{eq:fourfermiontypeI}. They are given by
\begin{eqnarray}
&&
\omega^{uu ~\!\! (0)}_{ijmn}
=2~ \! \frac{1}{{\tilde{m}_2}^2}\delta_{ij}\delta_{mn} \frac{m^{u}_{i} m^{u*}_{m}}{v^2} \cot^2 \beta 
\quad \quad
\omega^{uu\pm ~\!\! (0)}_{ijmn}
=2~ \! \frac{1}{{\tilde{m}_2}^2} \frac{V^T_{ij} m^{u}_{j} m^{u*}_{m}V^*_{mn}}{v^2} \cot^2 \beta 
\nonumber \\ & &
\omega^{dd ~\!\! (0)}_{ijmn}
=2 ~ \! \frac{1}{{\tilde{m}_2}^2}\delta_{ij}\delta_{mn}  \frac{m^{d*}_{i} m^{d}_{m}}{v^2} \tan^2 \beta 
\quad \quad
\omega^{dd\pm ~\!\! (0)}_{ijmn}
=2 ~ \! \frac{1}{{\tilde{m}_2}^2} \frac{V^*_{ij}m^{d*}_{j} m^{d}_{m}V^T_{mn}}{v^2} \tan^2 \beta  
\nonumber \\ & &
 \omega^{\ell\ell ~\!\! (0)}_{ijmn}= \omega^{\ell\ell \pm~\!\! (0)}_{ijmn}
 =
 2~  \! \frac{1}{{\tilde{m}_2}^2}\delta_{ij}\delta_{mn}  \frac{m^{\ell*}_{i} m^{\ell}_{m}}{v^2} \tan^2 \beta 
\quad \quad
%
% =
% 2~  \! \frac{1}{{\tilde{m}_2}^2}\delta_{ij}\delta_{mn}  \frac{m^{\ell*}_{i} m^{\ell}_{m}}{v^2} \tan^2 \beta 
\nonumber \\ & &
\omega^{d\ell~\!\! (0)}_{ijmn}
=
 2 ~ \! \frac{1}{{\tilde{m}_2}^2} \delta_{ij}\delta_{mn}  \frac{m^{d*}_{i} m^{\ell}_{m}}{v^2} \tan^2 \beta  
\quad \quad
\omega^{d\ell\pm ~\!\! (0)}_{ijmn}
=
 2 ~ \! \frac{1}{{\tilde{m}_2}^2} \delta_{mn} \frac{V^*_{ij}m^{d*}_{j} m^{\ell}_{m}}{v^2} \tan^2 \beta 
\nonumber \\ & &
 \omega^{ud ~\!\! (2)}_{ijmn}=
 -2 ~  \! \frac{1}{{\tilde{m}_2}^2}\delta_{ij}\delta_{mn}  \frac{m^{u}_{i} m^{d*}_{m}}{v^2}
\quad \quad \quad  \quad
 \omega^{ud\pm ~\!\! (2)}_{ijmn}=
 2 ~  \! \frac{1}{{\tilde{m}_2}^2} \frac{V^T_{ij} m^{u}_{j} V_{mn}^*m^{d*}_{n}}{v^2}
\nonumber \\ & &
 \omega^{u\ell ~\!\! (2)}_{ijmn}=
 -2 ~  \! \frac{1}{{\tilde{m}_2}^2} \delta_{ij}\delta_{mn} \frac{m^{u}_{i} m^{\ell*}_{m}}{v^2}
\quad \quad \quad \quad 
 \omega^{u\ell\pm ~\!\! (2)}_{ijmn}=
 2 ~  \! \frac{1}{{\tilde{m}_2}^2} \delta_{mn} \frac{V^T_{ij} m^{u}_{j} m^{\ell*}_{m}}{v^2}
\label{eq:fourfermiontypeII2}
\end{eqnarray}
where $\pm$ superindex indicates four fermion operators generated by charged Higgs exchange and $V$ is the CKM matrix. Note that some of the four fermion operators in \eqref{eq:fourfermiontypeII2} are $\tan\beta$ independent.
%\textit{Note for myself: the limits from $b\rightarrow s \gamma$ are $\tan\beta$ independent. I think that the reason is the following. The one loop diagram at zero photon momentum vanishes since the photon mass is protected by gauge invariance. At first order in the photon momentum, the chirality of the $b$ and $s$ external quarks are inverted. The only four fermion operator that leads to this diagram is $ \omega^{ud\pm ~\!\! (2)}_{ijmn}$, which is $\tan\beta$ independent. This is actually the piece of the calculation that is most relevant for the 1 loop computation, except at very low $\tan\beta$, in which case the  }

\subsection{ Type III}
\label{sec:typeIII}

The type III 2HDM is defined by setting  $\lambda_1^u=\lambda_1^d=\lambda_2^{\ell}=0$. The Yukawa matrices for the Higgs basis doublets are
\begin{eqnarray}
 \tilde{\lambda}^{u,d}_{1ij}   \! \! \! &=& \! \! \! 
 e^{{i\xi \over 2}} \lambda_{2ij}^{u,d}\sin\beta
 \nonumber \\
 \tilde{\lambda}^{\ell}_{1ij}   \! \! \! &=& \! \! \! 
 e^{-{i\xi \over 2}} \lambda_{1ij}^{\ell}\cos\beta
 \label{eq:UVYukawasIII0}
 \end{eqnarray}
 \begin{eqnarray}
\tilde{\lambda}^{u,d}_{2ij}  \! \! \! &=& \! \! \! 
\lambda_{2ij}^{u,d}\cos\beta  \nonumber \\
\tilde{\lambda}^{\ell}_{2ij}  \! \! \! &=& \! \! \! 
-e^{-i\xi}\lambda_{1ij}^{\ell}\sin\beta
\label{eq:UVYukawasIII}
\end{eqnarray}
Using \eqref{eq:kUkD}, \eqref{eq:UVYukawasIII0} and \eqref{eq:UVYukawasIII} we rewrite  $\lambda_{2ij}^{u,d,\ell}$ in terms of the fermion mass matrices
\begin{eqnarray}
\nonumber \lambda_{2ij}^{u,d} \! \! \! &=& \! \! \!  e^{{-i\xi \over 2}}  \frac{\sqrt{2}m^{u,d}_{ij} }{v } \csc \beta \\
\lambda_{2ij}^{\ell}\! \! \! &=& \! \! \!  e^{{i\xi \over 2}}  \frac{\sqrt{2}m^{\ell}_{ij} }{v } \sec \beta 
\label{eq:secondyukawaIII}
\end{eqnarray}
Using \eqref{eq:UVYukawasIII} and \eqref{eq:secondyukawaIII} the Yukawas for the heavy doublet can be expressed in terms of the quark mass matrices as
\begin{eqnarray}
&& \tilde{\lambda}^{u,d}_{2ij} =  \sqrt{2}e^{-{i\xi \over 2}} \cot\beta ~\!\frac{m^{u,d}_{ij}}{v}
\nonumber \\
&& \tilde{\lambda}^{\ell}_{2ij} =  -\sqrt{2}e^{-{i\xi \over 2}} \tan\beta ~\!\frac{m^{\ell}_{ij}}{v}
\label{eq:heavyyukIII}
\end{eqnarray}

Using \eqref{eq:heavyyukIII} in \eqref{eq:varphiffEFT}, the resulting couplings of the physical Higgs to up and down type quarks  in the low energy theory are
\begin{eqnarray}
 \lambda^{u,d}_{\varphi ij}
  \! \! \! &=& \! \! \!
\frac{m_{ij}^{u,d}}{v}
 \bigg[1
 -{\tilde{\lambda}}_6^* e^{-i\xi /2 }\cot\beta
 ~\! \frac{v^2}{{\tilde{m}_2}^2}
  +{\cal O}\bigg(\frac{v^4}{{\tilde{m}_2}^4}\bigg)  
 \bigg]
  \nonumber \\
v  \lambda^{u,d}_{\varphi^2 ij}  \! \! \! &=& \! \! \! 
  - \frac{m^{u,d}_{ij}}{v} ~\! 
 \bigg[ 3  {\tilde{\lambda}}_6^* e^{-{i\xi \over 2}} \cot\beta 
~\! \frac{v^2}{{\tilde{m}_2}^2}
+{\cal O}\bigg(\frac{v^4}{{\tilde{m}_2}^4}\bigg) 
\bigg]
\nonumber \\
v^2\lambda^{u,d}_{\varphi^3 ij}
\! \! \! &=& \! \! \! 
  - \frac{m^{u,d}_{ij}}{v} ~\! 
 \bigg[ 3  {\tilde{\lambda}}_6^* e^{-{i\xi \over 2}} \cot\beta 
~\! \frac{v^2}{{\tilde{m}_2}^2}
+{\cal O}\bigg(\frac{v^4}{{\tilde{m}_2}^4}\bigg) 
\bigg]
\end{eqnarray}
while the lepton Yukawas are
\begin{eqnarray}
 \lambda^{\ell}_{\varphi ij}
  \! \! \! &=& \! \! \!
\frac{m_{ij}^{\ell}}{v}
 \bigg[1
 +{\tilde{\lambda}}_6^* e^{-i\xi /2 }\tan\beta
 ~\! \frac{v^2}{{\tilde{m}_2}^2}
  +{\cal O}\bigg(\frac{v^4}{{\tilde{m}_2}^4}\bigg)  
 \bigg]
  \nonumber \\
v  \lambda^{\ell}_{\varphi^2 ij}  \! \! \! &=& \! \! \! 
  \frac{m^{\ell}_{ij}}{v} ~\! 
 \bigg[ 3  {\tilde{\lambda}}_6^* e^{-{i\xi \over 2}} \tan\beta 
~\! \frac{v^2}{{\tilde{m}_2}^2}
+{\cal O}\bigg(\frac{v^4}{{\tilde{m}_2}^4}\bigg) 
\bigg]
\nonumber \\
v^2\lambda^{\ell}_{\varphi^3 ij}
 \! \! \! &=& \! \! \!
   \frac{m^{\ell}_{ij}}{v} ~\! 
 \bigg[ 3  {\tilde{\lambda}}_6^* e^{-{i\xi \over 2}} \tan\beta 
~\! \frac{v^2}{{\tilde{m}_2}^2}
+{\cal O}\bigg(\frac{v^4}{{\tilde{m}_2}^4}\bigg) 
\bigg]
\end{eqnarray}

The  four fermion interactions \eqref{eq:fourfermion} are 
\begin{eqnarray}
\Omega^{uu~\!\! (0)}_{ijmn }  
  \! \! \! &=& \! \! \!
 2 ~ 
 \! \frac{1}{{\tilde{m}_2}^2} \frac{m^{u}_{ij} m^{u\dagger}_{mn}}{v^2} \cot^2 \beta 
 \nonumber \\
 \Omega^{dd ~\!\! (0)}_{ijmn } 
  \! \! \! &=& \! \! \!
 2 ~ 
 \! \frac{1}{{\tilde{m}_2}^2} \frac{m^{d\dagger}_{ij} m^{d}_{mn}}{v^2} \cot^2 \beta  
  \nonumber \\
 \Omega^{\ell\ell ~\!\! (0)}_{ijmn } 
   \! \! \! &=& \! \! \!
 2 ~ 
 \! \frac{1}{{\tilde{m}_2}^2} \frac{m^{\ell\dagger}_{ij} m^{\ell}_{mn}}{v^2} \tan^2 \beta 
 \nonumber \\
\Omega^{d\ell ~\!\! (0)}_{ijmn}
  \! \! \! &=& \! \! \!
 - 2 ~ 
 \! \frac{1}{{\tilde{m}_2}^2} \frac{m^{d\dagger}_{ij} m^{\ell}_{mn}}{v^2} 
 \nonumber \\
 \Omega^{u d ~\!\! (2)}_{ijmn}
  \! \! \! &=& \! \! \!
 2 ~ 
 \! \frac{1}{{\tilde{m}_2}^2} \frac{m^{u}_{ij} m^{d\dagger}_{mn}}{v^2} \cot^2 \beta 
 \nonumber \\
\Omega^{u \ell ~\!\! (2)}_{ijmn}
  \! \! \! &=& \! \! \!
- 2 ~ 
 \! \frac{1}{{\tilde{m}_2}^2} \frac{m^{u}_{ij} m^{\ell\dagger}_{mn}}{v^2}
\end{eqnarray}

We now rotate to the quark mass eigenbasis by using \eqref{eq:flavortransf}. The Yukawas are again diagonal in flavor space, and given by 
\begin{eqnarray}
 \lambda^{u,d}_{\varphi ij}
  \! \! \! &=& \! \! \!
 \delta_{ij}~\!\frac{m_{i}^{u,d}}{v}
 \bigg[1
 -{\tilde{\lambda}}_6^* e^{-i\xi /2 }\cot\beta
 ~\! \frac{v^2}{{\tilde{m}_2}^2}
  +{\cal O}\bigg(\frac{v^4}{{\tilde{m}_2}^4}\bigg)  
 \bigg]
  \nonumber \\
v  \lambda^{u,d}_{\varphi^2 ij}  \! \! \! &=& \! \! \! 
  -\delta_{ij}~\! \frac{m^{u,d}_{i}}{v} ~\! 
 \bigg[ 3  {\tilde{\lambda}}_6^* e^{-{i\xi \over 2}} \cot\beta 
~\! \frac{v^2}{{\tilde{m}_2}^2}
+{\cal O}\bigg(\frac{v^4}{{\tilde{m}_2}^4}\bigg) 
\bigg]
\nonumber \\
v^2\lambda^{u,d}_{\varphi^3 ij}
\! \! \! &=& \! \! \! 
  -\delta_{ij}~\! \frac{m^{u,d}_{i}}{v} ~\! 
 \bigg[ 3  {\tilde{\lambda}}_6^* e^{-{i\xi \over 2}} \cot\beta 
~\! \frac{v^2}{{\tilde{m}_2}^2}
+{\cal O}\bigg(\frac{v^4}{{\tilde{m}_2}^4}\bigg) 
\bigg]
\label{eq:YukawastypeIIIA}
\end{eqnarray}
\begin{eqnarray}
 \lambda^{\ell}_{\varphi ij}
  \! \! \! &=& \! \! \!
 \delta_{ij}~\!\frac{m_{i}^{\ell}}{v}
 \bigg[1
 +{\tilde{\lambda}}_6^* e^{-i\xi /2 }\tan\beta
 ~\! \frac{v^2}{{\tilde{m}_2}^2}
  +{\cal O}\bigg(\frac{v^4}{{\tilde{m}_2}^4}\bigg)  
 \bigg]
  \nonumber \\
v  \lambda^{\ell}_{\varphi^2 ij}  \! \! \! &=& \! \! \! 
\delta_{ij}  ~\! \frac{m^{\ell}_{i}}{v} ~\! 
 \bigg[ 3  {\tilde{\lambda}}_6^* e^{-{i\xi \over 2}} \tan\beta 
~\! \frac{v^2}{{\tilde{m}_2}^2}
+{\cal O}\bigg(\frac{v^4}{{\tilde{m}_2}^4}\bigg) 
\bigg]
\nonumber \\
v^2\lambda^{\ell}_{\varphi^3 ij}
 \! \! \! &=& \! \! \! 
\delta_{ij}  ~\! \frac{m^{\ell}_{i}}{v} ~\! 
 \bigg[ 3  {\tilde{\lambda}}_6^* e^{-{i\xi \over 2}} \tan\beta 
~\! \frac{v^2}{{\tilde{m}_2}^2}
+{\cal O}\bigg(\frac{v^4}{{\tilde{m}_2}^4}\bigg) 
\bigg]
\label{eq:YukawastypeIIIB}
\end{eqnarray}

The four fermion interactions in the quark mass  eigenbasis are defined as in \eqref{eq:fourfermiontypeI}. They are given by
\begin{eqnarray}
&&
\omega^{uu ~\!\! (0)}_{ijmn}
=2~ \! \frac{1}{{\tilde{m}_2}^2}\delta_{ij}\delta_{mn} \frac{m^{u}_{i} m^{u*}_{m}}{v^2} \cot^2 \beta 
\quad \quad
\omega^{uu\pm ~\!\! (0)}_{ijmn}
=2~ \! \frac{1}{{\tilde{m}_2}^2} \frac{V^T_{ij} m^{u}_{j} m^{u*}_{m}V^*_{mn}}{v^2} \cot^2 \beta 
\nonumber \\ 
&&
\omega^{dd ~\!\! (0)}_{ijmn}
=2 ~ \! \frac{1}{{\tilde{m}_2}^2}\delta_{ij}\delta_{mn}  \frac{m^{d*}_{i} m^{d}_{m}}{v^2} \cot^2 \beta 
\quad \quad
\omega^{dd\pm ~\!\! (0)}_{ijmn}
=2 ~ \! \frac{1}{{\tilde{m}_2}^2} \frac{V^*_{ij}m^{d*}_{j} m^{d}_{m}V^T_{mn}}{v^2} \cot^2 \beta  
\nonumber \\
&&
 \omega^{\ell\ell ~\!\! (0)}_{ijmn}
 = \omega^{\ell\ell \pm~\!\! (0)}_{ijmn}=
 2~  \! \frac{1}{{\tilde{m}_2}^2}\delta_{ij}\delta_{mn}  \frac{m^{\ell*}_{i} m^{\ell}_{m}}{v^2} \tan^2 \beta 
\quad \quad
%
% =
% 2~  \! \frac{1}{{\tilde{m}_2}^2}\delta_{ij}\delta_{mn}  \frac{m^{\ell*}_{i} m^{\ell}_{m}}{v^2} \tan^2 \beta 
\nonumber \\ 
&&
\omega^{d\ell~\!\! (0)}_{ijmn}
=
 -2 ~ \! \frac{1}{{\tilde{m}_2}^2} \delta_{ij}\delta_{mn}  \frac{m^{d*}_{i} m^{\ell}_{m}}{v^2} 
\quad \quad \quad \quad 
\omega^{d\ell\pm ~\!\! (0)}_{ijmn}
=
 -2 ~ \! \frac{1}{{\tilde{m}_2}^2} \delta_{mn} \frac{V^*_{ij}m^{d*}_{j} m^{\ell}_{m}}{v^2} 
\nonumber \\ 
&&
 \omega^{ud ~\!\! (2)}_{ijmn}=
 -2 ~  \! \frac{1}{{\tilde{m}_2}^2}\delta_{ij}\delta_{mn}  \frac{m^{u}_{i} m^{d*}_{m}}{v^2} \cot^2 \beta
\quad ~
 \omega^{ud\pm ~\!\! (2)}_{ijmn}=
 2 ~  \! \frac{1}{{\tilde{m}_2}^2} \frac{V^T_{ij} m^{u}_{j} V_{mn}^*m^{d*}_{n}}{v^2} \cot^2 \beta
\nonumber \\ 
&&
 \omega^{u\ell ~\!\! (2)}_{ijmn}=
 -2 ~  \! \frac{1}{{\tilde{m}_2}^2} \delta_{ij}\delta_{mn} \frac{m^{u}_{i} m^{\ell*}_{m}}{v^2}
\quad \quad \quad \quad \,
 \omega^{u\ell\pm ~\!\! (2)}_{ijmn}=
 2 ~  \! \frac{1}{{\tilde{m}_2}^2} \delta_{mn} \frac{V^T_{ij} m^{u}_{j} m^{\ell*}_{m}}{v^2}
\label{eq:fourfermiontypeIII2}
\end{eqnarray}
where $\pm$ superindex indicates four fermion operators generated by charged Higgs exchange and $V$ is the CKM matrix. 
\subsection{Type IV}
\label{sec:typeIV}

The type IV 2HDM is defined by setting  $\lambda_1^u=\lambda_2^d=\lambda_1^{\ell}=0$. The Yukawa matrices for the Higgs basis doublets are
\begin{eqnarray}
 \tilde{\lambda}^{u,\ell}_{1ij}   \! \! \! &=& \! \! \! 
 e^{{i\xi \over 2}} \lambda_{2ij}^{u,\ell}\sin\beta
 \nonumber \\
 \tilde{\lambda}^{d}_{1ij}   \! \! \! &=& \! \! \! 
 e^{-{i\xi \over 2}} \lambda_{1ij}^{d}\cos\beta
 \label{eq:UVYukawasIV0}
\end{eqnarray}
\begin{eqnarray}
\tilde{\lambda}^{u,\ell}_{2ij}  \! \! \! &=& \! \! \! 
\lambda_{2ij}^{u,\ell}\cos\beta  \nonumber \\
\tilde{\lambda}^{d}_{2ij}  \! \! \! &=& \! \! \! 
-e^{-i\xi}\lambda_{1ij}^{d}\sin\beta
\label{eq:UVYukawasIV}
\end{eqnarray}
Using \eqref{eq:kUkD}, \eqref{eq:UVYukawasIV0} and \eqref{eq:UVYukawasIV} we rewrite  $\lambda_{2ij}^{u,d,\ell}$ in terms of the fermion mass matrices
\begin{eqnarray}
\nonumber \lambda_{2ij}^{u,d} \! \! \! &=& \! \! \!  e^{{-i\xi \over 2}}  \frac{\sqrt{2}m^{u,d}_{ij} }{v } \csc \beta \\
\lambda_{2ij}^{\ell}\! \! \! &=& \! \! \!  e^{{i\xi \over 2}}  \frac{\sqrt{2}m^{\ell}_{ij} }{v } \sec \beta 
\label{eq:secondyukawaIV}
\end{eqnarray}

Using  \eqref{eq:UVYukawasIV} and \eqref{eq:secondyukawaIV}, the Yukawas for the heavy doublet can be expressed in terms of the quark mass matrices as
\begin{eqnarray}
&& \tilde{\lambda}^{u,d}_{2ij} =  \sqrt{2}e^{-{i\xi \over 2}} \cot\beta ~\!\frac{m^{u,d}_{ij}}{v}
\nonumber \\
&& \tilde{\lambda}^{\ell}_{2ij} =  -\sqrt{2}e^{-{i\xi \over 2}} \tan\beta ~\!\frac{m^{\ell}_{ij}}{v}
\label{eq:heavyyukIV}
\end{eqnarray}

Using \eqref{eq:heavyyukIV} in \eqref{eq:varphiffEFT}, the resulting couplings of the physical Higgs to up type quarks and leptons in the low energy theory are
\begin{eqnarray}
 \lambda^{u,\ell}_{\varphi ij}
  \! \! \! &=& \! \! \!
\frac{m_{ij}^{u,\ell}}{v}
 \bigg[1
 -{\tilde{\lambda}}_6^* e^{-i\xi /2 }\cot\beta
 ~\! \frac{v^2}{{\tilde{m}_2}^2}
  +{\cal O}\bigg(\frac{v^4}{{\tilde{m}_2}^4}\bigg)  
 \bigg]
  \nonumber \\
v  \lambda^{u,\ell}_{\varphi^2 ij}  
\! \! \! &=& \! \! \! 
 -\frac{m^{u,\ell}_{ij}}{v} ~\! 
 \bigg[ 3  {\tilde{\lambda}}_6^* e^{-{i\xi \over 2}} \cot\beta 
~\! \frac{v^2}{{\tilde{m}_2}^2}
+{\cal O}\bigg(\frac{v^4}{{\tilde{m}_2}^4}\bigg) 
\bigg]
\nonumber \\
v^2\lambda^{u,\ell}_{\varphi^3 ij}
\! \! \! &=& \! \! \! 
  -\frac{m^{u,\ell}_{ij}}{v} ~\! 
 \bigg[ 3  {\tilde{\lambda}}_6^* e^{-{i\xi \over 2}} \cot\beta 
~\! \frac{v^2}{{\tilde{m}_2}^2}
+{\cal O}\bigg(\frac{v^4}{{\tilde{m}_2}^4}\bigg) 
\bigg]
\label{eq:YukawastypeIVA}
\end{eqnarray}
while the down type quark Yukawas are
\begin{eqnarray}
 \lambda^{d}_{\varphi ij}
  \! \! \! &=& \! \! \!
 \frac{m_{ij}^{d}}{v}
 \bigg[1
 +{\tilde{\lambda}}_6^* e^{-i\xi /2 }\tan\beta
 ~\! \frac{v^2}{{\tilde{m}_2}^2}
  +{\cal O}\bigg(\frac{v^4}{{\tilde{m}_2}^4}\bigg)  
 \bigg]
  \nonumber \\
v  \lambda^{d}_{\varphi^2 ij}  \! \! \! &=& \! \! \! 
 \frac{m^{d}_{ij}}{v} ~\! 
 \bigg[ 3  {\tilde{\lambda}}_6^* e^{-{i\xi \over 2}} \tan\beta 
~\! \frac{v^2}{{\tilde{m}_2}^2}
+{\cal O}\bigg(\frac{v^4}{{\tilde{m}_2}^4}\bigg) 
\bigg]
\nonumber \\
v^2\lambda^{d}_{\varphi^3 ij}
\! \! \! &=& \! \! \! 
  \frac{m^{d}_{ij}}{v} ~\! 
 \bigg[ 3  {\tilde{\lambda}}_6^* e^{-{i\xi \over 2}} \tan\beta 
~\! \frac{v^2}{{\tilde{m}_2}^2}
+{\cal O}\bigg(\frac{v^4}{{\tilde{m}_2}^4}\bigg) 
\bigg]
\label{eq:YukawastypeIVB}
\end{eqnarray}

The  four fermion interactions \eqref{eq:fourfermion} are 
\begin{eqnarray}
\Omega^{uu~\!\! (0)}_{ijmn }  
  \! \! \! &=& \! \! \!
 2~ 
 \! \frac{1}{{\tilde{m}_2}^2} \frac{m^{u}_{ij} m^{u\dagger}_{mn}}{v^2} \cot^2 \beta 
 \nonumber \\
 \Omega^{dd ~\!\! (0)}_{ijmn } 
  \! \! \! &=& \! \! \!
 2~ 
 \! \frac{1}{{\tilde{m}_2}^2} \frac{m^{d\dagger}_{ij} m^{d}_{mn}}{v^2} \tan^2 \beta  
  \nonumber \\
 \Omega^{\ell\ell ~\!\! (0)}_{ijmn } 
   \! \! \! &=& \! \! \!
 2~ 
 \! \frac{1}{{\tilde{m}_2}^2} \frac{m^{\ell\dagger}_{ij} m^{\ell}_{mn}}{v^2} \cot^2 \beta 
 \nonumber \\
\Omega^{d\ell ~\!\! (0)}_{ijmn}
  \! \! \! &=& \! \! \!
 - 2~ 
 \! \frac{1}{{\tilde{m}_2}^2} \frac{m^{d\dagger}_{ij} m^{\ell}_{mn}}{v^2}
 \nonumber \\
 \Omega^{u d ~\!\! (2)}_{ijmn}
  \! \! \! &=& \! \! \!
- 2~ 
 \! \frac{1}{{\tilde{m}_2}^2} \frac{m^{u}_{ij} m^{d\dagger}_{mn}}{v^2}
 \nonumber \\
\Omega^{u \ell ~\!\! (2)}_{ijmn}
  \! \! \! &=& \! \! \!
 2~ 
 \! \frac{1}{{\tilde{m}_2}^2} \frac{m^{u}_{ij} m^{\ell\dagger}_{mn}}{v^2} \cot^2 \beta 
\end{eqnarray}
We now rotate to the quark mass eigenbasis by using \eqref{eq:flavortransf}. The Yukawas are again diagonal in flavor space. The up type and lepton Yukawas are
\begin{eqnarray}
 \lambda^{u,\ell}_{\varphi ij}
  \! \! \! &=& \! \! \!
 \delta_{ij}~\!\frac{m_{i}^{u,\ell}}{v}
 \bigg[1
 -{\tilde{\lambda}}_6^* e^{-i\xi /2 }\cot\beta
 ~\! \frac{v^2}{{\tilde{m}_2}^2}
  +{\cal O}\bigg(\frac{v^4}{{\tilde{m}_2}^4}\bigg)  
 \bigg]
  \nonumber \\
v  \lambda^{u,\ell}_{\varphi^2 ij}  \! \! \! &=& \! \! \!
  -\delta_{ij}~\! \frac{m^{u,\ell}_{i}}{v} ~\! 
 \bigg[ 3  {\tilde{\lambda}}_6^* e^{-{i\xi \over 2}} \cot\beta 
~\! \frac{v^2}{{\tilde{m}_2}^2}
+{\cal O}\bigg(\frac{v^4}{{\tilde{m}_2}^4}\bigg) 
\bigg]
\nonumber \\ 
 v^2\lambda^{u,\ell}_{\varphi^3 ij}
  \! \! \! &=& \! \! \!
  -\delta_{ij}~\! \frac{m^{u,\ell}_{i}}{v} ~\! 
 \bigg[ 3  {\tilde{\lambda}}_6^* e^{-{i\xi \over 2}} \cot\beta 
~\! \frac{v^2}{{\tilde{m}_2}^2}
+{\cal O}\bigg(\frac{v^4}{{\tilde{m}_2}^4}\bigg) 
\bigg]
\end{eqnarray}
while the down quark Yukawas are
\begin{eqnarray}
 \lambda^{d}_{\varphi ij}
  \! \! \! &=& \! \! \!
 \delta_{ij}~\!\frac{m_{i}^{d}}{v}
 \bigg[1
 +{\tilde{\lambda}}_6^* e^{-i\xi /2 }\tan\beta
 ~\! \frac{v^2}{{\tilde{m}_2}^2}
  +{\cal O}\bigg(\frac{v^4}{{\tilde{m}_2}^4}\bigg)  
 \bigg]
  \nonumber \\
v  \lambda^{d}_{\varphi^2 ij}  \! \! \! &=& \! \! \! 
\delta_{ij}  ~\! \frac{m^{d}_{i}}{v} ~\! 
 \bigg[ 3  {\tilde{\lambda}}_6^* e^{-{i\xi \over 2}} \tan\beta 
~\! \frac{v^2}{{\tilde{m}_2}^2}
+{\cal O}\bigg(\frac{v^4}{{\tilde{m}_2}^4}\bigg) 
\bigg]
\nonumber \\
v^2\lambda^{d}_{\varphi^3 ij}
  \! \! \! &=& \! \! \!
\delta_{ij}  ~\! \frac{m^{d}_{i}}{v} ~\! 
 \bigg[ 3  {\tilde{\lambda}}_6^* e^{-{i\xi \over 2}} \tan\beta 
~\! \frac{v^2}{{\tilde{m}_2}^2}
+{\cal O}\bigg(\frac{v^4}{{\tilde{m}_2}^4}\bigg) 
\bigg]
\label{eq:YukawastypeIV}
\end{eqnarray}

The four fermion interactions in the quark mass  eigenbasis are defined as in \eqref{eq:fourfermiontypeI}. They are given by
\begin{eqnarray}
&&
\omega^{uu ~\!\! (0)}_{ijmn}
=2~ \! \frac{1}{{\tilde{m}_2}^2}\delta_{ij}\delta_{mn} \frac{m^{u}_{i} m^{u*}_{m}}{v^2} \cot^2 \beta 
\quad \quad \quad
\omega^{uu\pm ~\!\! (0)}_{ijmn}
=2~ \! \frac{1}{{\tilde{m}_2}^2} \frac{V^T_{ij} m^{u}_{j} m^{u*}_{m}V^*_{mn}}{v^2} \cot^2 \beta 
\nonumber \\ 
&&
\omega^{dd ~\!\! (0)}_{ijmn}
=2 ~ \! \frac{1}{{\tilde{m}_2}^2}\delta_{ij}\delta_{mn}  \frac{m^{d*}_{i} m^{d}_{m}}{v^2} \tan^2 \beta 
\quad \quad \quad \!
\omega^{dd\pm ~\!\! (0)}_{ijmn}
=2 ~ \! \frac{1}{{\tilde{m}_2}^2} \frac{V^*_{ij}m^{d*}_{j} m^{d}_{m}V^T_{mn}}{v^2} \tan^2 \beta  
\nonumber \\
&&
 \omega^{\ell\ell ~\!\! (0)}_{ijmn}
 = \omega^{\ell\ell \pm~\!\! (0)}_{ijmn}=
 2~  \! \frac{1}{{\tilde{m}_2}^2}\delta_{ij}\delta_{mn}  \frac{m^{\ell*}_{i} m^{\ell}_{m}}{v^2} \cot^2 \beta 
\quad \quad \quad  \,
%
% =
% 2~  \! \frac{1}{{\tilde{m}_2}^2}\delta_{ij}\delta_{mn}  \frac{m^{\ell*}_{i} m^{\ell}_{m}}{v^2} \cot^2 \beta 
\nonumber \\ 
&&
\omega^{d\ell~\!\! (0)}_{ijmn}
=
 -2 ~ \! \frac{1}{{\tilde{m}_2}^2} \delta_{ij}\delta_{mn}  \frac{m^{d*}_{i} m^{\ell}_{m}}{v^2} 
\quad \quad \quad \quad \quad
\omega^{d\ell\pm ~\!\! (0)}_{ijmn}
=
 -2 ~ \! \frac{1}{{\tilde{m}_2}^2} \delta_{mn} \frac{V^*_{ij}m^{d*}_{j} m^{\ell}_{m}}{v^2} 
\nonumber \\ 
&&
 \omega^{ud ~\!\! (2)}_{ijmn}=
 -2 ~  \! \frac{1}{{\tilde{m}_2}^2}\delta_{ij}\delta_{mn}  \frac{m^{u}_{i} m^{d*}_{m}}{v^2} 
\quad \quad \quad \quad \quad
 \omega^{ud\pm ~\!\! (2)}_{ijmn}=
 2 ~  \! \frac{1}{{\tilde{m}_2}^2} \frac{V^T_{ij} m^{u}_{j} V_{mn}^*m^{d*}_{n}}{v^2}
\nonumber \\  \nonumber
&&
 \omega^{u\ell ~\!\! (2)}_{ijmn}=
 2 ~  \! \frac{1}{{\tilde{m}_2}^2} \delta_{ij}\delta_{mn} \frac{m^{u}_{i} m^{\ell*}_{m}}{v^2}\cot^2 \beta
\quad \quad \quad~
 \omega^{u\ell\pm ~\!\! (2)}_{ijmn}=
 -2 ~  \! \frac{1}{{\tilde{m}_2}^2} \delta_{mn} \frac{V^T_{ij} m^{u}_{j} m^{\ell*}_{m}}{v^2}\cot^2 \beta \\
\label{eq:fourfermiontypeIV2}
\end{eqnarray}
where $\pm$ superindex indicates four fermion operators generated by charged Higgs exchange and $V$ is the CKM matrix. 
\subsection{T violation in types I-IV 2HDM}
\label{sec:Tviolationtypes}

In this section we discuss T violation in the types I-IV 2HDM. We restrict ourselves to effective dimension six effects. First, recall that in section \ref{sec:physicalHiggs2HDMEFT} we concluded that at effective dimension six, T violation is contained only in the four fermion operators or in the Higgs Yukawas.  In the types I-IV 2HDM, T violation is further restricted. 

From equations \eqref{eq:fourfermiontypeI2}, \eqref{eq:fourfermiontypeII2}, \eqref{eq:fourfermiontypeIII2} and \eqref{eq:fourfermiontypeIV2}, we note that in the types I-IV 2HDM, the only T violation in four fermion operators is due to the known CKM phase. These T violating terms originate from integrating out the charged Higgs, arise only in flavor violating processes with CKM matrix insertions, and are present even if all the T violating phases of the 2HDM vanish.

On the other hand, the Higgs Yukawas in the types I-IV 2HDM, equations \eqref{eq:YukawastypeI}, \eqref{eq:YukawastypeIIA}, \eqref{eq:YukawastypeIIB},  \eqref{eq:YukawastypeIIIA},  \eqref{eq:YukawastypeIIIB},  \eqref{eq:YukawastypeIVA} and  \eqref{eq:YukawastypeIVB}, contain a universal T violating phase due to the $U(1)_{\textrm{PQ}}$ invariant term $\tilde{\lambda}_6^* e^{-{i\xi \over 2}}$, which introduces a phase between the Higgs Yukawas and the quark mass matrix. 
It is convenient to define the T violating phase $\delta_{\textrm{2HDM}}$
\begin{equation}
\sin \delta_{\textrm{2HDM}}=\frac{v^2}{{\tilde{m}_2}^2} \cot\beta \abs{\tilde{\lambda}_6} \sin\bigg[{\rm Arg}\left({{\tilde{\lambda}}_6^* e^{-i\xi /2 }}\right)\bigg] \bigg[1+{\cal O}\bigg(\frac{v^2}{{\tilde{m}_2}^2}\bigg)\bigg]
\label{eq:Tviolatingphase2HDMGW}
\end{equation}
Using the definition \eqref{eq:Tviolatingphase2HDMGW} in the Yukawas \eqref{eq:YukawastypeI}, \eqref{eq:YukawastypeIIA}, \eqref{eq:YukawastypeIIB},  \eqref{eq:YukawastypeIIIA},  \eqref{eq:YukawastypeIIIB},  \eqref{eq:YukawastypeIVA} and  \eqref{eq:YukawastypeIVB}, we find that the imaginary part of the Yukawas relative to the fermion mass matrices are given parametrized by $\delta_{\textrm{2HDM}}$
\begin{eqnarray}
 \textrm{Type I}:&& 
 \textrm{Im} \big[ \, \lambda^{u,d,\ell}_{\varphi ik} \big(m^{u,d,\ell}\big)^{-1}_{kj} \,\big] 
 =
  -\delta_{ij}\, \sin\delta_{\textrm{2HDM}} 
\nonumber \\[20pt]
\nonumber
%%%%%
 \textrm{Type II} :&& 
 \textrm{Im} \big[ \, \lambda^{u}_{\varphi ik} \big(m^{u\,*}\big)_{kj}^{-1} \,\big] 
 =
  -\delta_{ij}\,  \sin\delta_{\textrm{2HDM}} 
\nonumber  \\
  &&
   \textrm{Im} \big[ \, \lambda^{d,\ell}_{\varphi ik} \big(m^{d,\ell\,*}\big)_{kj}^{-1} \,\big] 
 =
\delta_{ij}\,  \tan^2\beta ~\! \sin\delta_{\textrm{2HDM}} 
  \nonumber \\[20pt]
   %%%%%
    \textrm{Type III} :&&   
 \textrm{Im} \big[ \, \lambda^{u,d}_{\varphi ik} \big(m^{u,d\,*}\big)_{kj}^{-1} \,\big] 
 =
  -\delta_{ij}\, \sin\delta_{\textrm{2HDM}} 
\nonumber  \\
  &&
   \textrm{Im} \big[ \, \lambda^{\ell}_{\varphi ik} \big(m^{\ell\,*}\big)_{kj}^{-1} \,\big] 
 =
 \delta_{ij}\, \tan^2\beta ~\! \sin\delta_{\textrm{2HDM}} 
  \nonumber \\[20pt]
   %%%%%
       \textrm{Type IV} :&&   
 \textrm{Im} \big[ \, \lambda^{u,\ell}_{\varphi ik} \big(m^{u,\ell\,*}\big)_{kj}^{-1} \,\big] 
 =
  -\delta_{ij}\, \sin\delta_{\textrm{2HDM}} 
\nonumber  \\
  &&
   \textrm{Im} \big[ \, \lambda^{d}_{\varphi ik} \big(m^{d\,*}\big)_{kj}^{-1} \,\big] 
 =
\delta_{ij}\,  \tan^2\beta ~\! \sin\delta_{\textrm{2HDM}} 
   \end{eqnarray}
   Note that $\delta_{\textrm{2HDM}}$ vanishes in the decoupling limit. The T violating phase $\delta_{\textrm{2HDM}}$ leads to an EDM through Barr-Zee diagrams \cite{Barr:1990vd}, that sets strong constraints on it.

In the quark mass eigenbasis it is clear that no flavor violation is needed for $\delta_{\textrm{2HDM}}$ to be observable. Also, in this basis the phase of the heavy doublet Yukawa $ \tilde{\lambda}^f_{2ij}$ is  $ e^{-{i\xi \over 2}}$, so in terms of the complete set of the T violating phases of the 2HDM \eqref{eq:completesetCP}, $\textrm{Arg}\big(\tilde{\lambda}_6^* e^{-{i\xi \over 2}}\big)$ corresponds to the phase $\textrm{Arg} \big({\tilde{\lambda}}_6^* \tilde{\lambda}^f_{2ij}\big)$. This phase is independent of the phases $\theta_1, \theta_2$ associated exclusively with the 2HDM potential. This was to be expected, since in section \ref{sec:2HDMEFT} we concluded that $\textrm{Arg} \big({\tilde{\lambda}}_6^* \tilde{\lambda}^f_{2ij}\big)$ is the only T violating phase of the 2HDM which shows up at effective dimension six. We stress out that these conclusions are valid for any type I-IV 2HDM, with the most general Higgs potential at the renormalizable level.

\section{Conclusions}

In this paper we studied and organized the low energy phenomenology of the SHSM and 2HDM near the decoupling limit using EFT. We worked at tree level. In the SHSM we worked up to effective dimension six, and in the 2HDM we worked up to effective dimension six in interactions involving fermions, and eight in purely bosonic interactions. The main output of this exercise is a general map between experimental signatures and theory. We summarize some of its main features in table \ref{tab:summarymap}. This map is a valuable, simple tool for interpreting experimental data. 

Several observations can be made thanks to the organization of the phenomenology, that will be studied in follow up papers. First, we point out that the main difference of extensions with singlets and doublets, is that couplings of the Higgs to fermions and to massive gauge bosons are modified at different effective dimension. For this reason, a well motivated quantity to study at LHC are ratios of the type
\begin{equation}
\bigg|
\frac{\lambda^f_{\varphi ij}}
{g_{\varphi VV}}
\,
\frac{g^{\textrm{\tiny{SM}}}_{\varphi VV}}
{\lambda^{f\, \textrm{\tiny{SM}}}_{\varphi ij}}
\bigg|
\end{equation}
Measurements of these ratios have been recently presented by ATLAS and CMS \cite{ATLASCMSCOMB}. In the SHSM these ratios can be obtained from \eqref{eq:HiggsgaugecouplingssingletEFT} and \eqref{eq:HiggsfermioncouplingssingletEFT}, and should be close to one, if radiative effects are small. In the 2HDM they are obtained from \eqref{eq:varphiVVEFT} and \eqref{eq:varphiffEFT} and should be generically different from one. These observations remain valid away from the decoupling limit. Ratios of different couplings of the Higgs are one of the main tools to discern between the different extensions of the Higgs sector. In the ratios, theoretical and/or experimental uncertainties might cancel. Ratios of Higgs Yukawas or Higgs couplings to massive gauge bosons over Higgs self couplings might also be interesting observables at colliders.

Moreover, the deviations of the couplings with respect to their SM values are controlled by a small subset of parameters of the UV completions. This leads to correlations between the deviations. The simplest case is in the SHSM, where both the Higgs couplings to two massive gauge bosons \eqref{eq:HiggsgaugecouplingssingletEFT} and the Higgs Yukawas \eqref{eq:HiggsfermioncouplingssingletEFT} suffer the same modification as pointed out above. In the 2HDM, the deviations of these couplings are not directly correlated in general. However, in the particular limit in which the Yukawas of the heavy doublet vanish, as in the type I 2HDM at large $\tan\beta$, both deviations show up first at effective dimension eight and are equal, mimicking the effective theory of the SHSM. In this case, both deviations can be understood as dilution by the complex alignment parameter. 

Regarding T violation, in the 2HDM EFT we identified the $U(1)_{\textrm{PQ}}$ invariant T violating phases of the full 2HDM which are most relevant at low energies. We showed that only relative phases between the Higgs potential coupling $\tilde{\lambda}_6$ and the heavy doublet Yukawas appear at effective dimension six. In particular, in types I-IV 2HDM we showed that there is only one such phase and it appears only in the Higgs Yukawas. This phase can be constrained by EDM experiments. 

We also organized all the effective dimension six flavor violating effects in the 2HDM EFT. All the four fermion operators were derived, and the flavor violating Higgs Yukawas were presented. For the general 2HDM, these results are $\tan\beta$ independent. Direct constraints on both the four fermion operator coefficients and on the Higgs Yukawas can be placed, and efforts have already been carried out in the literature \cite{Crivellin:2013wna}. Moreover, recent anomalies on flavor physics \cite{Lees:2012xj,Lees:2013uzd,Huschle:2015rga,Aaij:2015yra,Aaij:2013qta,Aaij:2014ora,Khachatryan:2015kon} provide strong motivation to study models with novel sources of flavor violation. These anomalies might be explained with tree level flavor violation \cite{Freytsis:2015qca}, and it remains interesting to perform a detailed study of all the alternatives within the 2HDM. 

\begin{table}[h!]
$
\begin{array}{|ccc|} 
\hline
  &  \textrm{ \large{The SHSM}} & \textrm{\large{The 2HDM}}   \\
  \hline
  \multirow{2}{*}{\parbox[c]{.3\linewidth}{\centering \textit{Couplings to gauge bosons \\ $g_{\varphi VV},g_{\varphi^2 VV} ~\! , ~\! V=W,Z $}}}
 & 
\multirow{2}{*}{\parbox[c]{.3\linewidth}{\centering ED 6 \eqref{eq:HiggsgaugecouplingssingletEFT} \\ \small{Always smaller than SM}}}
 &
 \multirow{2}{*}{\parbox[c]{.3\linewidth}{\centering ED 8 \eqref{eq:varphiVVEFT},\eqref{eq:varphivarphiVVEFT} \\ \small{Always smaller than SM}}}
\\

&
&
\\ 
\hdashline[0.5pt/5pt]
\multirow{2}{*}{\parbox[c]{.3\linewidth}{\centering \textit{Fermionic couplings \\ $\lambda^f_{\varphi ij} ~\! ,~\! f=u,d,\ell$}}}
&  
\multirow{2}{*}{\parbox[c]{.3\linewidth}{\centering ED 6 \eqref{eq:HiggsfermioncouplingssingletEFT} \\ \small{Always smaller than SM}}}
& 
\multirow{2}{*}{\parbox[c]{.3\linewidth}{\centering ED 6 \eqref{eq:varphiffEFT}
}}
\\
  & & 
  \\
  \hdashline[0.5pt/5pt]
  \multirow{2}{*}{\parbox[c]{.3\linewidth}{\centering \textit{Higgs self-couplings \\ $g_{\varphi^3}, g_{\varphi^4}$}}}
   &  
     \multirow{2}{*}{\parbox[c]{.3\linewidth}{\centering ED 6 \eqref{eq:selfcouplings1}}}
   &
\multirow{2}{*}{\parbox[c]{.3\linewidth}{\centering ED 6 \eqref{eq:Higgsself2EFT}, \eqref{eq:Higgsself1EFT}\\ \small{Always smaller than SM}}}
    \\
      & 
      &
       \\
  \hdashline[0.5pt/5pt]
  \multirow{3}{*}{\parbox[c]{.3\linewidth}{\centering    \textit{Flavor violation} }}
   &  
     \multirow{3}{*}{\parbox[c]{.3\linewidth}{\centering   \ding{55}}}
   &
     \multirow{3}{*}{\parbox[c]{.3\linewidth}{\centering    ED 6 \eqref{eq:fourfermion},\eqref{eq:varphiffEFT} \\ \small{$\Delta F=1, \Delta F=2$}. Chirality violating and preserving. }}
     \\
      &
  &
  \\
   &
  &
    \\
    \hdashline[0.5pt/5pt]
      \multirow{4}{*}{\parbox[c]{.3\linewidth}{\centering    \textit{T violation} }}
&
     \multirow{4}{*}{\parbox[c]{.3\linewidth}{\centering   \ding{55}}}
 &
       \multirow{4}{*}{\parbox[c]{.3\linewidth}{\centering   ED 6  \eqref{eq:fourfermion},\eqref{eq:varphiffEFT} \\ \small{Only in fermionic interactions.\\ Only one phase $\delta_{\textrm{2HDM}}$ \eqref{eq:Tviolatingphase2HDMGW} in types I-IV.}}}
  \\
    &
  &
   \\
    &
  &
\\
  &
  &
\\
       \hdashline[0.5pt/5pt]
 \multirow{2}{*}{\parbox[c]{.3\linewidth}{\centering \textit{Modifications correlated and controlled mostly by}}}
 &  
 \multirow{2}{*}{ \textrm{$\frac{\xi^2}{\mu^2}$ \eqref{eq:singletLagrangian} (or $\cos\gamma$ \eqref{eq:mixingmatrixsinglet})}}
 &
 \multirow{2}{*}{\parbox[c]{.3\linewidth}{\centering $\tilde{\lambda}_6$ \eqref{eq:2HDMLagrangian} (or $\Xi$ \eqref{eq:Vh}) and $\tilde{\lambda}^f_{2ij} \eqref{eq:actionHiggsbasis} ~\! , ~\! f=u,d,\ell$}}
  \\
  &
  &
  \\
\hline
\end{array}
$
\caption{Summary table of the main features of the SHSM and 2HDM effective field theories. ED stands for effective dimension. Each equation reference after ED 6 or ED 8 points to the coupling or operator where the corresponding effect can be read off. The rest of the equation references point to definitions of parameters.}
\label{tab:summarymap}
\end{table}

\section{Acknowledgments}
We thank Nathaniel Craig for discussions at the initial stage of this project. We thank Marco Farina for comments on the manuscript. This work was supported by the US Department of Energy under grant DOE-SC0010008. The work of D.E at the early stage of this project was supported by CONICYT Becas Chile and the Fulbright Program.

\appendix

\bibliography{treeA_bib}

%% edm electron leigh paban xu EDM gunion vega
\end{document}